\documentclass{aa}  

\usepackage{graphicx}
\usepackage{xcolor}
\usepackage{txfonts}
\begin{document}

   \title{Chemical compositions of five Planck cold clumps}


   \author{V. Wakelam
          \inst{1}
          \and
          P. Gratier \inst{1}
          \and 
          M. Ruaud\inst{2}
          \and R. Le Gal\inst{3}
          \and  L. Majumdar\inst{4}
          J.-C. Loison\inst{5} 
          \and K. M. Hickson\inst{5}
          }

   \institute{Laboratoire d'astrophysique de Bordeaux, Univ. Bordeaux, CNRS, B18N, all\'ee Geoffroy Saint-Hilaire, 33615 Pessac, France
              \email{valentine.wakelam@u-bordeaux.fr}
         \and
             NASA Ames Research Center, Moffett Field, CA, USA
\and 
Harvard-Smithsonian Center for Astrophysics, 60 Garden St., Cambridge, MA 02138, USA
\and
School of Earth and Planetary Sciences, National Institute of Science Education and Research, HBNI, Jatni 752050, Odisha, India
\and
Institut des Sciences Mol\'eculaires (ISM), CNRS, Univ. Bordeaux, 351 cours de la Lib\'eration, 33400 Talence, France\\
             }

   \date{Received xxxx; accepted xxxx}

 
  \abstract
   {}
   {Interstellar molecules form early in the evolutionary sequence of interstellar material that eventually forms stars and planets.  To understand this evolutionary sequence, it is important to characterize the chemical composition of its first steps.}
   {In this paper, we present the result of a 2 and 3 mm survey of five cold clumps identified by the Planck mission. We carried out a radiative transfer analysis on the detected lines in order to put some constraints on the physical conditions within the cores and on the molecular column densities. We also performed chemical models to reproduce the observed abundances in each source using the gas-grain model Nautilus.} 
   {Twelve molecules were detected: H$_2$CO, CS, SO, NO, HNO, HCO$^+$, HCN, HNC, CN, CCH, CH$_3$OH, and CO. Here, CCH is the only carbon chain we detected in two sources. Radiative transfer analyses of HCN, SO, CS, and CO were performed to constrain the physical conditions of each cloud with limited success. The sources have a density larger than $10^4$~cm$^{-3}$ and a temperature lower than 15~K. The derived species column densities are not very sensitive to the uncertainties in the physical conditions, within a factor of 2. The different sources seem to present significant chemical differences with species abundances spreading over one order of magnitude. The chemical composition of these clumps is poorer than the one of Taurus Molecular Cloud 1 Cyanopolyyne Peak (TMC-1 CP) cold core. Our chemical model reproduces the observational abundances and upper limits for 79 to 83\% of the species in our sources. The 'best' times for our sources seem to be smaller than those of TMC-1, indicating that our sources may be less evolved and explaining the smaller abundances and the numerous non-detections. Also, CS and HCN are always overestimated by our models.}
   {}


   \maketitle
%


\section{Introduction}

Star and planet formation is the final step in a long sequence of interstellar matter evolution that starts from the diffuse medium. The first step, that  of condensation, which sees the formation of cold cores, is not a well-constrained phase. The formation time and the evolution of the physical conditions are probably very variable, depending on the processes at work. From an observational point of view, cold cores are small (0.03 - 0.2 pc), dense ($10^4$-$10^5$~cm$^{-3}$), and cold (8-12~K) starless sources \citep{2007ARA&A..45..339B}. They are located within clumps (typical sizes of 0.3-3 pc), which are themselves located within clouds (2-15 pc). In these shielded regions, the gas and dust temperatures are expected to be below 20 K and the UV photons produced by surrounding massive stars cannot penetrate. Consequently, chemistry produces molecules that can survive, although they will be mostly  trapped in icy grain mantles. One key issue is determining what level of complexity  interstellar molecules are able to reach at this stage. A recent observation of complex organic molecules \citep{2012A&A...541L..12B,2012ApJ...759L..43C,2014ApJ...795L...2V} in the gas-phase in these objects indicates that molecular complexification could begin much earlier than previously thought. Reproducing this chemical complexity is still challenging for astrochemical models in which many chemical parameters, but also the processes themselves, are not well-constrained \citep{2010SSRv..156...13W}. Sometimes, even the abundances of the more basic species can be difficult to understand, particularly when the observational values published in the literature are in disagreement. Taking the example of CN, there are two abundances for this molecule in TMC-1 (CP) reported in the literature with a ratio of nearly 40 between them \citep{1984ApJ...283..668C,1997ApJ...486..862P}. Such disagreements can be attributed to the use of different telescopes, spectroscopic data, radiative transfer assumptions (local thermodynamic equilibrium, optical depth) etc.

From a general point of view, few sources exist for which a complete chemical survey, using coherent methods, have been performed. As far as we know, only two starless cold cores have been extensively studied thus far: TMC-1 (CP) and L134N (N). TMC-1 (CP) refers to the 'Cyanopolyyne Peak' within the Taurus molecular cloud \citep{1997ApJ...486..862P,1998FaDi..109..205O} while L134N (N) refers to the 'North Peak' in the isolated core L134N (also called L183) \citep{2000ApJ...542..870D}. Abundances observed in these two clouds have been compiled from the literature by \citet{2013ChRv..113.8710A} . These two cold cores present significant chemical differences, particularly with respect to carbon chains, which are much more abundant in TMC-1 than in L134N. The nature itself of L134N is, however, unclear as it might already be in the pre-stellar phase \citep{2004A&A...417..605P}. 

In this study, we present a 2 and 3 mm spectral survey of a selection of five cold clumps from the Planck Early Release Cold Core  \citep[ECC, ][]{2011A&A...536A..22P} catalogue. Using the detected lines, we tried to constrain the physical conditions within the observed sources and determine the molecular column densities (Section~\ref{obs}). In Section~\ref{diss}, we compare the observed values within the different cores and with the TMC-1 (CP) values. Section 4 presents a chemical model for each source. Finally, we offer a summary and our conclusions.

\section{Sources and observations}\label{obs}

\subsection{Selected sources}

To select the regions we intended to observe for the purposes of this study, we used the Planck Early Core catalogue \citep{2011A&A...536A..22P} and the follow-up mapping by \citet{2013ApJS..209...37M} of many of these sources in $^{12}$CO, $^{13}$CO, and C$^{18}$O. The sources presented in the Planck catalogue are quite large, appearing more similar to clumps rather than individual cores due to the Planck spatial resolution ($\sim$ 4.4$^{\prime}$ FWHM). They are, however, very likely to be cold and quiescent \citep{2012ApJ...756...76W}. 
The clumps were selected within the sample from \citeauthor{2013ApJS..209...37M} sample using several different criteria: large densities (a few $10^3$~cm$^{-3}$), large masses, low temperatures, originating from various regions of the sky, and be detected in the three CO isotopologues. The selected clumps are G153.34-08.00, G156.92-09.72, G157.12-11.56, G160.53-19.72, and G173.60-17.89 \citep[see Table 1 of][]{2013ApJS..209...37M}. Within these clumps, \citet{2013ApJS..209...37M}  found several sub-structures in CO. Here, we do not use the exact positions of the sub-cores identified by \citet{2013ApJS..209...37M}. Instead, for each of the clumps, we used the maximum $^{13}$CO(1-0) integrated intensity maps. 
The pointing positions for each of our clumps are summarised in Table~\ref{Table_sources}. For simplicity, we have labelled the observed sources from C1 to C5. The first three clumps (C1 to C3) are located in the California Molecular Cloud (CMC) \citep{2009ApJ...703...52L}, C4 is in the Perseus Molecular Cloud (PMC), while C5 is in the Taurus Molecular Cloud. These three clouds are located at approximately 140~pc for Taurus \citep{1987ApJ...322..706D}, 235~pc for Perseus \citep{2010A&A...512A..67L}, and 450~pc for California \citep{2009ApJ...703...52L}. 

After  our observations were complete, the Planck catalogue was revised (using more sophisticated numerical modelling and combining other highest frequency channels of Planck) and the Planck Catalogue of Galactic Cold Clumps (PGCC) was published \citep{2016A&A...594A..28P}. The central positions of the clumps were modified and some of the previously identified clumps were removed because they did not satisfy the compactness criterion. In particular, G173.60-17.89 is among the clumps that are not present in this catalogue. These clumps, however, have sizes of a few arc minutes, which is much larger than the difference between our position and the new Planck catalogue positions. Herschel observations are not available for these positions. So we could not better identify  the positions of the cores, if there were any at all.  

\begin{table*}
\caption{Observed sources}
\begin{center}
\begin{tabular}{lccccccccc}
\hline
\hline
Planck Clump's name & PGCC Ra \& Dec & Observed Ra \& Dec & labelled name  \\
& (J2000)&  (J2000) & \\
\hline
G153.34-08.00 & 03:48:41.80 +44:09:14.0 & 03:48:43.30 +44:07:45.0 & C1  \\
G156.92-09.72 & 03:57:30.00 +40:33:52.0 & 03:57:30.82 +40:34:22.0 & C2  \\
G157.12-11.56 & 03:51:58.80  +39:02:20.0 & 03:51:58.12  +39:00:31.0 & C3  \\
G160.53-19.72 & 03:38:57.70 +30:39:17.0 & 03:39:02.50 +30:41:30.0 & C4  \\
G173.60-17.89 & 04:32:50.26 +23:21:57.8 $^a$& 04:24:38.79 +23:23:51.0 & C5  \\
\hline
\end{tabular}
\end{center}
$^a$ This source is not in the PGCC catalogue, so we report the coordinates from the ECC Planck catalogue here. 
\label{Table_sources}
\end{table*}%

\subsection{Observations}

The observations were taken with the Institut de RadioAstronomie Millim\'etrique (IRAM) 30m telescope, where the data were acquired in a single 55 h observing run in April-May 2016. The Eight MIxer Receiver (EMIR) 3 and 2 mm receivers were tuned at 28 distinct frequencies to enable a nearly continuous coverage of the 3 mm (between 71.8 GHz and 116.2 GHz) and 2 mm bands (between 126.8 GHz and 164 GHz) at a frequency channel spacing of 48 kHz (corresponding to velocity channel spacing ranging from 0.09 to 0.2 km/s). The observing conditions were very good, with typical  system temperatures of 120 K at 3 mm and 150 K at 2 mm, which enabled us to reach a typical noise level of 8 mK at 3 mm and 18 mK at 2 mm.

The five sources were observed using position switching with an OFF position selected for each source  having minimal emission in the \citet{2013ApJS..209...37M} CO maps. Contamination from the 'off' is only present for the $^{12}$CO lines in sources C2, C3, and C4 but at different velocities. Primary pointing and focus were done on Mercury and secondary focus was achieved on quasars 0316+413, 0355+508, 0430+052. The average pointing corrections were between 3" and 4" (4 times smaller than the 15'' beam at 164GHz). The beam size was between 35'' (at 72~GHz) and 15'' (at 164~GHz). %

\subsection{Data reduction}

 Data reduction, line identification, and line fitting were carried out using the CLASS/GILDAS package \citep{2005sf2a.conf..721P}\footnote{http://www.iram.fr/IRAMFR/GILDAS/}. The spectroscopic catalogue used for  line identification is the JPL database\footnote{https://spec.jpl.nasa.gov/}  and the CDMS database\footnote{https://cdms.astro.uni-koeln.de/classic/} \citep{2005JMoSt.742..215M}. Within the observed frequency ranges, 12 molecules were detected in at least one of the sources, plus one isotopic line of CS and of HCO$^+$, and three isotopic lines of CO. The list of detected molecules and lines (with spectroscopic information and the critical density of the lines at 15~K) is given in Table~\ref{detect_lines}. The observed spectra for these lines are given in Figs.~\ref{CO_spec} to \ref{sulphur_spec}. The Gaussian line fit parameters for all detected lines in each of the five sources are given in Tables \ref{lineC1} to \ref{lineC5}. An absence of line parameters means that the lines were not detected and in that case, we just give the noise level. The 3$\sigma$ upper limits on the integrated intensities were computed assuming a gaussian shape and a line width of 1~km.s$^{-1}$  similar to the typical line width detected for the other molecules in these sources with the CASSIS software\footnote{http://cassis.irap.omp.eu} (developed by IRAP-UPS/CNRS), providing the produced line does not go above three times the rms. All temperatures given here are main beam temperatures. In the observed spectral range, we expected to detect more types of molecules, such as carbon chains. Table~\ref{lineparam_nondetected} lists these molecules and the spectroscopic information of the lower energetic transition present in the observed frequency range.  \\
Looking at the number of detected lines and line intensities, the five clouds present significant differences. In C2, for instance, only half of the molecules were detected, whereas in C3, there is only one molecule that was not detected. CCH is detected only in C1 and C4, while NO and HNO are only detected in C3. CS, SO, HCN, HNC, and CO are detected in all five sources.  The lines in C4 have a larger width than in the other sources and present a double peak in $^{13}$CO, CS, H$_2$CO, HCN, HCO$^+$, HNC, and SO. The $^{12}$CO line in C4 does not have a double peak but shows a strong redshifted wing. The presence of wings or multiple peaks in the emission lines of Planck's cold clumps have been discussed by \citet{2013ApJS..209...37M} and \citet{2019A&A...622A..32L}, where  it is often interpreted as a signature of multi-components in the line of sight, which may then be the case for C4. Furthermore, $^{12}$CO in C3 and C5 also present a small double peak profile but since none of the other species (including isotopic CO) have it, we assume that this is due to an optical depth effect. Molecules such as HC$_3$N and C$_3$H$_2$ were not detected in any of the observed sources. For the lines presenting a double peak, we checked that the integrated intensity obtained with a single Gaussian profile does not change significantly as compared to a fit with two Gaussians.

\subsection{Method of analysis and observational constraints}

\subsubsection{Physical conditions in the clumps}\label{temperature}

Our analysis of the molecular content of these clumps is limited by the small number of lines detected for each species. As an attempt to constrain the physical conditions in each of the clumps, we first use the $^{12}$CO observed line intensity to determine the gas kinetic temperature assuming that the line is optically thick and the source is filled in the beam \citep[see e.g. equation 1 of][]{2013ApJS..209...37M}. The assumption that the sources fill the beam may not be correct as we will discuss later in the paper. Table \ref{Table_physical} contains our computed kinetic temperatures derived from $^{12}$CO. When compared to the study of  \citet{2013ApJS..209...37M}, we found that the observed peak intensities of CO are larger in our sample in many cases and sometimes by a factor of 3. This could be an effect due to beam dilution if the sources do not fill the beam as our beam is smaller than theirs. This could also be the consequence of a different selected position of the sources. For C2 however, we have the same position as their G156.92-09.72C1 (see their Table 4). At this position, our $^{12}$CO intensity is 21\% smaller than theirs and we have a similar $^{13}$CO intensity, however, our C$^{18}$O intensity is twice as large. Because we have different CO intensities, our gas kinetic temperatures derived from CO are different -- and in most cases, smaller.

\begin{figure*}
\includegraphics[width=0.27\linewidth]{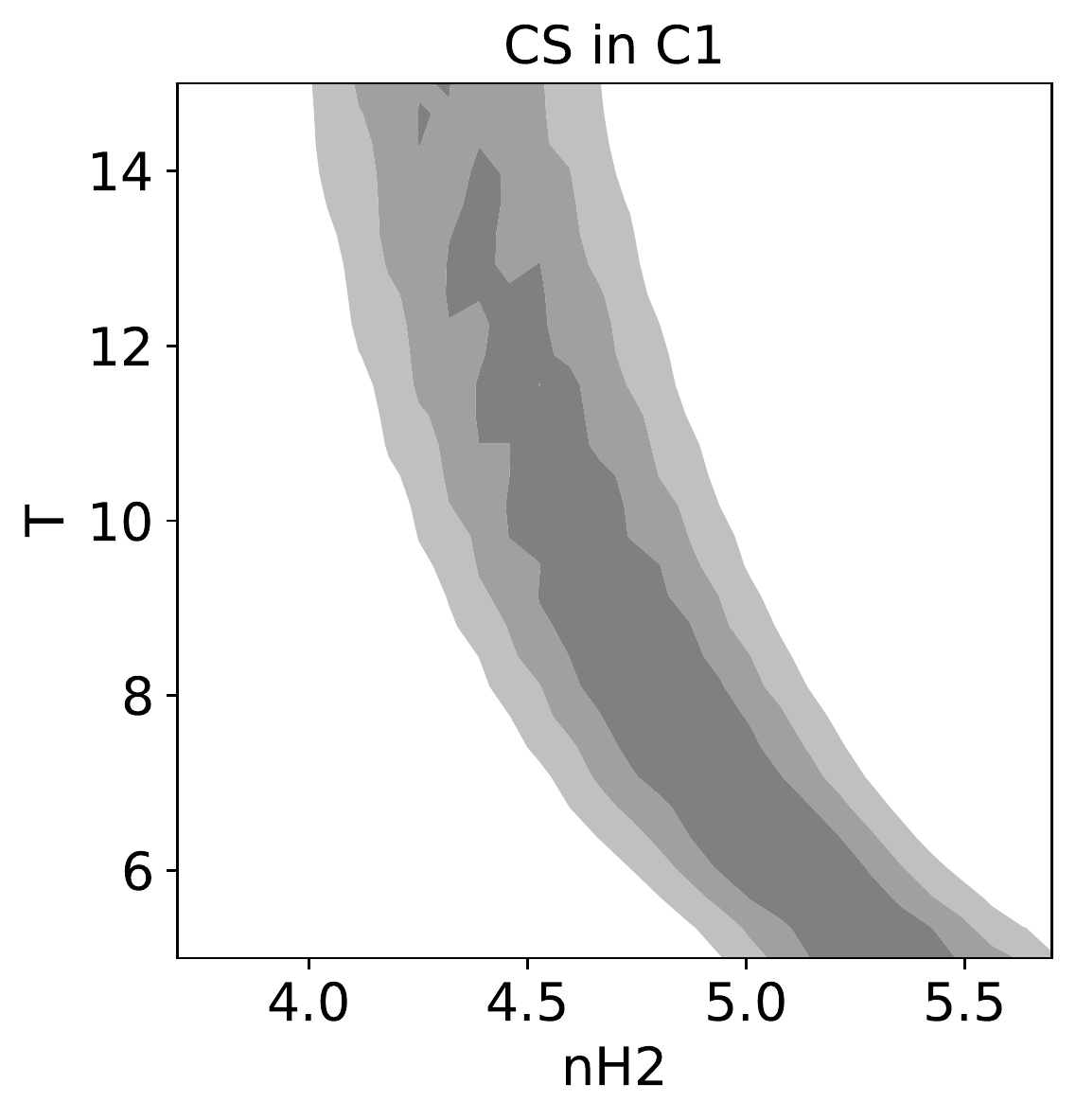}
\includegraphics[width=0.27\linewidth]{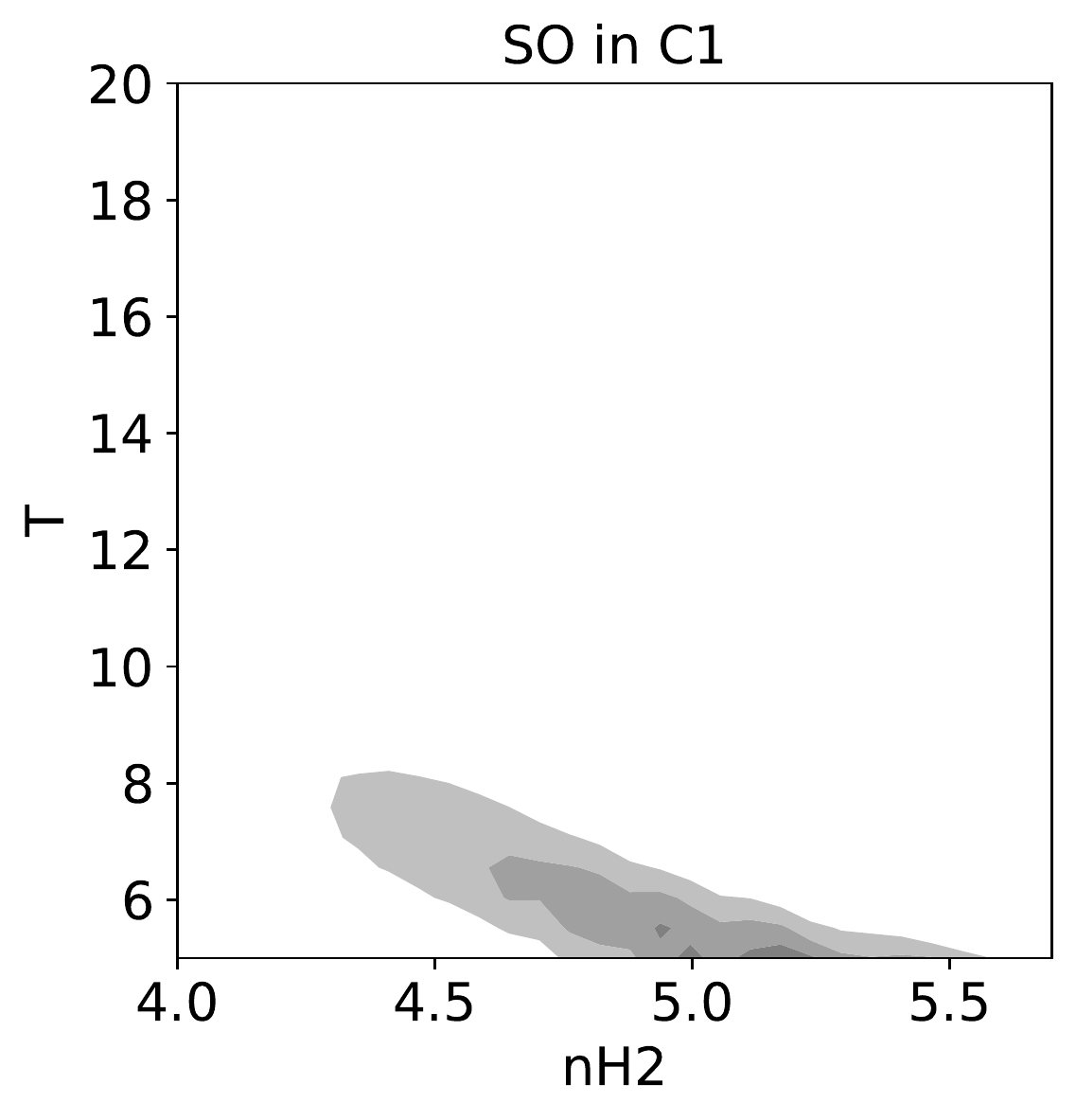}
\includegraphics[width=0.27\linewidth]{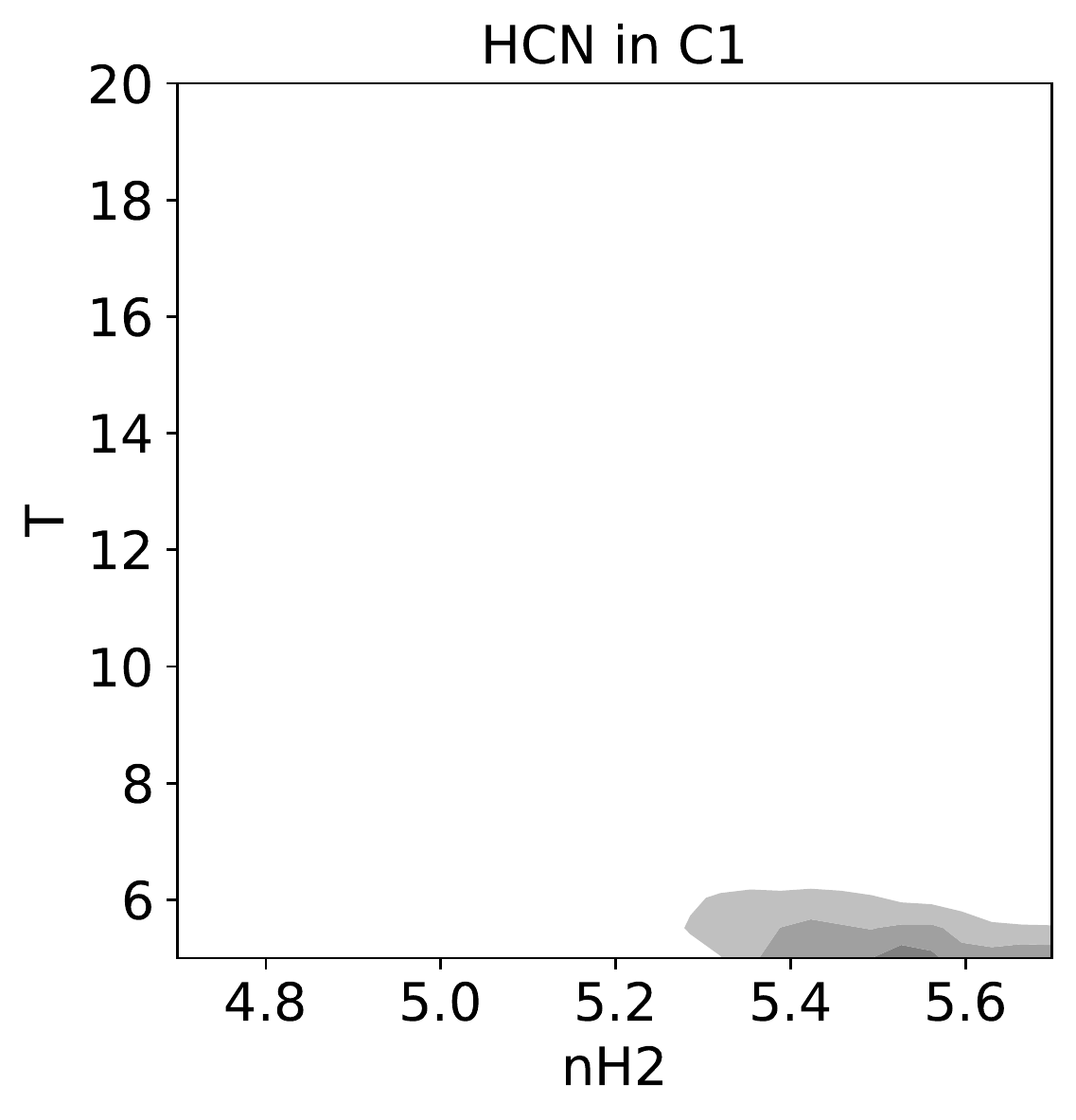}
\includegraphics[width=0.27\linewidth]{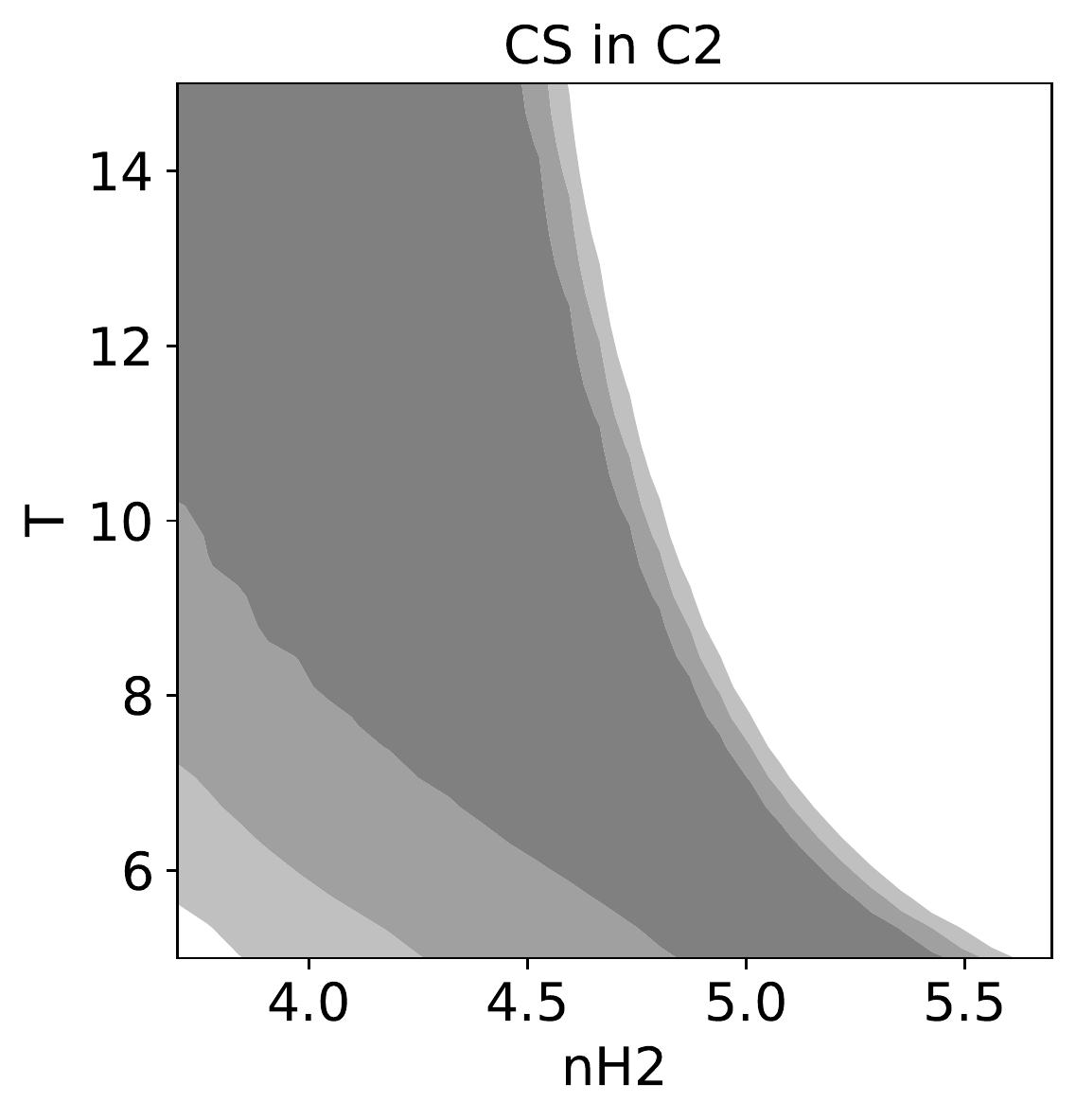}
\includegraphics[width=0.27\linewidth]{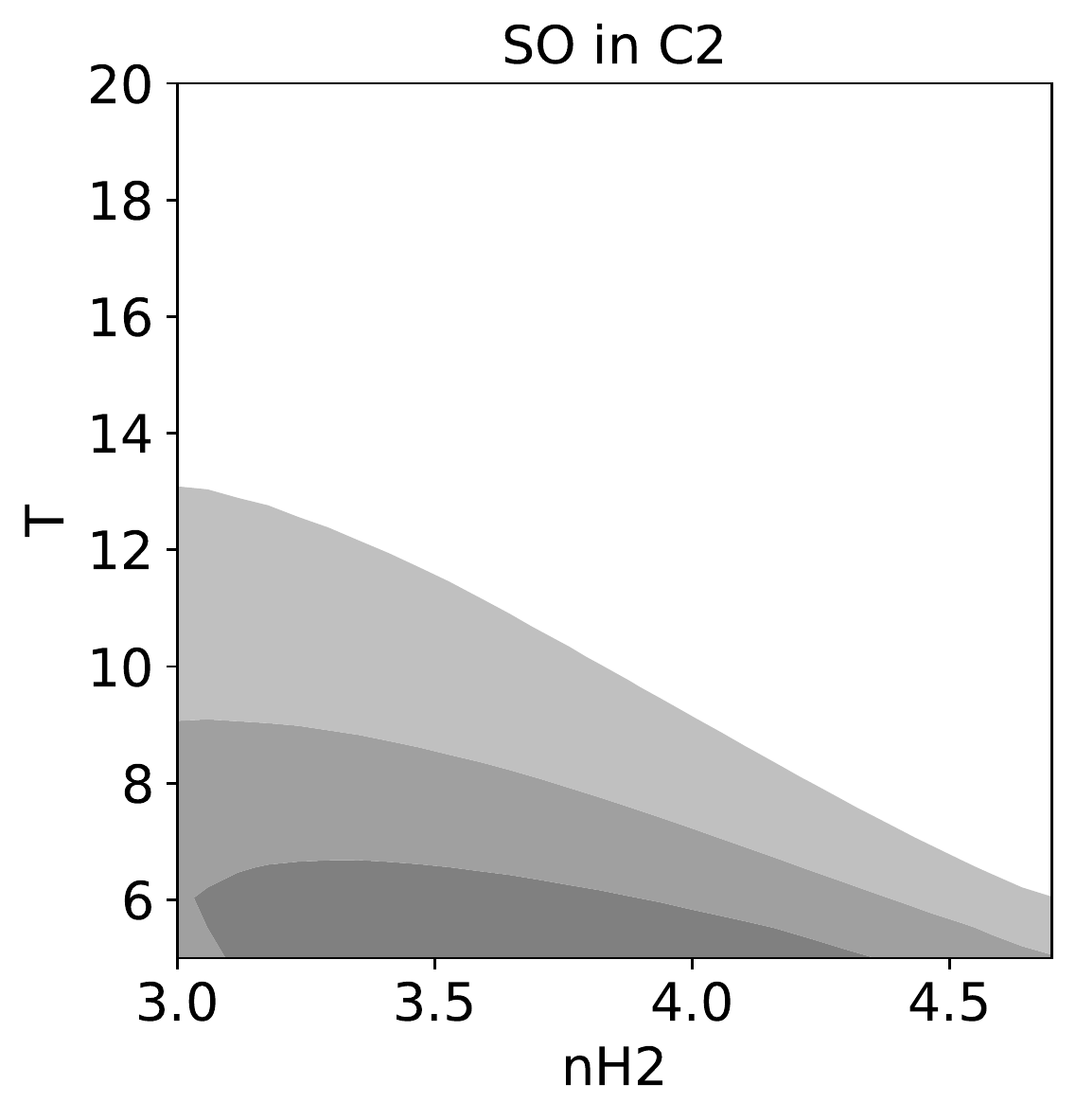}
\includegraphics[width=0.27\linewidth]{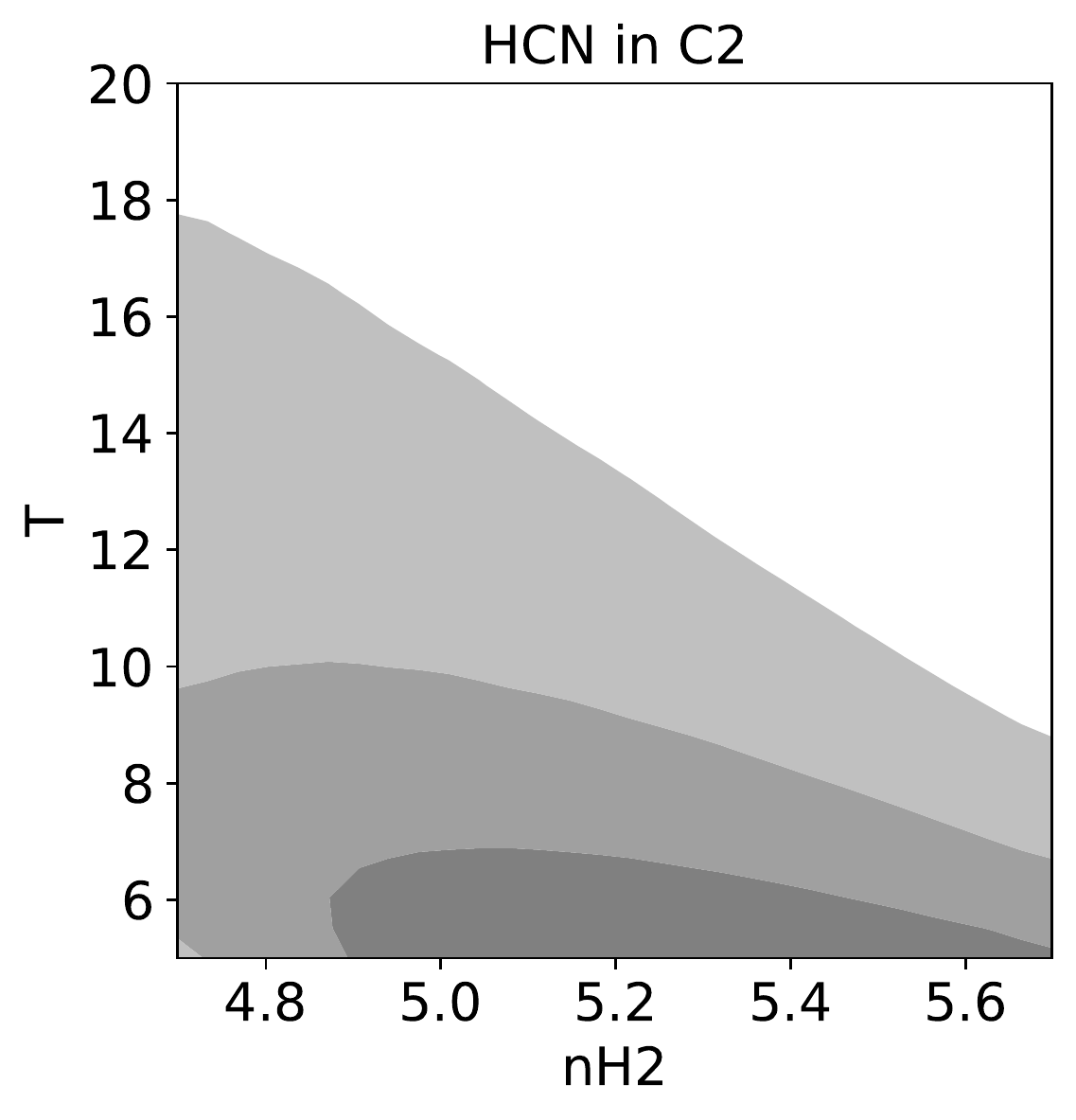}
\includegraphics[width=0.27\linewidth]{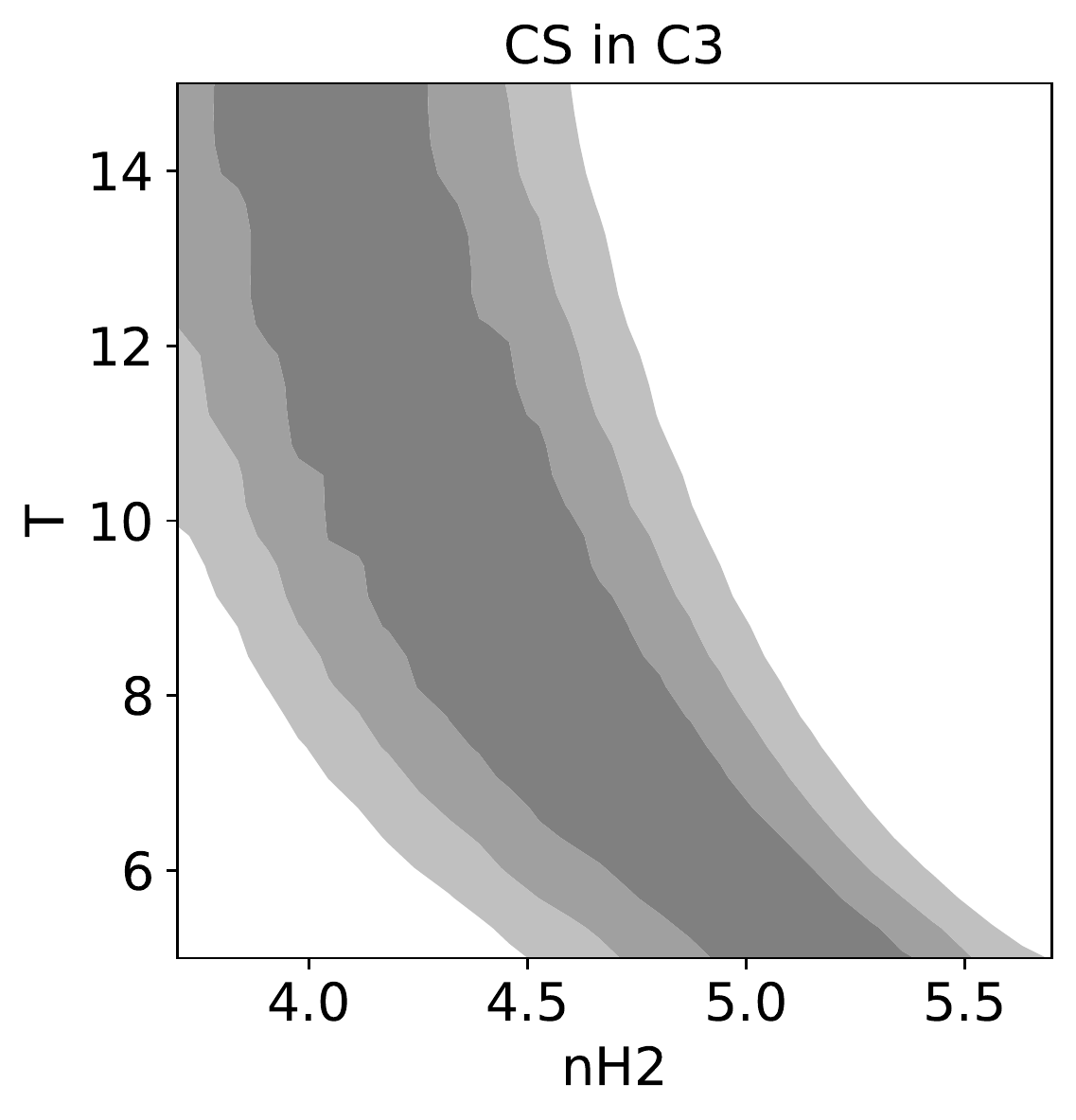}
\includegraphics[width=0.27\linewidth]{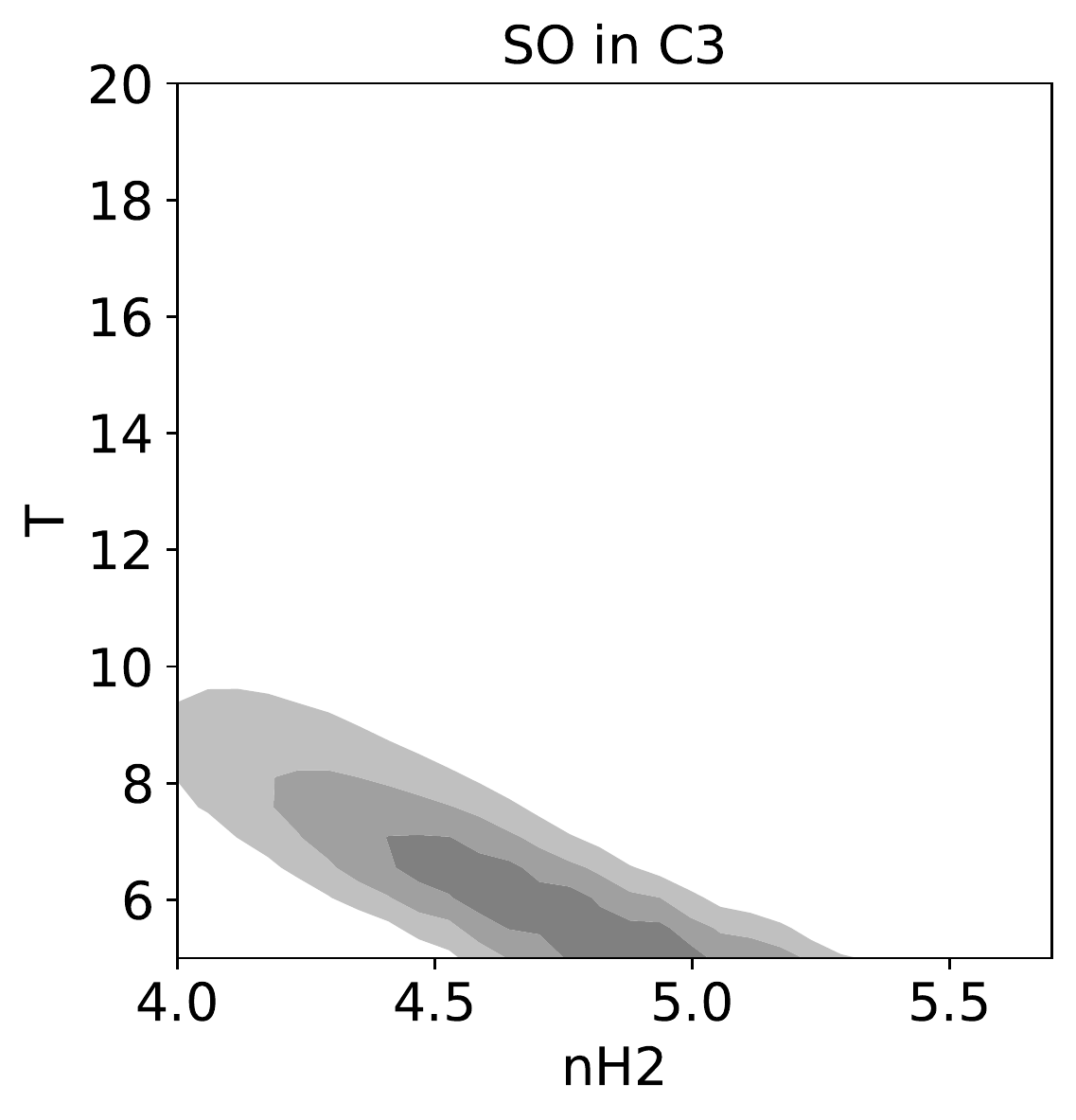}
\includegraphics[width=0.27\linewidth]{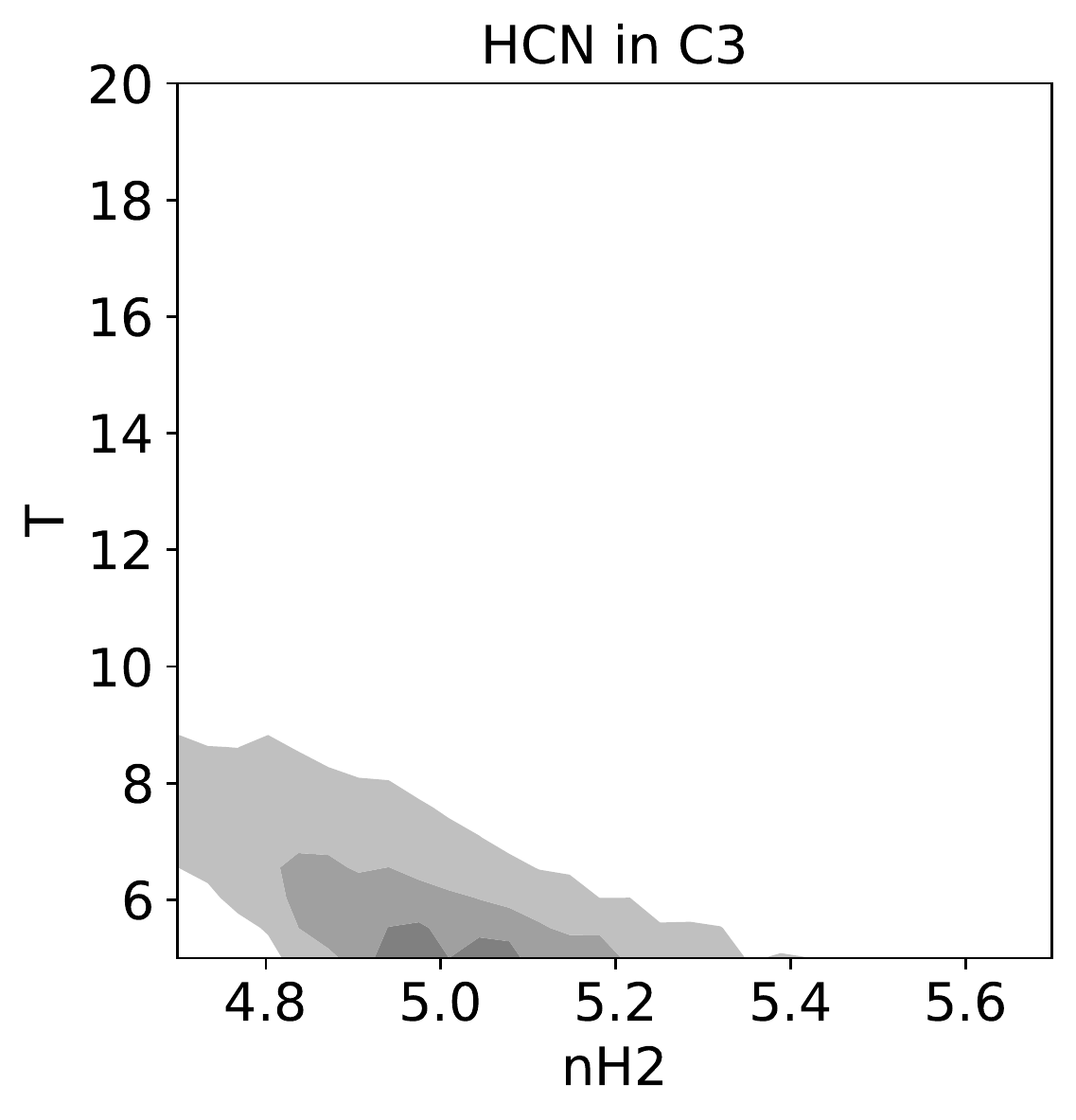}
\includegraphics[width=0.27\linewidth]{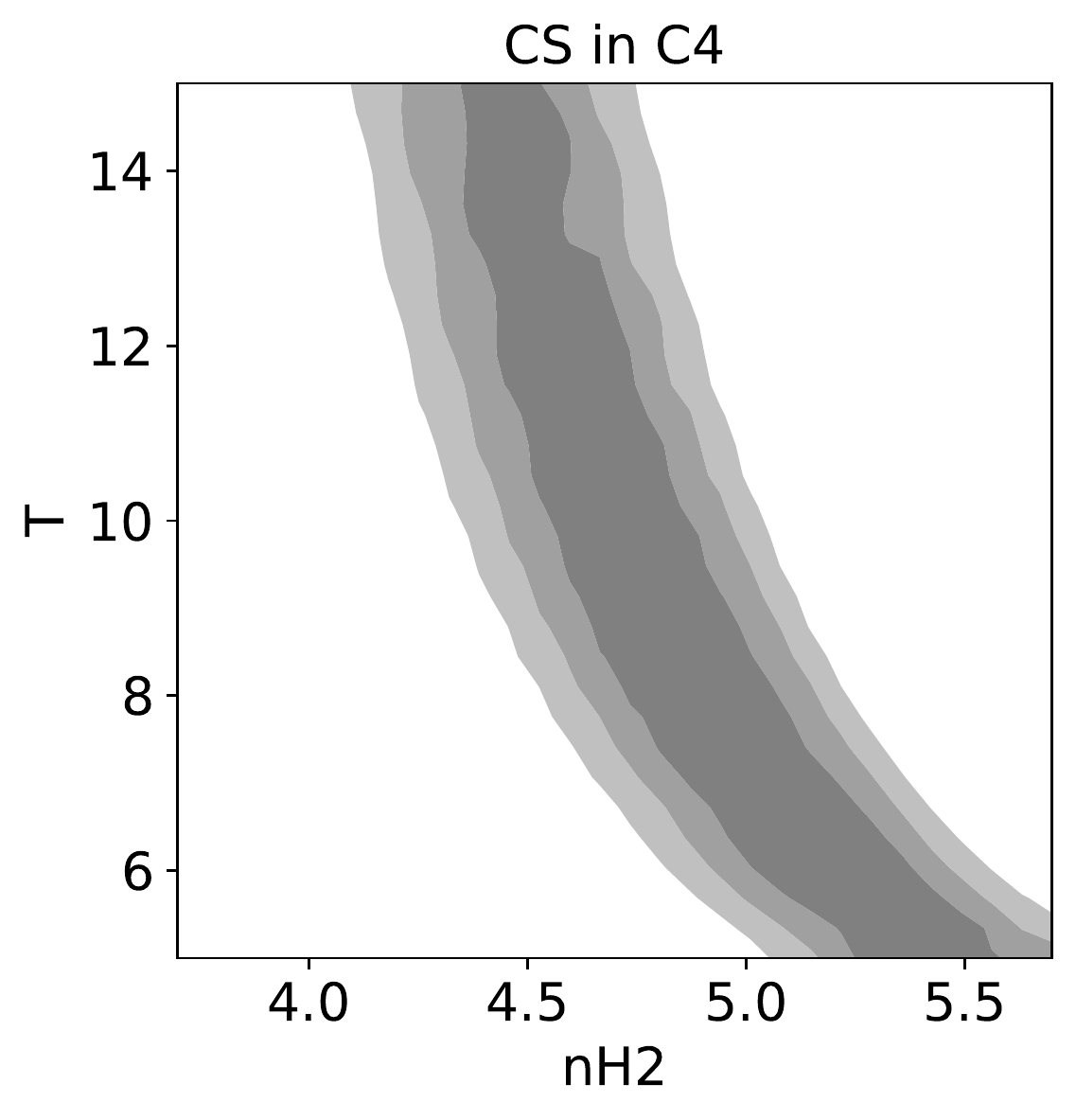}
\includegraphics[width=0.27\linewidth]{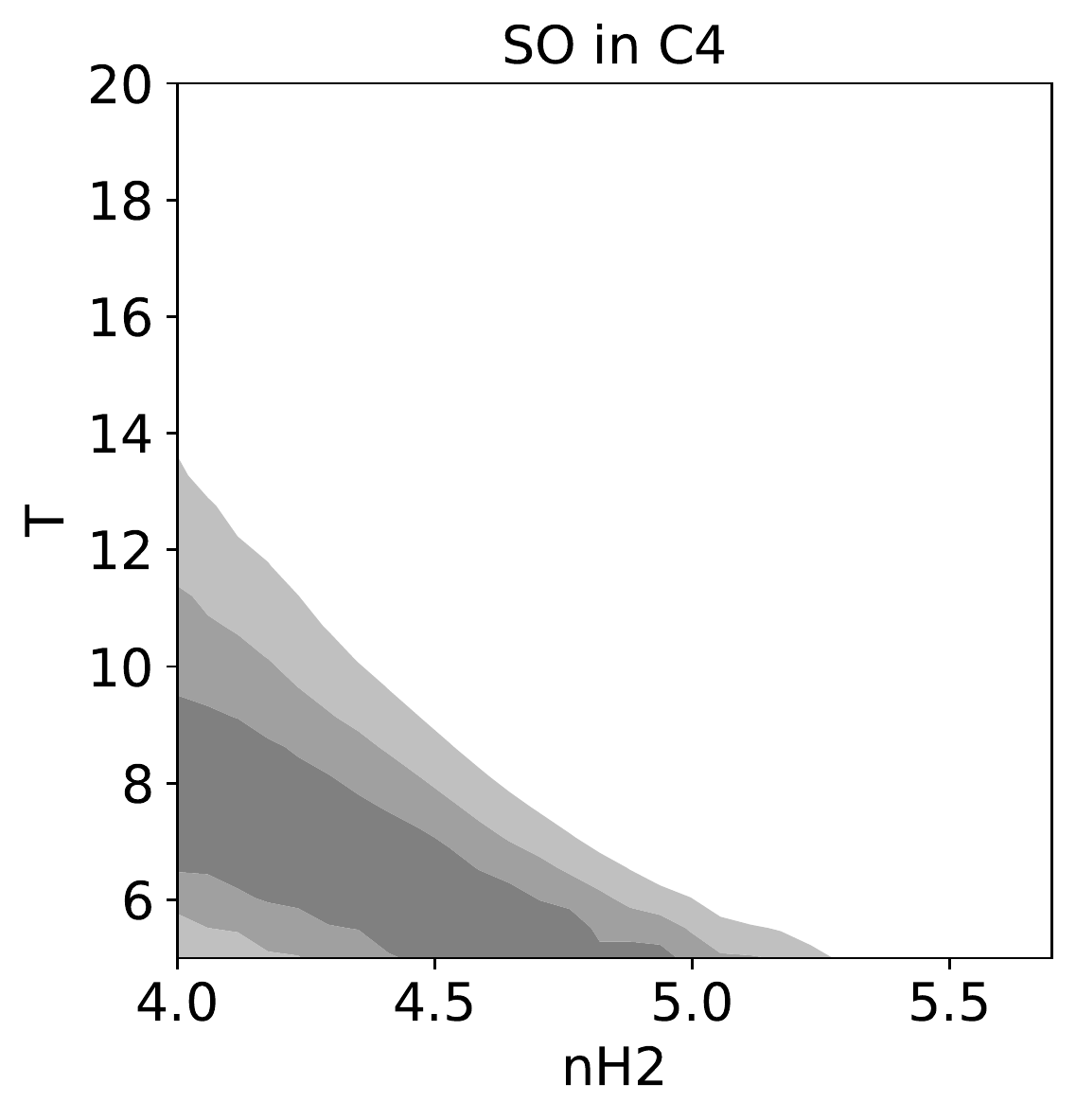}
\includegraphics[width=0.27\linewidth]{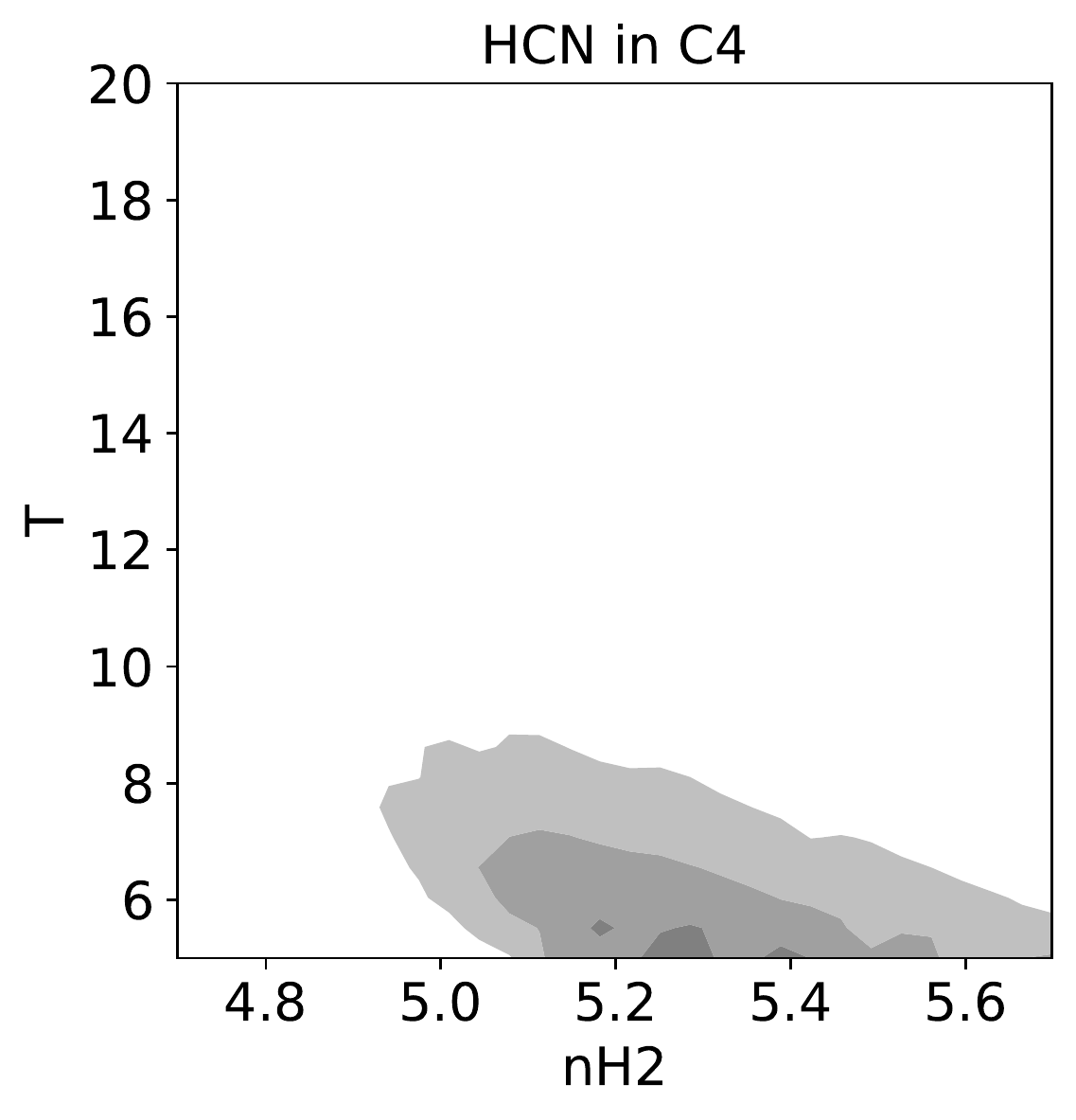}
\includegraphics[width=0.27\linewidth]{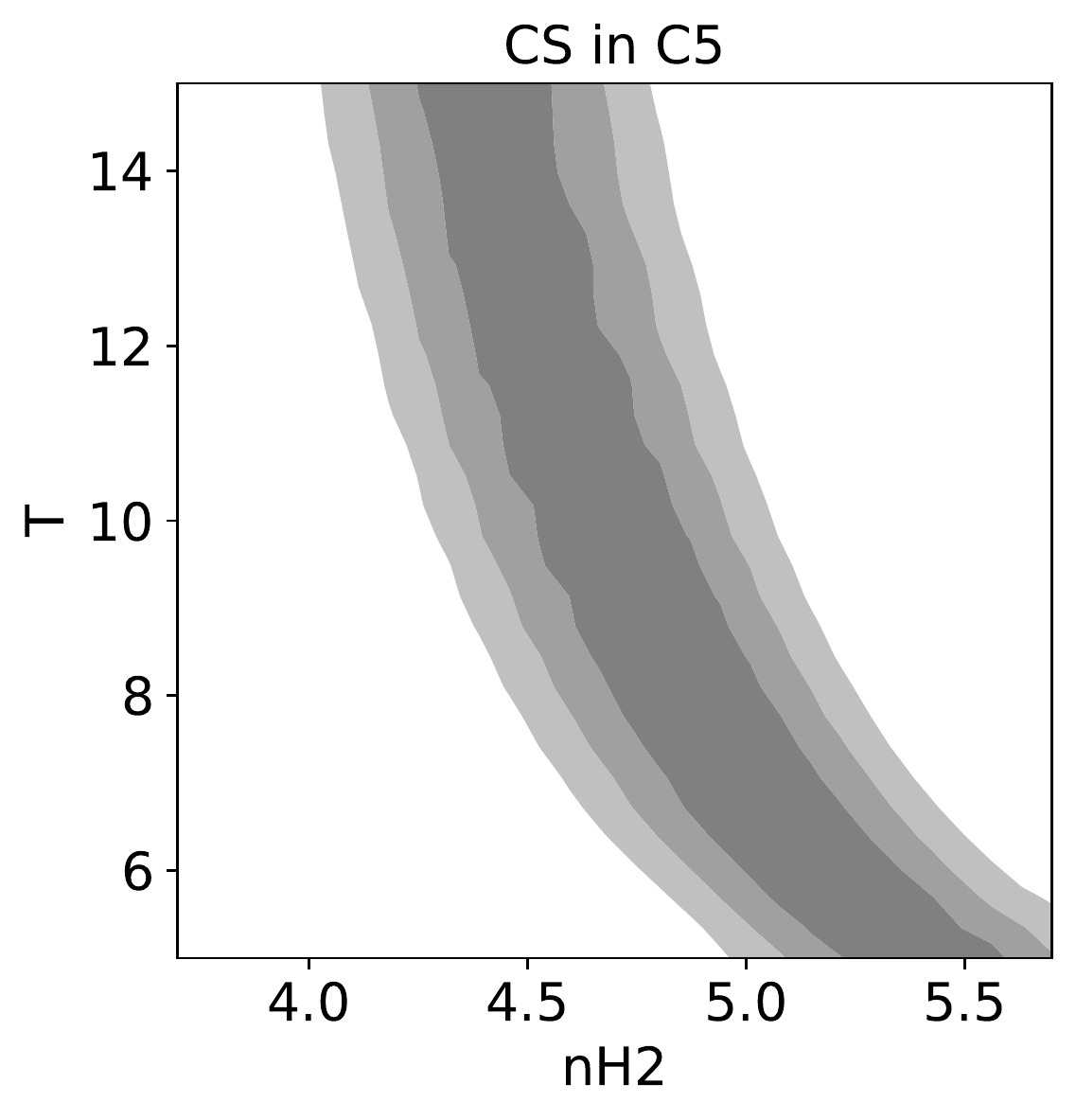}
\includegraphics[width=0.27\linewidth]{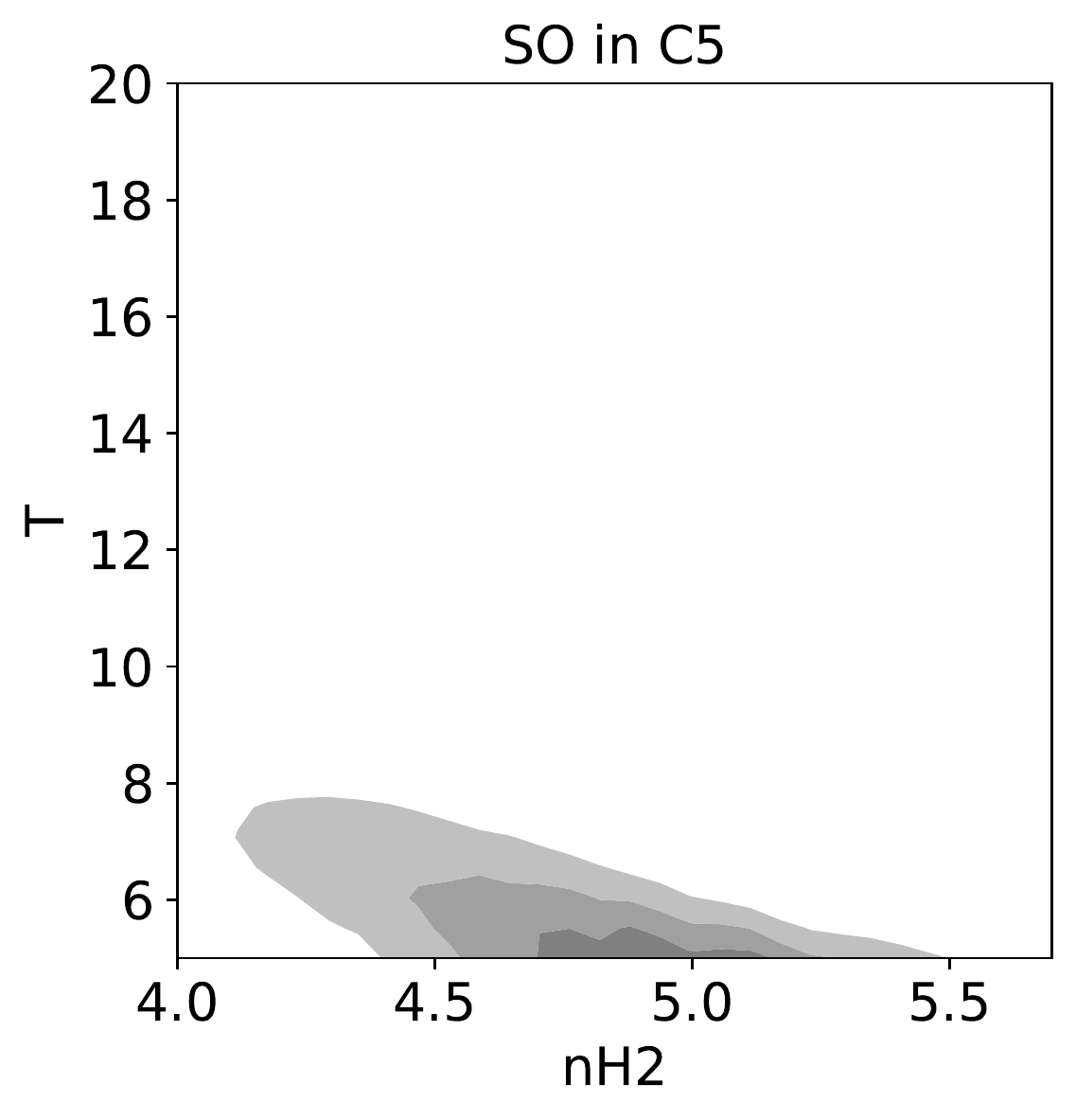}
\includegraphics[width=0.27\linewidth]{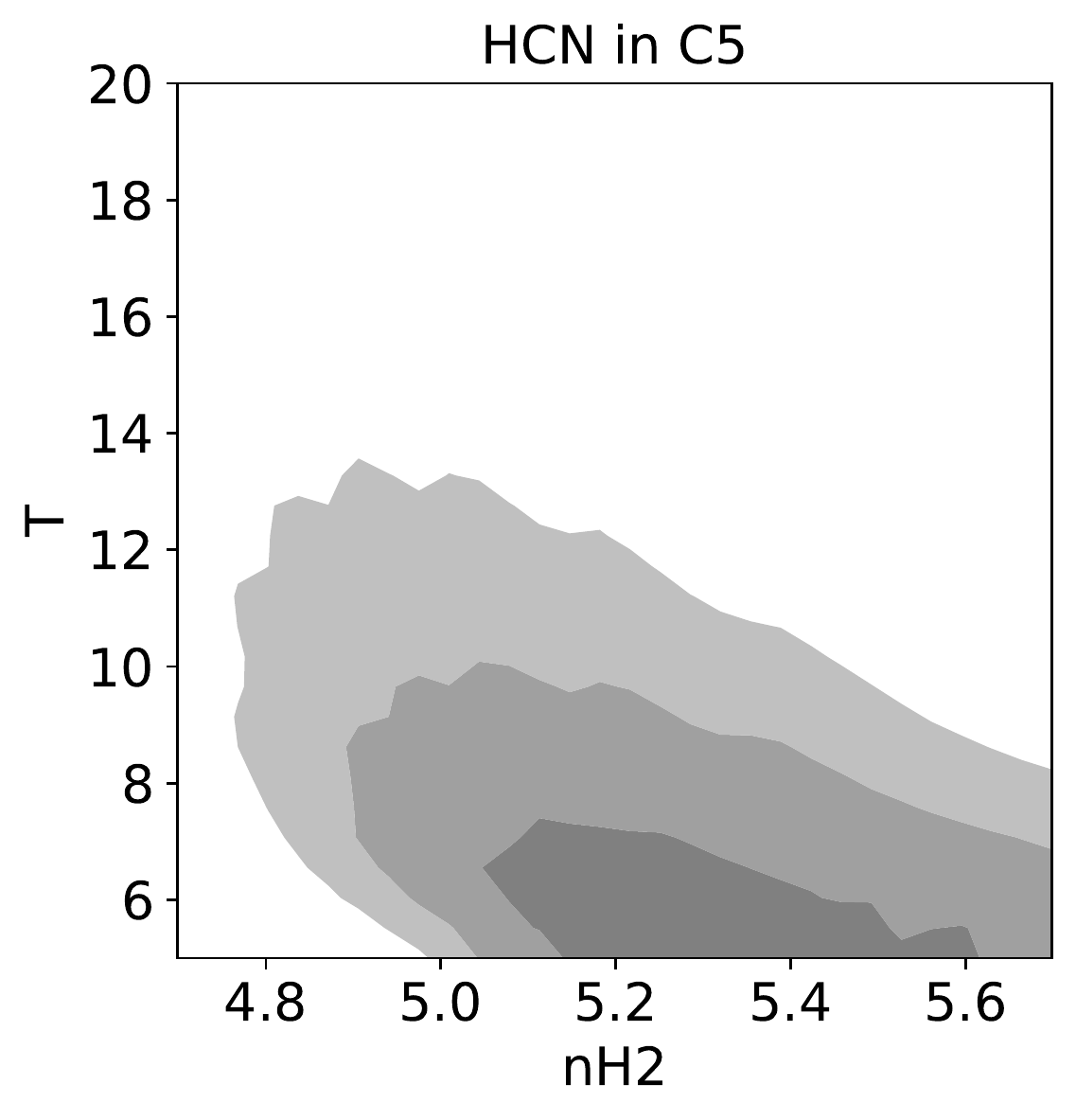}
\caption{$\chi^2$ contours ($1\sigma$, $2\sigma$, and $3\sigma$ confidence intervals) projected over the column density axis.\label{chi2_projN}}
\end{figure*}

To  independently derive a kinetic gas temperature and a total hydrogen density, we studied the excitation conditions of HCN, SO, and CS using $\chi^2$ minimisation scripts in CASSIS with the RADEX\footnote{http://www.strw.leidenuniv.nl/~moldata/radex.html} model \citep{2007A&A...468..627V}, assuming non-LTE conditions and a homogeneous slab. The $\chi^2$ values were computed by fitting the full line profiles to synthetic spectra. Non-detected lines are also considered for the $\chi^2$ calculation. We varied the gas density, the kinetic temperature, and the species column densities in the ranges ($5\times 10^3$ - $5\times 10^5$ ~cm$^{-3}$), (5 - 15 or 20~K), and ($10^{12}$ - $5\times 10^{14}$ cm$^{-2}$),  respectively, with regular grids of 30 points each. We used the collisional coefficients from \citet{2017MNRAS.468.1084H} for HCN, \citet{2007JChPh.126p4312L} for SO, and \citet{2013JChPh.139t4304D,2018MNRAS.478.1811D} for CS. All three are defined down to 5~K. 
For HCN, the detection of the hyperfine component of the 1-0 line and the non-detection of the 2-1 line provide strong constraints. For CS, two lines were detected in almost all sources, while for SO, only two lines (namely, the less energetic ones) were detected but there are seven lines in our sample (with upper energies up to 30~K). For each molecule, we obtained a 3D matrix of $\chi^2$ for a grid of 30x30x30 values of temperature, gas density, and molecular column density. The 'best' physical conditions (and column density for these molecules) are determined by taking the position in the three-dimensional (3D) parameter space for which the global $\chi^2$ is the smallest (i.e. the likelihood is maximal). The determined physical conditions obtained with the three molecules in the five sources are summarised in Table~\ref{Table_physical}. In order to devise a visualisation of the constraints on the physical conditions given by the line analysis, we summed the likelihood over all column densities (N) to get a two-dimensional (2D) map of the $\chi^2$ for (n$_{H_2}$, T) taking into account the uncertainty on N. Fig.~\ref{chi2_projN} shows the 1, 2, and 3 $\sigma$ confidence levels as a function of the gas density and temperature. Similarly, Figs. \ref{chi2_projT} and \ref{chi2_projnH2} show the $\chi^2$ contours projected over the temperatures and gas density respectively (see Appendix~\ref{Appendix_chi2}). We note that when summing the likelihood over one the parameters, the maximum likelihood (or minimum $\chi^2$) may correspond to different values of parameters than the global optimum over the full parameter space spanned by the 3D matrix. When we sum the values over the column densities axis, however, the values of the best parameters are identical for the 2D and 3D case.
 For comparison, the dust temperatures from the PGCC catalogue are listed in the last column of Table~\ref{Table_physical}. 

\newpage

\begin{table*}
\caption{Physical properties of the sources.}
\begin{center}
\begin{tabular}{lccccccccc}
\hline
\hline
 &  CO constraints & \multicolumn{2}{c}{HCN constraints} & \multicolumn{2}{c}{SO constraints} &  \multicolumn{2}{c}{CS constraints}  & \\
Sources & T$_{\rm kin}$ (K) & T$_{\rm kin}$ (K) & n$_{\rm H_2}$ (cm$^{-3}$) & T$_{\rm kin}$ (K) & n$_{\rm H_2}$ (cm$^{-3}$)  & T$_{\rm kin}$ (K) & n$_{\rm H_2}$ (cm$^{-3}$) & T$_{\rm dust, Planck}$ (K) \\
\hline
 C1 &  12  & 5  & $3.3\times 10^5$ & 5 & $1.5\times 10^5$ & 7 & $8.7\times 10^4$ & 12.2\\
C2 &  7  & 5  &  $2.1\times 10^5$ & 5 & $5.2\times 10^3$ & 12 & $1.8\times 10^4$ &  11.6\\
C3 & 12  & 5  & $1.1\times 10^5$& 5.5 & $5.7\times 10^4$ & 7.4 & $4.6\times 10^4$ & 10.6 \\
C4 & 15  & 5  & $1.9\times 10^5$ & 6 & $3\times 10^4$ & 5.7 & $1.6\times 10^5$ & 11.5\\
C5 &  9  & 5  & $1.9\times 10^5$ & 5 & $8.6\times 10^4$ & 7 & $1\times 10^5$ & 12\\
\hline
\end{tabular}
\end{center}
\label{Table_physical}
\end{table*}%

\begin{figure}[htbp]
\includegraphics[width=1\linewidth]{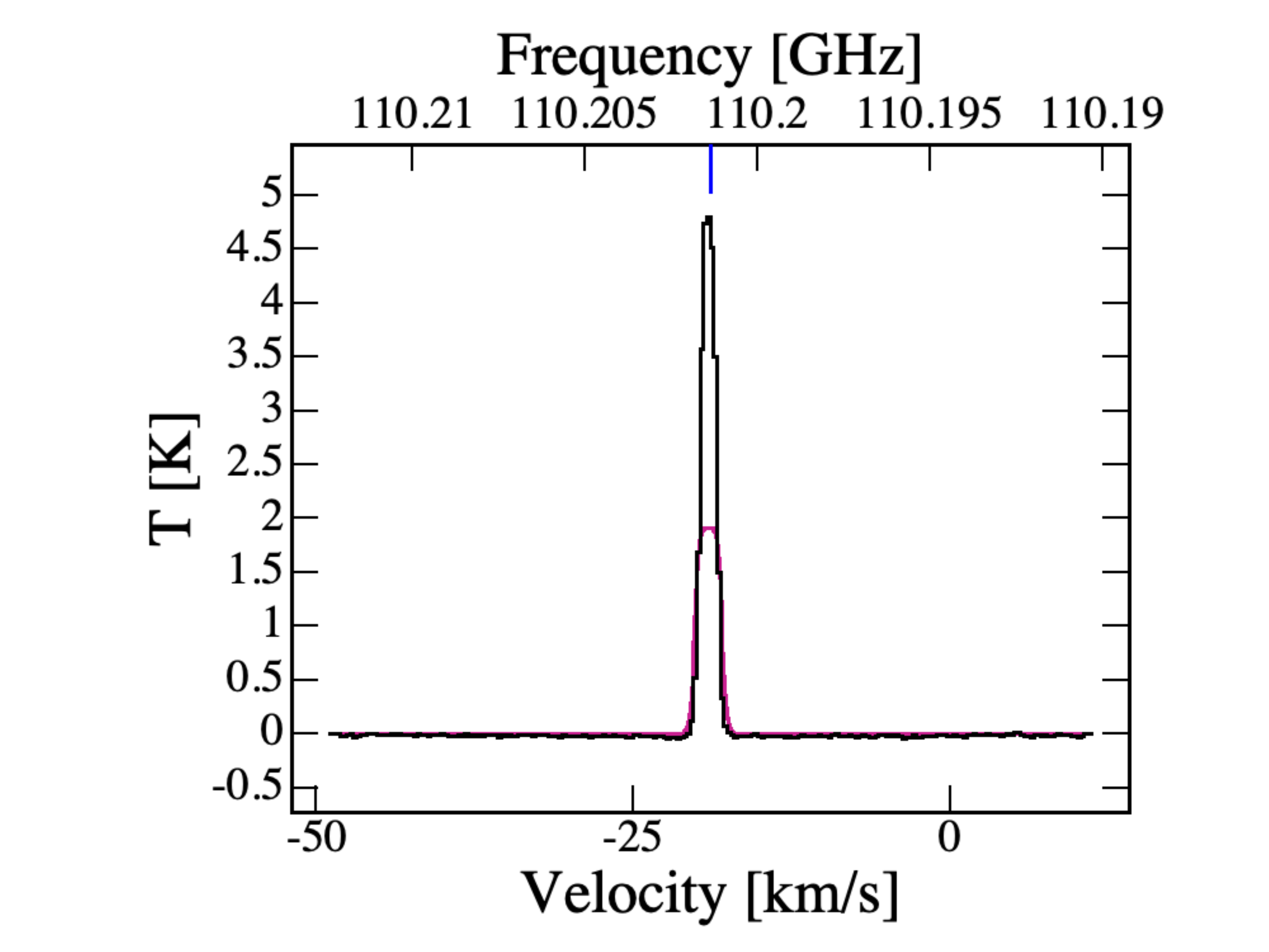}
\caption{$^{13}$CO spectrum (in main beam temperature) observed in C1 in black and theoretical spectra at LTE in red for a kinetic temperature of 5~K and a column density of $2\times 10^{16}$~cm$^{-2}$.  \label{CO_5K}}
\end{figure}

\begin{figure}[htbp]
\includegraphics[width=0.5\linewidth]{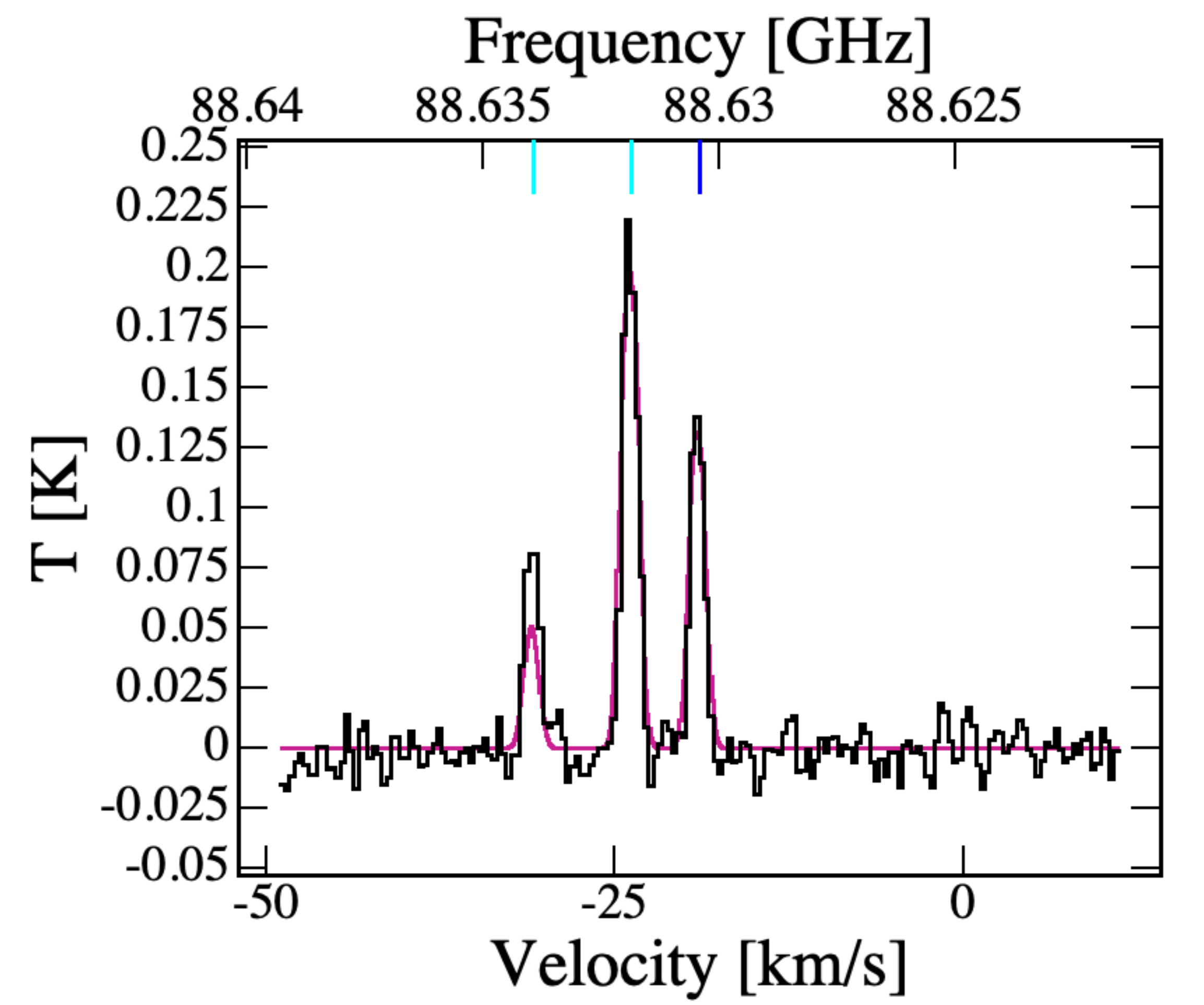}
\includegraphics[width=0.5\linewidth]{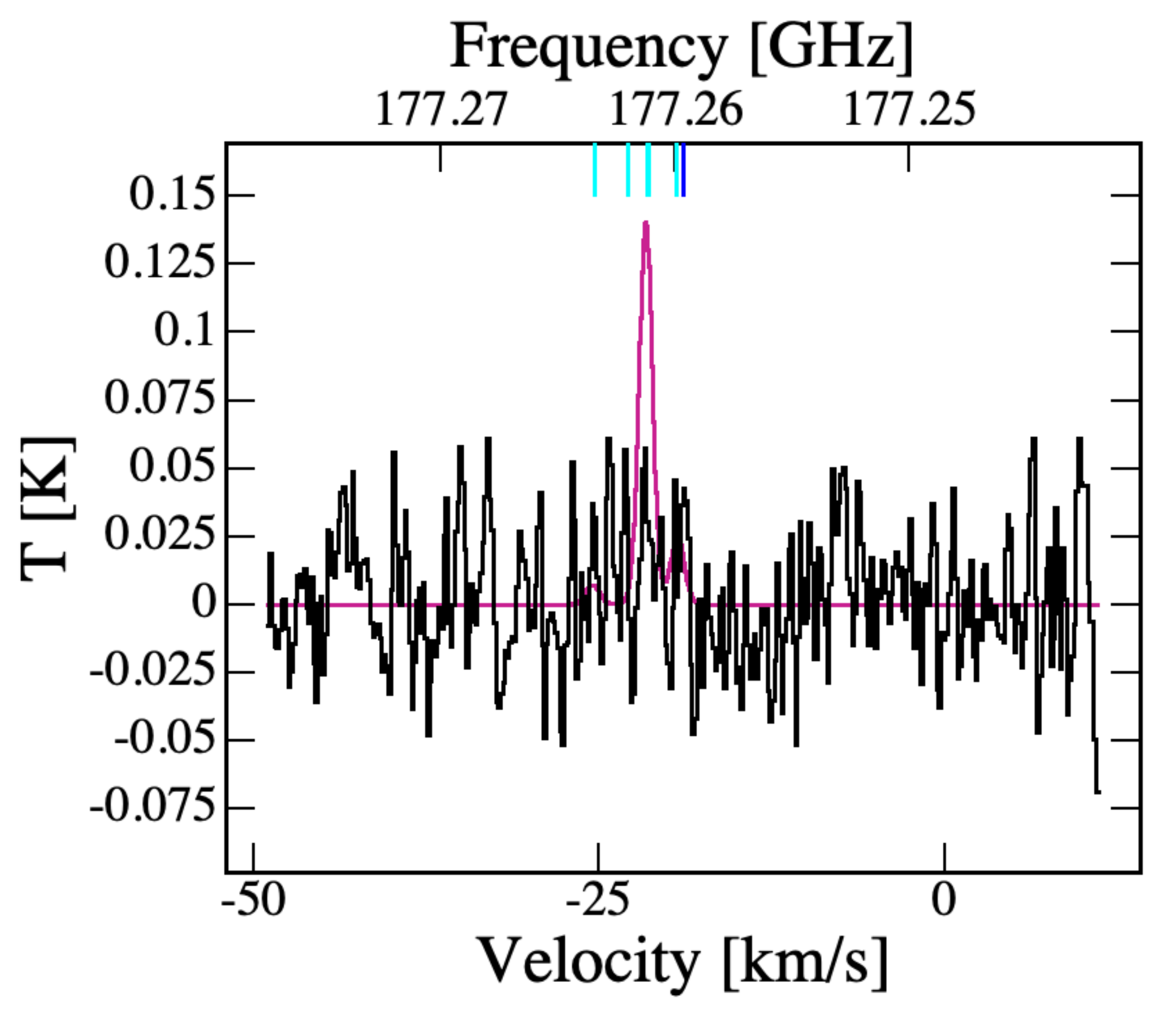}
\caption{HCN 1-0 (left) and 2-1 (right) spectra (in main beam temperature) observed in C1 in black and theoretical spectra at non-LTE in red for a kinetic temperature of 12~K, a column density of $4\times 10^{12}$~cm$^{-2}$, and a H$_2$ density of $3\times 10^4$~cm$^{-3}$.  \label{HCN_12K}}
\end{figure}

Temperature constraints given by SO and HCN are rather strong towards very low temperatures while the constraints given by CS are weaker because of the degeneracy between temperature and density (see Fig.~\ref{chi2_projN}).  Although the $\chi^2$ contours seem to indicate that temperatures could be smaller than 5~K, the collisional rates of these molecules are not defined below this temperature, so it would be risky to extrapolate them. Such low temperatures have already been found by \citet{2011A&A...534A..77P} and \citet{2017MNRAS.470.3194Q} in pre-stellar cores using HCN lines. We do not expect our sources to be pre-stellar cores but our analysis could be biased by optical depth effects and an overly simplified radiative transfer analysis \citep{2017MNRAS.468.1084H}. 
For SO and HCN, the non-detection of the slightly more energetic lines puts strong constraints on the maximum gas temperature. These constraints are, however, obtained with a small number of detected lines, particularly for C2, where only a single line for each of CS and SO are detected. These very low temperatures are not compatible with the bright $^{12}$CO and $^{13}$CO emission lines. In fact, if we assume kinetic temperatures close to 5~K, the  $^{12}$CO and $^{13}$CO lines are systematically underestimated. Figure~\ref{CO_5K} shows the theoretical $^{13}$CO (1-0) line (fitted over the observed one in C1) for a kinetic temperature of 5~K and a column density of $2\times 10^{16}$~cm$^{-2}$ under LTE. Increasing the column density does not increase the line intensity as the line becomes optically thick. Similarly, Fig. \ref{HCN_12K} shows the theoretical HCN (1-0) and (2-1) lines for a kinetic temperature of 12~K (fitted over the observed one in C1), a gas density of $3\times 10^4$~cm$^{-3}$, and a column density of $4\times 10^{12}$~cm$^{-2}$. The 2-1 line is clearly overestimated. Decreasing the gas density or the HCN column density will decrease the 1-0 line intensity before the 2-1 line reaches the noise level. The incompatibility between the kinetic temperatures traced by CO and the other molecules (HCN and SO) seems to indicate that the molecules do not trace the same layer of material.  The $^{12}$CO and $^{13}$CO (1-0) lines in our sample are optically thick, so their emissions are likely to come from the outer layer of the clouds, while the other molecules may originate from more dense and cold material. 



\subsubsection{Deriving molecular column densities}

\begin{table*}
\caption{Sets of physical conditions used to derive the molecular column densities. Warm conditions: Temperature (K) from the CO lines and H$_2$ density (cm$^{-3}$) from CS for the CO temperature. Cold conditions:  Best fits for HCN lines.  }
\begin{center}
\begin{tabular}{lllllll}
\hline
\hline
Source & Warm conditions & Cold conditions \\
\hline
C1 & 12~K, $2.9\times 10^4$~cm$^{-3}$ & 5~K, $3.3\times 10^5$~cm$^{-3}$ \\
C2 & 7~K, $8.7\times 10^4$~cm$^{-3}$ & 5~K, $2.1\times 10^5$~cm$^{-3}$ \\
C3 & 12~K, $1.7\times 10^4$~cm$^{-3}$ & 5~K, $1.1\times 10^5$~cm$^{-3}$ \\
C4 & 15~K, $2.9\times 10^4$~cm$^{-3}$ & 5~K, $1.9\times 10^5$~cm$^{-3}$ \\
C5 & 9~K, $6.0\times 10^4$~cm$^{-3}$ & 5~K, $2.1\times 10^5$~cm$^{-3}$ \\
TMC1 & 10~K, $1.5\times 10^4$~cm$^{-3}$ & 8~K, $3.4\times 10^4$~cm$^{-3}$\\
\hline
\end{tabular}
\end{center}
\label{models}
\end{table*}%

Although the constraints on the kinetic temperatures and gas densities are quite uncertain, the derived column densities are not overly sensitive to them. To get an idea of the values of the species column densities, we used two different values of (T, n$_{\rm H_2}$). The first one is the 'best' fit given by HCN for each cloud, which is a very cold and dense solution. The second one is the kinetic temperature given by $^{12}$CO and the gas density given by CS for this temperature, which corresponds to a somewhat less dense and warmer solution. These sets of physical conditions, called 'warm conditions' and 'cold conditions', are listed in Table~\ref{models}. The column densities for each molecule are then derived using the $\chi^2$ minimisation scripts provided by CASSIS with the Radex radiative transfer code to find the best fit of the observed line profiles. The species column densities are varied in a regular grid of 30 points between $10^{12}$ and $5\times 10^{14}$ cm$^{-2}$. The line widths and positions are set by the observations. Non-LTE conditions are assumed for all species, except for HNO as no collisional rates exist for this molecule. Considering the Einstein coefficient and an approximation of the excitation coefficients, the critical density should be lower than our typical densities of the detected line (Francois Lique, private communication). For the other molecules, the collisional coefficients used are the following: \citet{2014MNRAS.441..664Y} for HCO$^+$, \citet{2012MNRAS.422..812K} for CN, \citet{2010MNRAS.406...95R} for CH$_3$OH, \citet{2012MNRAS.421.1891S} for CCH, \citet{2011PCCP...13.8204D} for HNC, \citet{2009A&A...493..557L} for NO, \citet{2013MNRAS.432.2573W} for H$_2$CO, and \citet{2010ApJ...718.1062Y} for CO. The collisional coefficients for CCH, CN, HNC, and H$_2$CO are defined down to 5~K, while the ones for H$_2$CO, NO, HCO$^+$, and CH$_3$OH only go down to 10~K. The files in the Radex format were downloaded from the LAMDA database\footnote{https://home.strw.leidenuniv.nl/$\sim$moldata/} and the CASSIS website.  The species column densities obtained in the five sources and for the two physical conditions are listed in Tables \ref{col_dens_C1} to \ref{col_dens_C5}, with the computed reduced $\chi^2$. Due to the low temperatures in the clumps, ortho and para forms of H$_2$CO have been fitted separately and summed. The same was done for the A and E forms of CH$_3$OH. For undetected molecules, we have  determined the upper limits on the column densities to reach the noise level with the physical parameters previously mentioned. For the molecules detected in none of the sources and listed in Table~\ref{lineparam_nondetected}, we estimated the upper limits on the column density assuming LTE, optically thin lines, and a beam-filling factor of 1. The considered upper limits on the integrated intensities are 0.06~K.km.s$^{-1}$ for sources C1 to C3, 0.03~K.km.s$^{-1}$ for C4, and 0.05~K.km.s$^{-1}$ for C5.

\begin{table*}
\caption{Species column densities (in cm$^{-2}$) in C1. See Table~\ref{models} for the corresponding 'warm' and 'cold' conditions. Boldface indicates the smallest $\chi^2$ if both conditions do not give the same value. $\chi^2_{\rm red}$ stands for reduced $\chi^2$.}
\begin{center}
\begin{tabular}{l|cc|cc}
\hline
\hline
& \multicolumn{2}{c|}{Warm conditions} & \multicolumn{2}{c}{Cold conditions} \\
Molecule & N (cm$^{-2}$) & $\chi^2_{\rm red}$ & N (cm$^{-2}$) & $\chi^2_{\rm red}$ \\
\hline
H$_2$CO & $5.4\times 10^{12}$ & {\bf 2.0} & $3.5\times 10^{12}$ & 2.3 \\
CS & $6.9\times 10^{12}$  & {\bf 1.2}  & $4.5\times 10^{12}$ & 1.3\\
SO &$5.9\times 10^{12}$ & 3.0 & $9\times 10^{12}$ & {\bf 2.9}\\
NO & $\le 6.0\times 10^{13}$ & - & $\le 8.0\times 10^{13}$ & - \\
HNO &$\le 7.0\times 10^{11}$ & - & $\le 4.0\times 10^{11}$ & - \\
HCO$^+$ & $1.1\times 10^{12}$ & 1.5 & $9.0\times 10^{11}$ & {\bf 1.4} \\
HCN &$4.7\times 10^{12}$ & 1.4 & $2.5\times 10^{12}$ & {\bf 1.3} \\
HNC & $9.9\times 10^{11}$ & 1.5 & $5.8\times 10^{11}$ & {\bf 1.1}\\
CN &  $9.9\times 10^{12}$ & 1.8 & $5.8\times 10^{12}$ & 1.8 \\
CCH & $2.9\times 10^{12}$ & 1.0 & $2.9\times 10^{12}$ & 1.0 \\
CH$_3$OH & $4.7\times 10^{12}$ & 1.2 &  $6.2\times 10^{12}$  & {\bf 1.1} \\
C$^{18}$O & $8.7\times 10^{14}$ & 1.4 & $1.1\times 10^{15}$ & 1.4 \\
c-C$_3$H$_2$ & $\le 1.2\times 10^{12}$ & - & $\le 1\times 10^{12}$ & - \\
l-C$_3$H$_2$  & $\le 2.6\times 10^{11}$ & - & $\le 1\times 10^{10}$ & - \\
CH$_3$CN & $\le  1.9\times 10^{11}$ & - & $\le 3\times 10^{11}$ & - \\
C$_3$N & $\le  4.2\times 10^{12}$ & - & $\le 2\times 10^{13}$ & - \\
c-C$_3$H  & $\le 4.9\times 10^{12}$ & - & $\le 6\times 10^{12}$ & - \\
l-C$_5$H$_2$  & $\le 8.7\times 10^{11}$ & - & $\le 1\times 10^{11}$ & - \\
HNCCC & $\le  2.7\times 10^{11}$ & - & $\le 1\times 10^{12}$ & - \\
HCCNC & $\le  9.9\times 10^{11}$ & - & $\le 4\times 10^{12}$ & - \\
H$_2$CCN & $\le  3.0\times 10^{11}$ & - & $\le 2\times 10^{9}$ & - \\
HCCCN & $\le  2.4\times 10^{11}$ & - & $\le 9\times 10^{11}$ & - \\
l-C$_4$H$_2$& $\le   6.0\times 10^{11}$ & - & $\le  4\times 10^{10}$ & - \\
N$_2$H$^+$ & $\le  6.6\times 10^{11}$ & - & $\le 7\times 10^{11}$ & - \\
H$_2$CS & $\le  1.0\times 10^{12}$ & - & $\le 1\times 10^{12}$ & - \\
\hline
\end{tabular}
\end{center}
\label{col_dens_C1}
\end{table*}%

\begin{table*}
\caption{Species column densities (in cm$^{-2}$) in C2. See Table~\ref{models} for the corresponding 'warm' and 'cold' conditions. Boldface indicates the smallest $\chi^2$ if both conditions do not give the same value. $\chi^2_{\rm red}$ stands for reduced $\chi^2$.}
\begin{center}
\begin{tabular}{l|cc|cc}
\hline
\hline
& \multicolumn{2}{c|}{Warm conditions} & \multicolumn{2}{c}{Cold conditions} \\
Molecule & N (cm$^{-2}$) & $\chi^2_{\rm red}$ & N (cm$^{-2}$) & $\chi^2_{\rm red}$ \\
\hline
H$_2$CO & $\le 7.0\times 10^{11}$ &- & $\le 7.0\times 10^{11}$ & -\\
CS & $1.0\times 10^{12}$ & 1.9 & $1.1\times 10^{12}$ & 1.9 \\
SO & $8.5\times 10^{11}$ &  3.9  & $1.3\times 10^{12}$ & 3.9 \\
NO & $\le 5.0\times 10^{13}$ &-&$\le 7.0\times 10^{13}$ &-\\
HNO &$\le 3.0\times 10^{11}$ & - & $\le 3.0\times 10^{11}$ & - \\
HCO$^+$ & $2.5\times 10^{11}$ & {\bf 1.7} & $2.5\times 10^{11}$ & 1.8\\
HCN &$6.9\times 10^{11}$ & 0.6 & $5.5\times 10^{11}$ & {\bf 0.5}\\
HNC & $1.5\times 10^{11}$  & 1.2 & $1.6\times 10^{11}$ & 1.2\\
CN & $\le 6.0\times 10^{12}$   &- &$\le 5.0\times 10^{12}$& -\\
CCH &$\le 2.0\times 10^{12}$ &- &$\le 2.0\times 10^{12}$ &- \\
CH$_3$OH & $\le 3.0\times 10^{12}$  & -& $\le 3.0\times 10^{12}$ & -\\
C$^{18}$O & $7.6\times 10^{14}$  & 4.3 & $9.9\times 10^{14}$ & {\bf 4.1} \\
c-C$_3$H$_2$ & $ \le 1\times 10^{12}$ & - & $\le 1\times 10^{12}$ & - \\
l-C$_3$H$_2$  & $\le 4\times 10^{10}$ & - & $\le 1\times 10^{10}$ & - \\
CH$_3$CN  & $\le 2\times 10^{11}$ & - & $\le 3\times 10^{11}$ & - \\
C$_3$N  & $\le 8\times 10^{12}$ & - & $\le 2\times 10^{13}$ & - \\
c-C$_3$H  & $\le 5\times 10^{12}$ & - & $\le 5\times 10^{12}$ & - \\
l-C$_5$H$_2$  & $\le 2\times 10^{11}$ & - & $\le 1\times 10^{11}$ & - \\
HNCCC  & $\le 5\times 10^{11}$ & - & $\le 1\times 10^{12}$ & - \\
HCCNC  & $\le 2\times 10^{12}$ & - & $\le 4\times 10^{12}$ & - \\
H$_2$CCN  & $\le 2\times 10^{10}$ & - & $\le 2\times 10^{09}$ & - \\
HCCCN  & $\le 4\times 10^{11}$ & - & $\le 9\times 10^{11}$ & - \\
l-C$_4$H$_2$  & $\le 1\times 10^{11}$ & - & $ \le 4\times 10^{10}$ & - \\
N$_2$H$^+$  & $\le 6\times 10^{11}$ & - & $\le 7\times 10^{11}$ & - \\
H$_2$CS  & $\le 1\times 10^{12}$ & - & $\le 1\times 10^{12}$ & - \\
\hline
\end{tabular}
\end{center}
\label{col_dens_C2}
\end{table*}%

\begin{table*}
\caption{Species column densities (in cm$^{-2}$) in C3. See Table~\ref{models} for the corresponding 'warm' and 'cold' conditions. Boldface indicates the smallest $\chi^2$ if both conditions do not give the same value. $\chi^2_{\rm red}$ stands for reduced $\chi^2$.}
\begin{center}
\begin{tabular}{l|cc|cc}
\hline
\hline
& \multicolumn{2}{c|}{Warm conditions} & \multicolumn{2}{c}{Cold conditions} \\
Molecule & N (cm$^{-2}$) & $\chi^2_{\rm red}$ & N (cm$^{-2}$) & $\chi^2_{\rm red}$ \\
\hline
H$_2$CO & $8.0\times 10^{12}$  &  2.9 & $6.4\times 10^{12}$ & {\bf 2.8} \\
CS & $4.7\times 10^{12}$ & 1.5 & $3.8\times 10^{12}$ & 1.5 \\
SO & $9.0\times 10^{12}$ & 2.8 & $1.4\times 10^{13}$ & {\bf 2.6} \\
NO & $1.7\times 10^{14}$ & 0.4 & $3.3\times 10^{14}$ & 0.4 \\
HNO &$1.3\times 10^{12}$  & {\bf 0.8} & $2.2\times 10^{12}$ & 0.9 \\
HCO$^+$ & $1.7\times 10^{12}$ & {\bf 4.0} & $1.4\times 10^{12}$ & 4.1\\
HCN & $7.3\times 10^{12}$ & 1.1 & $3.8\times 10^{12}$ & {\bf 0.9} \\
HNC & $1.5\times 10^{12}$ & 2.9 & $8.7\times 10^{11}$ & 2.9 \\
CN & $2.5\times 10{13}$  & 0.5 & $2.2\times 10^{13}$ & 0.5 \\
CCH &$\le 3.0\times 10^{12}$ & - & $\le 2.0\times 10^{12}$ & -  \\
CH$_3$OH & $5.2\times 10^{12}$ & 1.2 & $7.2\times 10^{12}$ & {\bf 1.2} \\
C$^{18}$O & $3.9\times 10^{14}$ & 2.2 & $5.0\times 10^{14}$ & 2.2 \\
c-C$_3$H$_2$ & $\le 1.2\times 10^{12}$ & - & $\le 1\times 10^{12}$ & - \\
l-C$_3$H$_2$  & $\le 2.6\times 10^{11}$ & - & $\le 1\times 10^{10}$ & - \\
CH$_3$CN & $\le  1.9\times 10^{11}$ & - & $\le 3\times 10^{11}$ & - \\
C$_3$N & $\le  4.2\times 10^{12}$ & - & $\le 2\times 10^{13}$ & - \\
c-C$_3$H  & $\le 4.9\times 10^{12}$ & - & $\le 5\times 10^{12}$ & - \\
l-C$_5$H$_2$  & $\le 8.7\times 10^{11}$ & - & $\le 1\times 10^{11}$ & - \\
HNCCC & $\le  2.7\times 10^{11}$ & - & $\le 1\times 10^{12}$ & - \\
HCCNC & $\le  9.9\times 10^{11}$ & - & $\le 4\times 10^{12}$ & - \\
H$_2$CCN & $\le  3.0\times 10^{11}$ & - & $\le 2\times 10^{09}$ & - \\
HCCCN & $\le  2.4\times 10^{11}$ & - & $\le 9\times 10^{11}$ & - \\
l-C$_4$H$_2$& $\le   6.0\times 10^{11}$ & - & $\le  4\times 10^{10}$ & - \\
N$_2$H$^+$ & $\le  6.6\times 10^{11}$ & - & $\le 7\times 10^{11}$ & - \\
H$_2$CS & $\le  1.0\times 10^{12}$ & - & $\le 1\times 10^{12}$ & - \\
\hline
\end{tabular}
\end{center}
\label{col_dens_C3}
\end{table*}%

\begin{table*}
\caption{Species column densities (in cm$^{-2}$) in C4. See Table~\ref{models} for the corresponding 'warm' and 'cold' conditions. Boldface indicates the smallest $\chi^2$ if both conditions do not give the same value. $\chi^2_{\rm red}$ stands for reduced $\chi^2$.}
\begin{center}
\begin{tabular}{l|cc|cc}
\hline
\hline
& \multicolumn{2}{c|}{Warm conditions} & \multicolumn{2}{c}{Cold conditions} \\
Molecule & N (cm$^{-2}$) & $\chi^2_{\rm red}$ & N (cm$^{-2}$) & $\chi^2_{\rm red}$ \\
\hline
H$_2$CO & $3.6\times 10^{12}$ & 1.7 & $3.7\times 10^{12}$ & {\bf 1.3} \\
CS & $7.3\times 10^{12}$ & 1.7 & $9.0\times 10^{12}$ & 1.7 \\
SO &$3.1\times 10^{12}$ & 1.6 & $5.9\times 10^{12}$ & 1.6 \\
NO & $\le 5.0\times 10^{13}$ & - &$\le 8.0\times 10^{13}$ & - \\
HNO & $\le 8.0\times 10^{11}$ &- & $\le 3.0\times 10^{11}$ & - \\
HCO$^+$ &$5.9\times 10^{11}$ &1.2 & $5.8\times 10^{11}$ & 1.2 \\
HCN & $7.3\times 10^{12}$ & 0.7 & $5.9\times 10^{12}$  & {\bf 0.6}\\
HNC & $8.6\times 10{11}$ & 1.0 & $6.6\times 10^{11}$ & {\bf 0.9}\\
CN &  $\le 1.0\times 10^{13}$ &  - &$\le 1.0\times 10^{13}$ & - \\
CCH & $6.6\times 10^{12}$ & 0.9 & $9.9\times 10^{12}$ & 0.9 \\
CH$_3$OH & $2.4\times 10^{12}$ & 3.3 & $4.7\times 10^{12}$ & {\bf 3.2}\\
C$^{18}$O & $1.5\times 10^{15}$ & 6.2 & $2.2\times 10^{15}$ & {\bf 5.5} \\
c-C$_3$H$_2$  & $\le 8\times 10^{11}$ & - & $\le 6\times 10^{11}$ & - \\
l-C$_3$H$_2$  & $\le 2\times 10^{11}$ & - & $\le 7\times 10^{09}$ & - \\
CH$_3$CN  & $\le 1\times 10^{11}$ & - & $\le 1\times 10^{11}$ & - \\
C$_3$N  & $\le 2\times 10^{12}$ & - & $\le 9\times 10^{12}$ & - \\
c-C$_3$H & $ \le 3\times 10^{12}$ & - & $\le 3\times 10^{12}$ & - \\
l-C$_5$H$_2$ & $\le 4\times 10^{11}$ & - & $\le  5\times 10^{10}$ & - \\
HNCCC  & $\le 1\times 10^{11}$ & - & $\le 5\times 10^{11}$ & - \\
HCCNC  & $\le 5\times 10^{11}$ & - & $\le 2\times 10^{12}$ & - \\
H$_2$CCN  & $\le 2\times 10^{11}$ & - & $\le 9\times 10^{08}$ & - \\
HCCCN  & $\le 1\times 10^{11}$ & - & $\le 4\times 10^{11}$ & - \\
l-C$_4$H$_2$  & $\le 3\times 10^{11}$ & - & $\le 2\times 10^{10}$ & - \\
N$_2$H$^+$  & $\le 4\times 10^{11}$ & - & $\le 3\times 10^{11}$ & - \\
H$_2$CS  & $\le 6\times 10^{11}$ & - & $\le 7\times 10^{12}$ & - \\
\hline
\end{tabular}
\end{center}
\label{col_dens_C4}
\end{table*}%

\begin{table*}
\caption{Species column densities (in cm$^{-2}$) in C5. See Table~\ref{models} for the corresponding 'warm' and 'cold' conditions. Boldface indicates the smallest $\chi^2$ if both conditions do not give the same value. $\chi^2_{\rm red}$ stands for reduced $\chi^2$.}
\begin{center}
\begin{tabular}{l|cc|cc}
\hline
\hline
& \multicolumn{2}{c|}{Warm conditions} & \multicolumn{2}{c}{Cold conditions} \\
Molecule & N (cm$^{-2}$) & $\chi^2_{\rm red}$ & N (cm$^{-2}$) & $\chi^2_{\rm red}$ \\
\hline
H$_2$CO & $2.8\times 10^{12}$ & 2.3 & $3.2\times 10^{12}$ & 2.3 \\
CS & $4.7\times 10^{12}$ & 0.9 & $5.9\times 10^{12}$ & 0.9 \\
SO & $3.1\times 10^{12}$ & 2 & $5.9\times 10^{12}$ & {\bf 1.8} \\
NO &$\le 7.0\times 10^{13}$ &-&$\le 9.0\times 10^{13}$ & - \\
HNO & $\le 4.0\times 10^{11}$ & - & $\le 4.0\times 10^{11}$ & - \\
HCO$^+$ & $3.8\times 10^{11}$ &1.1 & $4.7\times 10^{11}$ & 1.1 \\
HCN & $3.1\times 10^{12}$  & 0.5 & $2.5\times 10^{12}$  & {\bf 0.4} \\
HNC & $5.0\times 10^{11}$ & 0.7 & $5.0\times 10^{11}$ & 0.7 \\
CN &$\le 7.0\times 10^{12}$ & - & $\le 8.0\times 10^{12}$ & -\\
CCH &$\le 1.0\times 10^{12}$ & -& $\le 2.0\times 10^{12}$&-\\
CH$_3$OH & $2.2\times 10^{12}$ & 2.5 & $3.7\times 10^{12}$ & {\bf 2.4} \\
C$^{18}$O & $5.0\times 10^{14}$ & 2.8 & $7.6\times 10^{14}$ & 2.8 \\
c-C$_3$H$_2$  & $\le  9\times 10^{11}$ & - & $\le 1\times 10^{12}$ & - \\
l-C$_3$H$_2$   & $\le 1\times 10^{11}$ & - & $\le 1\times 10^{10}$ & - \\
CH$_3$CN   & $\le 1\times 10^{11}$ & - & $\le 2\times 10^{11}$ & - \\
C$_3$N   & $\le 4\times 10^{12}$ & - & $\le 1\times 10^{13}$ & - \\
c-C$_3$H   & $\le 3\times 10^{12}$ & - & $\le 4\times 10^{12}$ & - \\
l-C$_5$H$_2$   & $\le 7\times 10^{11}$ & - & $\le 8\times 10^{10}$ & - \\
HNCCC  & $\le  3\times 10^{11}$ & - & $\le 9\times 10^{11}$ & - \\
HCCNC   & $\le 1\times 10^{12}$ & - & $\le 3\times 10^{12}$ & - \\
H$_2$CCN  & $\le  1\times 10^{11}$ & - & $\le 1\times 10^{9}$ & - \\
HCCCN  & $\le  2\times 10^{11}$ & - & $\le 7\times 10^{11}$ & - \\
l-C$_4$H$_2$  & $\le  4\times 10^{11}$ & - & $ 3\le \times 10^{10}$ & - \\
N$_2$H$^+$   & $\le 5\times 10^{11}$ & - & $\le 6\times 10^{11}$ & - \\
H$_2$CS   & $\le 8\times 10^{11}$ & - & $\le 1\times 10^{12}$ & - \\
\hline
\end{tabular}
\end{center}
\label{col_dens_C5}
\end{table*}%

For HCN, SO, and CH$_3$OH, the colder and denser physical conditions produce the smallest $\chi^2$ values. For the other species, the conclusion is less clear as some molecules are either not sensitive to these parameters or the results vary from source to source. For instance, HCO$^+$  is better fitted by the physical conditions derived from HCN in C1, insensitive in C4 and C5, and better reproduced by the other physical conditions in C2 and C3. Whatever the physical conditions applied, the computed species column densities vary by less than a factor of 2.

\section{Comparing the chemical composition of the different clumps}\label{diss}

To compare the chemical composition of the different clumps, we computed mean species column densities using the two values listed in Tables~\ref{col_dens_C1} to \ref{col_dens_C5} and divided these values by the C$^{18}$O mean column density multiplied by a constant $^{12}$CO/C$^{18}$O ratio of 557 \citep{1999RPPh...62..143W}. We assume here that there is no oxygen isotopic fractionation based on \citet{2019MNRAS.485.5777L}. We chose CO as a reference because it was detected in all five sources. In addition, C$^{18}$O is optically thin and so it is very likely to originate from the same layer of material as the other molecules, which is contrary to the properties of $^{12}$CO and $^{13}$CO.
The CO abundance can be affected by depletion but its abundance is less affected by other physical parameters, such as density and temperature. For instance, \citet{1997ApJ...486..862P}  preferred to compare the abundances with respect to HCO$^+$. This molecule is less abundant than CO and its abundance is more affected by the physical conditions, as can be seen in Fig.~\ref{CO_HCOp_ab} in the appendix. 

\begin{figure}[htbp]
\includegraphics[width=1.0\linewidth]{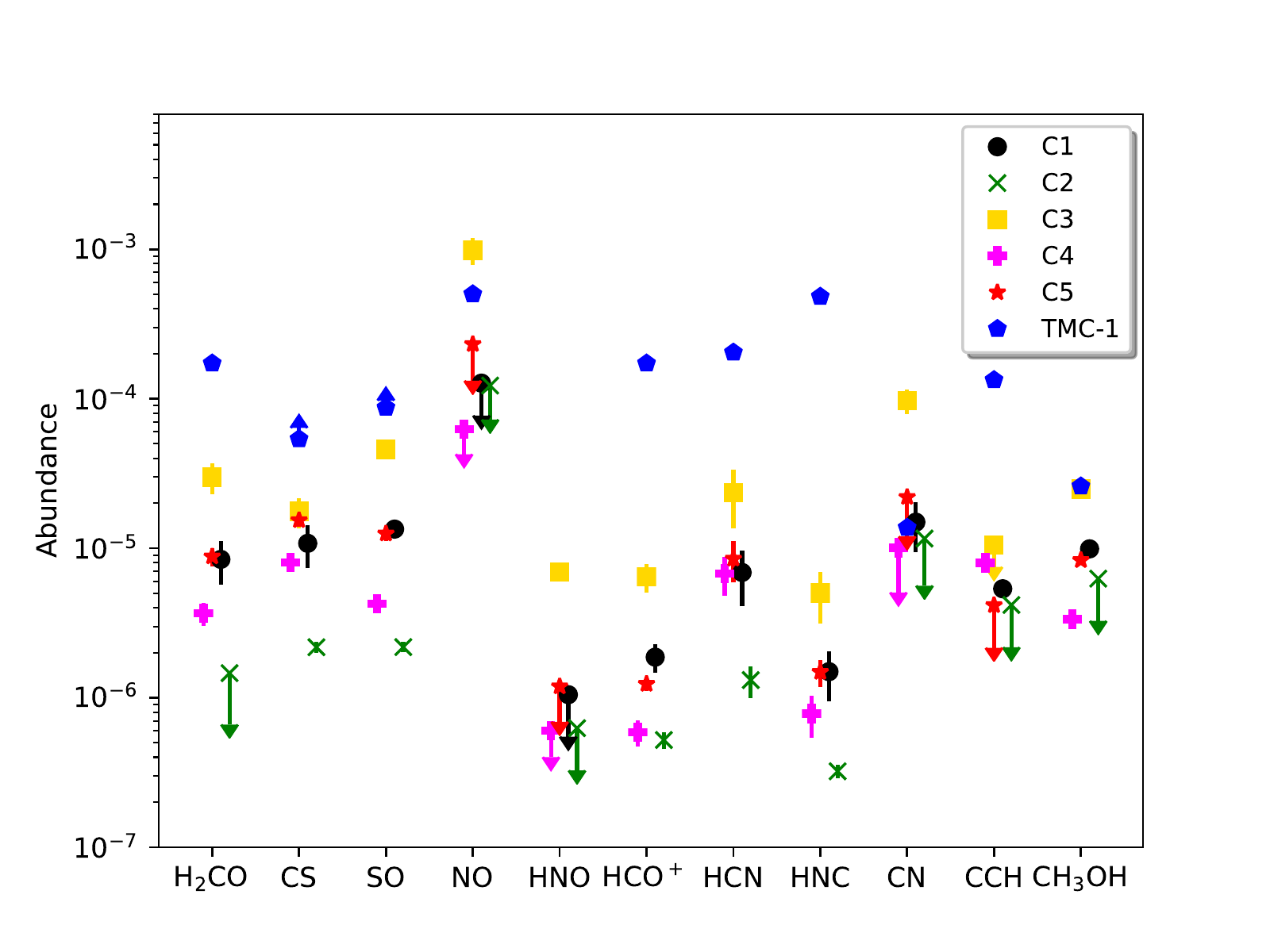}
\caption{Species abundances (of the molecules listed in Table~\ref{detect_lines}) with respect to CO in each of the clouds. Arrows represent upper or lower limits. Vertical lines represent error bars computed from the variation of species column densities due to the uncertainty in the physical conditions. Blue points are the values reported in the literature towards the cold core TMC-1 (CP).  \label{abundances}}
\end{figure}

\begin{figure}[htbp]
\includegraphics[width=1.0\linewidth]{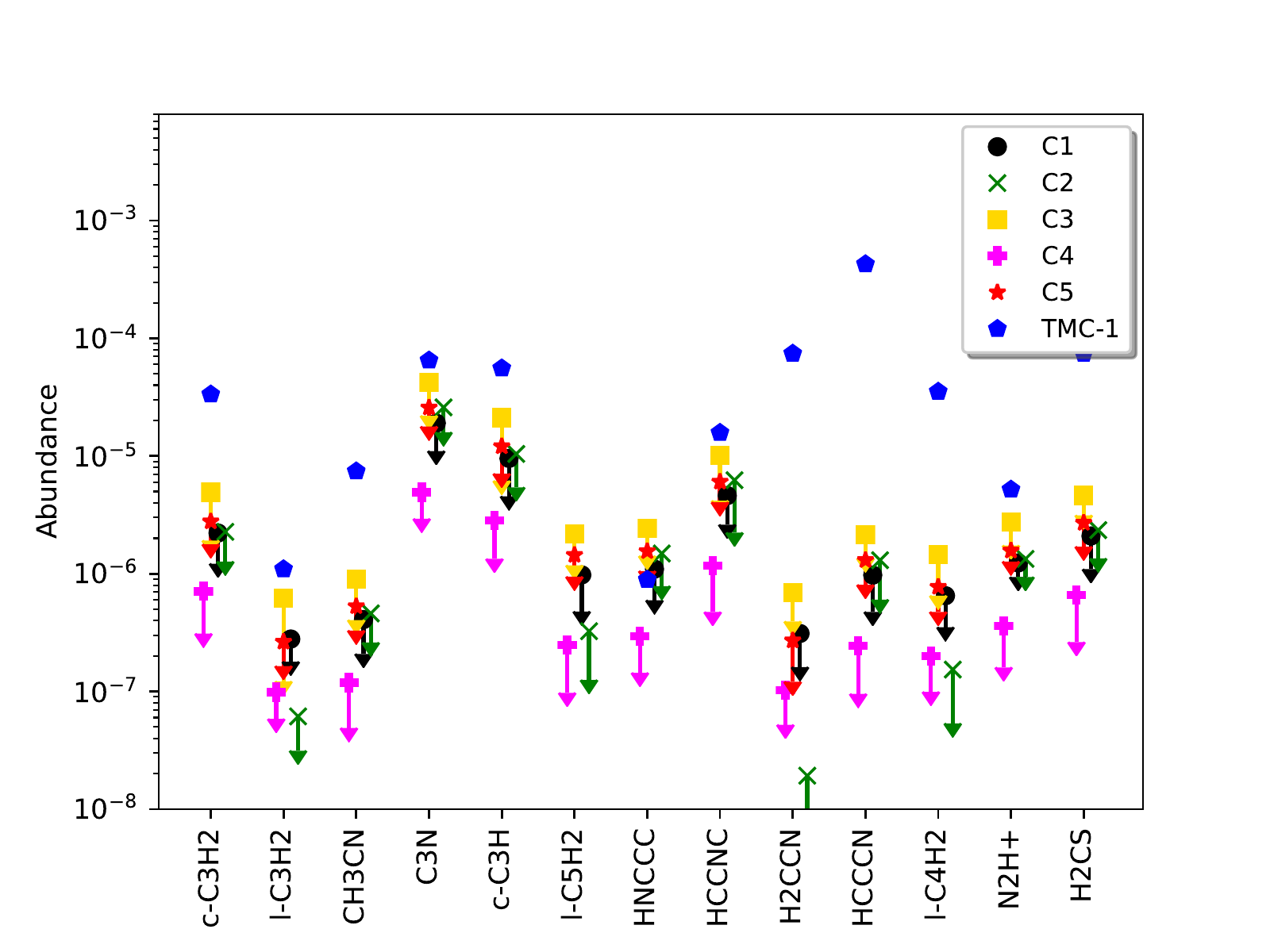}
\caption{Upper limits on the species abundances not detected in any of the sources (and listed in Table~\ref{lineparam_nondetected})
 with respect to CO in each of the clouds. Arrows represent upper or lower limits. Vertical lines represent error bars computed from the variation of species column densities due to the uncertainty in the physical conditions. Blue points are the values reported in the literature towards the cold core TMC-1 (CP). \label{abundances_nd}}
\end{figure}

The abundances with respect to CO are shown in Fig.~\ref{abundances} for molecules detected in at least one of the clouds (listed in Table~\ref{detect_lines}) and in Fig.~\ref{abundances_nd} for the molecules detected in none of the sources (listed in Table~\ref{lineparam_nondetected}). For comparison, we also report the values from the literature for the cold clump TMC-1 (CP): \citet{1997ApJ...486..862P} for HCO$^+$, HCN, HNC, CN, CCH, and N$_2$H$^+$ ; \citet{2016ApJS..225...25G} for SO, CS, CH$_3$OH, c-C$_3$H$_2$, l-C$_3$H$_2$, CH$_3$CN, C$_3$N, c-C$_3$H, HNCCC, HCCNC, H2CCN, HC$_3$N, l-C$_4$H$_2$, and H$_2$CS ; \citet{1998FaDi..109..205O} for H$_2$CO ; and \citet{1993A&A...268..212G} for NO. As
far as we know for this source, HNO and l-C$_5$H$_2$ have not been reported in the literature.  For CO, the abundance reported by \citet{1997ApJ...486..862P} is very high ($1.4\times 10^{-4}$ with respect to H$_2$). \citet{1992IAUS..150..171O} reported a lower abundance of $8\times 10^{-5}$. A more recent study by 
\citet{2019A&A...624A.105F} obtained $9.7\times 10^{-5}$ from C$^{18}$O observations. We chose to use this last value to compute the species abundances with respect to CO in TMC-1, but the overall results are not significantly affected if alternative values are used. We note that the H$_2$ column density used for the CO abundance is also different from the other references ($1.8\times 10^{22}$~cm$^{-2}$ from Fuente et al. 2019, instead of the generally assumed value of $10^{22}$~cm$^{-2}$), so we scaled the abundance using the same H$_2$ column densities before dividing the species abundances.

Among our observed clouds, C3 presents clearly larger abundances with respect to CO while C2 seems to have smaller abundances. As compared to TMC-1 (CP), all our clumps have smaller abundances, except for the molecules CN, SO, CS, and NO; CN seems to be less abundant in TMC-1 (CP) than in C1 and C3, while CH$_3$OH and NO seem to be less abundant in TMC-1 (CP) than in C3. Overall, all molecular abundances vary by nearly a factor of 10 or more from source to source. For the molecules that are not detected in any of our sources, half of them present an upper limit at the same level as the observed value in TMC-1 (CP) and half of them are below. As an example, the upper limit on N$_2$H$^+$ is not a strong constraint while the HC$_3$N upper limits are well below what is observed in TMC-1 (CP).

Looking at the CS/SO abundance ratios in the different clumps, we find a small variation around 1: from 0.4 to 1.9. The ratio of HCN/HNC is always larger than 1: from 4 to 8.6. A ratio of HCN/HNC that is larger than one may indicate a non negligible abundance of atomic carbon in the gas phase as HNC is quickly converted to HCN through its reaction with C \citep{2014MNRAS.443..398L}. 
NO/HNO is larger than 100 in C3 where both molecules were detected. The CN/NO abundance ratio is 0.1 in C3 and the HCN/CN ratio is 0.4 in C1 and 0.2 in C3. The o/p ratio of H$_2$CO is between 2.2 and 2.5 in all sources where it was detected. 

\section{Chemical modelling}\label{chem_mod}

To chemically characterise the observed cores, we used the gas-grain chemical code Nautilus \citep{2016MNRAS.459.3756R}. This model follows the chemical evolution of the gas and ices surrounding interstellar grains by solving a set of differential equations. The various gas, gas-grain, and grain-surface processes included in the model are described in \citet{2016MNRAS.459.3756R} while all the chemical parameters are the ones described in \citet{2019MNRAS.486.4198W}. For each source, we have run two models employing the sets of physical conditions used to derive the observed column densities (see Table~\ref{models} for the corresponding physical conditions). We computed the chemistry for $10^7$~yr starting from an atomic composition (except for hydrogen, assumed to be entirely molecular) with the same initial abundances as in Table 1 of \citet{2018A&A...611A..96R}. The elemental abundances for chemical models are the abundances of elements that remain in the gas-phase and available for chemistry, while the rest are locked into refractory compounds. The choice of elemental abundances for the chemical modelling of cold cores is somewhat arbitrary as elemental depletion is observed to increase with the density in the diffuse medium but uncertainties remain on what happens for densities larger than 10 cm$^{-3}$ (conditions under which the atomic lines cannot be observed any longer). So the constraints we have on these parameters are based on the observation of atomic abundances in the diffuse medium (that have also been reviewed over time with successive generations of telescopes). These observed abundances are very often modified to reduce the abundance of metals and sulphur. Many different values have been used in the literature, sometimes even tuned to reproduce one specific observation \citep[see for instance discussions and references in ][]{2008ApJ...680..371W,2020MNRAS.497.2309W}.   For this work, we used the  abundances observed towards the diffuse region $\zeta$Oph (v = -15 km.s$^{-1}$) listed in \citet{2009ApJ...700.1299J} without additional depletion, especially for sulphur, whose abundance is then fairly high compared to what is usually used for dense sources.
For all sources, we assumed a 'standard' cosmic-ray ionisation rate of $10^{-17}$~s$^{-1}$. We   discuss the elemental abundances and the cosmic-ray ionisation rate further in the remainder of this section. The visual extinction is taken to be 15 to limit the effect of direct UV photons as there is no indication that these are illuminated regions.

\subsection{Comparing the model predictions with the observations}

From the observations, we have constraints on 24 species (not including CO). Among them, 5 to 11 are detected in our sources. To compare the model predictions with the observations, we have followed the same approach as in \citet{2006A&A...451..551W} by counting the number of species that are reproduced by our model at each time. We assume an agreement if a species is detected and its abundance is within a factor of 10 of the modelled abundance.  If the species is not detected, we have an upper limit. In that case, we assume that the modelled abundance has to be below the observed upper limit. Since the factor of 10 is an approximate uncertainty, we assume that the observed upper limit needs to be larger than the modelled abundance divided by 10 (meaning that the ratio between the modelled and observed upper limit needs to be smaller than 10). If we had chosen a more conservative approach, assuming for instance that the observed upper limit has to be larger than ten times the modelled abundance, then our agreement between modelled and observed abundances would be less than the results shown in this section. These indicators of agreement need to be taken with caution and are not a real indication of quality of model but are, rather, used to compare one model with the other.
For TMC-1, we have two lower limits (due to optical depth effects); thus, for these points to get an agreement, the observed lower limit needs to be smaller than the modelled abundance multiplied by ten.
If these criteria are not fulfilled, then we do not have agreement. Since we do not have any estimate on the H$_2$ column densities, we again use CO as a reference and compare the observed and modelled species abundances with respect to CO. We also ran two models for TMC-1 using the physical conditions determined by \citet{2006ApJ...653.1342L} (a temperature of 8~K for a H$_2$ density of $3\times 10^4$~cm$^{-3}$) and by \citet{2019A&A...624A.105F} (10~K for a H$_2$ density of $1.5\times 10^4$~cm$^{-3}$). These conditions are very similar to what we found in our sample. 

\begin{figure*}[htbp]
\centering
\includegraphics[width=0.48\linewidth]{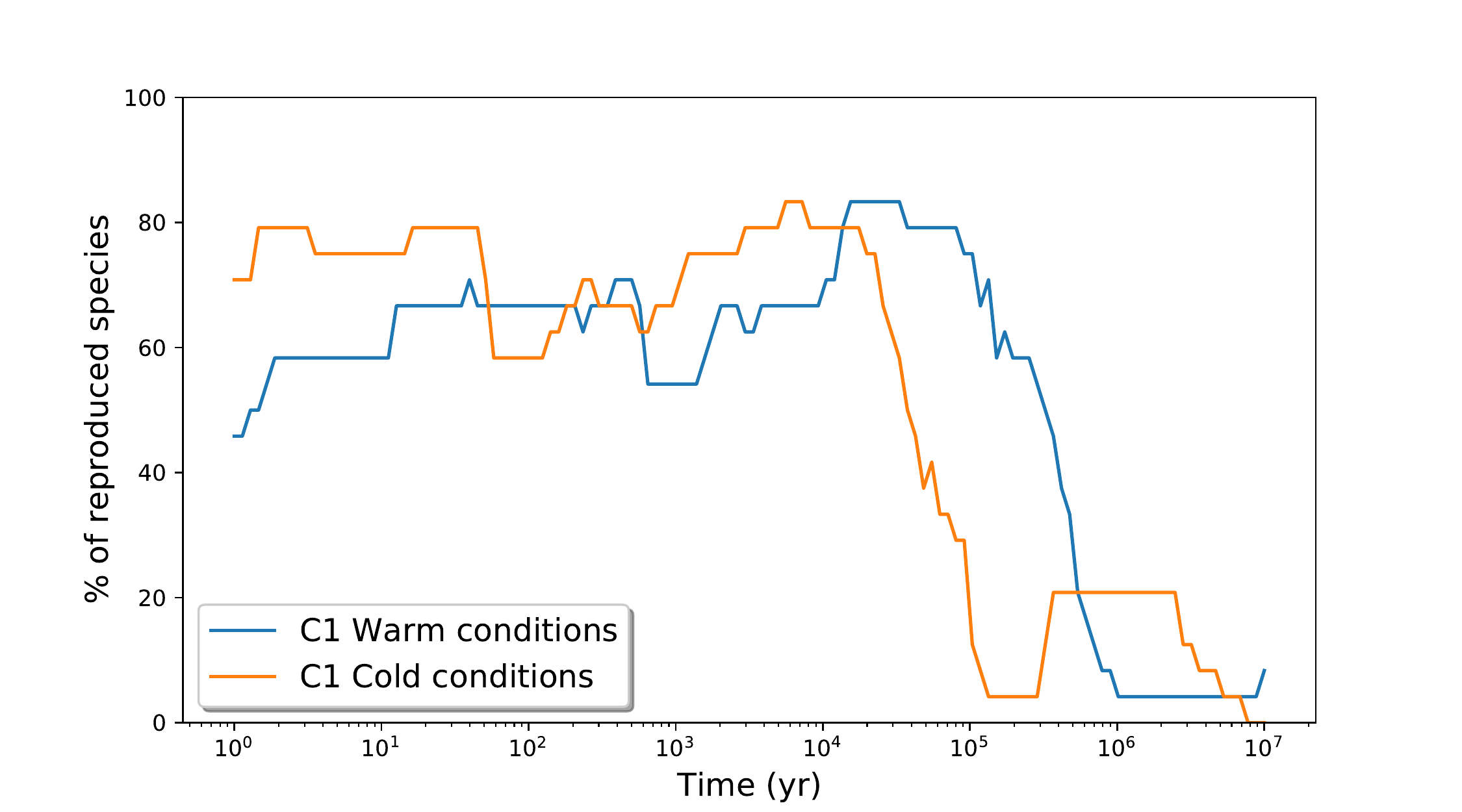}
\includegraphics[width=0.48\linewidth]{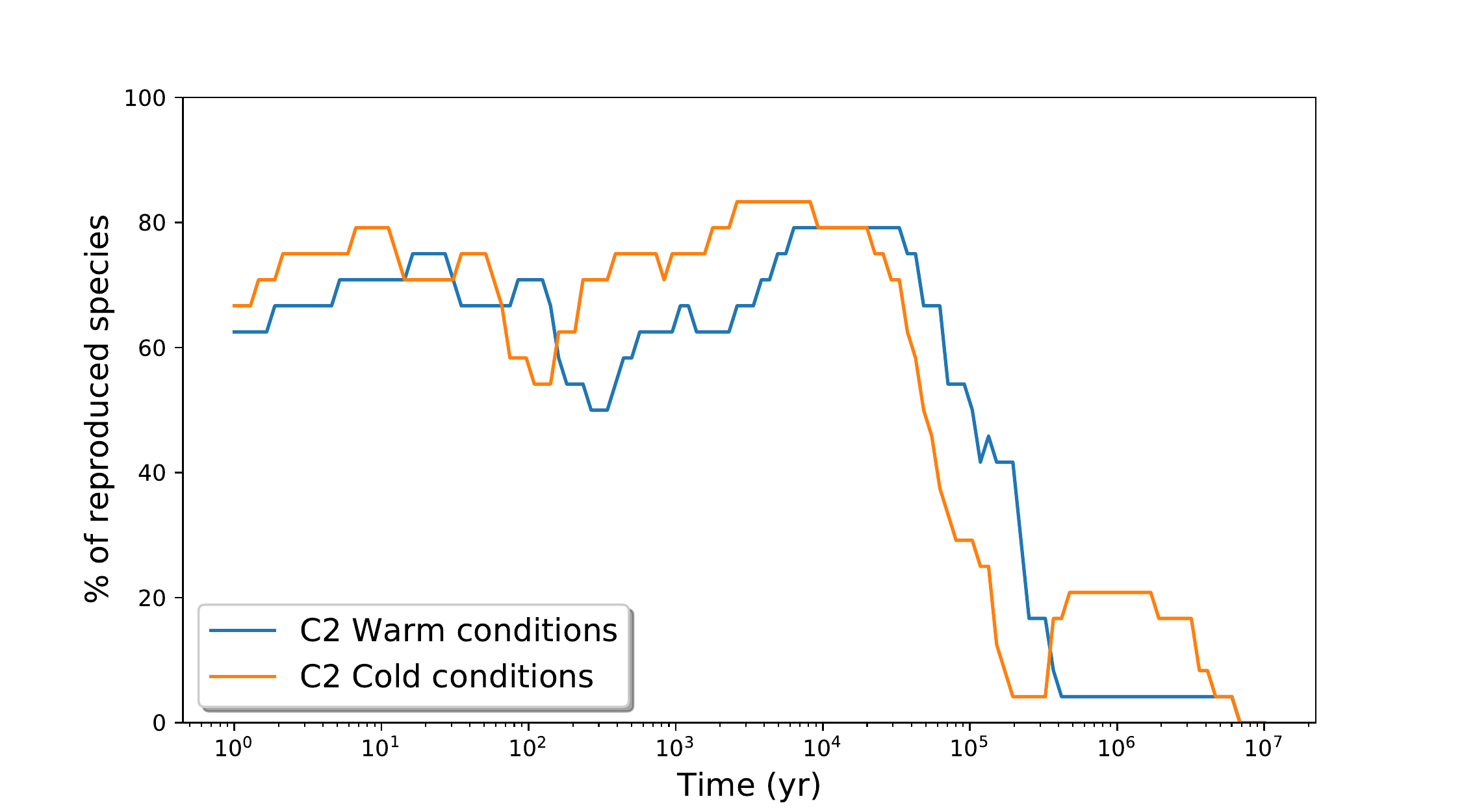}
\includegraphics[width=0.48\linewidth]{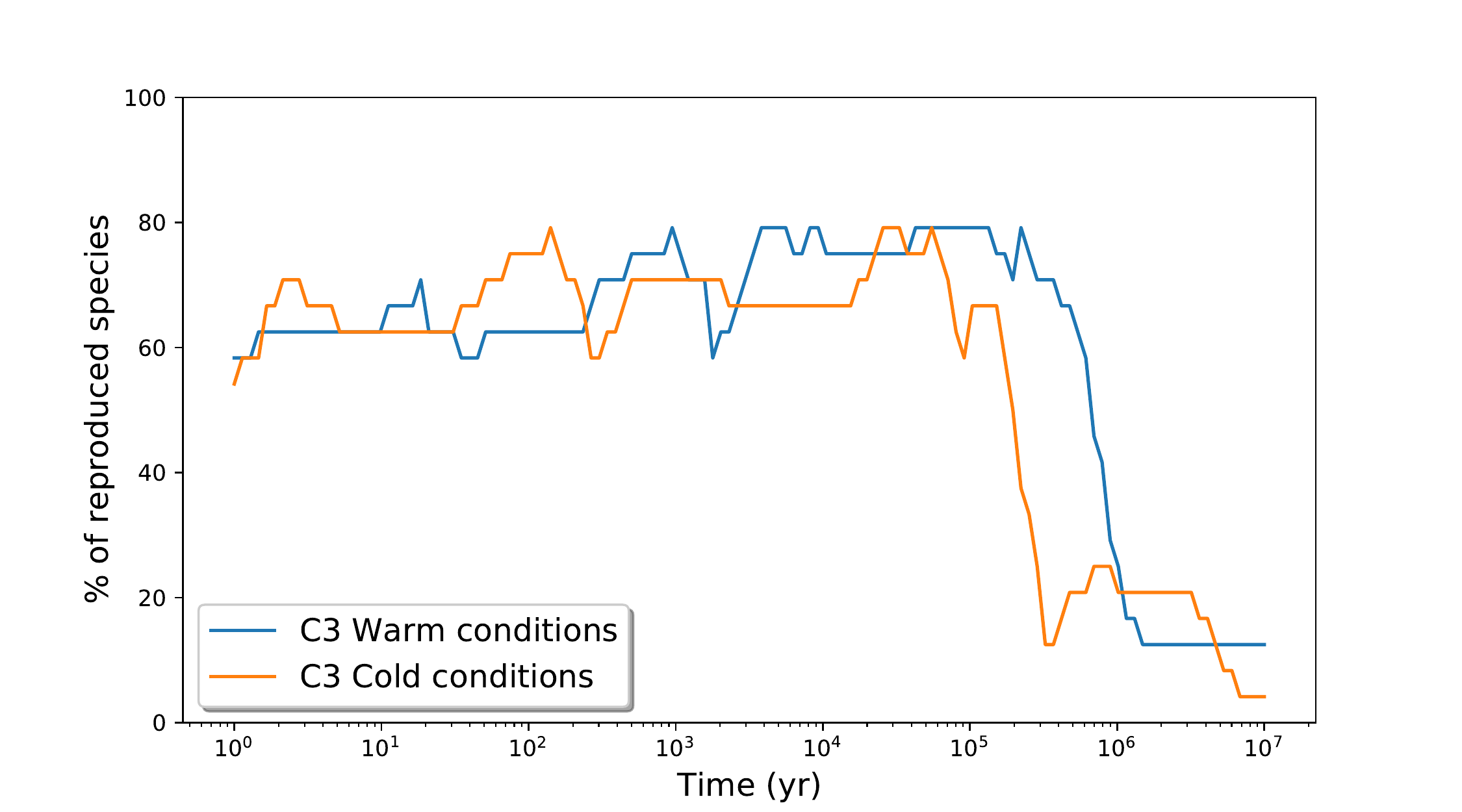}
\includegraphics[width=0.48\linewidth]{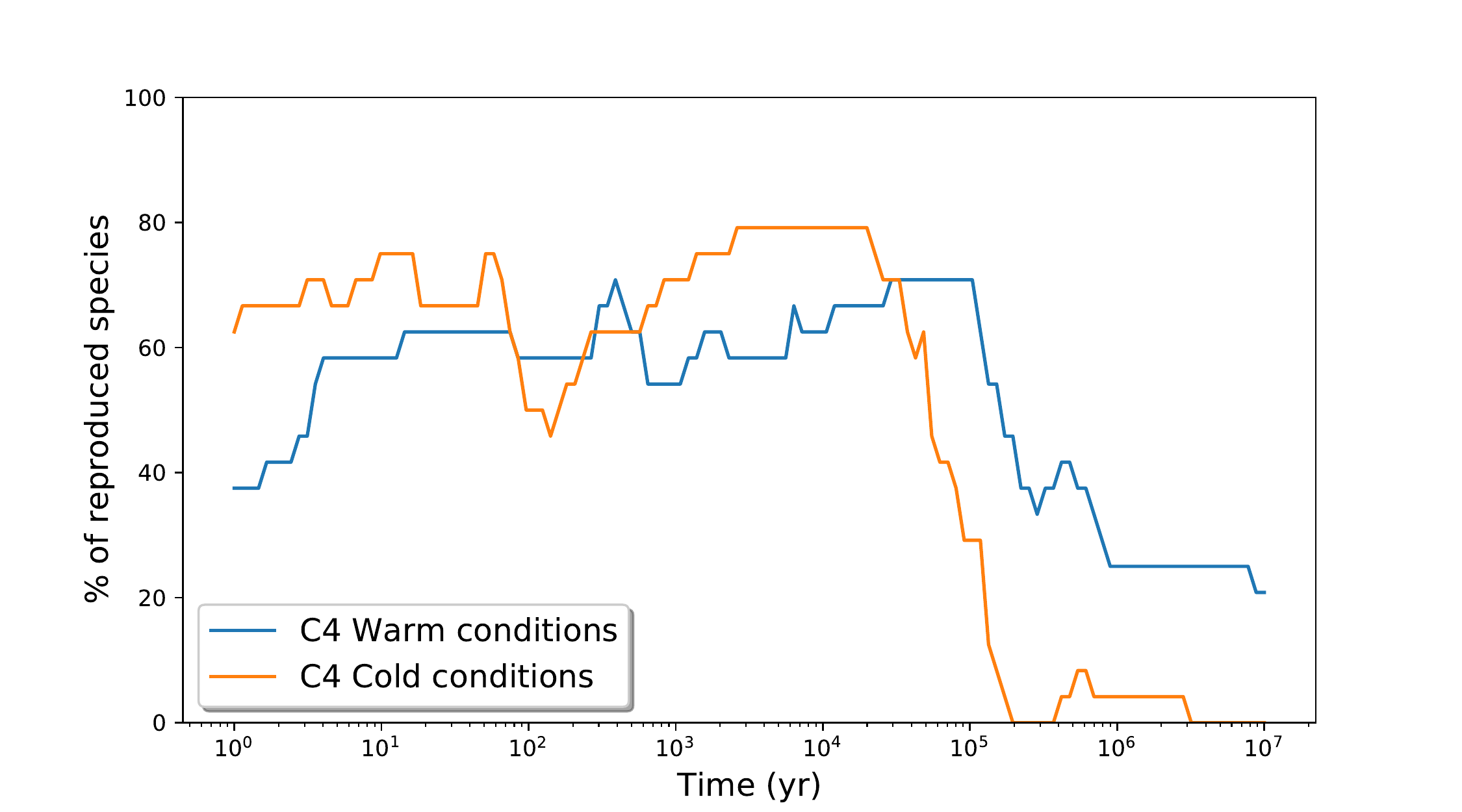}
\includegraphics[width=0.48\linewidth]{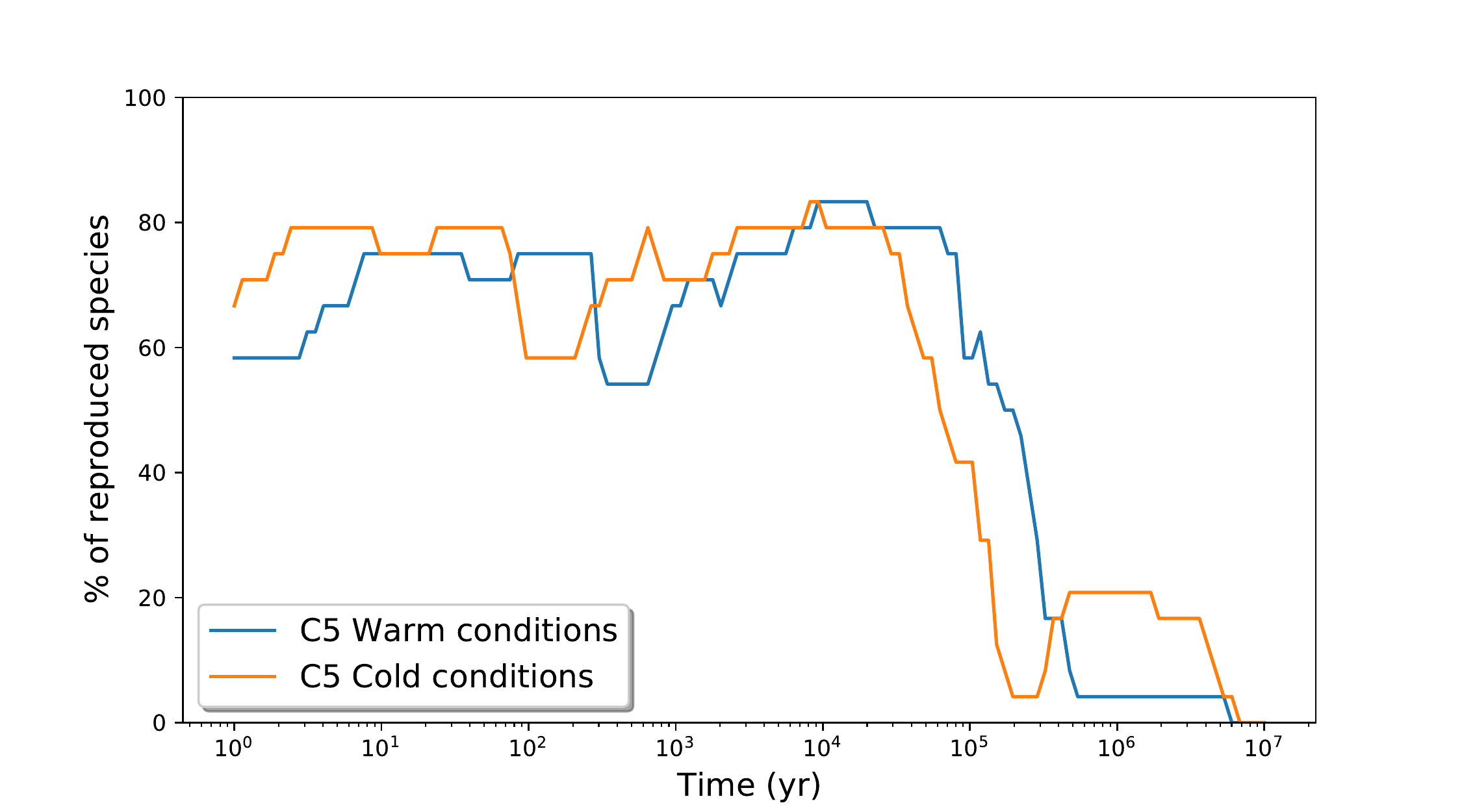}
\includegraphics[width=0.48\linewidth]{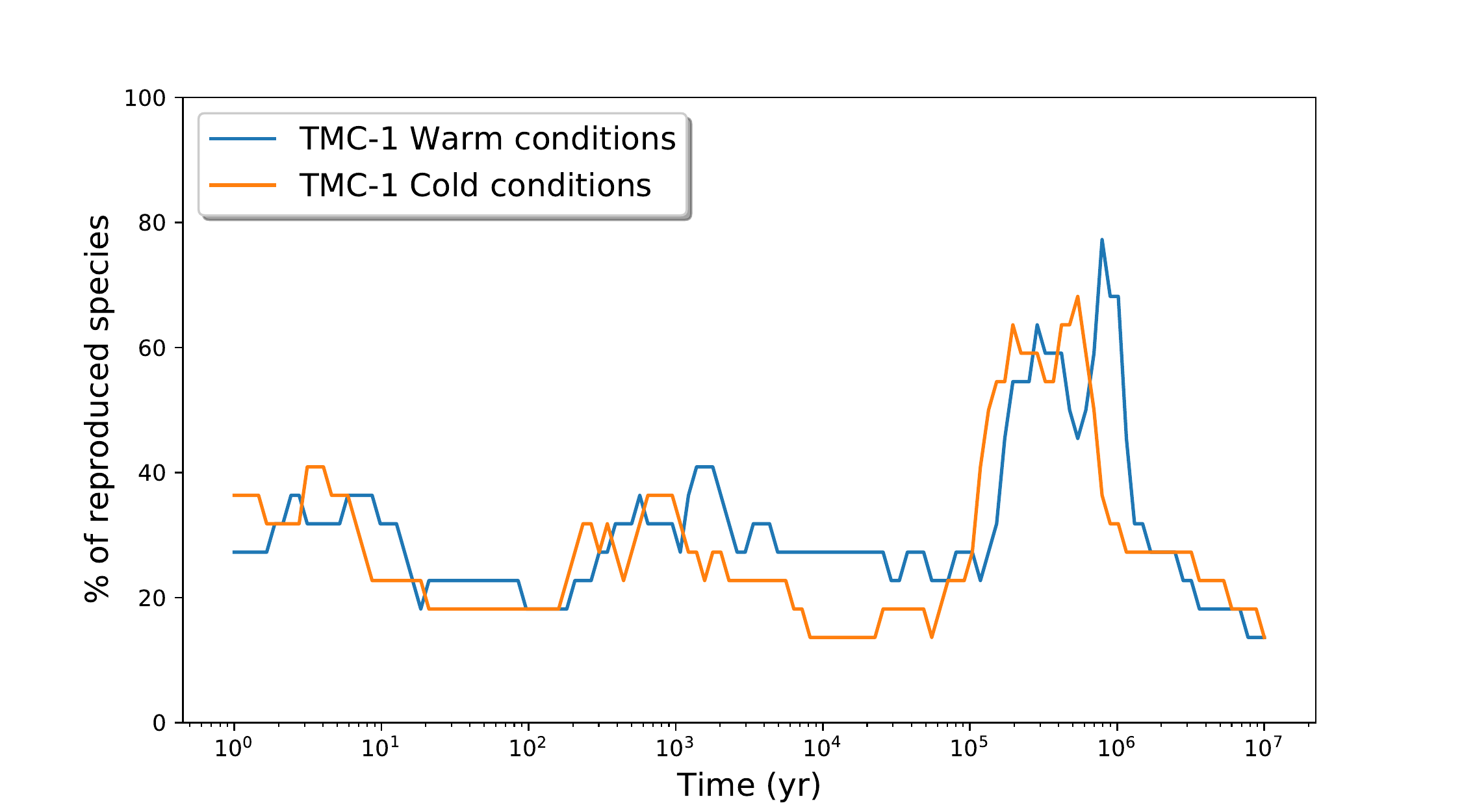}
\caption{Percentage of species reproduced by the different models for each source as a function of time. 'Warm' conditions and 'cold' conditions refer to the set of physical conditions as listed in Table~\ref{models}.\label{agreement}}
\end{figure*}

Figure~\ref{agreement} shows the percentage of reproduced species in each source for each model (see Table~\ref{models} for a summary of the physical conditions). Overall, we find a good agreement, with between 79 and 83\% of the molecules reproduced by the models. For TMC-1, the agreement is still good (68 to 77\%). The percentage of reproduced species might seem smaller than for our sources, however, it is important to keep in mind that in TMC-1, all considered molecules are detected while in our sources, we mostly have upper limits, which makes the comparison with the model less robust. In Table~\ref{agreement_table}, we report some constraints (best percentage and time) derived from the model/observation comparison. For each source, we have two different physical conditions. 
The last column of the table lists the species that are not reproduced by the model at the 'best time'. A graphic view of the model agreement is shown in Appendix~\ref{appendixD}. In Figs. \ref{modelled_observed_1} and \ref{modelled_observed_2}, we show for each model the ratio between the modelled and observed abundances at the best times. 

The first obvious result is that the strongest constraints on the time are upper limits as the agreement for all sources drops sharply after a few $10^4$ or $10^5$~yr, depending on the source. The maximum time is higher for lower-density models. Indeed, the model needs more time to achieve a similar chemical stage if the density is lower. Here, C4 is the only source for which one of the two sets of physical conditions (the 'cold' ones) seems to better agree with the observations. 
Both physical conditions for TMC-1 seem to indicate a more evolved source, which is in agreement with the fact that a lot of molecules are detected (and these are more abundant) in this source as opposed to the other ones. \\
Looking at the species that are not reproduced by the models at the best times in C1 to C5 sources, most of them are among the detected ones (in particular, CS, HCN, and CH$_3$OH), which weakens our analysis. The smaller times that we obtain (as compared to TMC-1) can be explained by the fact that molecules will build up with time slower than CO. They reach a maximum at times when CO has already started to deplete. So the model can only reproduce the small observed abundances (and the non-detections) at early times. 
For CS and HCN, whatever the time, the model overestimates the abundance. The fact that chemical models overestimate the CS abundance in cold cores has also been found in previous studies \citep{2020A&A...637A..39N}. The possible failure of the model for this molecule will be discussed in Bulut et al. (accepted in A\&A). For HCN, this may indicate that our observed abundance is underestimated. In fact, in our observations, the ratio between the three hyperfine lines of the 1-0 HCN line is not 1:5:3, as expected from the line parameters, but 2:5:3 (Fig. \ref{HCN_12K}). Anomalies in the HCN hyperfine structures are frequent in cold cores and their origin has been the subject of several studies \citep{1981A&A....97..213G,1993A&A...279..506G}. Among the causes, optical depth effects, overlap effects of 2-1 hyperfine transition to the 1-0 hyperfine intensity ratio, and the density and velocity structures of cold cores have been found to affect this ratio. All these effects would not necessarily affect the overall derived abundance, except for optical depth effects. Beam dilution may on the contrary produce an underestimation of the real abundance if the emitting structure were smaller than our beam. These could also explain the very low temperatures found from the radiative transfer analysis of the 1-0 HCN line. In addition, CS is reproduced in the TMC-1 model because the observations only give a lower limit due to opacity effects. 

\begin{table*}
\caption{Percentage of species reproduced by the models in each source, corresponding best times and species not reproduced. See text for details. 'Warm' and 'cold' refer to the set of physical conditions as listed in Table~\ref{models}. C3 Warm$^a$ and $^b$ are two different periods of time giving the same agreement.}
\begin{center}
\begin{tabular}{l|ccc}
\hline\hline
Sources & Best times (yr) & Best \% & Non reproduced species \\
\hline
C1 Warm & $(1.5-3.3)\times 10^4$ & 83\% & CS, HCN, CH$_3$OH, CH$_3$CN \\
C1 Cold & $(5.6-7.2) 10^3$ & 83\% & CS, HCN, CH$_3$OH, CH$_3$CN \\
C2 Warm & $(6.3-33)\times 10^3$  & 79\% & CS, HCN, HNC, CH$_3$CN, H$_2$CCN\\
C2 Cold & $(2.6-8.2) \times 10^3$ & 83\% & CS, HCN, HNC, CH$_3$CN\\
C3 Warm$^a$& $(3.8-9.3)\times 10^3$  & 79\% & CS, NO, HCN, CCH, CH$_3$OH  \\
C3 Warm$^a$ & $(4.2-22)\times 10^4$  & 79\% & CN, NO, HCN, CH$_3$OH, CH$_3$CN  \\
C3 Cold & $(2.5-5.5) \times 10^4$ & 79\% & CS, NO, HCN, CH$_3$OH, CH$_3$CN  \\
C4 Warm & $(2.9-10) \times 10^4$ & 71\% & CS, SO, HCN, HNC, CH$_3$OH, CH$_3$CN, HC$_3$N, C$_4$H$_2$ \\
C4 Cold & $(2.6-20) \times 10^4$ & 79\% & CS, HCN, HNC, CH$_3$OH, CH$_3$CN \\
C5 Warm& $(8.3 - 20) \times 10^4$ & 83\% & CS, HCN, CH$_3$OH, CH$_3$CN \\
C5 Cold & $(8.2-9.3) 10^4$ & 83\% & CS, HCN, CH$_3$OH, CH$_3$CN \\
TMC-1 8~K & $\sim 6\times 10^5$ & 68\% & H$_2$CO, HCN, CN, HCCNC, HC$_3$N, C$_4$H$_2$, H$_2$CS\\
TMC-1 10~K & $\sim 10^6$ & 77\% & CN, HCCNC, HC$_3$N, C$_4$H$_2$, H$_2$CS\\
\hline
\end{tabular}
\end{center}
$^a$ There are two different periods of time giving the same agreement for C3.
\label{agreement_table}
\end{table*}%

\subsection{Effect of the cosmic-ray ionisation rate and elemental abundances}

For the simulations presented above, we have considered a 'standard' cosmic-ray ionisation rate of $10^{-17}$~s$^{-1}$ \citep{1971ApJ...165...41S}. This value is highly uncertain. In TMC-1, \citet{1994Ap&SS.220..261N} estimated $\zeta$ to be between $3\times 10^{-17}$~s$^{-1}$ and $8\times 10^{-17}$~s$^{-1}$ while \citet{1997ApJ...486..862P} found a $\zeta$ of $6\times 10^{-17}$~s$^{-1}$. We used this later value and ran  all the models presented in the previous section once again. The comparison between these new models and the observations are presented in Fig.~\ref{agreement_highZ} in the appendix. The results are not significantly different and the list of species not reproduced given in Table~\ref{agreement_table} stands. The only real change is that the TMC-1 observations are better reproduced by the colder and denser model.

Another model parameter that is not well constrained is the elemental abundances. Atoms are known to deplete in the diffuse interstellar medium, a depletion that is not fully understood or quantified at this stage \citep{1998ApJ...499..267T,2010ApJ...710.1009W,2009ApJ...700.1299J}. As the initial conditions for our models, we chose to use the atomic abundances observed in the diffuse medium, while many chemical studies assume more depleted values that are usually refereed as 'low metal' conditions \citep{1982ApJS...48..321G}. Among the elements, sulphur is probably the one whose elemental abundance is the most varied in models in order to reproduce the low observed abundances of S-bearing species in cold sources \citep[see for instance][]{2017MNRAS.469..435V}. In our previous models, we have used the cosmic abundance of sulphur as its elemental abundance. To test the effect of this parameter, we have run the same models with a ten times smaller abundance (i.e. $3\times 10^{-6}$ with respect to H$_2$). The comparison of these models with the observations is shown in Fig.~\ref{agreement_depS} in the appendix. Overall, the results are not much changed. The agreement is similar or slightly worse with our sources. SO and CS are not better reproduced. The comparison with TMC-1 is slightly better but the best times do not change.\\
The C/O elemental ratio is very often varied (from smaller than 1 to larger than 1) to reproduce the large abundances of carbon bearing species \citep{1986MNRAS.222..689H,1995ApJ...443..664B,2011A&A...530A..61H}. We tested the effect of increasing this ratio to 1.2 by decreasing the oxygen elemental abundance. As expected, the observations in TMC-1 are better reproduced (see Fig.~\ref{agreement_C_O} in the appendix) at times between $1\times 10^5$ and $4\times 10^5$~yr with both physical conditions. The agreement is less good for the other sources. 

\section{Conclusion}

In this work, we observe two large spectral bands (71-116~GHz and 126-182~GHz) in five cold clumps of the  Planck Early Core catalogue, located in three different large molecular clouds (California, Perseus, and Taurus molecular clouds). Within these sources, only a small number of molecules were detected (H$_2$CO, CS, SO, NO, HNO, HCO$^+$, HCN, HNC, CN, CCH, CH$_3$OH, and CO), as compared to the long list of targeted molecules seen in cold cores such as TMC-1(CP). A study of the excitation conditions of molecules such as CO, HCN, SO, and CS has shown incompatibilities: HCN (1-0) observed lines can only be reproduced by very low temperature (of 5~K) while the $^{12}$CO and $^{13}$CO (1-0) lines indicate higher temperatures and lower densities. The $^{12}$CO and $^{13}$CO (1-0) lines are optically thick, contrary to the other observed lines. The $^{12}$CO and $^{13}$CO (1-0) lines very likely originate from the external warmer and less dense layers of the clumps while the other lines probe inner denser and colder regions. In addition, there may have some substructures within the antenna beam that could  explain why we have systematically larger CO fluxes as compared to \citet{2013ApJS..209...37M}, who mapped the region with a larger beam. \\
Based on two different assumed physical conditions (one constrained by HCN and the other constrained by CS and CO), we computed the species column densities. Overall, the computed column densities are not very sensitive to the assumed physical conditions. Assuming that the CO abundance (computed from optically thin C$^{18}$O) is the same in all sources (which may not be the case), the molecular abundances are spread over almost one order of magnitude in the different sources. 
Compared to TMC-1(CP), these clumps appear quite poor both in abundance levels and in molecular diversity. We did not detect any carbon chain molecules. \\
 With a full gas-grain model and the physical conditions derived in each source, we are able to reproduce between 79\% and 83\% of the observed species (including the upper limits). This is slightly better than in TMC-1, we note, however: 1) we do not reproduce CS and HCN, which are both overestimated by the models; and 2) we have mostly upper limits, a condition that provides fewer constraints for the model. The 'best' times for our sources seem to be smaller than for TMC-1, indicating that our sources may be less evolved; this allows us to explain the smaller abundances and the numerous non-detections. Considering a cosmic-ray ionisation rate that is larger than the standard $10^{-17}$~s$^{-1}$ one does not  significantly impact our results. If we deplete the sulphur 
 as compared to cosmic values the agreement between the model and the CS observed abundance is not improved. If we increase the C/O elemental ratio, we improve the model agreement with TMC-1, but worsening it for the sources presented here.


\begin{acknowledgements}
The authors acknowledge the CNRS program "Physique et Chimie du Milieu Interstellaire" (PCMI) co-funded by the Centre National d'Etudes Spatiales (CNES). 
M. Ruaud's research was supported by an appointment to the NASA Postdoctoral Program at NASA Ames Research Center, administered by Universities Space Research Association under contract with NASA. 
\end{acknowledgements}





   \bibliographystyle{aa} 
\bibliography{biblio} 
%


\appendix

\section{Line properties in the five sources.}\label{appendixA}

\begin{table*}
\caption{List of detected transitions.}
\begin{center}
\begin{tabular}{llllll|l}
\hline
\hline
Molecule & Frequency (MHz) & Eup (K) & Gup & Aij (s$^{-1}$) & transition  & n$_{\rm c}$ (cm$^{-3}$) \\
\hline
H$_2$CO    &       72837.948  &    3.5  &  3 & 8.15e-06    &      1 0 1 -- 0 0 0   & $2\times 10^5$  \\        
H$_2$CO     &     140839.502   &   21.9  & 15&  5.30e-05      &    2 1 2 -- 1 1 1 & $7\times 10^5$\\ 
H$_2$CO      &    145602.949   &    10.5 &   5 & 7.81e-05     &     2 0 2 -- 1 0 1 & $8\times 10^5$\\
H$_2$CO      &    150498.334   &    22.6  & 15 & 6.47e-05    &      2 1 1 -- 1 1 0 & $1\times 10^6$\\ 
\hline
CS         &    97980.950   &   7.1  &  5 & 1.69e-05      &        2 -- 1             & $4\times 10^5$    \\
CS         &   146969.033   &   14.1 &   7 & 6.11e-05        &      3 -- 2 & $1\times 10^6$\\
C$^{34}$S &          96412.940  &   6.9  &  5  &1.61e-05    &          2 -- 1 & - \\ 
\hline
 SO      &       99299.870  &    9.2  &  7 & 1.15e-05    &        2 3 -- 1 2   & $2\times 10^5$    \\        
SO     &       138178.600  &   15.9  &  9 & 3.23e-05   &         3 4 -- 2 3& $4\times 10^5$\\
\hline
NO     &       150176.480   &    7.2  &  6 & 3.31e-07    &    2-1 2 3 -- 1 1 1 2& $2\times 10^4$ \\
NO     &       150198.760   &    7.2  &  4  &1.84e-07     &   2-1 2 2 -- 1 1 1 1& $2\times 10^4$\\
NO     &       150218.730  &     7.2  &  4  &1.47e-07      &  2-1 2 2 -- 1 1 1 2& $1\times 10^4$\\
NO     &       150225.660  &     7.2  &  2 & 2.94e-07       & 2-1 2 1 -- 1 1 1 1 & $4\times 10^4$\\
NO    &        150439.120  &     7.2  &  4&  1.48e-07 &       2 1 2 2 -- 1-1 1 2& $1\times 10^4$\\
NO     &       150546.520 &    7.2  &  6  &3.33e-07  &      2 1 2 3 -- 1-1 1 2 & $2\times 10^4$\\
NO     &       150580.560  &     7.2  &  2 & 2.96e-07  &      2 1 2 1 -- 1-1 1 1 & $4\times 10^4$\\
NO     &       150644.340  &     7.2  &  4 & 1.85e-07   &     2 1 2 2 -- 1-1 1 1 & $2\times 10^4$\\
\hline
HNO  &          81477.490   &     3.9  &  5 & 2.23e-06      &  1 0 1 2 -- 0 0 0 1 & -   \\
HNO  &          81477.490  &    3.9  &  3 & 2.23e-06   &     1 0 1 1 -- 0 0 0 1   & -   \\
HNO    &        81477.490  &    3.9  &  1 & 2.23e-06    &    1 0 1 0 -- 0 0 0 1  & -    \\
\hline
HCO$^+$ &   89188.523   &   4.3  &  3 & 4.16e-05   &       1 0 0 -- 0 0 0    & $2\times 10^5$    \\     
HCO$^+$ & 178375.010  &   12.8  &  5 & 4.00e-04     &     2 0 0 -- 1 0 0 & $9\times 10^5$\\
H$^{13}$CO$^+$&       86754.288  &     4.2 &   3 & 3.83e-05   &           1 -- 0   & - \\
\hline
HCN    &        88630.416   &  4.3  & 3 & 2.43e-05      &      1 1 -- 0 1  & $1\times 10^6$   \\          
HCN    &        88631.847   &    4.3  &  5 & 2.43e-05    &        1 2 -- 0 1 & $1\times 10^6$   \\           
HCN    &        88633.936   &    4.3  &  1 & 2.43e-05      &      1 0 -- 0 1  & $1\times 10^6$    \\   
\hline     
HNC   &      90663.593   &     4.4  &  3 & 2.69e-05     &         1 -- 0  & $3\times 10^5$  \\
\hline
CN & 113144.190   &    5.4  &  2 & 1.05e-05   &  1 1 2 1 2 -- 0 1 2  3 2 & $2\times 10^6$  \\
CN &  113191.325  &     5.4 &   4 & 6.68e-06  &  1 1 2 1 2 -- 0 3 2 3 2 & $3\times 10^6$\\
CN &  113488.142  &     5.4  &  4  &6.73e-06  &     1 3 2 1 2 -- 0 3 2 1 2 & $3\times 10^6$\\
CN &  113490.985  &     5.4  &  6 & 1.19e-05   &   1 3 2 1 2 -- 0 5 2 3 2 & $3\times 10^6$\\
\hline
CCH     &       87316.925  &     4.2  &  5 & 1.65e-06   &       1 2 2 -- 0 1 1 & $1\times 10^5$ \\
CCH      &      87402.004  &    4.2 &   3 & 1.38e-06   &       1 1 1 -- 0 1 1 & $1\times 10^5$ \\
\hline
CH$_3$OH   &       96739.358  &    12.5  &  5 & 2.56e-06   &     2-1   0 -- 1-1   0 & $3\times 10^4$\\ 
CH$_3$OH    &      96741.371  &   7.0 &   5 & 3.41e-06    &    2 0 + 0 -- 1 0 + 0  & $3\times 10^4$\\
CH$_3$OH   &      145097.435  &  19.5  &  7 & 1.10e-05   &     3-1   0 -- 2-1   0 & $3\times 10^5$   \\
CH$_3$OH    &     145103.185 &  13.9  &  7 & 1.23e-05   &     3 0 + 0 -- 2 0 + 0  & $1\times 10^5$ \\
\hline
C$^{18}$O       &  109782.173  &     5.3 &   3 & 6.27e-08   &           1 -- 0 & $2\times 10^3$\\
$^{13}$CO   &     110201.354 &     5.3  &  3&  6.33e-08    &          1 -- 0 & $2\times 10^3$ \\
C$^{17}$O$^a$   &      112359.284  &     5.4 &   3 & 6.70e-08    &          1 -- 0 & $2\times 10^3$\\
CO       &     115271.202  &     5.5   & 3&  7.20e-08      &        1 -- 0 & $2\times 10^3$ \\
\hline
\end{tabular}
\end{center}
 $^a$ C$^{17}$O has a hyperfine structure with three lines at 112358.7770, 112358.9820, and 112360.0070 MHz according to the CMDS database. Both the JPL and LAMDA databases (which our CASSIS analysis is based on)  assume only one component at 112359.284 MHz. In our observations, the lines at 112358.7770 MHz and 112358.9820 MHz are blended.
\label{detect_lines}
\end{table*}%

\begin{table*}
\caption{Transitions used to determine upper limit for molecules detected in none of the sources.}
\begin{center}
\begin{tabular}{lllllll}
\hline
\hline
Molecule & Frequency (MHz) & Eup (K) & Gup & Aij (s$^{-1}$) & transition & n$_{\rm c}$ (cm$^{-3}$) \\
\hline
c-C$_3$H$_2$ & 82093.559 & 6.4 & 5 & $2.07\times 10^{-5}$ & 2 0 2 0 -- 1 1 1 0 & $1\times 10^6$ \\
l-C$_3$H$_2$ & 83165.345  &   10.0  &  9 & $5\times 10^{-5}$     &     4 0 4 -- 3 0 3 & -  \\
CH$_3$CN &73590.218 & 8.8  & 18 & $3.17\times 10^{-5}$   &         4 0 -- 3 0 & $2\times 10^5$ \\
C$_3$N & 79150.986 &  17.1 &  16 & $1.29\times 10^{-5}$   &       8 9 8 -- 7 8 7 & - \\
c-C$_3$H & 121273.585  &    6.5 &   3 & $2.05\times 10^{-5}$  &    2 1 1 2 1 -- 1 1 0 1 0 & - \\
l-C$_5$H$_2$ & 73447.066 &   30.0 &  33 & $7.78\times 10^{-5}$  &       16 016 -- 15 015  & -  \\
HNC$_3$ &  74692.276  &  16.1 &  19 &  $7.33\times 10^{-5}$    &        8 9 -- 7 8  & -\\
HCCNC & 79484.131 &  17.2 &  19 & $2.36\times 10^{-5}$     &       8 9 -- 7 8 &- \\
H$_2$CCN & 80479.940 &   9.7 &  90 & $3.30\times 10^{-5}$      &  4 0 4 5 -- 3 0 3 4 &- \\
HC$_3$N &  72783.822  &   15.7 &  51 & $2.93\times 10^{-5}$   &           8 -- 7 & $5\times 10^5$  \\
l-C$_4$H$_2$ & 80383.887  & 19.3 &  19 & $4.81\times 10^{-5}$    &      9 0 9 -- 8 0 8  & -\\
N$_2$H$^+$ & 93176.130  &   4.5  &  3 & $3.63\times 10^{-5}$    &        1 0 -- 0 1   & $1\times 10^5$  \\
H$_2$CS & 103040.220 &   9.9  &  7&  $1.48\times 10^{-5}$   &       3 0 3 -- 2 0 2 & $1\times 10^5$  \\
\hline
\end{tabular}
\end{center}
\label{lineparam_nondetected}
\end{table*}%

\begin{table*}
\caption{Results of line fitting in C1.}
\begin{center}
\begin{tabular}{ll|llllllllll|}
\hline
\hline
Molecule & Frequency & \multicolumn{5}{|c}{C1}  \\
 & MHz &  W    &               v$_{\rm lsr}$     &       FWHM      &       T$_{\rm peak}$  & rms \\
 & & K.km.s$^{-1}$ & km.s$^{-1}$ & km.s$^{-1}$ & K & K \\
\hline
H$_2$CO    &       72837.948  &    0.35$\pm$0.02  & -19.04$\pm$0.03 & 1.09$\pm$0.06&  0.30 & 0.02   \\        
H$_2$CO     &     140839.502   &  0.24$\pm$0.09  &   -19.22$\pm$0.02 &   0.99$\pm$0.04 &  0.22 & 0.01   \\ 
H$_2$CO      &    145602.949   &   0.08$\pm$0.02 &   -19.29$\pm$0.08 &    0.75$\pm$0.23 &  0.10& 0.02  \\
H$_2$CO      &    150498.334   &  0.13$\pm$0.01   &  -19.04$\pm$0.04 &   1.14$\pm$0.09 &  0.10 & 0.01   \\ 
\hline
CS         &    97980.950   &     0.59$\pm$0.01    &  -19.08$\pm$0.01 &  1.16$\pm$0.03 & 0.48 & 0.02     \\
CS         &   146969.033   &  0.19$\pm$0.01  &  -19.04$\pm$0.03 &     1.05$\pm$0.06 &  0.17& 0.01  \\
C$^{34}$S &          96412.940  &  $\le 0.06$ & -&- & -&  0.01 \\ 
\hline
 SO      &       99299.870  &   0.54$\pm$0.01  &  -19.15$\pm$0.01 & 0.99$\pm$0.02 &  0.51 & 0.01    \\        
SO     &       138178.600  & 0.18$\pm$0.01   &  -19.23$\pm$0.03 &     0.86$\pm$0.06 & 0.19 & 0.01   \\
\hline
NO     &       150176.480   &  $\le 0.06$ & -&- & -& 0.01  \\
NO     &       150198.760   &  $\le 0.06$ & -&- & -&   0.01 \\
NO     &       150218.730  &  $\le 0.06$ & -&- & -&   0.01\\
NO     &       150225.660  &  $\le 0.06$ & -&- & -&   0.01 \\
NO    &        150439.120  & $\le 0.06$ & -&- & -&   0.01 \\
NO     &       150546.520 &  $\le 0.06$ & -&- & -&   0.01 \\
NO     &       150580.560  & $\le 0.06$ & -&- & -&   0.01\\
NO     &       150644.340  &  $\le 0.06$ & -&- & -&   0.01\\
\hline
HNO  &          81477.490   &    $\le 0.02$ & -&- &  -& 0.007  \\
\hline
HCO$^+$ &   89188.523   &   0.66$\pm$0.01   &   -19.23$\pm$0.01 &    1.29$\pm$0.02 & 0.48 & 0.01   \\     
HCO$^+$ & 178375.010  & 0.27$\pm$0.03     &   -19.30$\pm$0.07 &     1.34 $\pm$0.21 & 0.19 &0.04  \\
H$^{13}$CO$^+$&       86754.288  &   $\le 0.06$ & -&- & -&   0.01 \\
\hline
HCN    &        88630.416   & 0.177$\pm$0.007    &   -19.08$\pm$0.02 &     1.12$\pm$0.04 & 0.15 & 0.007   \\          
HCN    &        88631.847   &  0.298$\pm$0.007    &   -19.09$\pm$0.06 &    1.24$\pm$0.03 & 0.22 & 0.007   \\           
HCN    &        88633.936   & 0.119$\pm$0.007     &   -19.01$\pm$0.22 &    1.22$\pm$0.08 &  0.09 & 0.007   \\   
\hline     
HNC   &      90663.593   & 0.23$\pm$0.01  &   -18.96$\pm$0.03 &   1.39$\pm$0.08 &  0.16 & 0.01  \\
\hline
CN & 113144.190   &  $\le 0.06$ & -&- &  -&   0.01   \\
CN &  113191.325  &  $\le 0.06$ & -&- &  -& 0.01 \\
CN &  113488.142  &  $\le 0.06$ & -&- & -&   0.01  \\
CN &  113490.985  & 0.06$\pm$0.01 & -18.98$\pm$0.05 & 0.76$\pm$0.12 & 0.07 & 0.01\\
\hline
CCH     &       87316.925  &  0.05$\pm$0.01 & -18.56$\pm$0.11&  1.06$\pm$0.22& 0.04 & 0.01  \\
CCH      &      87402.004  &   $\le 0.06$ & -&- &  - &   0.01  \\
\hline
CH$_3$OH   &       96739.358  &  0.11$\pm$0.03    &  -19.15$\pm$0.16 &     1.44$\pm$0.62 &  0.07 & 0.02   \\ 
CH$_3$OH    &      96741.371  &  0.10$\pm$0.02     &   -19.09$\pm$0.07 &   0.79$\pm$0.14 & 0.12 & 0.02  \\
CH$_3$OH   &      145097.435  &   0.052$\pm$0.009& -19.16$\pm$0.11&   1.19$\pm$0.19& 0.04  &   0.01 \\
CH$_3$OH    &     145103.185 &   $\le 0.06$ & -&- & - &  0.01 \\
\hline
C$^{18}$O       &  109782.173  &  0.899$\pm$0.006    &   -19.181$\pm$0.003 &     0.938$\pm$0.007 &  0.90 & 0.007    \\
$^{13}$CO   &     110201.354 &  6.969$\pm$0.002    &   -19.09$\pm$0.000 &     1.254$\pm$0.001 &   5.22 & 0.008  \\
C$^{17}$O$^a$   &      112359.284  &    0.182$\pm$0.008     &  -18.21$\pm$0.02 &    1.11$\pm$0.05 &  0.15 & 0.02  \\
 &  &    0.071$\pm$0.007 &   -21.09$\pm$0.03 &     0.77$\pm$0.07 & 0.08 & 0.02  \\
CO       &     115271.202  &  17.34$\pm$0.06   & -19.081$\pm$0.003 &    1.803$\pm$0.007 &   9.04 & 0.06   \\
\hline
\end{tabular}
\end{center}
 $^a$ C$^{17}$O has a hyperfine structure; see comment in Table~\ref{detect_lines}.\\
W is the integrated intensity, v$_{\rm lsr}$  the central velocity, FWHM  the line width, and T$_{\rm peak}$ the peak intensity of a gaussian fitting of each line. The rms is the noise level of each line.
\label{lineC1}
\end{table*}%

\begin{table*}
\caption{Results of the line fitting in C2.}
\begin{center}
\begin{tabular}{ll|llllllllll|}
\hline
\hline
Molecule & Frequency & \multicolumn{5}{|c}{C2}  \\
 & MHz &  W    &               v$_{\rm lsr}$     &       FWHM      &       T$_{\rm peak}$  & rms \\
  & & K.km.s$^{-1}$ & km.s$^{-1}$ & km.s$^{-1}$ & K & K \\
\hline
 H$_2$CO    &       72837.948  &  $\le 0.06$ & -&- & - & 0.02 \\        
H$_2$CO     &     140839.502   &  $\le 0.03$ & -&- &- & 0.01 \\ 
H$_2$CO      &    145602.949   &  $\le 0.06$ & -&- &- & 0.01 \\
H$_2$CO      &    150498.334    & $\le 0.06$ & -&- &- & 0.01 \\ 
\hline
CS         &    97980.950     &  0.11$\pm$0.01 &   -7.27$\pm$0.04 &  0.79$\pm$0.12 &  0.13 & 0.01   \\
CS         &   146969.033   &   $\le 0.03$ & -&- & - & 0.01  \\
C$^{34}$S &          96412.940  &  $\le 0.06$ & -&- & -&  0.02  \\ 
\hline
 SO      &       99299.870   &  0.08$\pm$0.01 &   -7.32$\pm$0.04 &   0.75$\pm$0.12 &  0.10 & 0.012 &  \\        
SO     &       138178.600  &  $\le 0.06$ & -&- & - &  0.01  \\
\hline
NO     &       150176.480   &  $\le 0.06$ & -&- & -&  0.02\\
NO     &       150198.760   &  $\le 0.03$ & -&- & -&   0.01 \\
NO     &       150218.730  &  $\le 0.03$ & -&- & -&   0.01 \\
NO     &       150225.660  &  $\le 0.06$ & -&- & -&  0.02 \\
NO    &        150439.120  &  $\le 0.06$ & -&- & -& 0.02 \\
NO     &       150546.520 &  $\le 0.06$ & -&- & -&  0.02 \\
NO     &       150580.560  &  $\le 0.06$ & -&- & -&  0.02 \\
NO     &       150644.340  &  $\le 0.06$ & -&- & -&  0.02\\
\hline
HNO  &          81477.490   &    $\le 0.03$ & -&- &  -&   0.009  \\
\hline
HCO$^+$ &   89188.523   &   0.18$\pm$0.01 &    -7.32$\pm$0.03 &     1.02$\pm$0.08 &  0.16 & 0.01  \\     
HCO$^+$ & 178375.010  &   $\le 0.09$ & -&- & - & 0.03  \\
H$^{13}$CO$^+$&       86754.288  &   $\le 0.03$ & -&- & -&   0.01 \\
\hline
HCN    &        88630.416   &  0.0496$\pm$0.007 &    -7.02$\pm$0.10 &    1.58$\pm$0.23 & 0.029 & 0.006  \\          
HCN    &        88631.847   &  0.0558$\pm$0.005 &   -7.35$\pm$0.06 &    0.89$\pm$0.10 & 0.059 & 0.006   \\           
HCN    &        88633.936   &   0.059$\pm$0.01 & -6.0$\pm$0.4 &  3.75$\pm$0.58  & 0.015 & 0.006  \\   
\hline     
HNC   &      90663.593   &  0.057$\pm$0.010 &    -7.30$\pm$0.10 &    1.17$\pm$0.29 & 0.0461 & 0.009   \\
\hline
CN & 113144.190   &  $\le 0.03$ & -&- &  -&  0.01 \\
CN &  113191.325  &  $\le 0.03$ & -&- &  -&   0.01  \\
CN &  113488.142  &   $\le 0.03$ & -&- & -&   0.01 \\
CN &  113490.985  & $\le 0.03$ & -&- & -&   0.01 \\
\hline
CCH     &       87316.925  &    $\le 0.02$ & -&- &  -& 0.008  \\
CCH      &      87402.004  &   $\le 0.02$ & -&- &  - &  0.008    \\
\hline
CH$_3$OH   &       96739.358  &     $\le 0.06$ & -&- & - & 0.02  \\ 
CH$_3$OH    &      96741.371  &     $\le 0.06$ & -&- & - & 0.02 \\
CH$_3$OH   &      145097.435  &   $\le 0.03$ & -&- &-  & 0.01 \\
CH$_3$OH    &     145103.185 &   $\le 0.03$ & -&- & - &  0.01 \\
\hline
C$^{18}$O       &  109782.173  &   0.898$\pm$0.005 &  -7.373$\pm$0.002 &  0.696$\pm$0.005 &  1.212 & 0.007  \\
$^{13}$CO   &     110201.354 &   4.31$\pm$0.01 &   -7.265$\pm$0.002 &   1.220$\pm$0.004 &  3.32 & 0.01  \\
C$^{17}$O$^a$   &      112359.284  &    0.201$\pm$0.008 &  -6.39$\pm$0.02 &  0.96$\pm$0.04 & 0.19 & 0.01 \\
 & &   0.103$\pm$0.007 &  -9.37$\pm$0.02 &   0.63$\pm$0.05 &  0.15 & 0.01 \\
CO       &     115271.202  &     8.88$\pm$0.44  &  -7.25$\pm$0.05 &   2.08$\pm$0.11 & 4.0 & 0.4  \\
\hline
\end{tabular}
\end{center}
 $^a$ C$^{17}$O has a hyperfine structure; see comment in Table~\ref{detect_lines}.\\
W is the integrated intensity, v$_{\rm lsr}$  the central velocity, FWHM  the line width, and T$_{\rm peak}$ the peak intensity of a gaussian fitting of each line. The rms is the noise level of each line.
\label{lineC2}
\end{table*}%

\begin{table*}
\caption{Results of the line fitting in C3.}
\begin{center}
\begin{tabular}{ll|lllll}
\hline
\hline
Molecule & Frequency & \multicolumn{5}{|c}{C3}  \\
 & MHz &  W    &               v$_{\rm lsr}$     &       FWHM      &       T$_{\rm peak}$  & rms \\
   & & K.km.s$^{-1}$ & km.s$^{-1}$ & km.s$^{-1}$ & K & K \\
\hline
 H$_2$CO    &       72837.948  &  0.378$\pm$0.02  &    -2.12$\pm$0.02 &   0.912$\pm$0.005 & 0.39 & 0.02    \\        
H$_2$CO     &     140839.502   & 0.200$\pm$0.007 &   -2.12$\pm$0.01 &  0.73$\pm$0.03 & 0.25 & 0.01 \\ 
H$_2$CO      &    145602.949   &   0.114$\pm$0.007 &   -2.17$\pm$0.01 &    0.57$\pm$0.04 &0.19 & 0.01 \\
H$_2$CO      &    150498.334   & 0.160$\pm$0.008  &   -2.14$\pm$0.01 &    0.72$\pm$0.04 &  0.21 & 0.01\\ 
\hline
CS         &    97980.950   &     0.270$\pm$0.009  &    -2.01$\pm$0.02 &     1.07$\pm$0.04 &0.23 & 0.01        \\
CS         &   146969.033   & 0.08$\pm$0.01 &  -2.05$\pm$0.06 &   1.05$\pm$0.14 & 0.08 & 0.01 \\
C$^{34}$S &          96412.940  &   $\le 0.06$ & -&- & - & 0.02 \\ 
\hline
 SO      &       99299.870  &   0.57$\pm$0.01  &   -2.210$\pm$0.005 &  0.63$\pm$0.01 & 0.85 & 0.01 \\        
SO     &       138178.600  & 0.16$\pm$0.01  & 27293.807$\pm$0.007 &   0.43$\pm$0.03 & 0.36 & 0.05\\
\hline
NO     &       150176.480   & 0.127$\pm$0.008     &   -2.11$\pm$0.01 &    0.50$\pm$0.04 &  0.24 & 0.01 \\
NO     &       150198.760   & 0.069$\pm$0.007 &    -2.23$\pm$0.04 &    0.764$\pm$0.091 &  0.08 & 0.01 \\
NO     &       150218.730  & 0.048$\pm$0.006 &    -2.13$\pm$0.03 &   0.45$\pm$0.06 &  0.10 & 0.01\\
NO     &       150225.660  & 0.036$\pm$0.006 &    -2.11$\pm$0.03 &  0.48$\pm$0.09 & 0.07 & 0.01 \\
NO    &        150439.120  & 0.053$\pm$0.008 &    -2.04$\pm$0.04 &   0.59$\pm$0.11 &  0.08 & 0.02\\
NO     &       150546.520 & 0.126$\pm$0.006     &    -2.04$\pm$0.01 &    0.49$\pm$0.03 & 0.24 & 0.01 \\
NO     &       150580.560  &  0.052$\pm$0.009 &   -2.06$\pm$0.07 &   0.71$\pm$0.18 &  0.06 & 0.01 \\
NO     &       150644.340  & 0.063$\pm$0.007 &   -2.16$\pm$0.04 &    0.68$\pm$0.09 &  0.08 & 0.01 \\
\hline
HNO  &          81477.490   &   0.051$\pm$0.005 &  -2.39$\pm$0.03 &   0.65$\pm$0.06 & 0.073  & 0.007 \\
\hline
HCO$^+$ &   89188.523   & 0.50$\pm$0.02     &    -1.75$\pm$0.02 &    1.50$\pm$0.05 &  0.31 & 0.02  \\     
HCO$^+$ & 178375.010  & $\le 0.06$ & -&- & - & 0.02 \\
H$^{13}$CO$^+$&       86754.288  &  0.06$\pm$0.01 &    -2.14$\pm$0.08 &   0.79$\pm$0.22 & 0.07 & 0.02 \\
\hline
HCN    &        88630.416   &  0.14$\pm$0.01  &  -1.95$\pm$0.04 &  1.36$\pm$0.10 & 0.098 & 0.008\\          
HCN    &        88631.847   &  0.177$\pm$0.009   &  -1.90$\pm$0.08 &  1.33$\pm$0.07 & 0.125 & 0.008 \\           
HCN    &        88633.936   &   0.08$\pm$0.01 & -1.82$\pm$0.30 &    1.53$\pm$0.19 &  0.050 & 0.008 \\   
\hline     
HNC   &      90663.593   & 0.22$\pm$0.01  & -1.961$\pm$0.044 &   1.450$\pm$0.103 & 0.14615 & 0.014 \\
\hline
CN & 113144.190   & 0.04$\pm$0.01 &  -1.71$\pm$0.13 &   1.03$\pm$0.29 &  0.04 & 0.01 \\
CN &  113191.325  &  0.07$\pm$0.01 &    -1.98$\pm$0.08 &     1.04$\pm$0.18 &  0.06 & 0.01 \\
CN &  113488.142  & 0.06$\pm$0.02 &   -2.01$\pm$0.17 &   1.46$\pm$0.55 &  0.04 & 0.02 \\
CN &  113490.985  & 0.07$\pm$0.01 &   -2.32$\pm$0.17 &     1.71$\pm$0.29 & 0.04 & 0.01\\
\hline
CCH     &       87316.925  &    $\le 0.03$ & -&- & -  & 0.01 \\
CCH      &      87402.004  &  $\le 0.03$ & -&- & -& 0.01 \\
\hline
CH$_3$OH   &       96739.358  & 0.07$\pm$0.01 &  -2.15$\pm$0.04 &    0.39$\pm$0.24 & 0.16 & 0.03 \\ 
CH$_3$OH    &      96741.371  &  0.099$\pm$0.014 &  -2.20$\pm$0.03 &    0.50$\pm$0.08 & 0.18 & 0.02 \\
CH$_3$OH   &      145097.435  & 0.042$\pm$0.004 &   -2.17$\pm$0.02 &   0.51$\pm$0.06 & 0.078  & 0.008 \\
CH$_3$OH    &     145103.185 &   0.052$\pm$0.004 &   -2.18$\pm$0.01 &   0.48$\pm$0.04 & 0.101 & 0.007 \\
\hline
C$^{18}$O       &  109782.173  & 0.471$\pm$0.007     &   -2.053$\pm$0.006 &   0.78$\pm$0.01 & 0.56  & 0.01 \\
$^{13}$CO   &     110201.354 & 7.11$\pm$0.05   &   -1.987$\pm$0.004 &    1.184$\pm$0.009 &  5.64 & 0.06 \\
C$^{17}$O$^a$   &112359.284 &   0.108$\pm$0.009     &   -1.00$\pm$0.04 &  1.13$\pm$0.10 & 0.09  & 0.01  \\
&     &  0.057$\pm$0.007 &   -3.91$\pm$0.05 &   0.760$\pm$0.10 &  0.07 & 0.01 \\
CO       &     115271.202  &  22.13$\pm$0.88   &  -1.82$\pm$0.04 &   2.22$\pm$0.10 &  9.35 & 0.79\\
\hline
\end{tabular}
\end{center}
  $^a$ C$^{17}$O has a hyperfine structure; see comment in Table~\ref{detect_lines}.\\
W is the integrated intensity, v$_{\rm lsr}$  the central velocity, FWHM  the line width, and T$_{\rm peak}$ the peak intensity of a gaussian fitting of each line. The rms is the noise level of each line.
\label{lineC3}
\end{table*}%

\begin{table*}
\caption{Results of the line fitting in C4.}
\begin{center}
\begin{tabular}{ll|llllllllll|}
\hline
\hline
Molecule & Frequency & \multicolumn{5}{|c}{C4}  \\
 & MHz &  W    &               v$_{\rm lsr}$     &       FWHM      &       T$_{\rm peak}$  & rms \\
  & & K.km.s$^{-1}$ & km.s$^{-1}$ & km.s$^{-1}$ & K & K \\
\hline
 H$_2$CO    &       72837.948  &    0.26 $\pm$0.02 &    3.05$\pm$0.07 &     1.80$\pm$0.15 &  0.14 &   0.02 \\        
H$_2$CO     &     140839.502   &   0.191$\pm$0.009& 3.35$\pm$0.05&   1.98$\pm$0.10& 0.09 & 0.01\\ 
H$_2$CO      &    145602.949   &   0.0160$\pm$0.004 &     0.52$\pm$0.04 &    0.30$\pm$0.08 &  0.05 & 0.01   \\
H$_2$CO      &    150498.334    &   0.15$\pm$0.02 &     3.06$\pm$0.13 &     2.29$\pm$0.33 &  0.06 & 0.02 \\ 
\hline
CS         &    97980.950     &   0.81$\pm$0.01 &     3.21$\pm$0.01 &    2.00$\pm$0.03 &  0.38 & 0.01  \\
CS         &   146969.033   &    0.29$\pm$0.01 &     3.19$\pm$0.05 &     2.10$\pm$0.12 &  0.13 & 0.02 \\
C$^{34}$S &          96412.940  &   0.11$\pm$0.01 &      3.07$\pm$0.19 &     3.74$\pm$0.57 &  0.029 & 0.007   \\ 
\hline
 SO      &       99299.870   &   0.29$\pm$0.01  &   3.37$\pm$0.03 &     1.87$\pm$0.06 & 0.148 & 0.009 \\        
SO     &       138178.600  &  $\le 0.03$ & - & - & - & 0.01  \\
\hline
NO     &       150176.480   & $\le 0.03$ & - & - & - & 0.01 \\
NO     &       150198.760   & $\le 0.03$ & - & - & - & 0.01  \\
NO     &       150218.730  & $\le 0.03$ & - & - & - & 0.01  \\
NO     &       150225.660  & $\le 0.03$ & - & - & - & 0.01   \\
NO    &        150439.120  &$\le 0.03$ & - & - & - & 0.01  \\
NO     &       150546.520 & $\le 0.03$& - & - & - & 0.01  \\
NO     &       150580.560  & $\le 0.03$ & - & - & - & 0.01  \\
NO     &       150644.340  &  $\le 0.03$ & - & - & - & 0.01 \\
\hline
HNO  &          81477.490   &   $\le 0.02$ & - & - & - & 0.008  \\
\hline
HCO$^+$ &   89188.523   &   0.37$\pm$0.01 &  3.32$\pm$0.03 &  2.10$\pm$0.06 &  0.166 & 0.008 \\     
HCO$^+$ & 178375.010  &  $\le 0.12$ & - & - & - & 0.04  \\
H$^{13}$CO$^+$&       86754.288  & $\le 0.03$ & - & - & - & 0.01  \\
\hline
HCN    &        88630.416   &   0.279$\pm$ 0.006 & 3.40$\pm$0.02 &    2.08$\pm$0.04 &  0.125 & 0.005  \\          
HCN    &        88631.847   &   0.445$\pm$0.006 &    3.36$\pm$0.11 &     2.24$\pm$0.03 &  0.186 & 0.005 \\           
HCN    &        88633.936   &   0.158$\pm$0.006 &    3.26$\pm$0.42 &    2.06$\pm$0.08 &  0.072 & 0.005 \\   
\hline     
HNC   &      90663.593   &    0.259$\pm$0.015 & 3.33$\pm$0.06 &   2.28$\pm$0.15 & 0.106   & 0.01 \\
\hline
CN & 113144.190   & $\le 0.03$ & - & - & - & 0.01  \\
CN &  113191.325  & $\le 0.03$ & - & - & - & 0.01  \\
CN &  113488.142  & $\le 0.03$ & - & - & - & 0.01 \\
CN &  113490.985  & $\le 0.03$ & - & - & - & 0.01 \\
\hline
CCH     &       87316.925  &  0.08$\pm$0.01 &    2.74$\pm$0.08 &   1.30$\pm$0.22 & 0.06 & 0.01 \\
CCH      &      87402.004  &  0.06$\pm$0.01 &   2.77$\pm$0.13 &    1.33$\pm$0.33 &  0.04 & 0.01  \\
\hline
CH$_3$OH   &       96739.358  &   0.050$\pm$0.008 &  9.04$\pm$0.12 &   1.50$\pm$0.24 & 0.03 & 0.008 \\ 
CH$_3$OH    &      96741.371  &  0.09$\pm$0.008 &     2.92$\pm$0.07 &  1.58$\pm$0.17 &  0.05 & 0.008 \\
CH$_3$OH   &      145097.435  &  $\le 0.02$&   -&  - &  - & 0.008 \\
CH$_3$OH    &     145103.185 & $\le 0.02$ &  -& - &  - & 0.008 \\
\hline
C$^{18}$O       &  109782.173  &  1.80$\pm$0.01&   3.245$\pm$0.006 &   1.86$\pm$0.01& 0.91 &0.01 \\
$^{13}$CO   &     110201.354 &    14.51$\pm$0.01&   3.218$\pm$0.001&  2.176$\pm$0.002&  6.26 & 0.01 \\
C$^{17}$O$^a$   &      112359.284  &  0.32$\pm$0.02&  4.12$\pm$0.04&  1.84$\pm$0.11& 0.17 &0.01 \\
 & &   0.15$\pm$0.02 &     1.04$\pm$0.01&   1.80$\pm$0.22& 0.008 & 0.01 \\
CO       &     115271.202  &   31.0 $\pm$0.2 &   3.10$\pm$0.01&  2.40$\pm$0.02 &  12.12  &0.05  \\
\hline
\end{tabular}
\end{center}
  $^a$ C$^{17}$O has a hyperfine structure; see comment in Table~\ref{detect_lines}.\\
W is the integrated intensity, v$_{\rm lsr}$  the central velocity, FWHM  the line width, and T$_{\rm peak}$ the peak intensity of a gaussian fitting of each line. The rms is the noise level of each line.
\label{lineC4}
\end{table*}%

\begin{table*}
\caption{Results of the line fitting in C5.}
\begin{center}
\begin{tabular}{ll|llllllllll|}
\hline
\hline
Molecule & Frequency & \multicolumn{5}{|c}{C5}  \\
 & MHz &  W    &               v$_{\rm lsr}$     &       FWHM      &       T$_{\rm peak}$  & rms \\
  & & K.km.s$^{-1}$ & km.s$^{-1}$ & km.s$^{-1}$ & K & K \\
\hline
 H$_2$CO    &       72837.948  &   0.24$\pm$ 0.01 &   3.89$\pm$0.01&   0.64$\pm$0.05 & 0.35 &0.01\\        
H$_2$CO     &     140839.502   &  0.158$\pm$0.006&   3.85$\pm$0.01& 0.71$\pm$0.03 & 0.21 & 0.01 \\ 
H$_2$CO      &    145602.949   &  0.070$\pm$0.008&  3.87$\pm$0.04& 0.69$\pm$0.08  &  0.10 &0.01  \\
H$_2$CO      &    150498.334    &  0.088$\pm$0.008&  3.94$\pm$0.03&   0.58$\pm$0.06 &  0.14  & 0.02 \\ 
\hline
CS         &    97980.950     &    0.491$\pm$0.008&     3.874$\pm$0.006&  0.75$\pm$0.01& 0.616 & 0.01 \\
CS         &   146969.033   &    0.172$\pm$0.009&  3.87$\pm$0.02&   0.63$\pm$0.04& 0.26 & 0.02 \\
C$^{34}$S &          96412.940  &     0.04$\pm$0.007&    3.85$\pm$0.05&   0.48$\pm$0.07& 0.08 &0.01 \\ 
\hline
 SO      &       99299.870   &   0.32431$\pm$0.008&     3.844$\pm$0.008&   0.66$\pm$0.02& 0.46& 0.01 \\        
SO     &       138178.600  &   0.12079 $\pm$0.008&  3.81$\pm$0.02&   0.60$\pm$0.05& 0.19  & 0.01\\
\hline
NO     &       150176.480   & $\le 0.05$ & -& -& -& 0.01 \\
NO     &       150198.760   & $\le 0.05$& -& -& -&0.01 \\
NO     &       150218.730  &  $\le 0.05$& -& -& -&0.01\\
NO     &       150225.660  &  $\le 0.05$& -& -& -&0.01 \\
NO    &        150439.120  &$\le 0.05$& -& -& -&0.01 \\
NO     &       150546.520 & $\le 0.05$& -& -& -&0.01 \\
NO     &       150580.560  & $\le 0.05$& -& -& -&0.01 \\
NO     &       150644.340  & $\le 0.05$& -& -& -& 0.01\\
\hline
HNO  &          81477.490   &  $\le 0.03$ & -& -& -& 0.007  \\
\hline
HCO$^+$ &   89188.523   &   0.27$\pm$0.01&  3.91$\pm$0.02&     0.88$\pm$0.04& 0.29 & 0.01 \\     
HCO$^+$ & 178375.010  &   $\le 0.19$ & -& -& -& 0.05  \\
H$^{13}$CO$^+$&       86754.288  & $\le 0.05$ & -& -& -& 0.01 \\
\hline
HCN    &        88630.416   &   0.151$\pm$0.008&    3.95$\pm$0.09 &    1.05$\pm$0.07&  0.134 & 0.008 \\          
HCN    &        88631.847   &      0.187$\pm$0.007&  3.89$\pm$0.02&   0.89$\pm$0.04&  0.196& 0.008 \\           
HCN    &        88633.936   &   0.07$\pm$0.007&   3.94$\pm$0.15 &   0.79$\pm$0.09& 0.08 & 0.008\\   
\hline     
HNC   &      90663.593   &  0.171$\pm$0.008&   3.94$\pm$0.02 &   0.97$\pm$0.06 &0.16 & 0.01 \\
\hline
CN & 113144.190   & $\le 0.05$& -& -& -& 0.01 \\
CN &  113191.325  & $\le 0.05$& -& -& -& 0.01 \\
CN &  113488.142  & $\le 0.05$& -& -& -&0.01 \\
CN &  113490.985  & $\le 0.05$& -& -& -&0.01 \\
\hline
CCH     &       87316.925  &$\le 0.05$& -& -& -& 0.01 \\
CCH      &      87402.004  &$\le 0.05$& -& -& -& 0.01  \\
\hline
CH$_3$OH   &       96739.358  &  0.03$\pm$0.009 &    3.85$\pm$0.06&    0.40$\pm$0.15&  0.07 & 0.01  \\ 
CH$_3$OH    &      96741.371  &   0.06$\pm$0.009&   3.92$\pm$0.04&    0.59$\pm$0.11 &  0.09 &0.01 \\
CH$_3$OH   &      145097.435  &   0.02$\pm$0.004&   4.00$\pm$0.05&    0.51$\pm$0.09 & 0.04 & 0.008 \\
CH$_3$OH    &     145103.185 &   0.02$\pm$ 0.004&   3.80$\pm$0.04&    0.51$\pm$0.09 &  0.05  & 0.009 \\
\hline
C$^{18}$O       &  109782.173  &    0.527$\pm$0.005&    3.958$\pm$0.002&   0.557$\pm$0.006&  0.889 & 0.007 \\
$^{13}$CO   &     110201.354 & 4.862$\pm$0.008&   3.961$\pm$0.001&    0.968$\pm$0.002&  4.72 & 0.01\\
C$^{17}$O$^a$   &      112359.284  &   0.109$\pm$0.009&    4.87$\pm$0.03&     0.79$\pm$0.07 & 0.13 & 0.01 \\
 & &   0.04$\pm$0.006 &     2.02$\pm$0.04&    0.41$\pm$0.05&  0.01 & 0.01  \\
CO       &     115271.202  &   14.67$\pm$0.07&    3.858$\pm$0.005&    2.08$\pm$0.01&   6.60 & 0.06 \\
\hline
\end{tabular}
\end{center}
 $^a$ C$^{17}$O has a hyperfine structure, see comment on Table~\ref{detect_lines}.\\
W is the integrated intensity, v$_{\rm lsr}$  the central velocity, FWHM  the line width, and T$_{\rm peak}$ the peak intensity of a gaussian fitting of each line. rms is the noise level of each line.
\label{lineC5}
\end{table*}%

\section{Spectra of the detected lines}\label{appendix_spectra}

\begin{figure*}[htbp]
\centering
\includegraphics[width=1\linewidth]{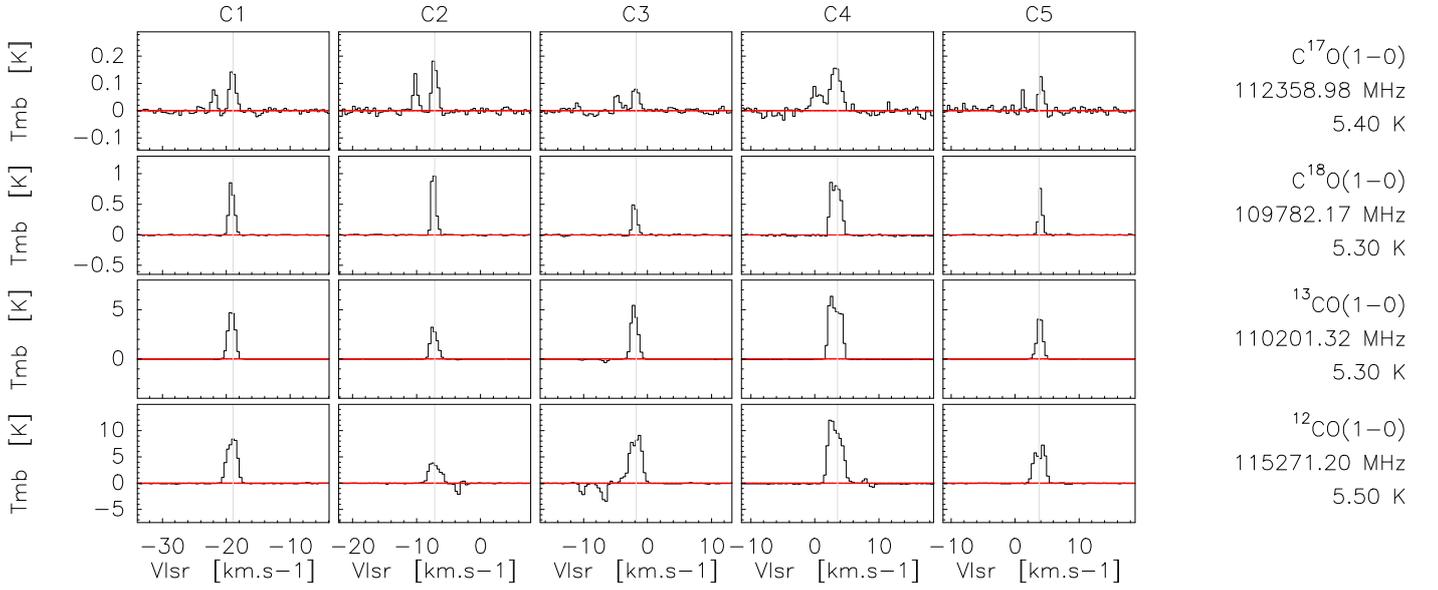}
\caption{Observed spectra (main beam temperature as a function of lsr velocity) of the CO lines in the five sources (each column). The name of the molecule, frequency, and upper energy level are
indicated to the right of each line. \label{CO_spec}}
\end{figure*}

\begin{figure*}[htbp]
\centering
\includegraphics[width=1\linewidth]{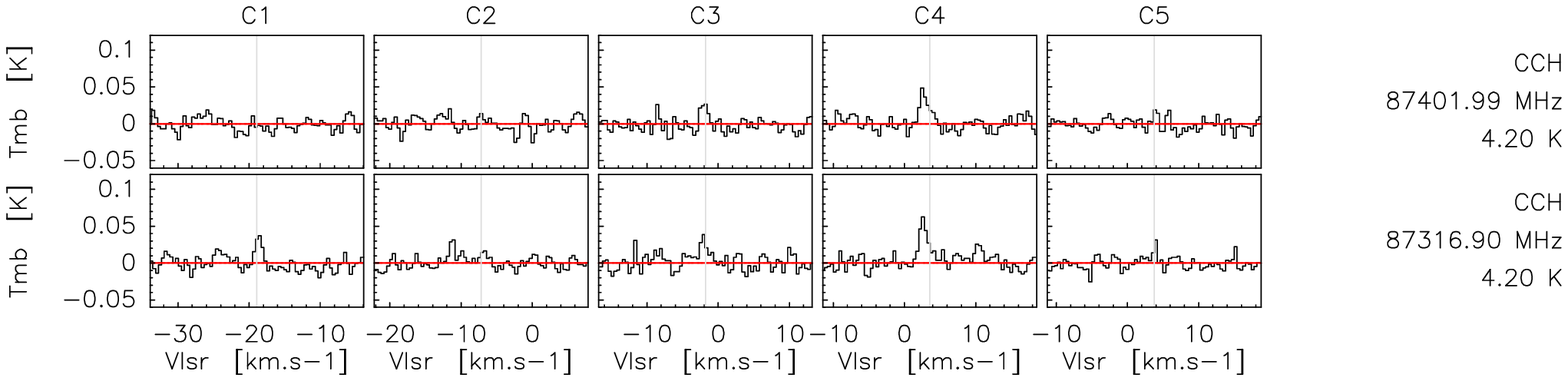}
\caption{Observed spectra (main beam temperature as a function of lsr velocity) of the CCH lines in the five sources (each column). The name of the molecule, frequency, and upper energy level are
indicated to the right of each line.  \label{CH_spec}}
\end{figure*}

\begin{figure*}[htbp]
\centering
\includegraphics[width=1\linewidth]{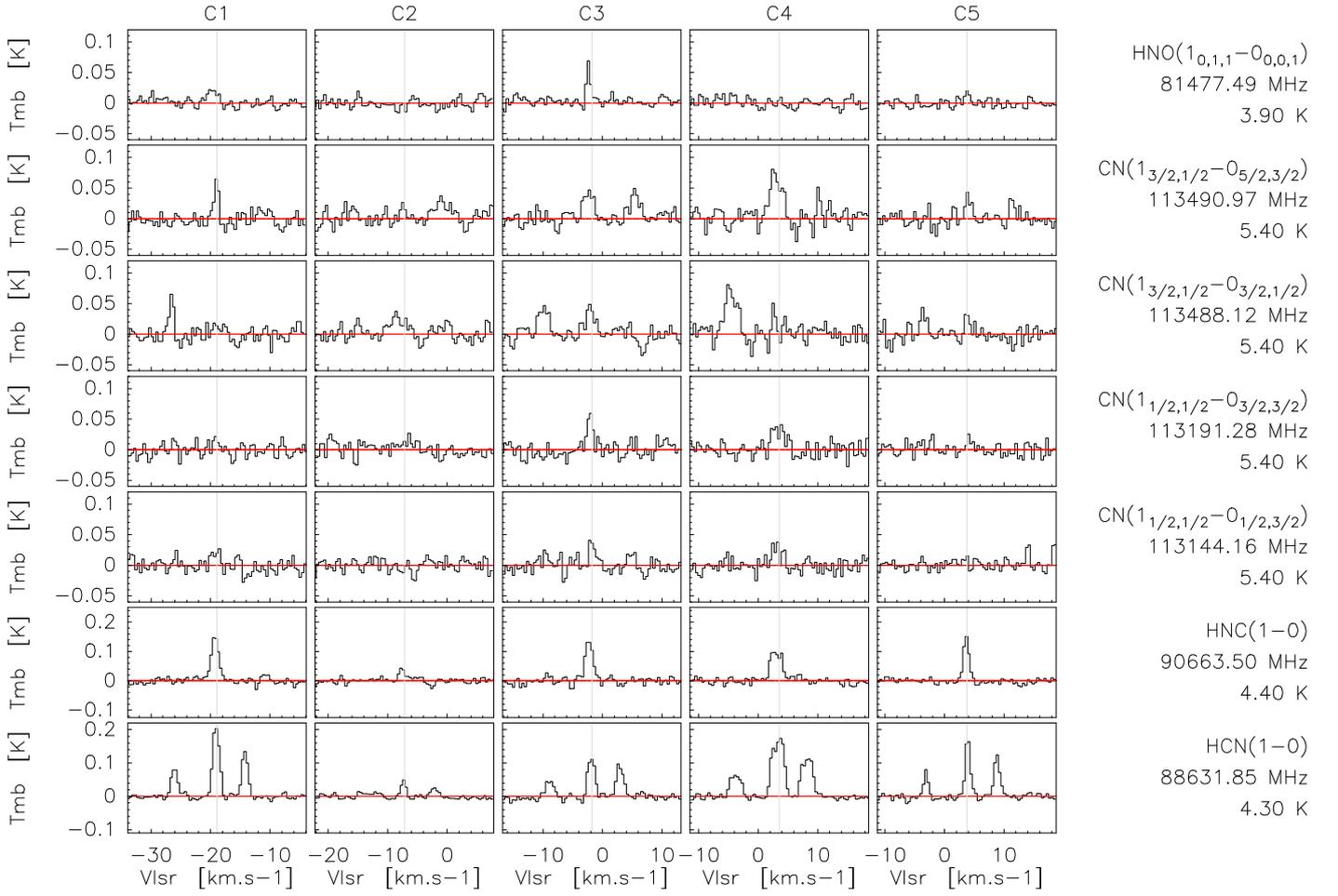}
\caption{Observed spectra (main beam temperature as a function of lsr velocity) of the detected lines of nitrogen bearing species (except NO) in the five sources (each column). The name of the molecule, frequency, and upper energy level are indicated on the right for each line. The HCN hyperfine structure is shown on one single figure, while we have split the CN lines in four different plots. \label{nitrogen_spec}}
\end{figure*}

\begin{figure*}[htbp]
\centering
\includegraphics[width=1\linewidth]{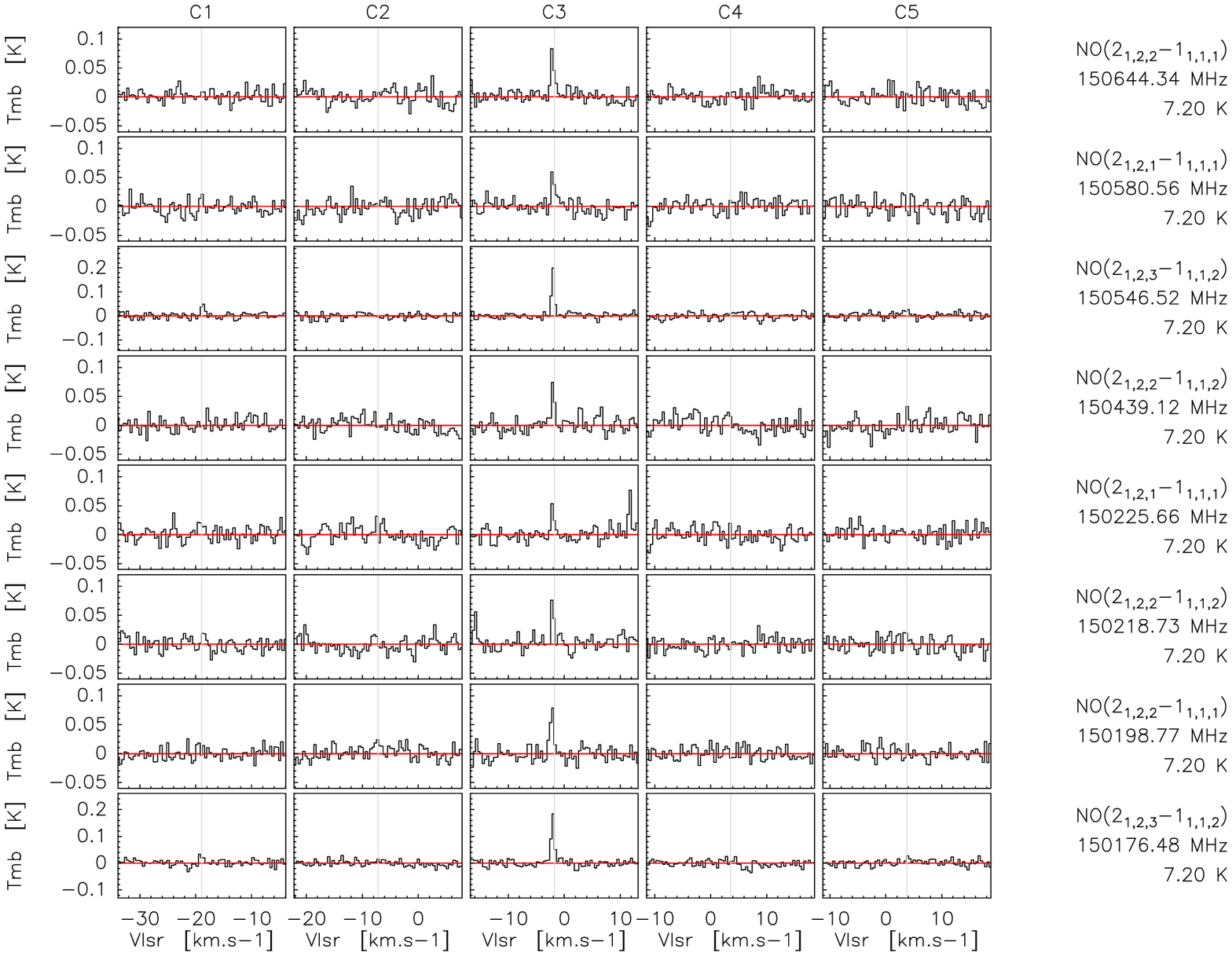}
\caption{Observed spectra (main beam temperature as a function of lsr velocity) of the detected lines of NO in the five sources (each column). The name of the molecule, frequency, and upper energy level are
indicated to the right of each line.  \label{NO_spec}}
\end{figure*}

\begin{figure*}[htbp]
\centering
\includegraphics[width=1\linewidth]{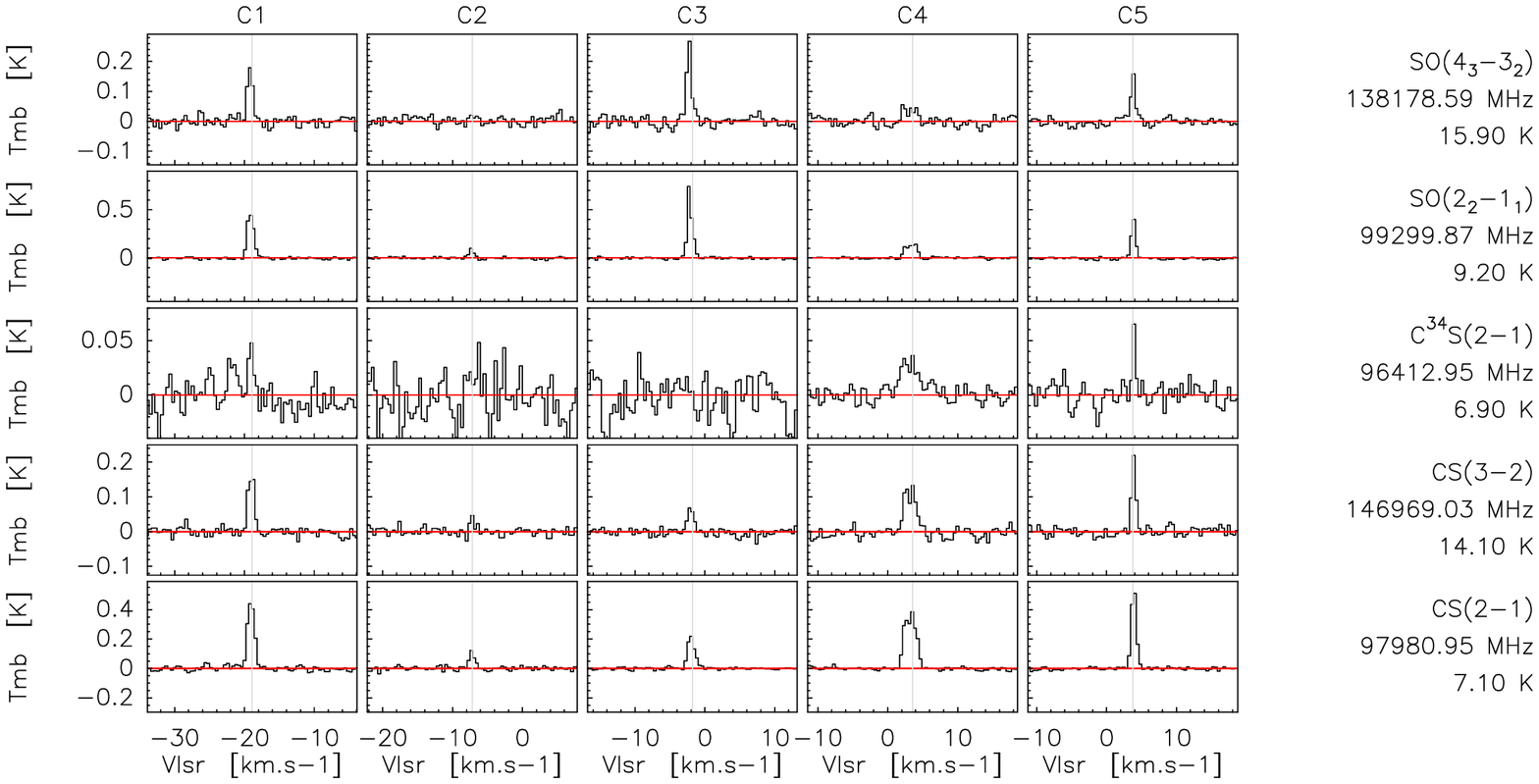}
\caption{Observed spectra (main beam temperature as a function of lsr velocity) of the detected lines of sulphur bearing species in the five sources (each column). The name of the molecule, frequency, and upper energy level are indicated to the right of each line.  \label{sulphur_spec}}
\end{figure*}

\section{$\chi^2$ analysis of the CS, HCN, and SO detected lines}\label{Appendix_chi2}


\begin{figure*}
\includegraphics[width=0.27\linewidth]{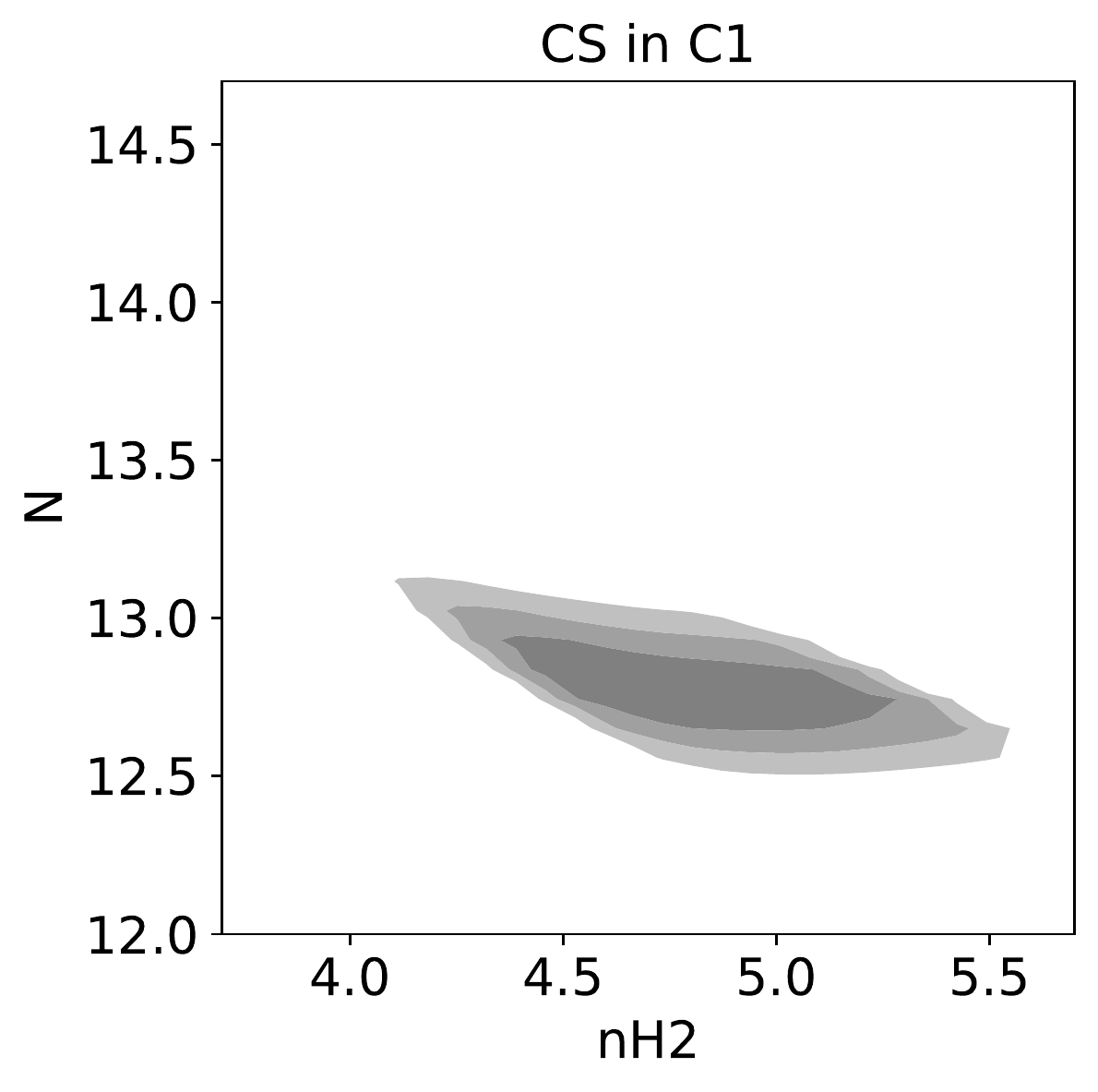}
\includegraphics[width=0.27\linewidth]{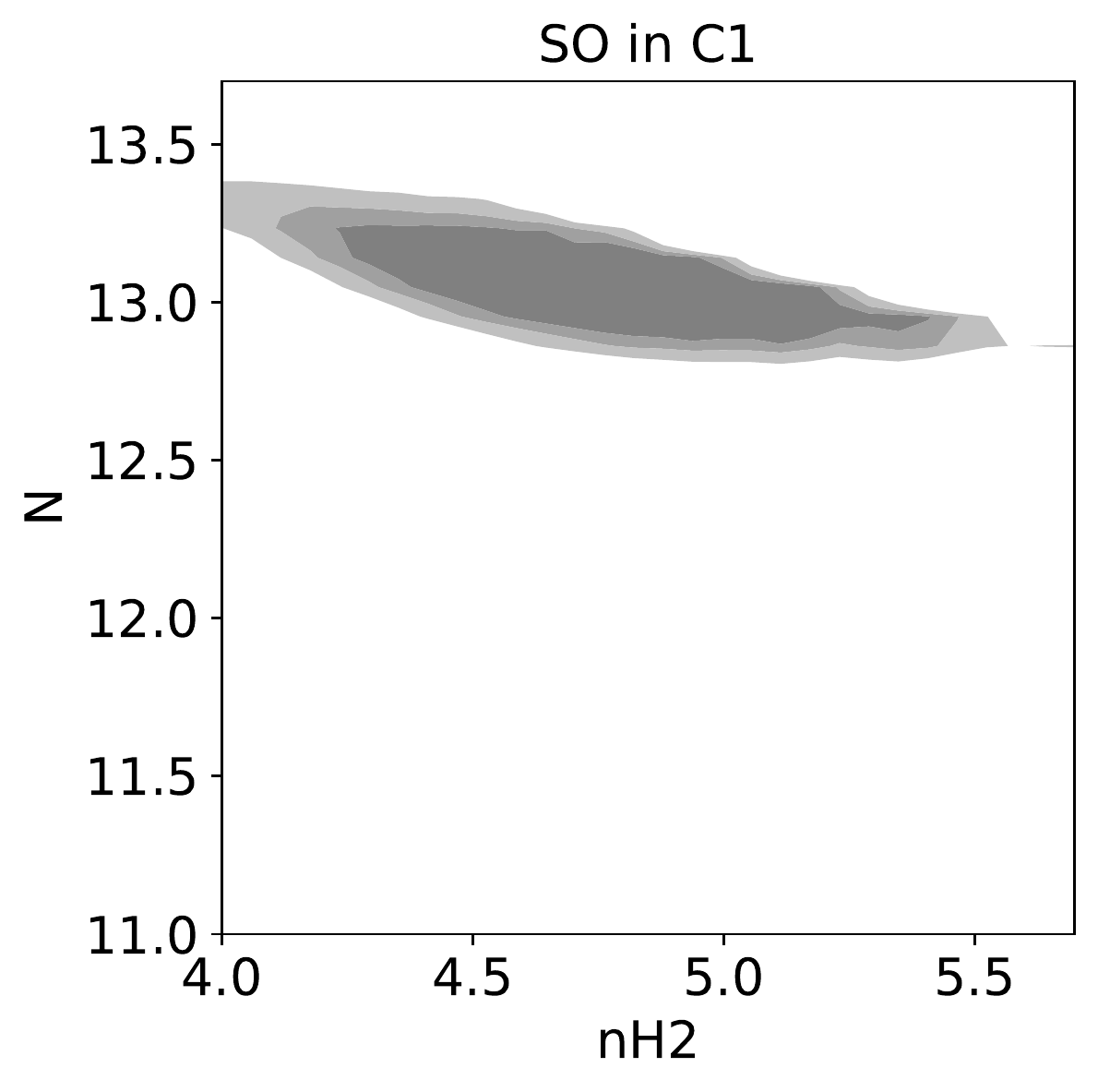}
\includegraphics[width=0.27\linewidth]{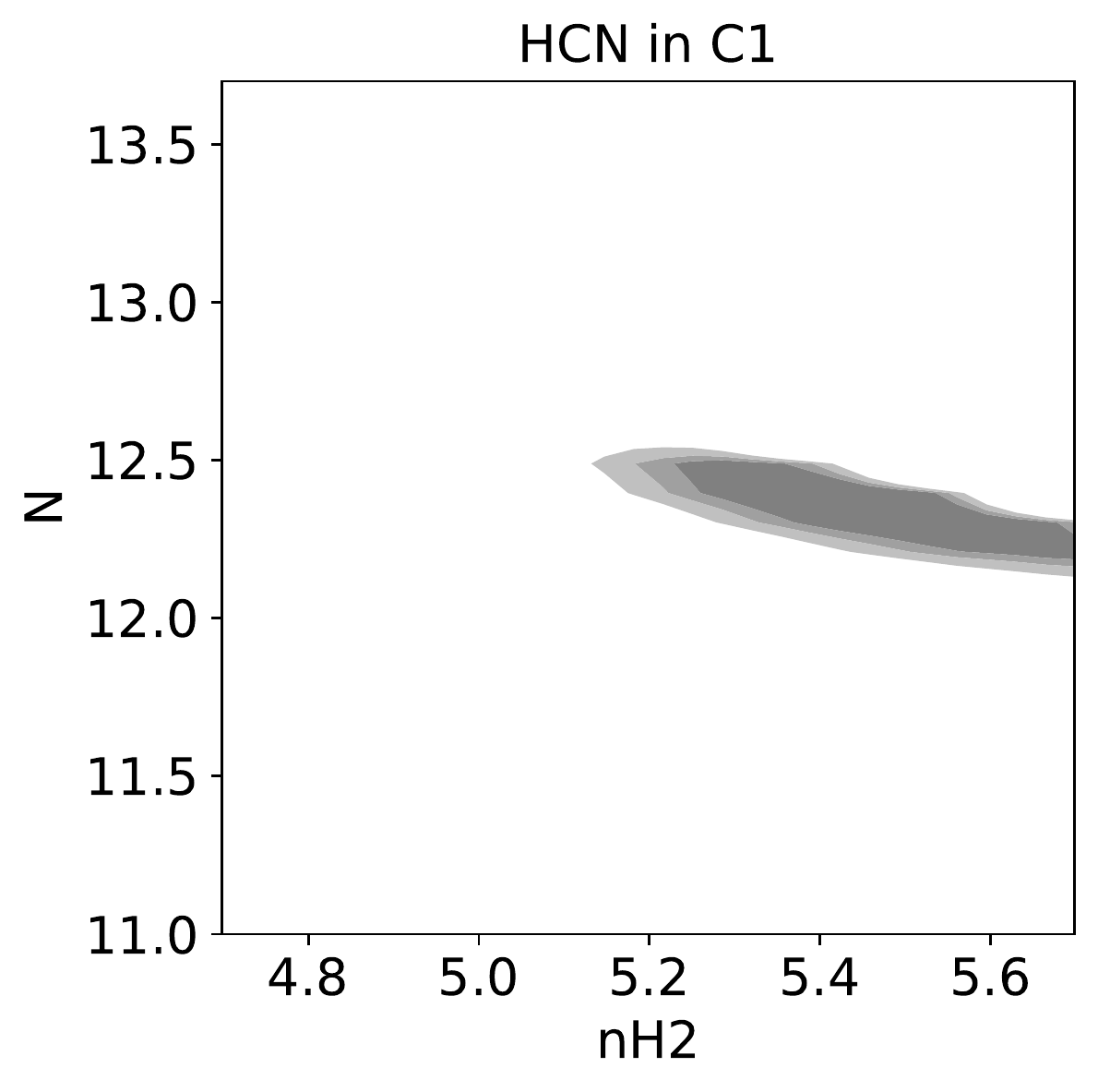}
\includegraphics[width=0.27\linewidth]{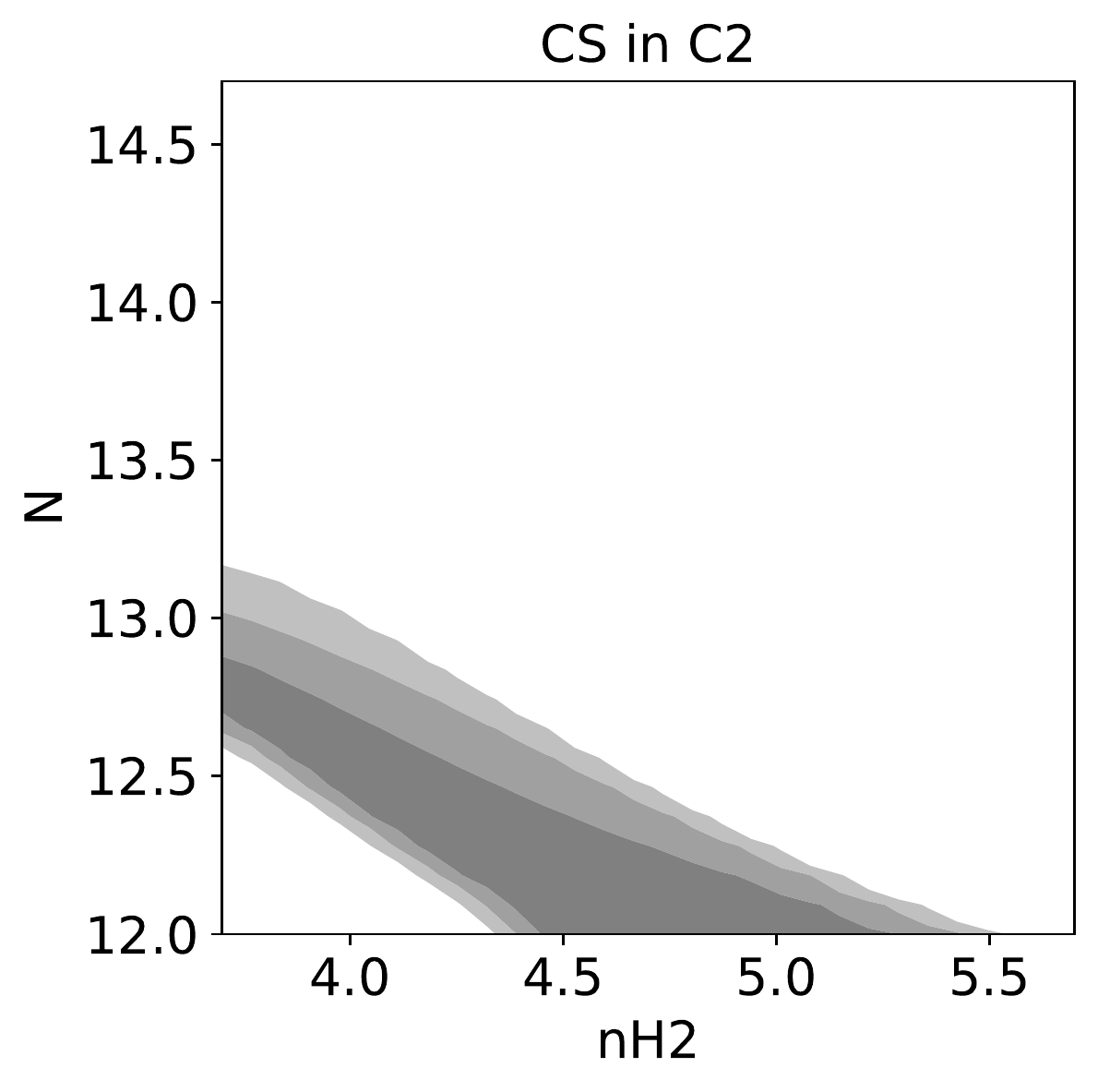}
\includegraphics[width=0.27\linewidth]{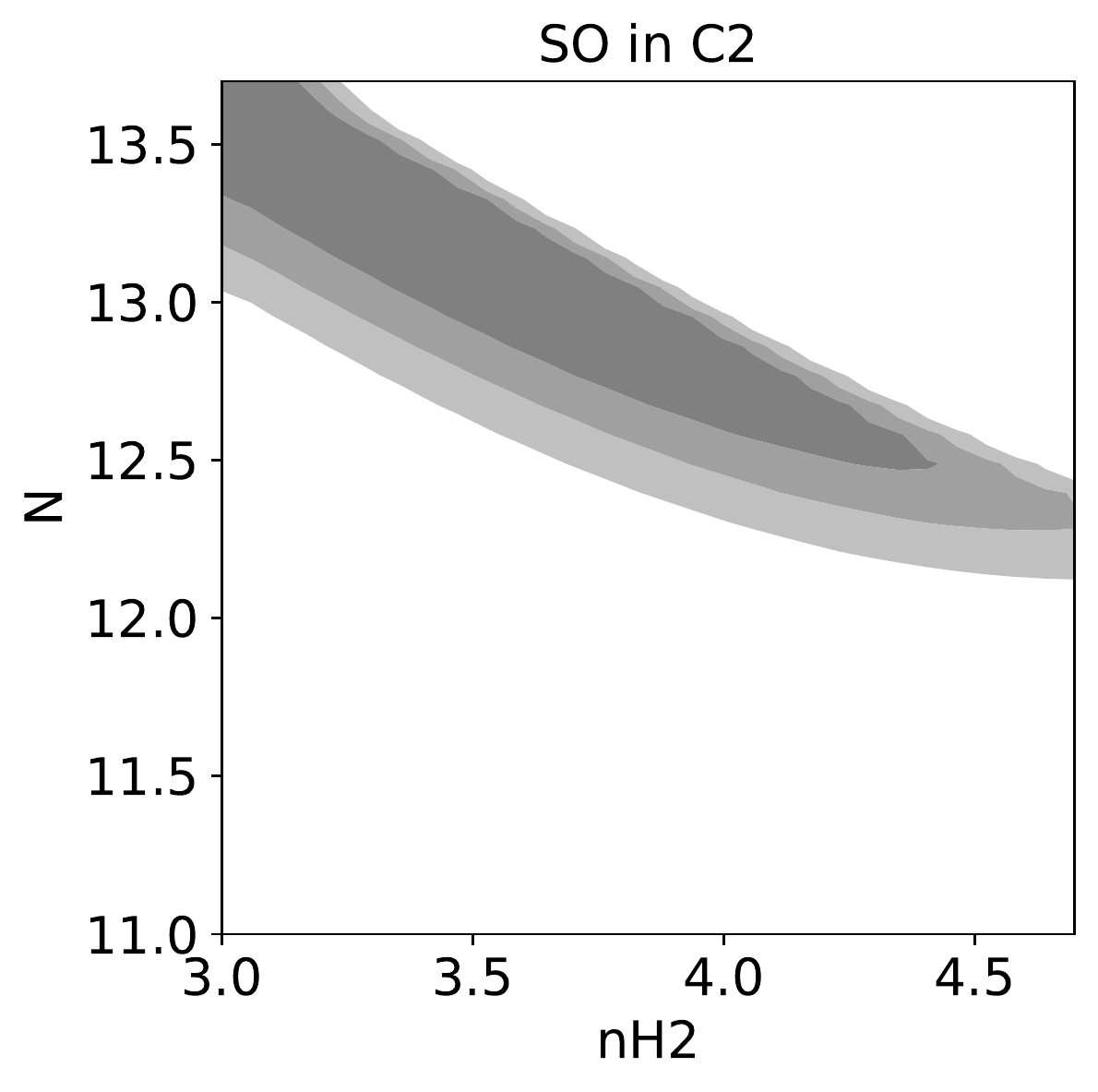}
\includegraphics[width=0.27\linewidth]{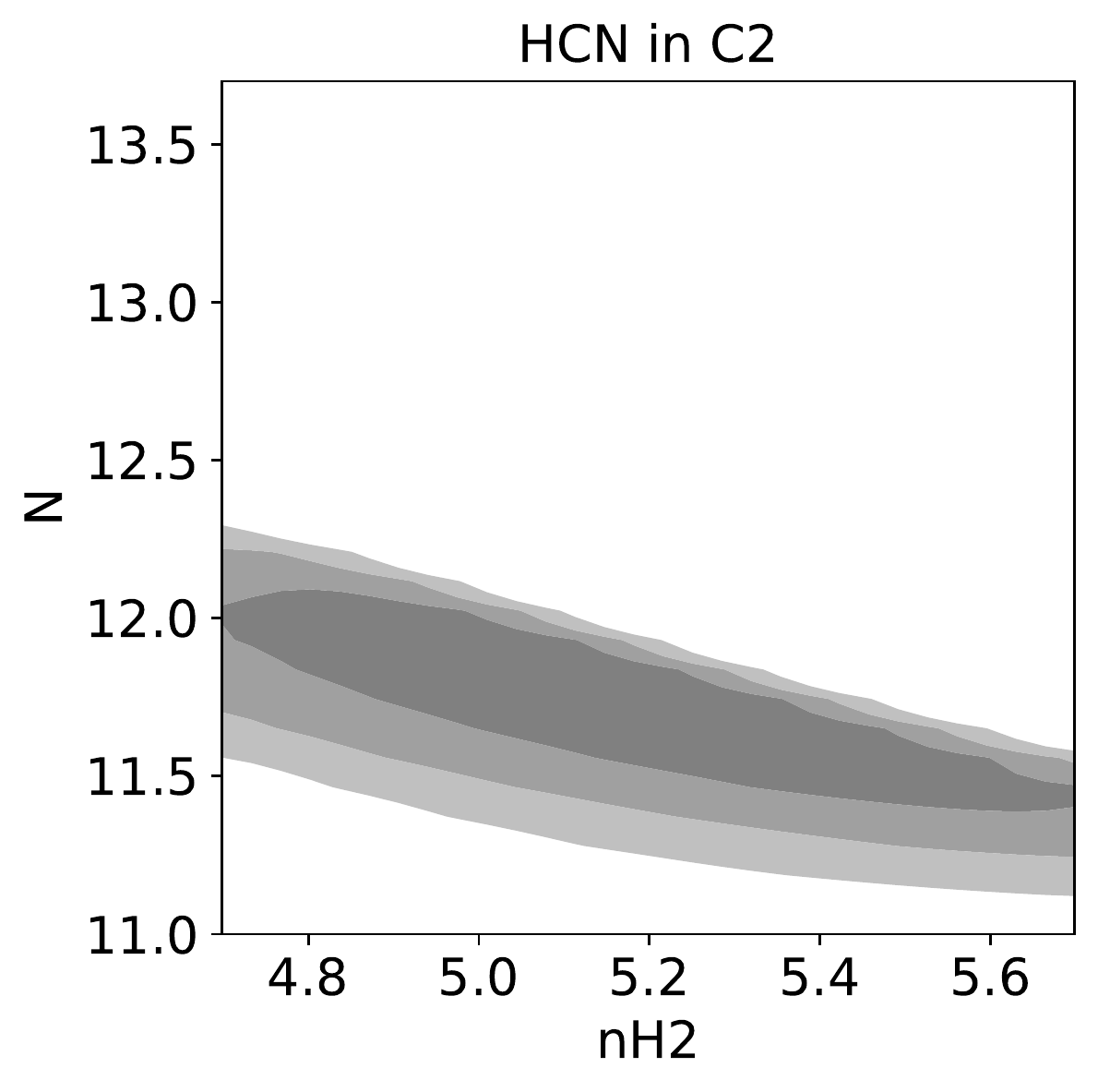}
\includegraphics[width=0.27\linewidth]{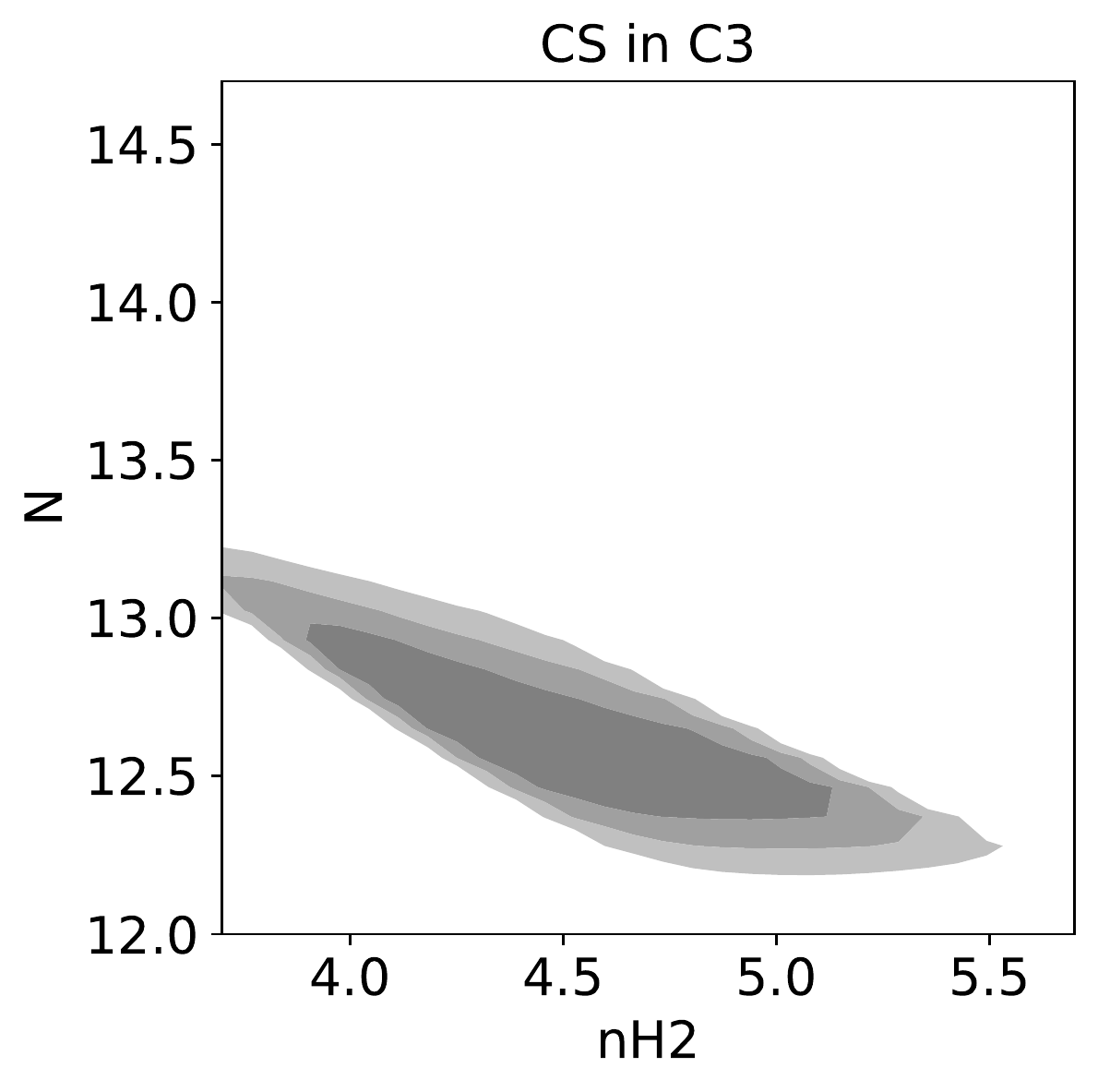}
\includegraphics[width=0.27\linewidth]{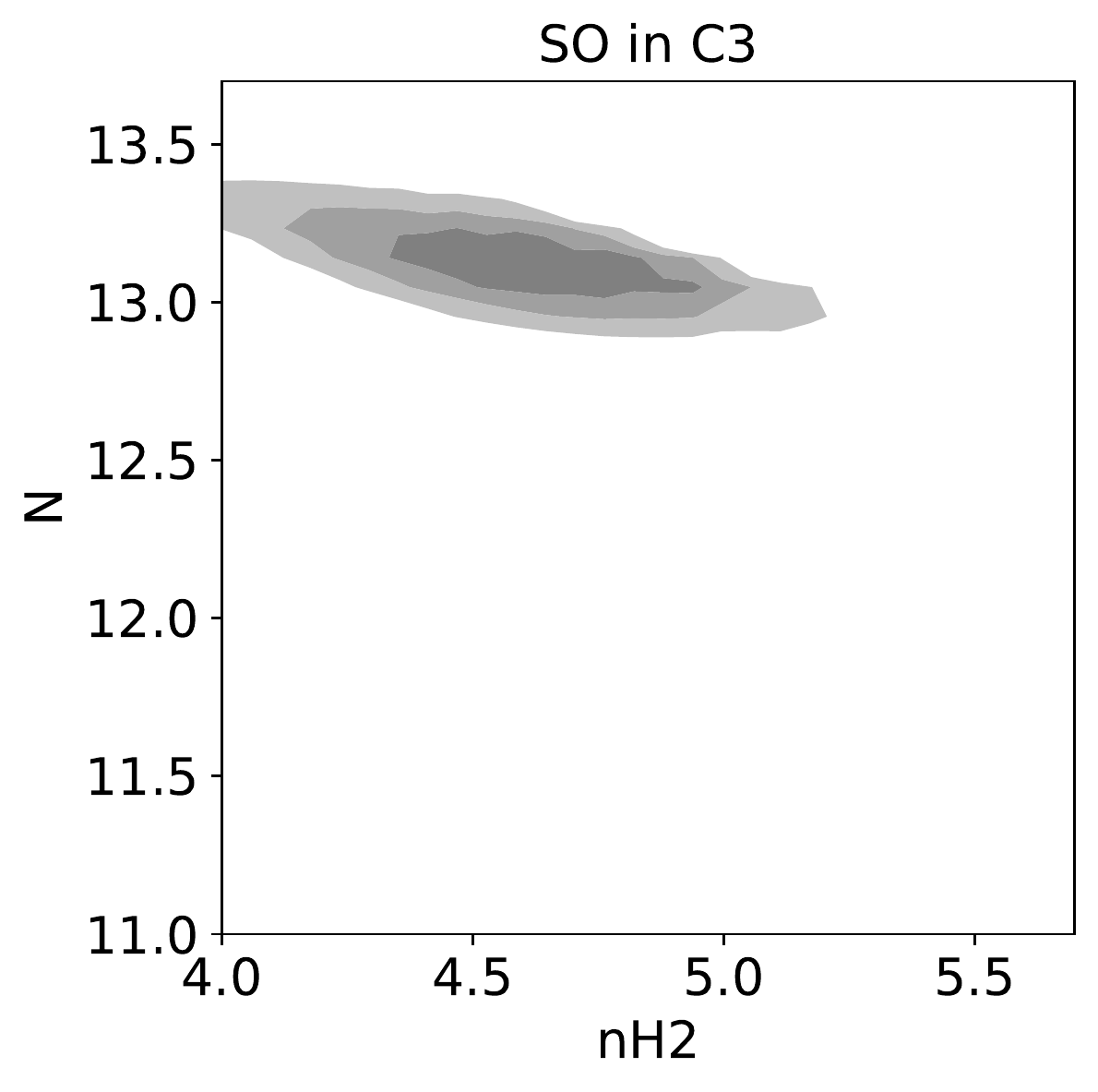}
\includegraphics[width=0.27\linewidth]{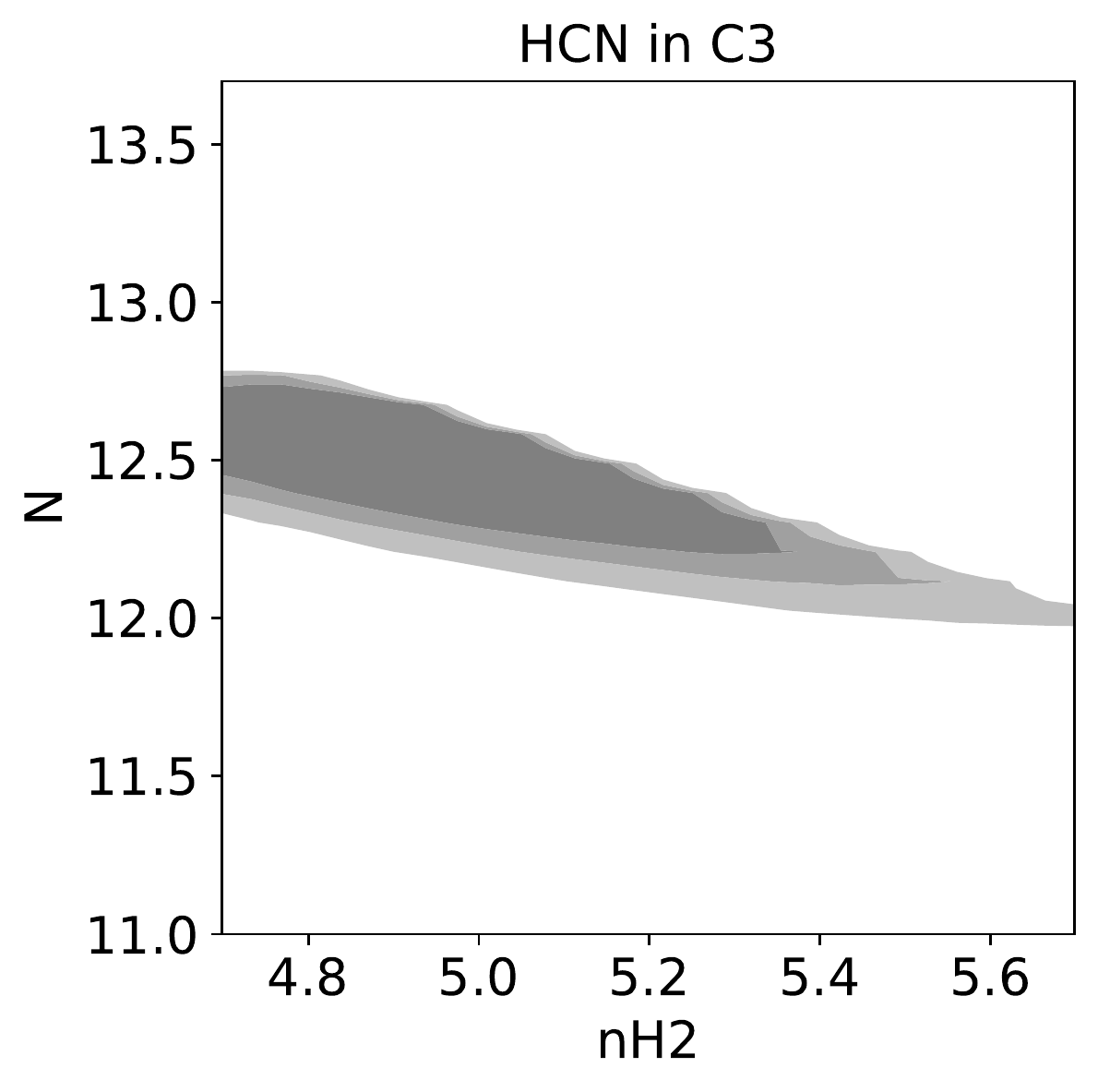}
\includegraphics[width=0.27\linewidth]{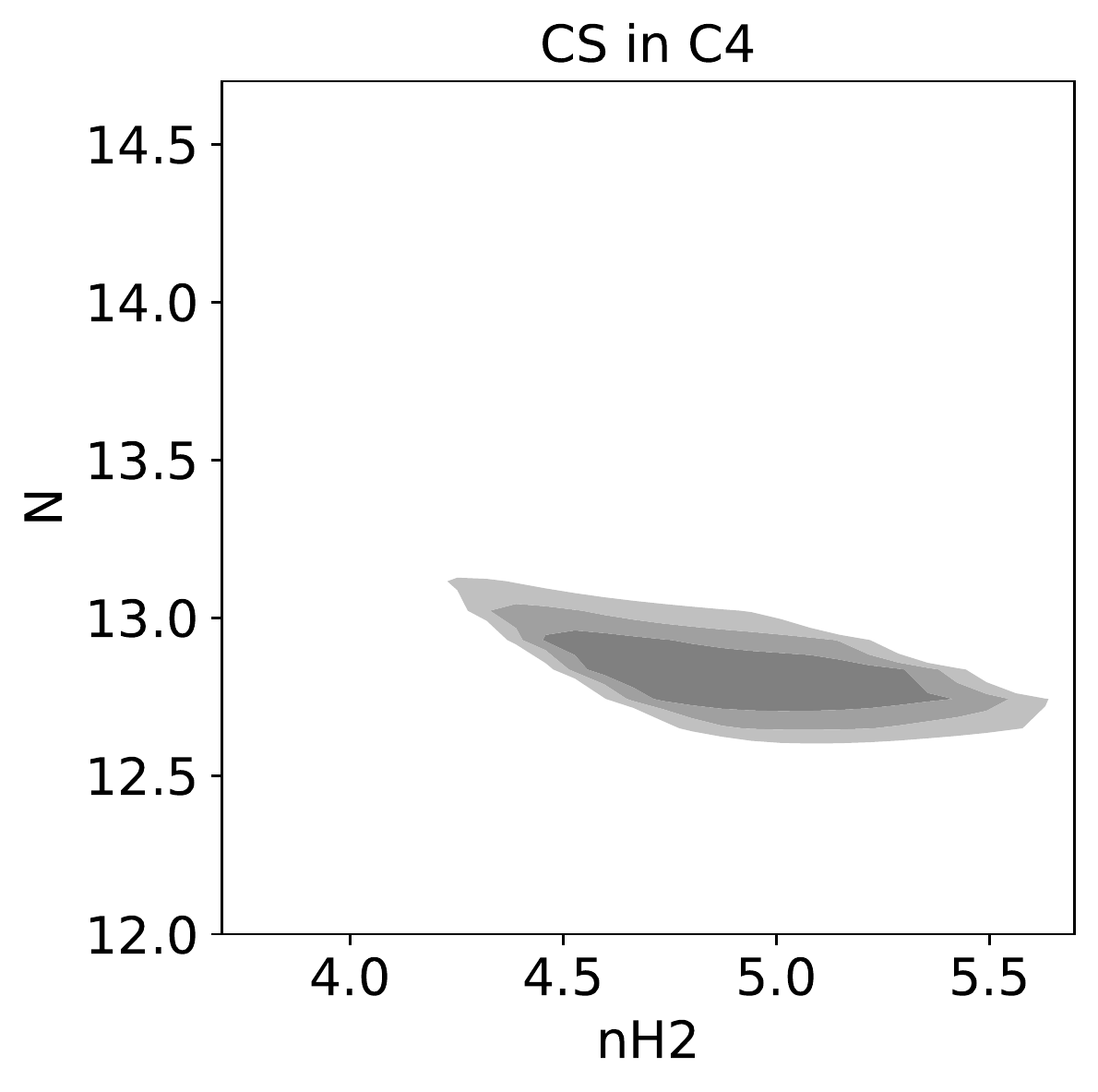}
\includegraphics[width=0.27\linewidth]{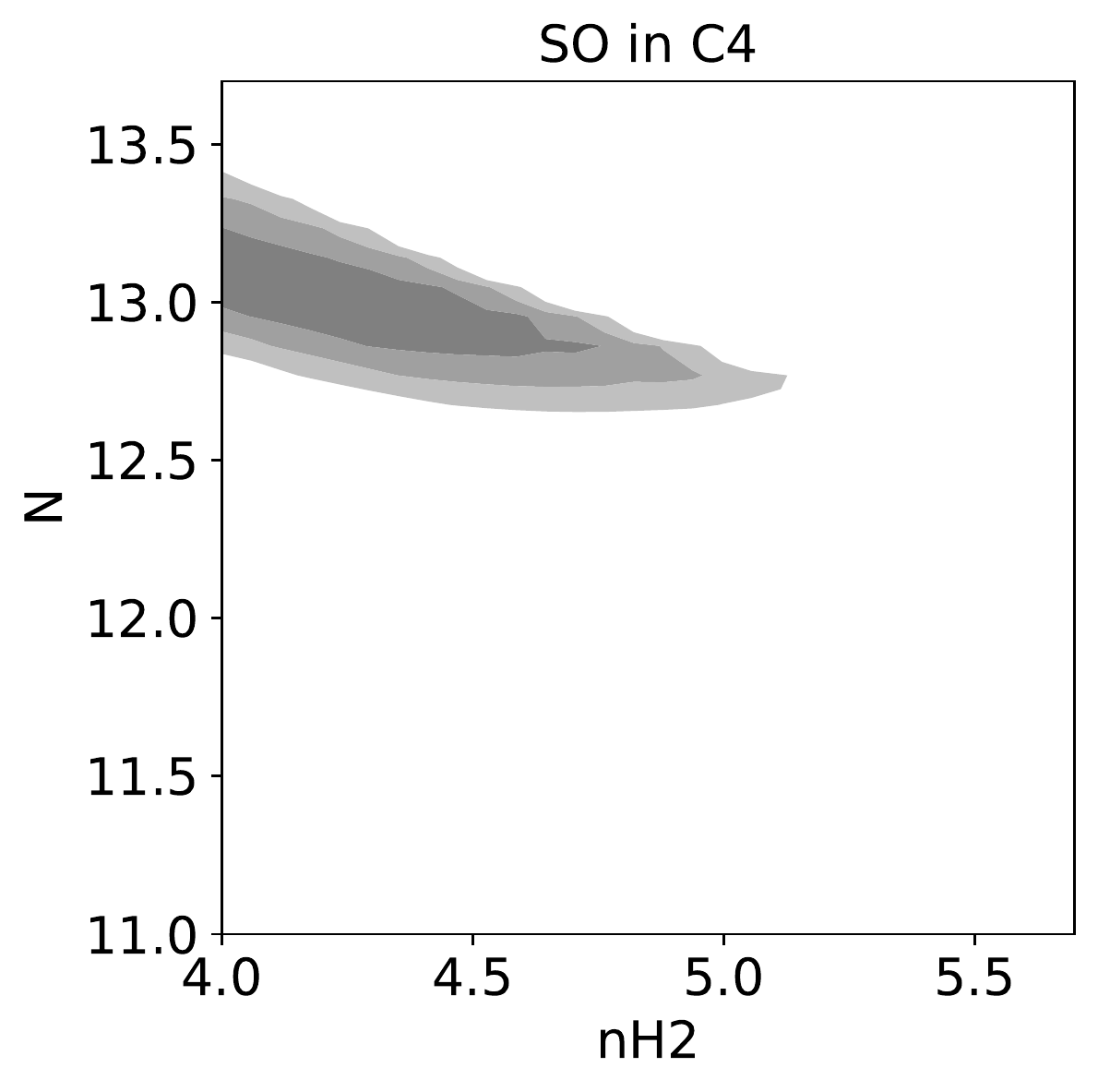}
\includegraphics[width=0.27\linewidth]{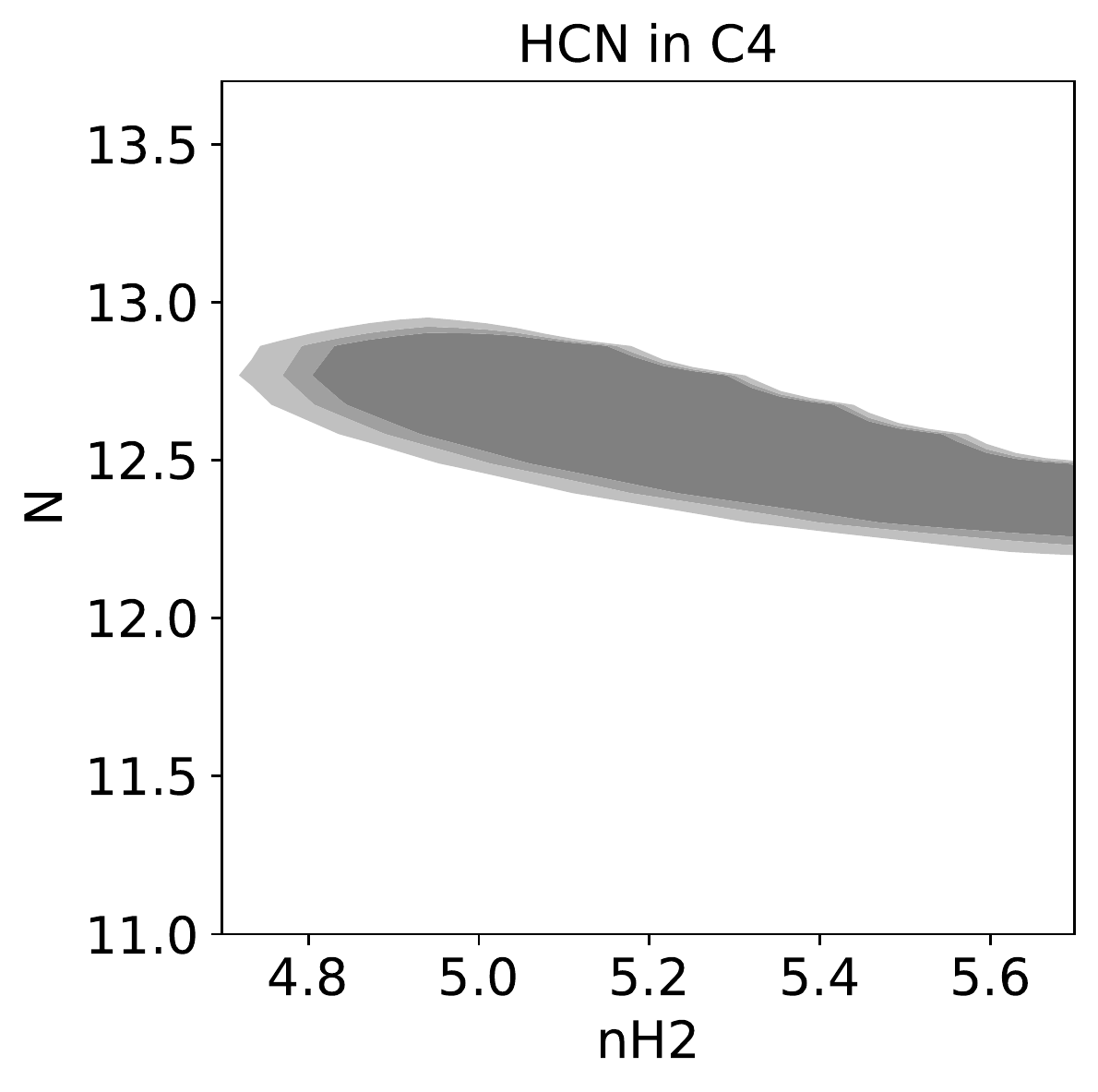}
\includegraphics[width=0.27\linewidth]{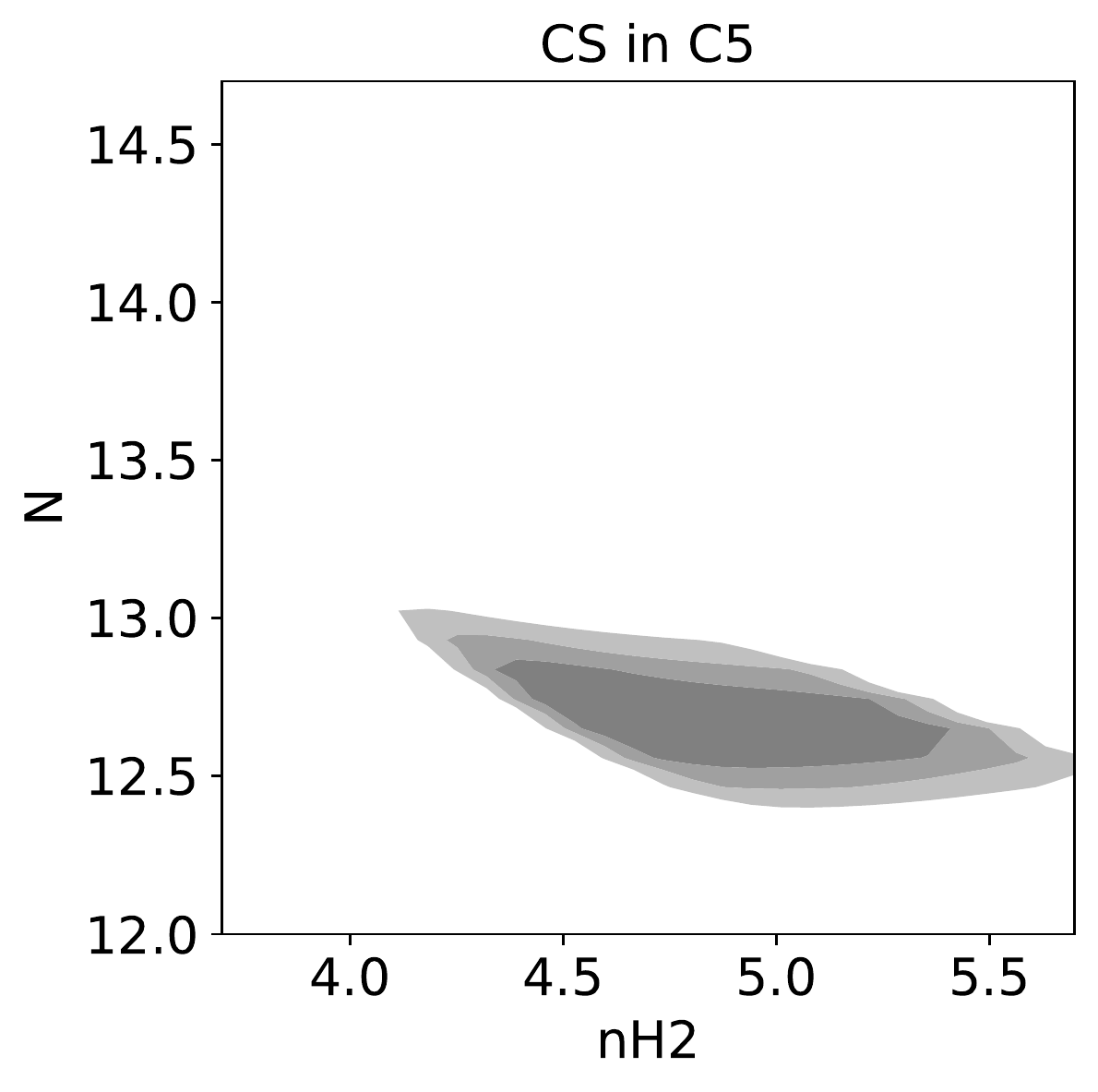}
\includegraphics[width=0.27\linewidth]{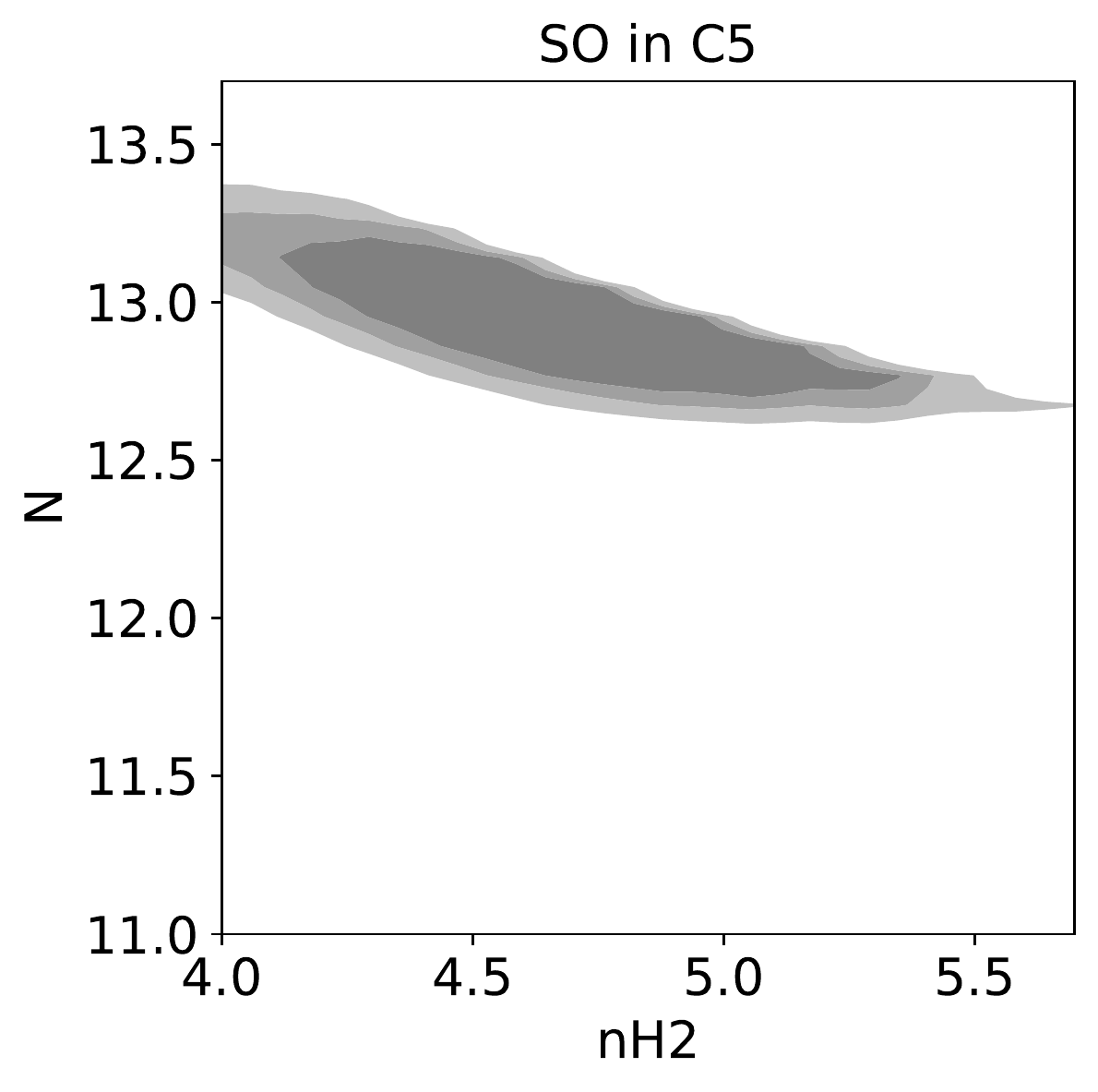}
\includegraphics[width=0.27\linewidth]{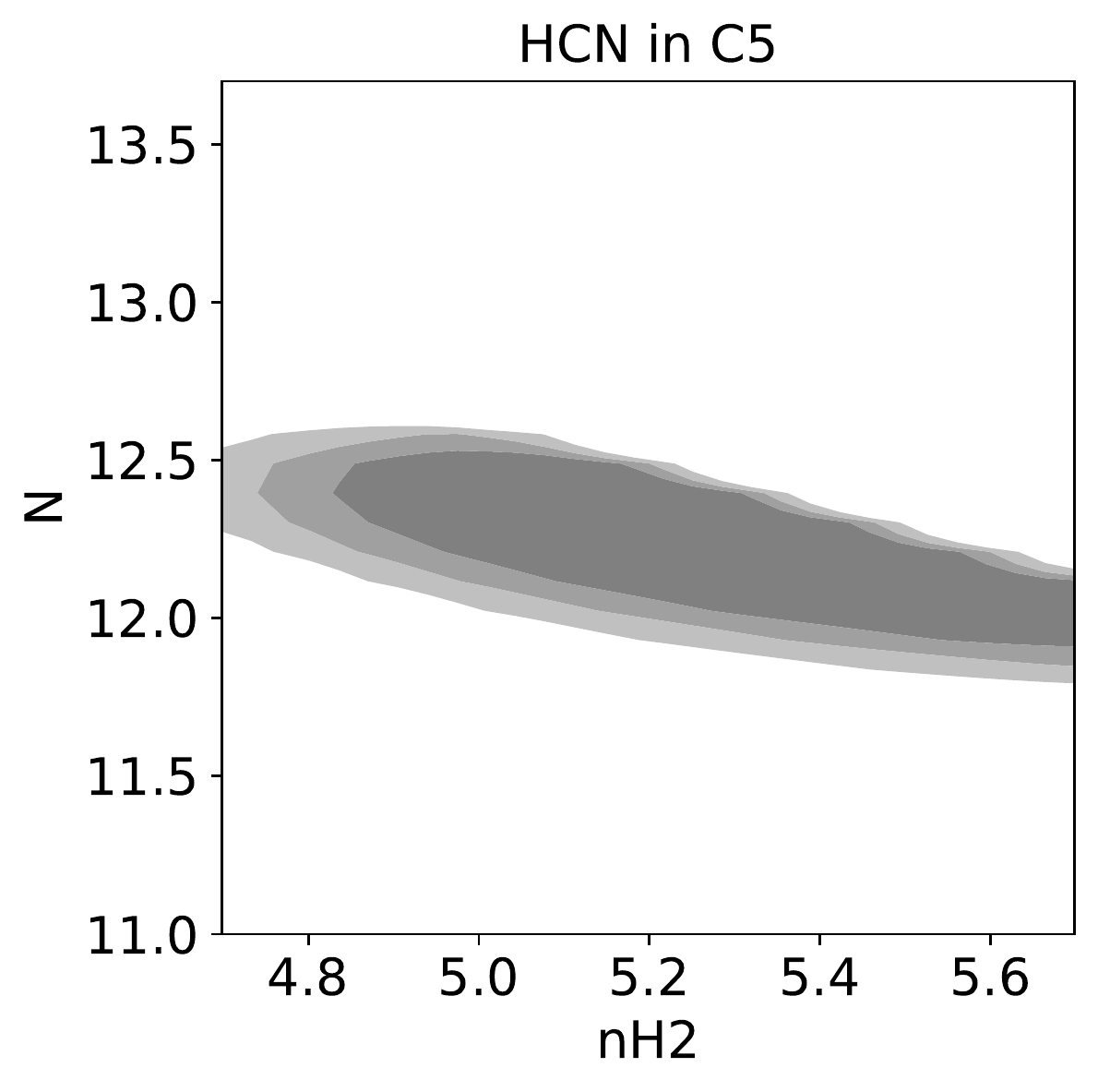}
\caption{$\chi^2$ contours ($1\sigma$, $2\sigma$, and $3\sigma$ confidence intervals) projected over the temperature axis. \label{chi2_projT}}
\end{figure*}

\begin{figure*}
\includegraphics[width=0.27\linewidth]{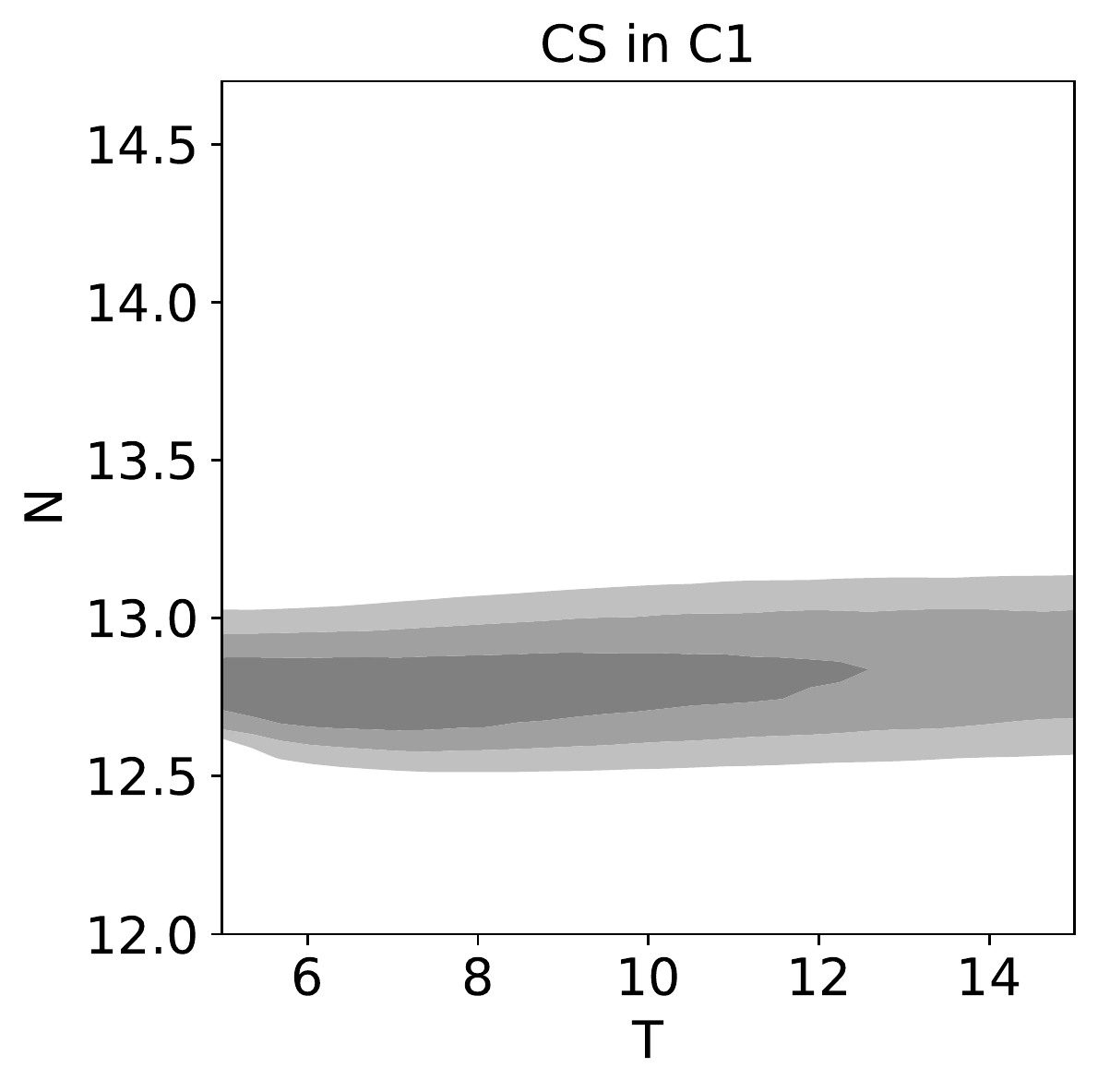}
\includegraphics[width=0.27\linewidth]{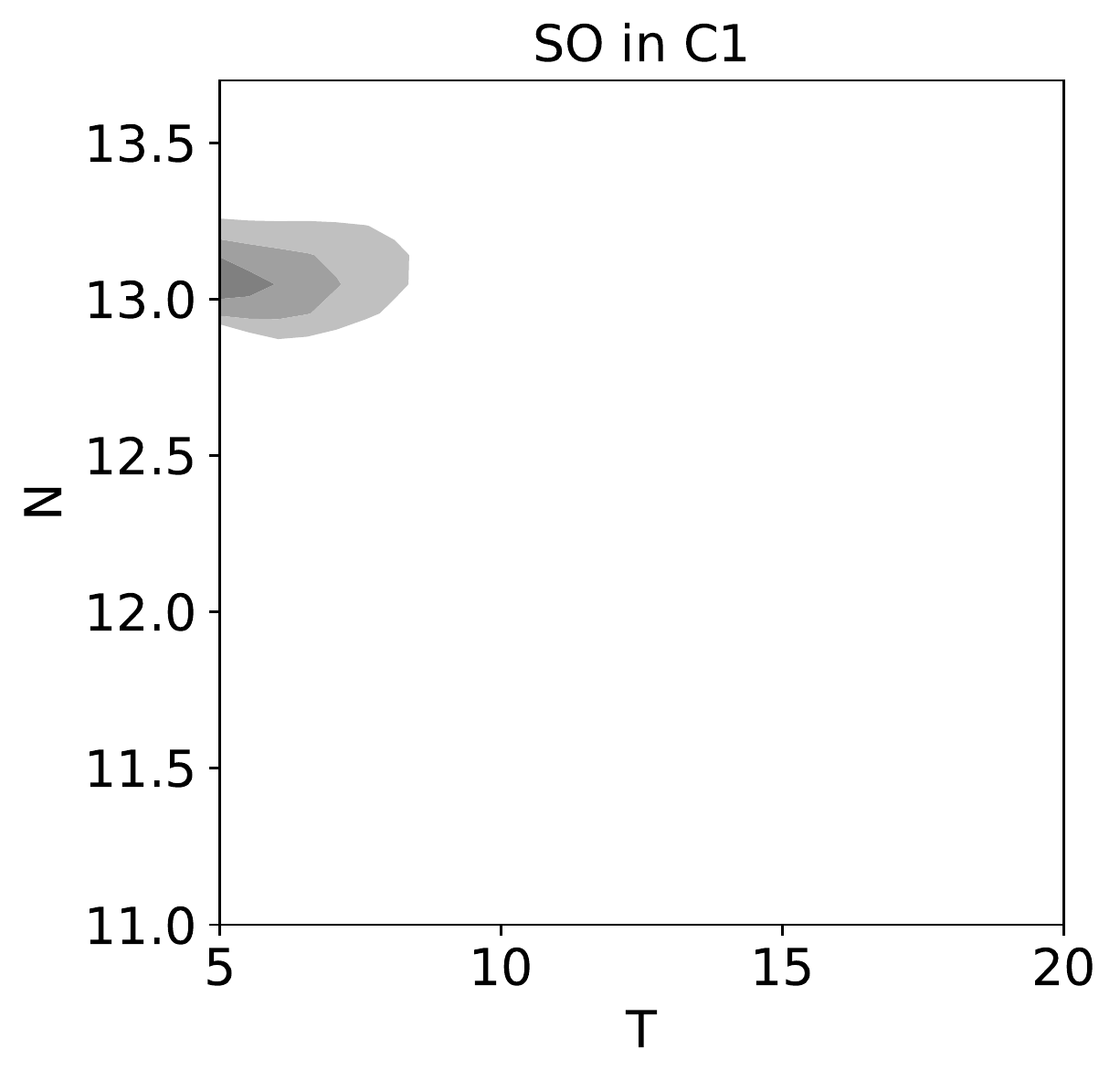}
\includegraphics[width=0.27\linewidth]{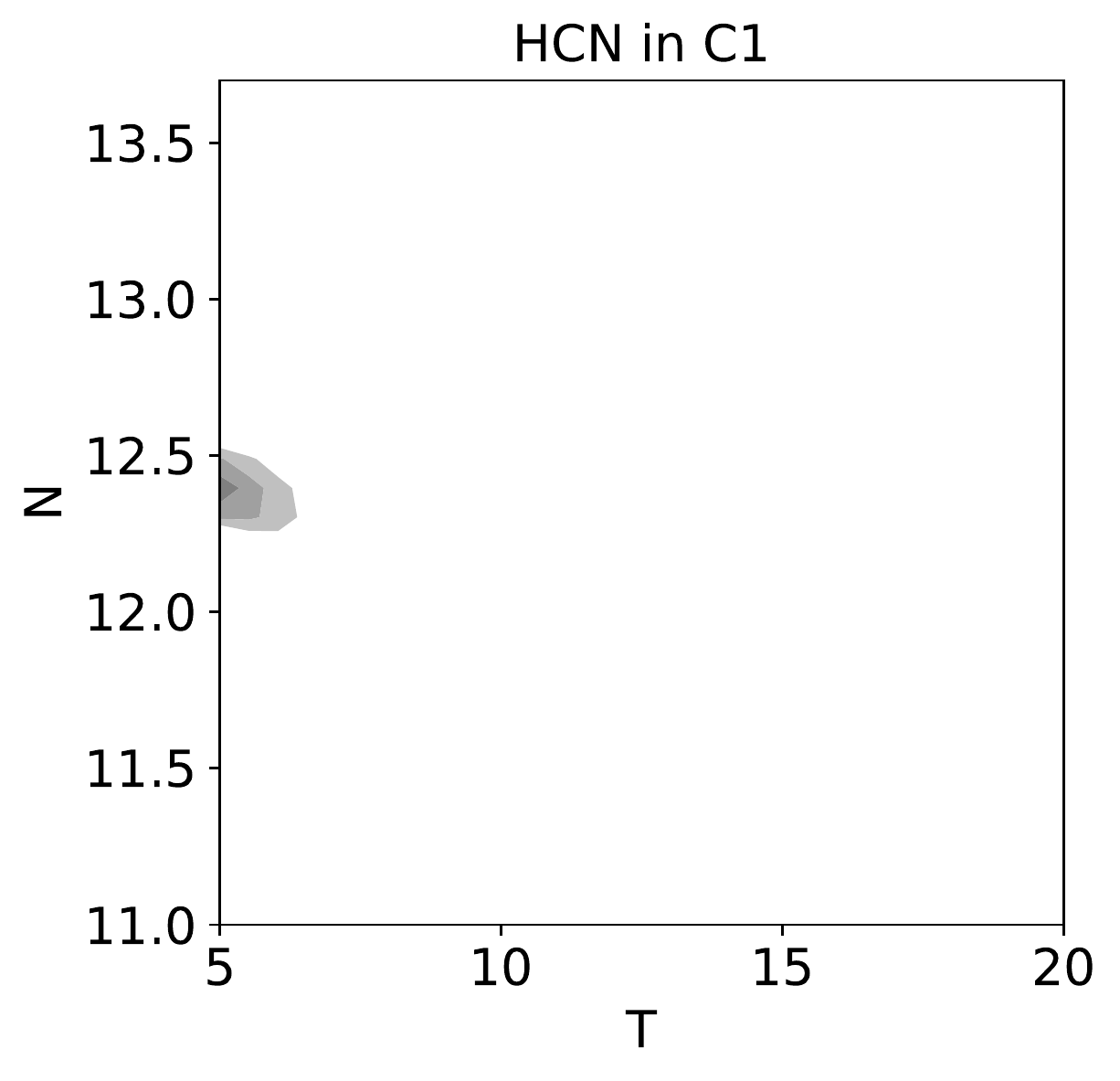}
\includegraphics[width=0.27\linewidth]{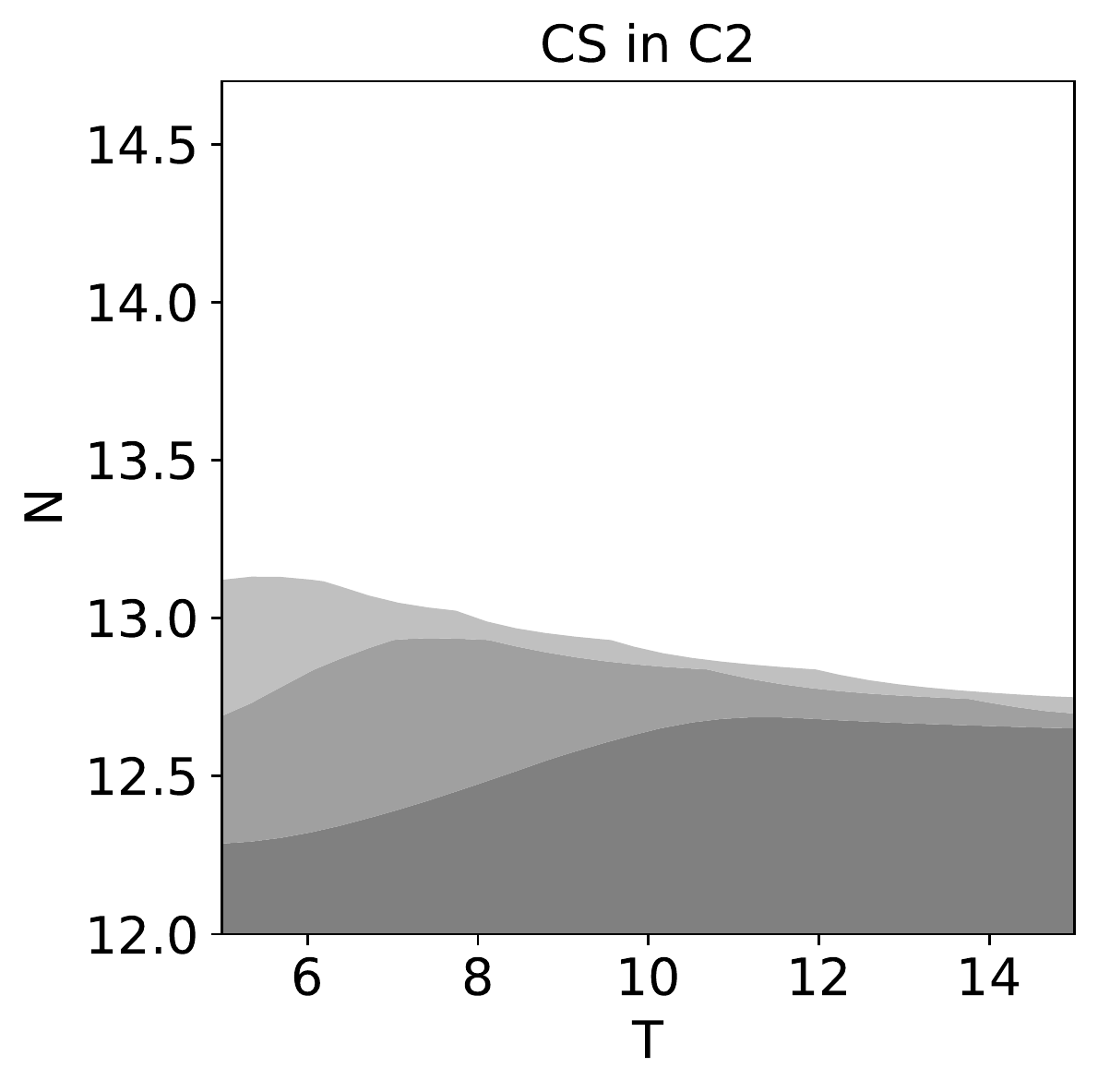}
\includegraphics[width=0.27\linewidth]{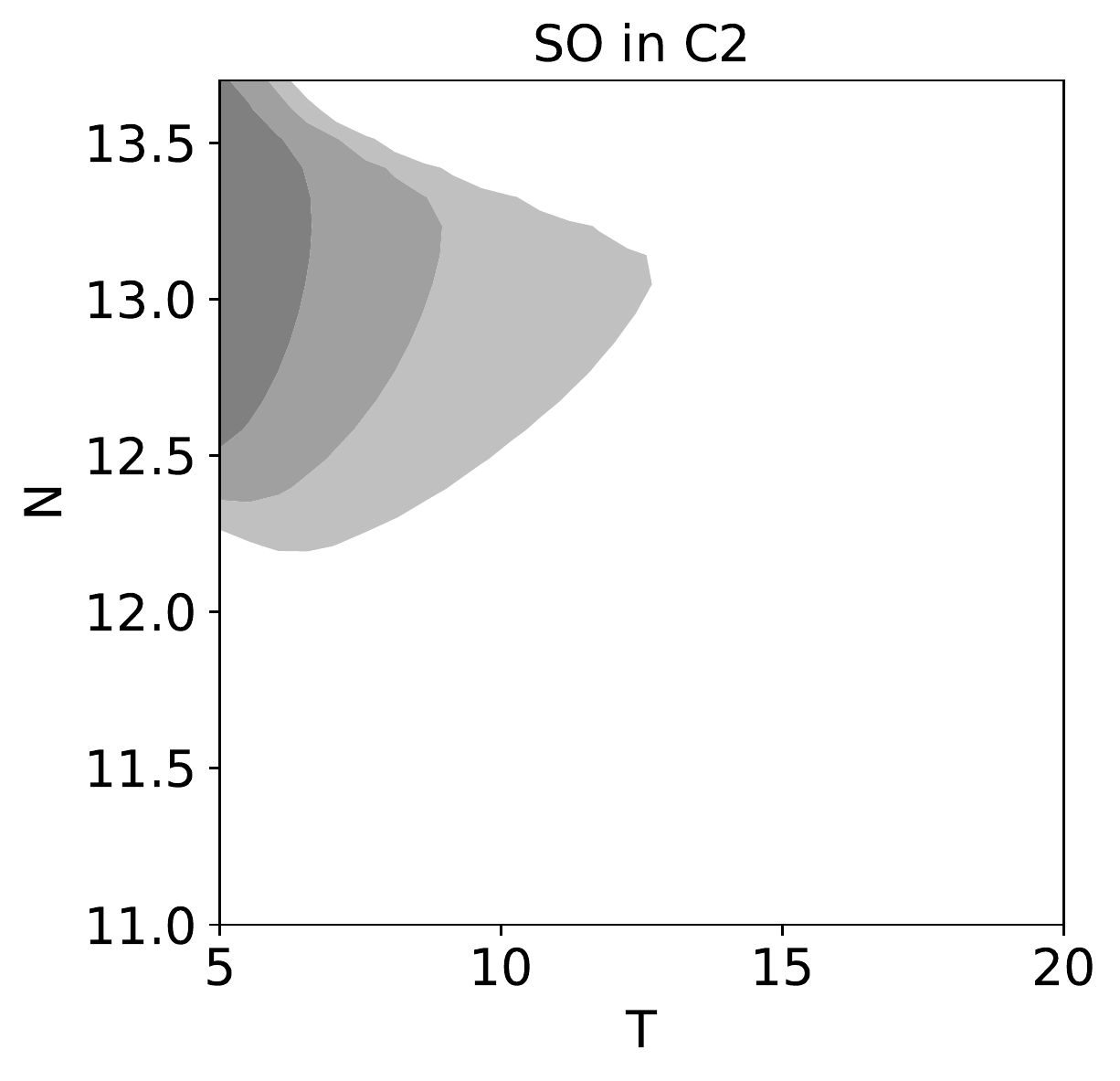}
\includegraphics[width=0.27\linewidth]{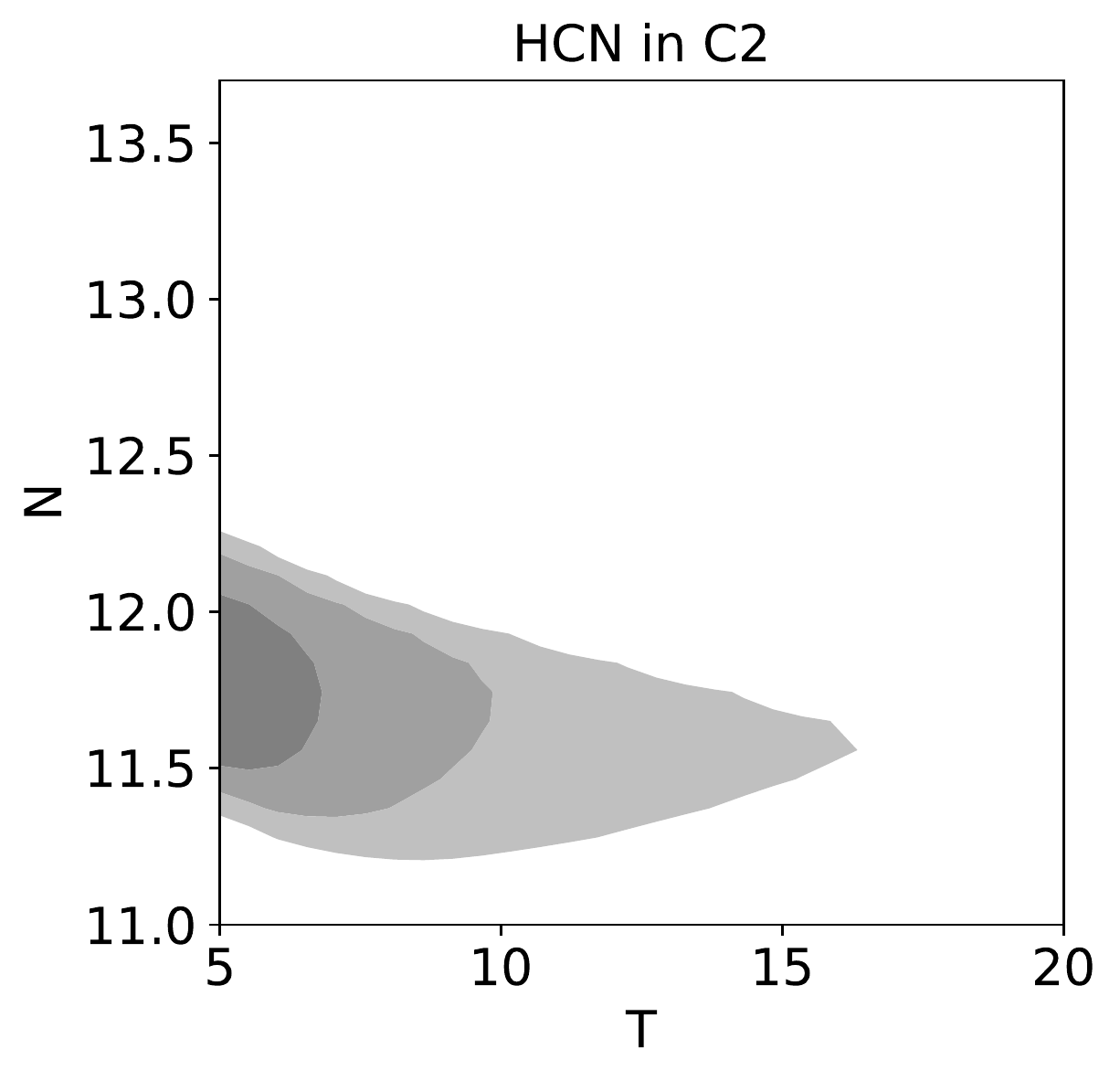}
\includegraphics[width=0.27\linewidth]{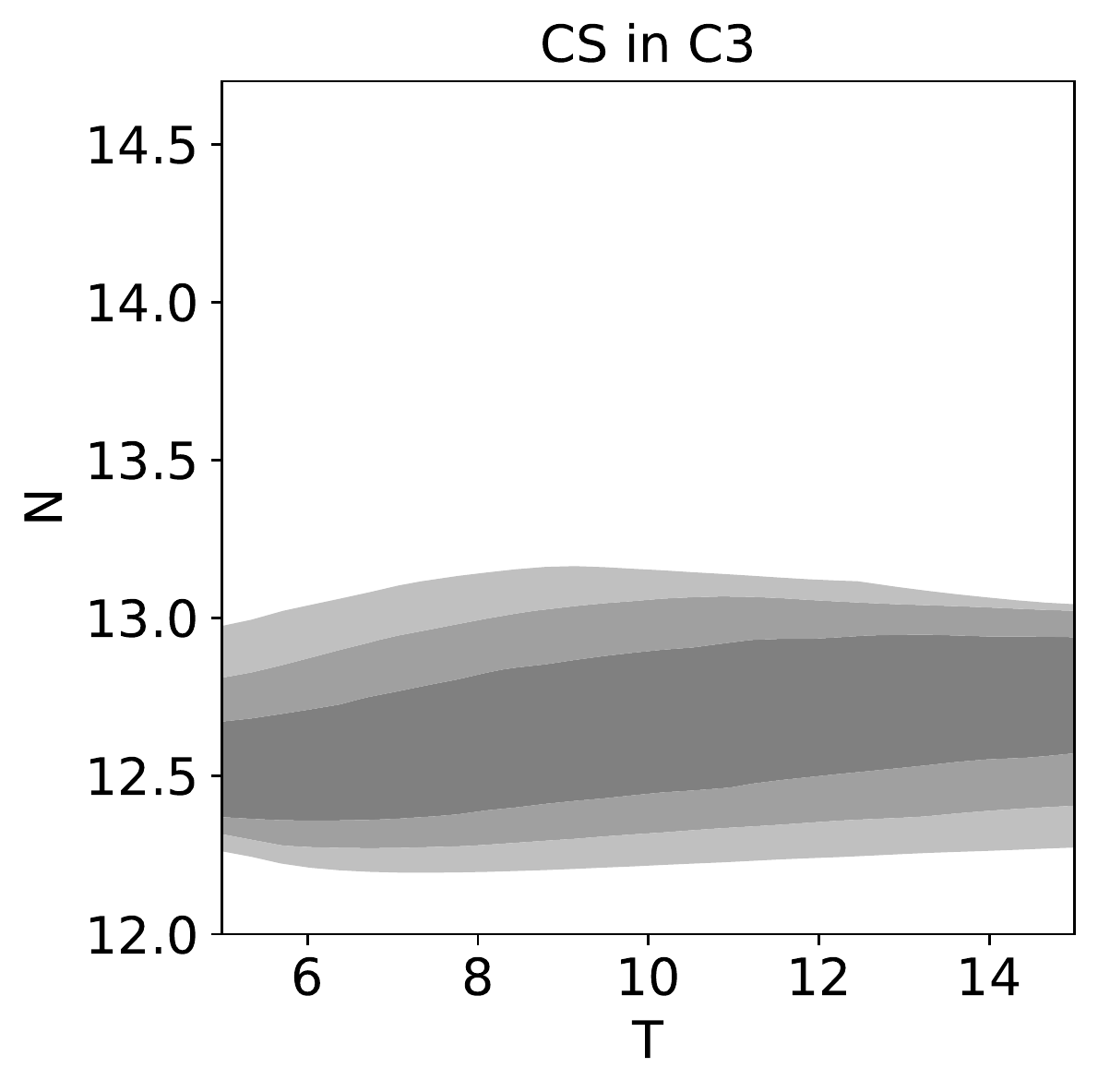}
\includegraphics[width=0.27\linewidth]{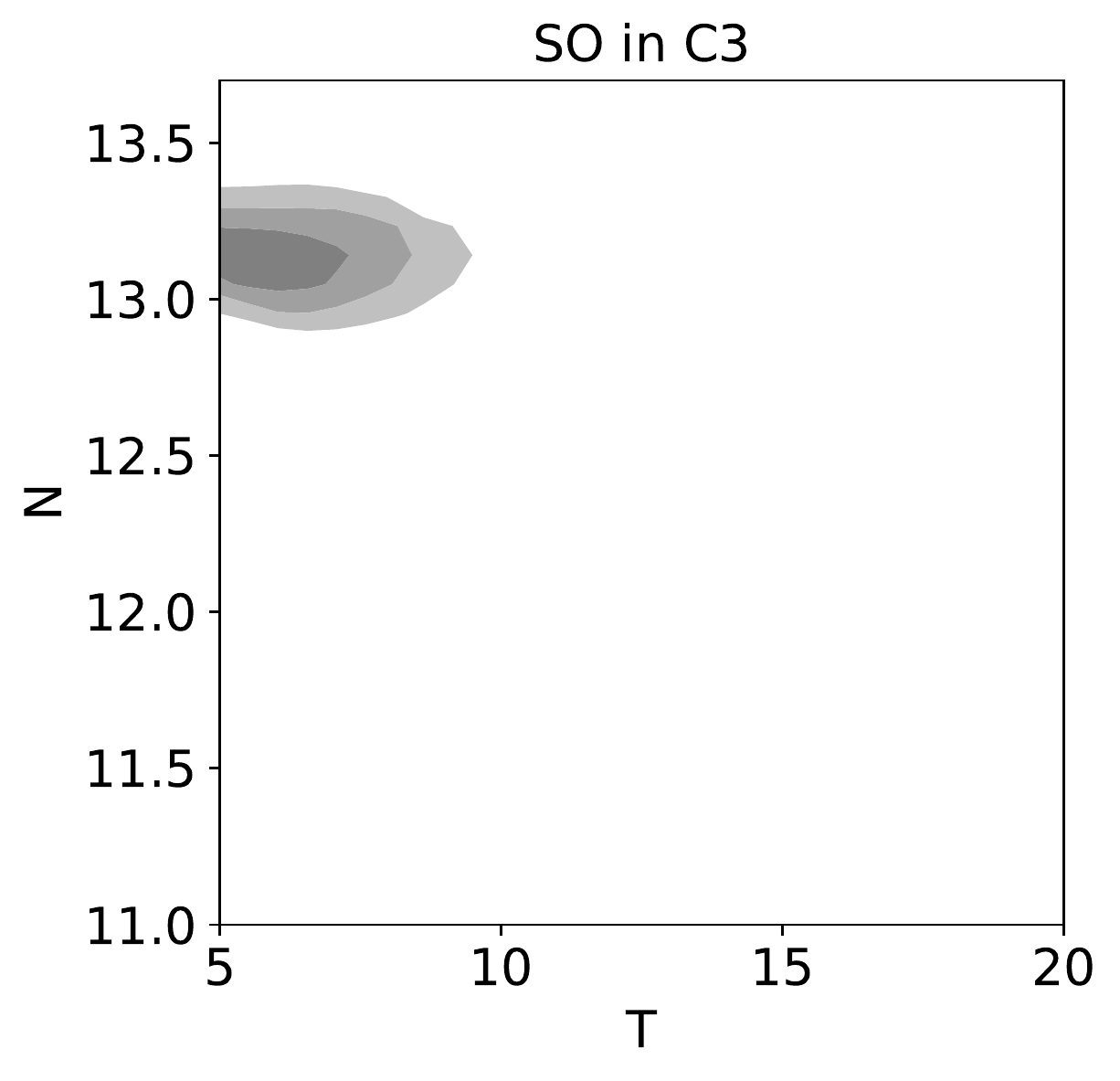}
\includegraphics[width=0.27\linewidth]{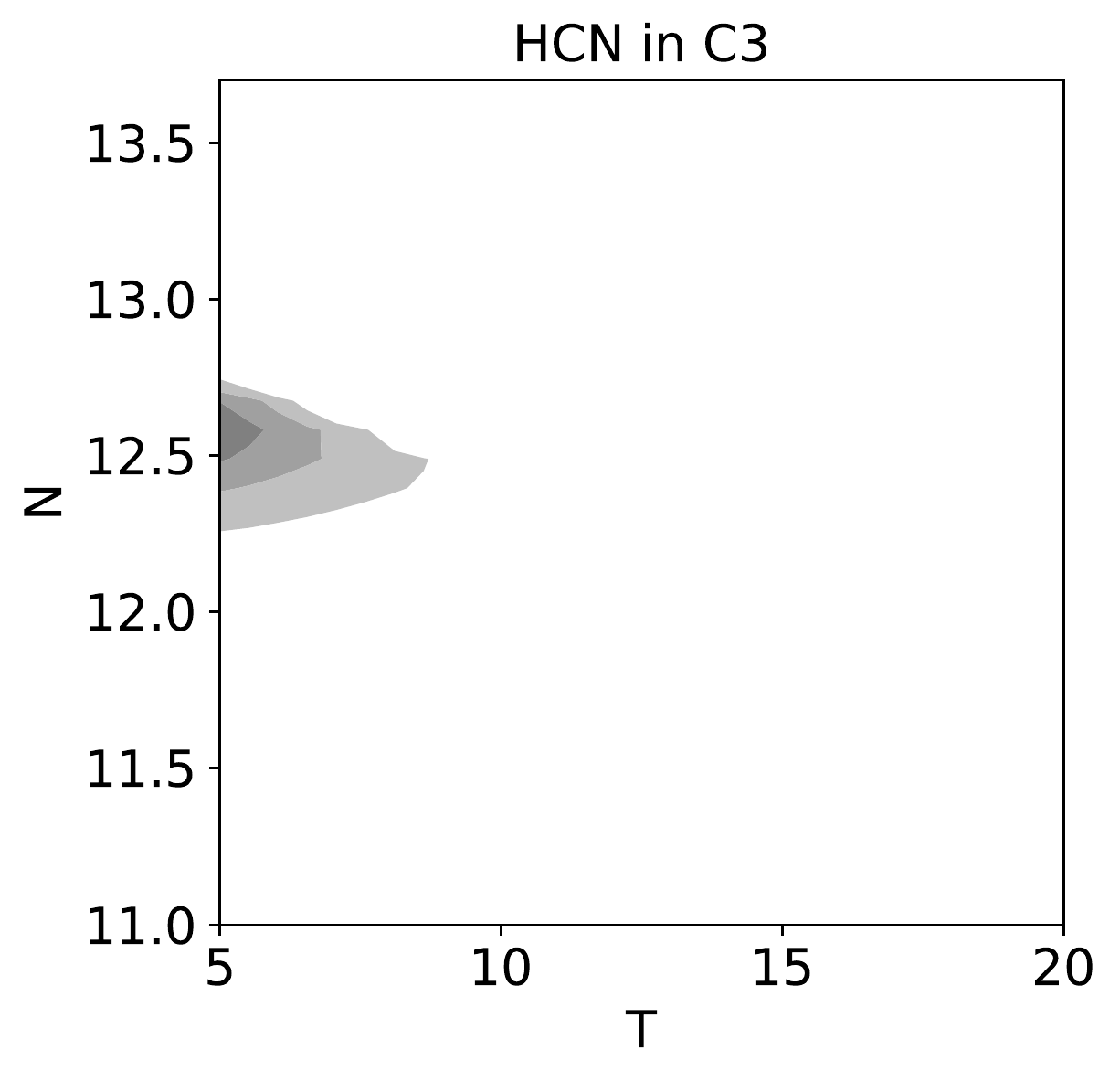}
\includegraphics[width=0.27\linewidth]{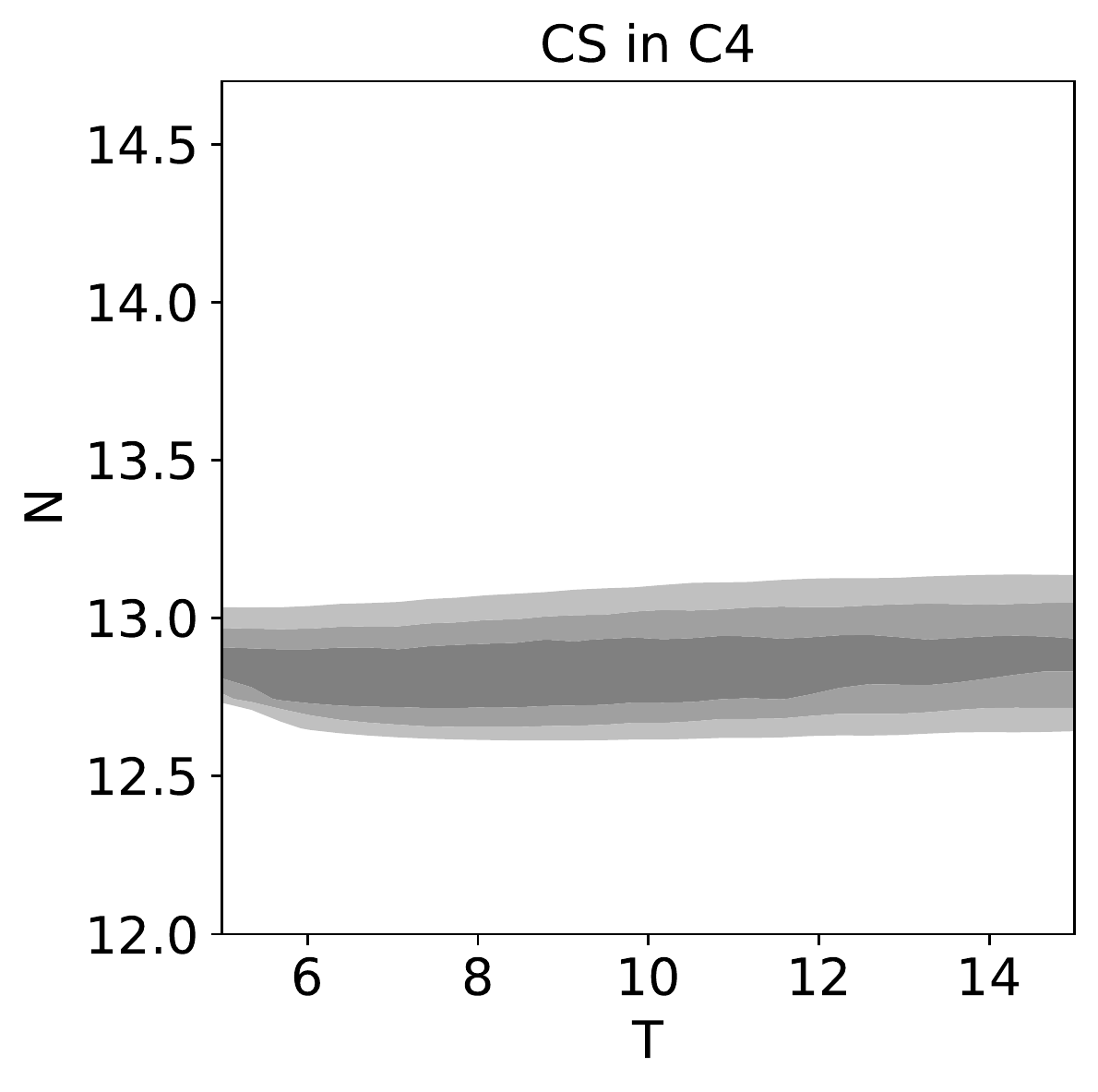}
\includegraphics[width=0.27\linewidth]{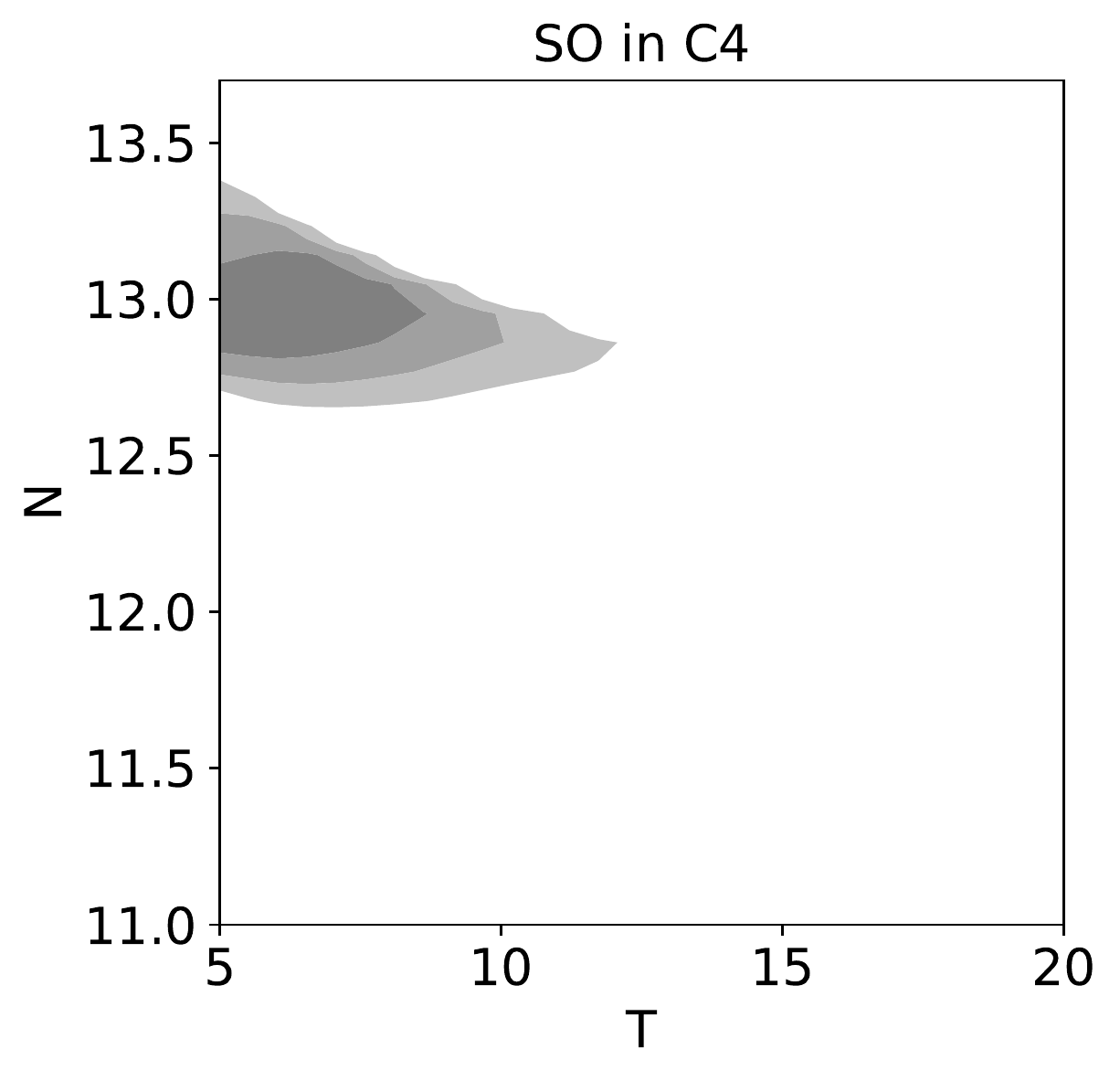}
\includegraphics[width=0.27\linewidth]{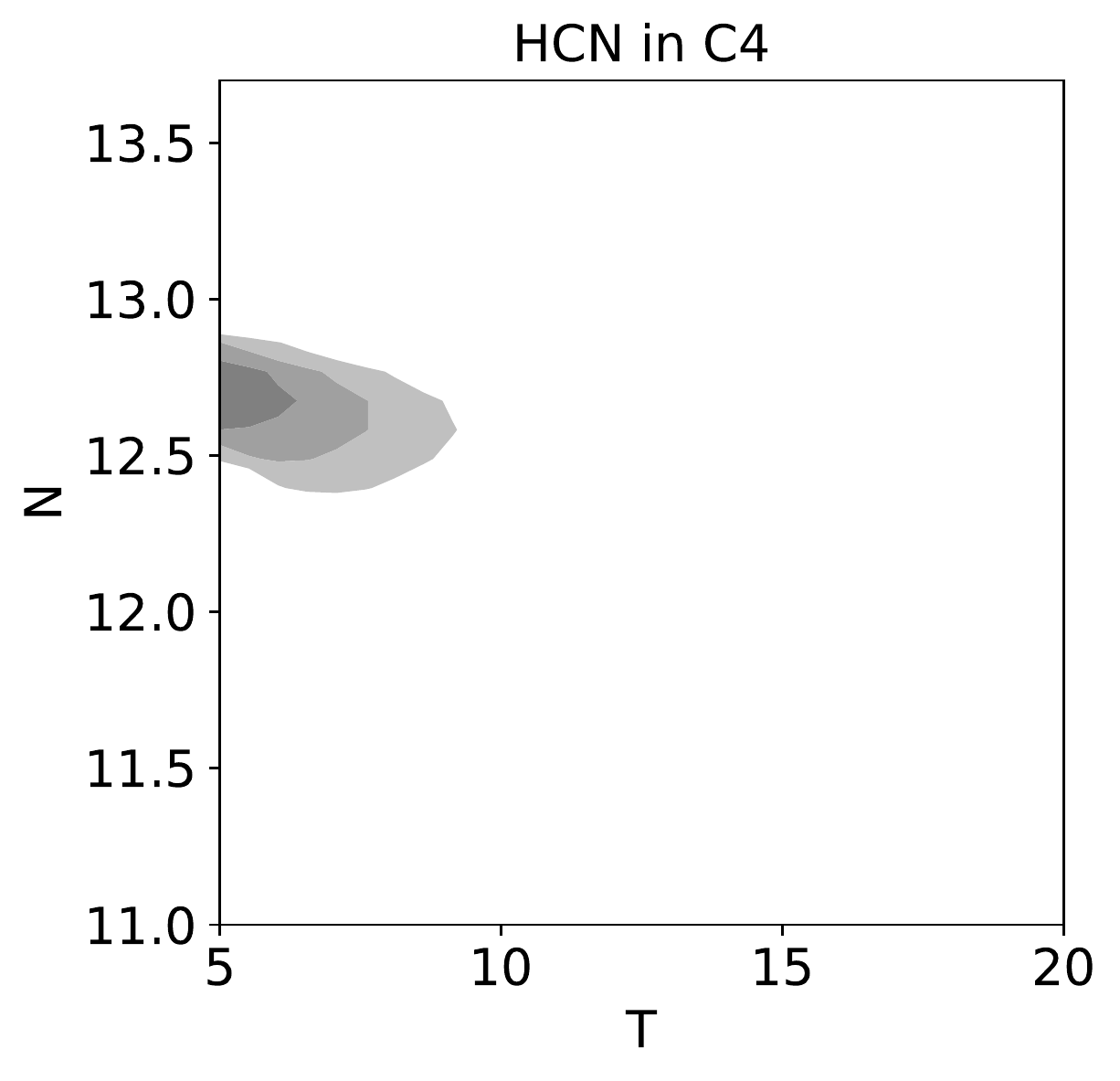}
\includegraphics[width=0.27\linewidth]{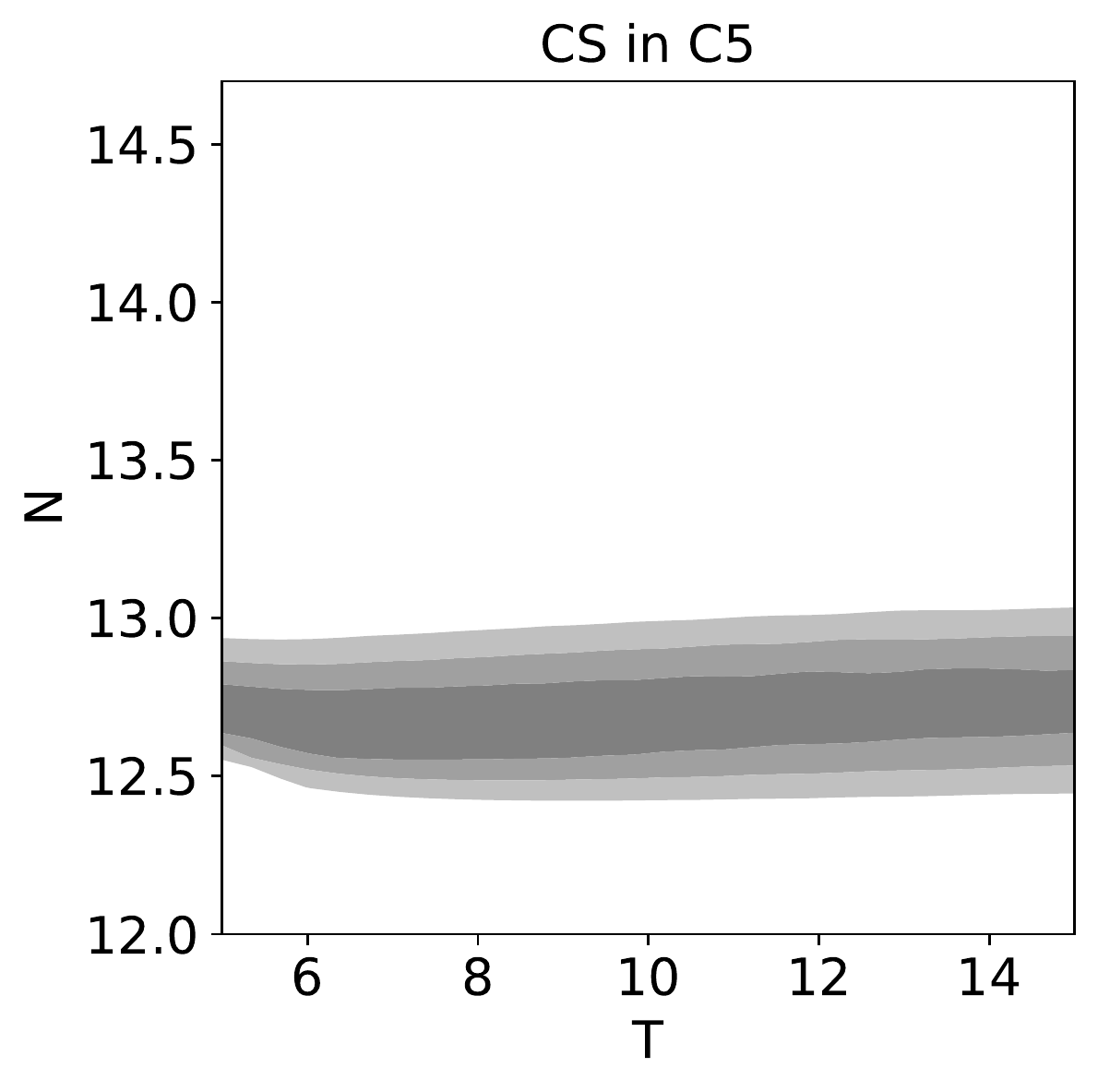}
\includegraphics[width=0.27\linewidth]{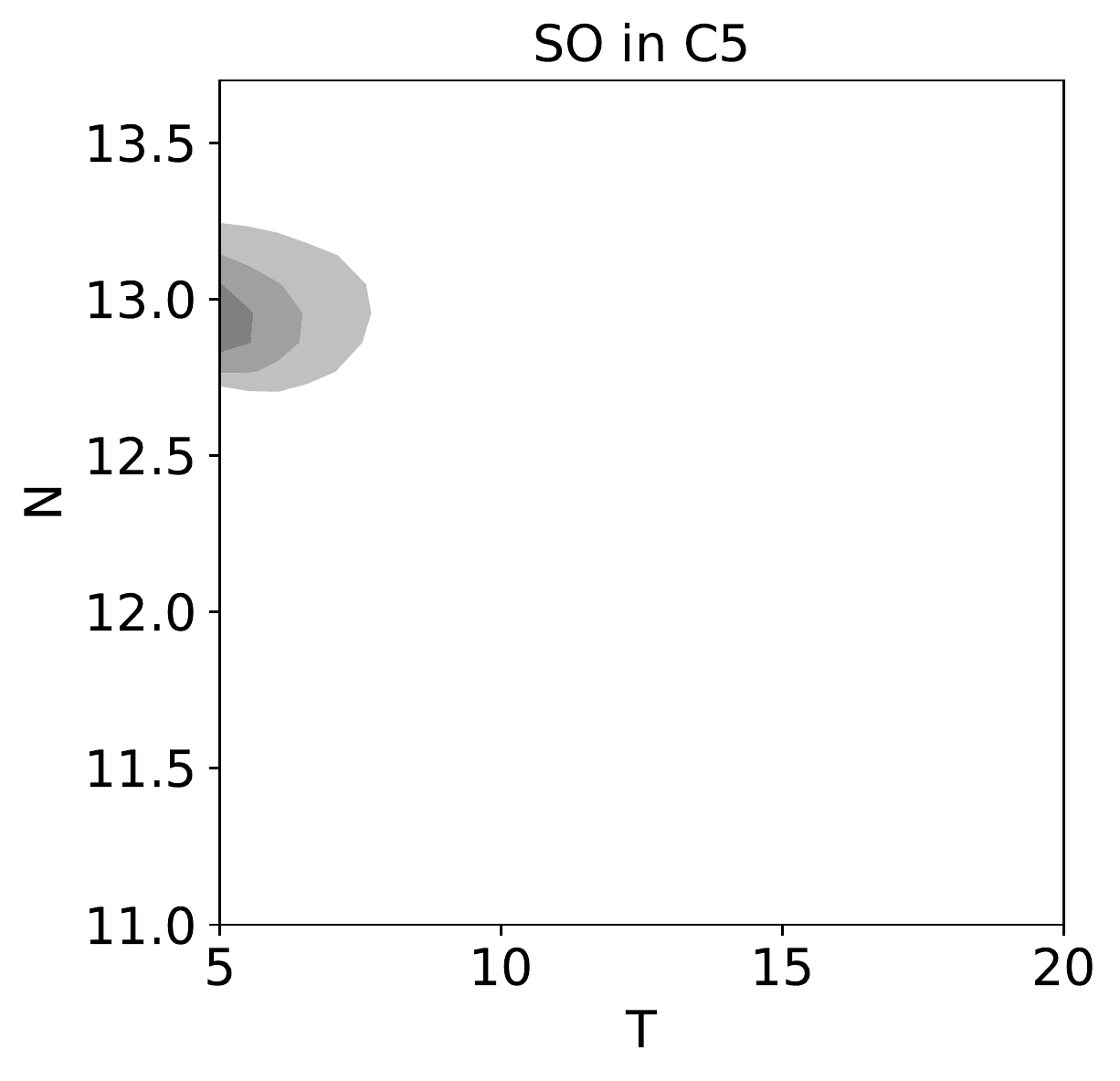}
\includegraphics[width=0.27\linewidth]{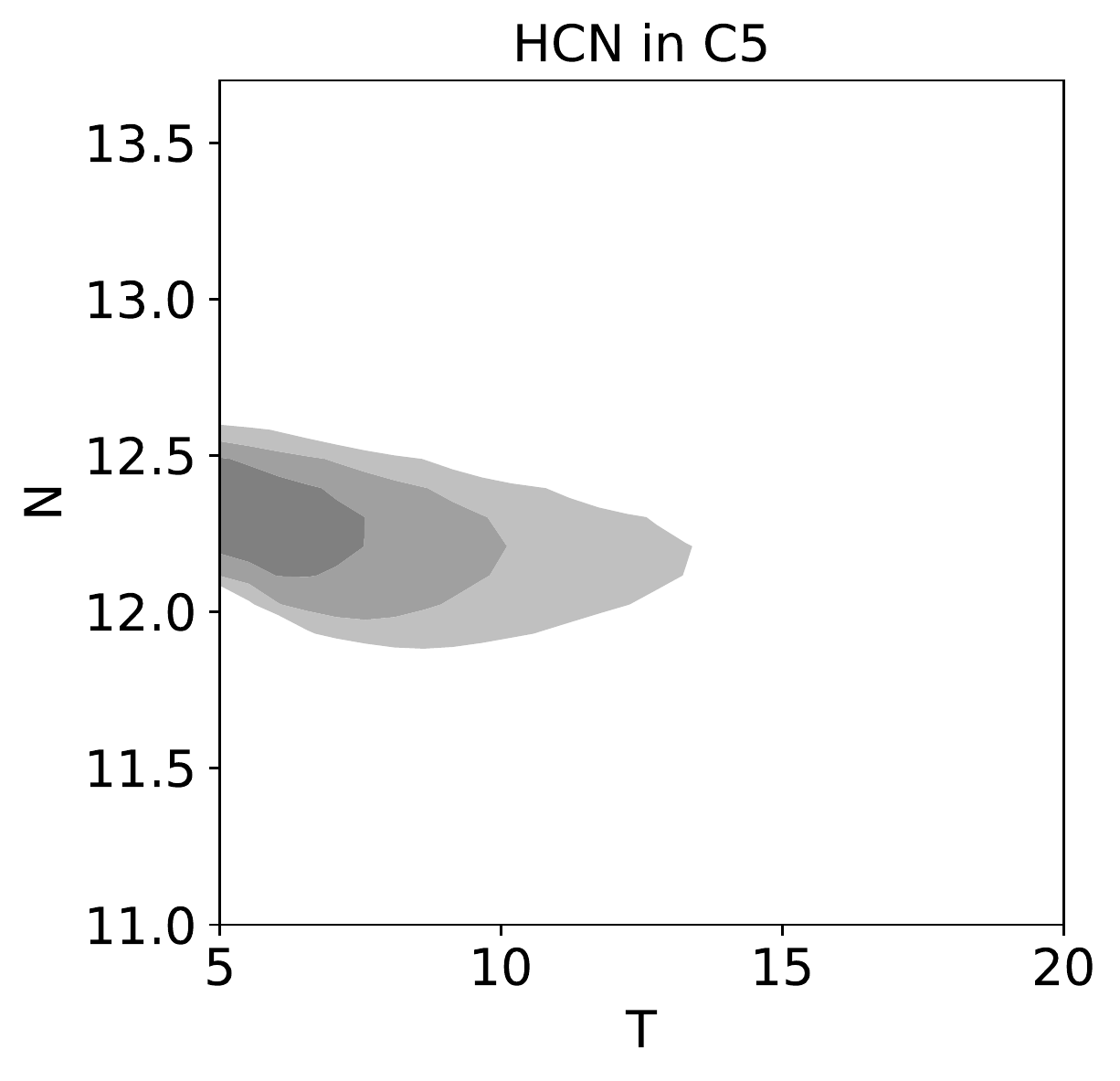}
\caption{$\chi^2$ contours ($1\sigma$, $2\sigma$, and $3\sigma$ confidence intervals) projected over the gas density axis. \label{chi2_projnH2}}
\end{figure*}

\section{Model predictions for CO and HCO$^+$ abundances as a function of time}\label{appendixB}

\begin{figure*}[htbp]
\centering
\includegraphics[width=0.48\linewidth]{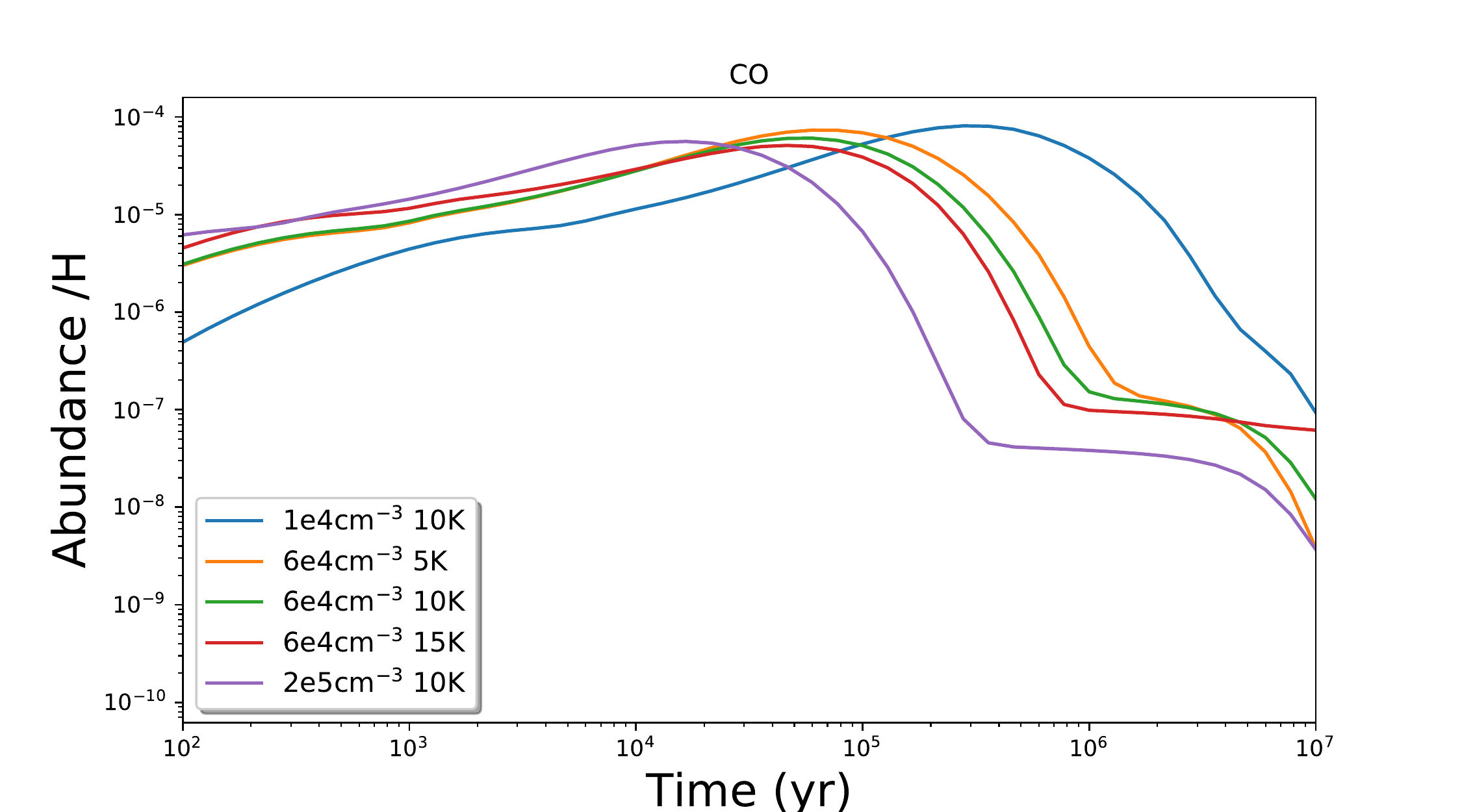}
\includegraphics[width=0.48\linewidth]{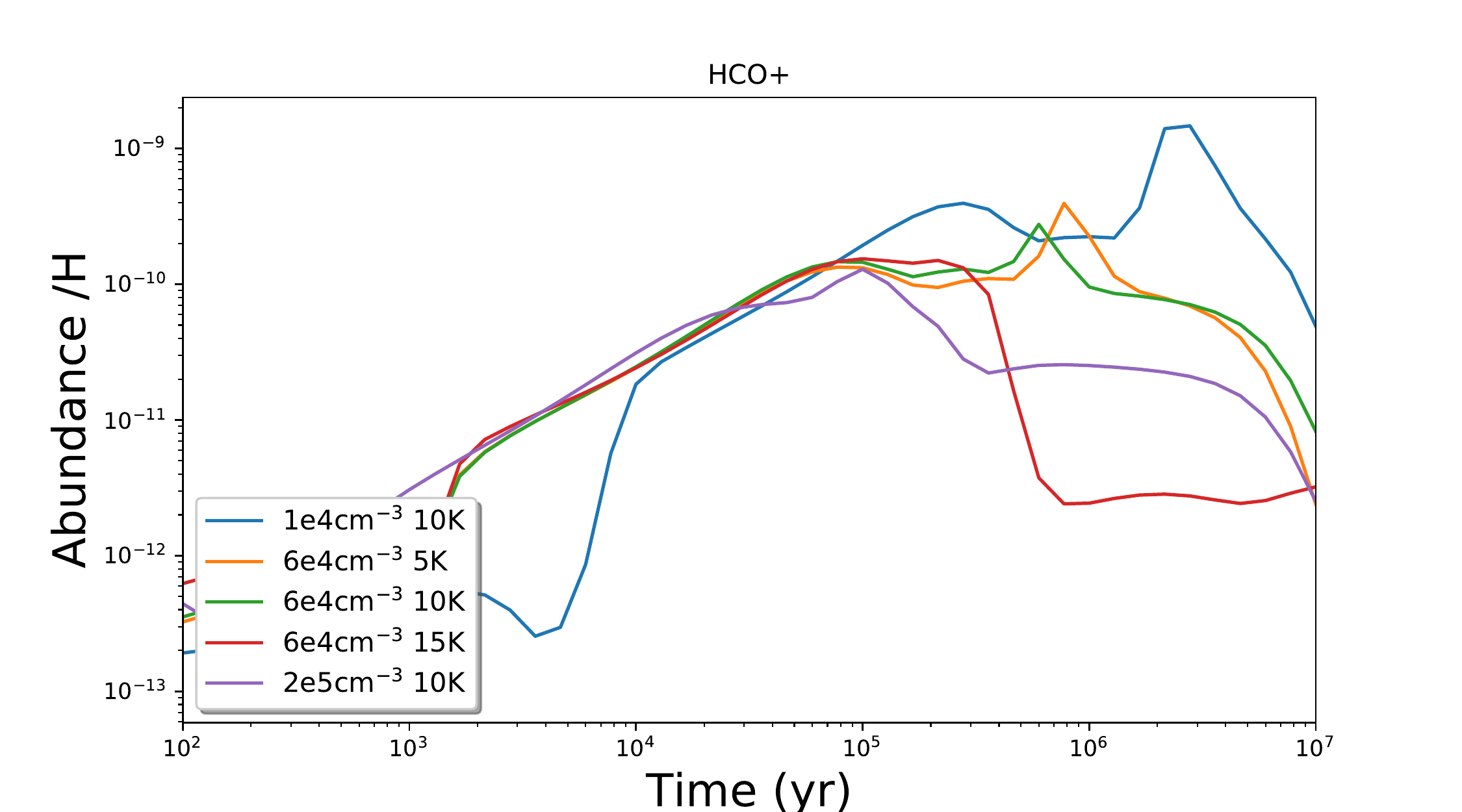}
\caption{Abundance of CO and HCO$^+$ as a function of time using different sets of physical conditions. \label{CO_HCOp_ab}}
\end{figure*}

\section{Percentage of reproduced species with a higher cosmic-ray ionisation rate and different elemental abundances.}\label{appendixC}

\begin{figure*}[htbp]
\centering
\includegraphics[width=0.48\linewidth]{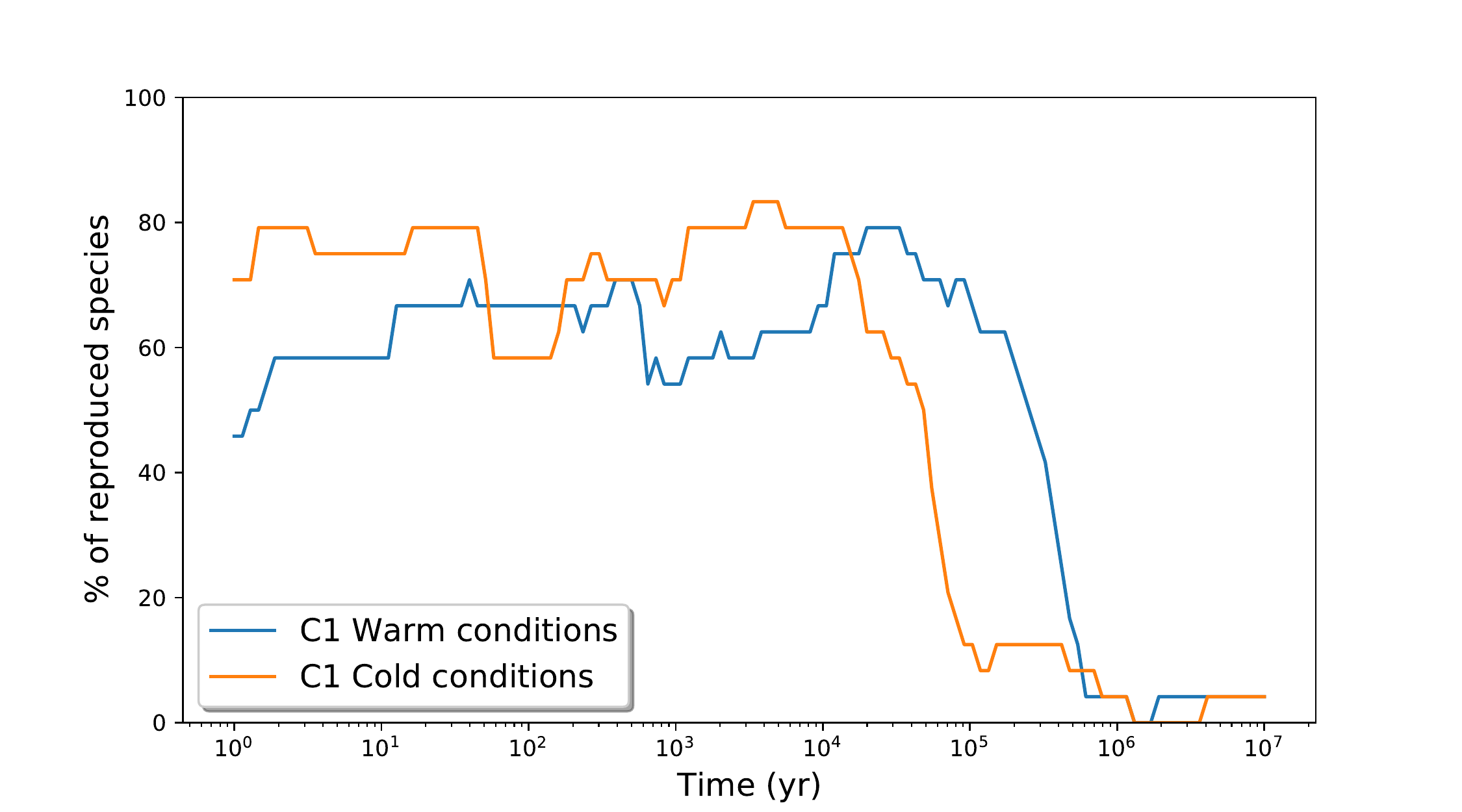}
\includegraphics[width=0.48\linewidth]{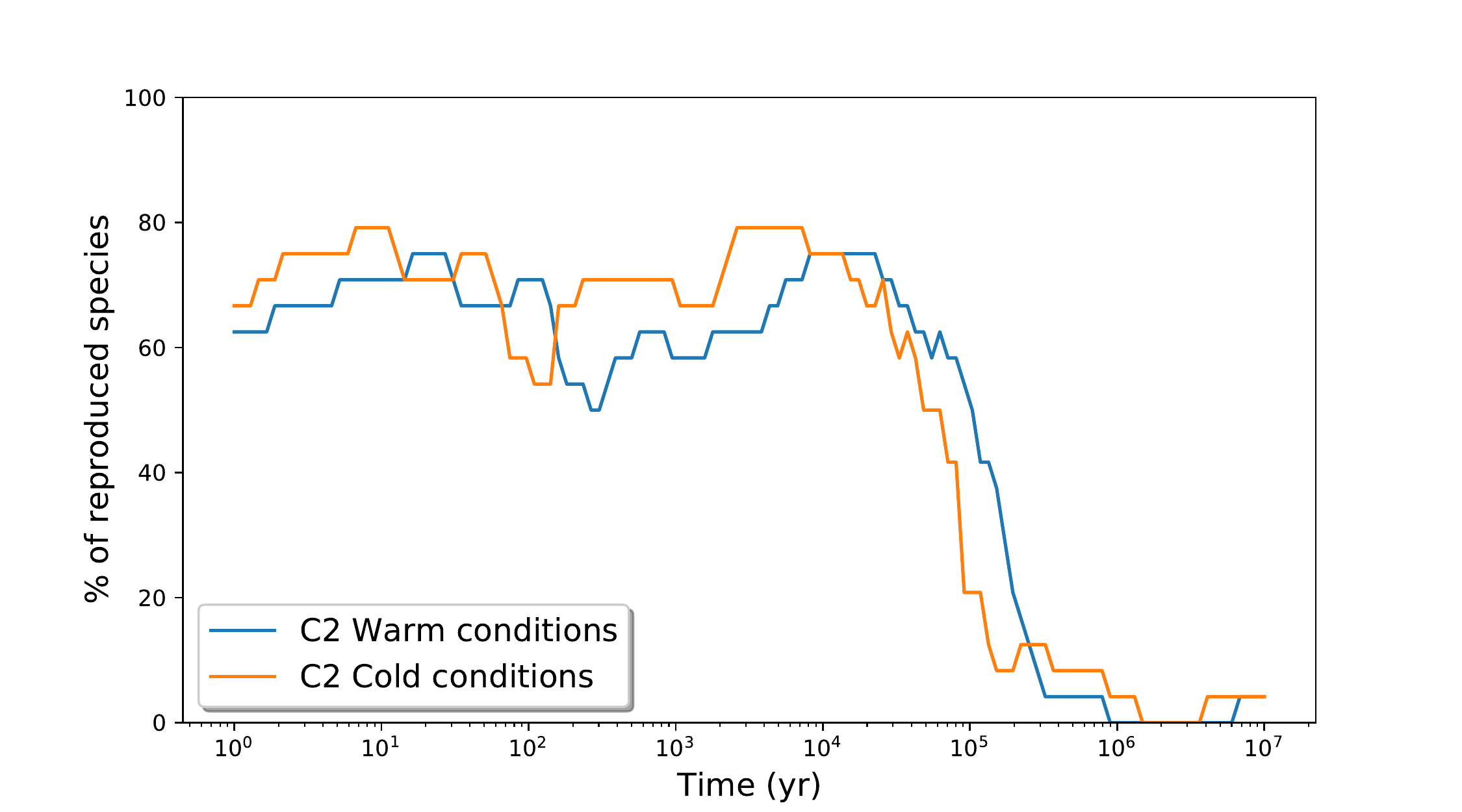}
\includegraphics[width=0.48\linewidth]{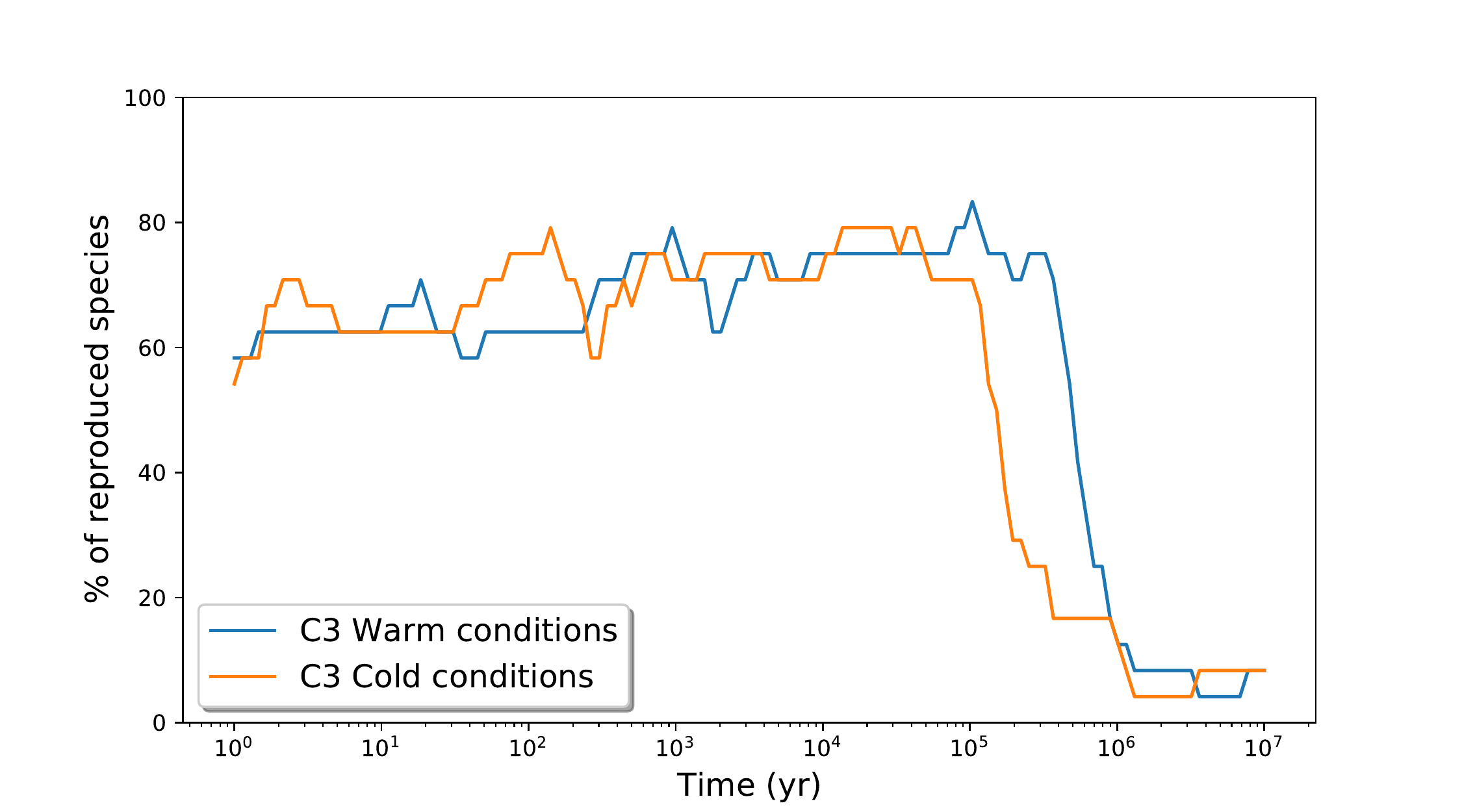}
\includegraphics[width=0.48\linewidth]{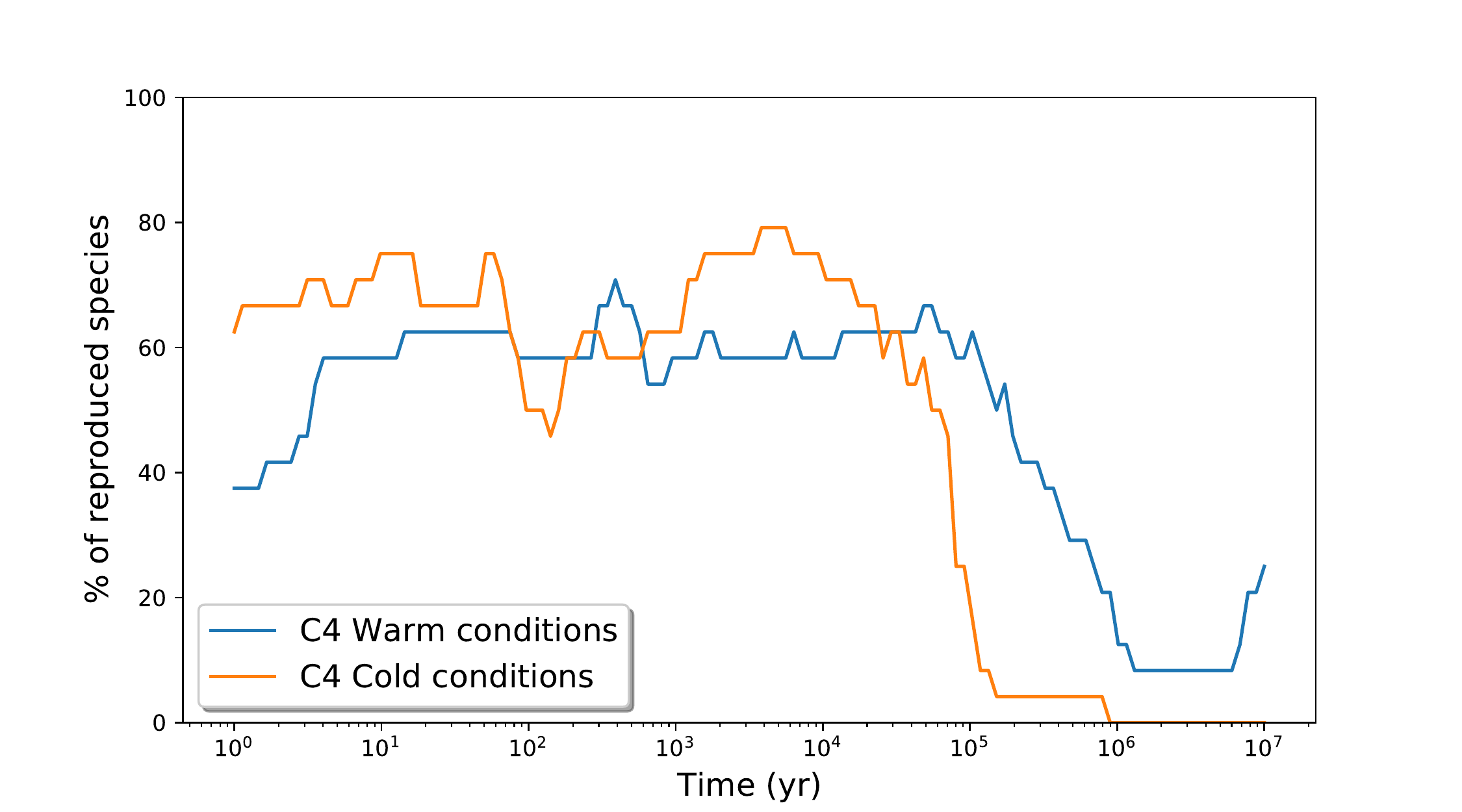}
\includegraphics[width=0.48\linewidth]{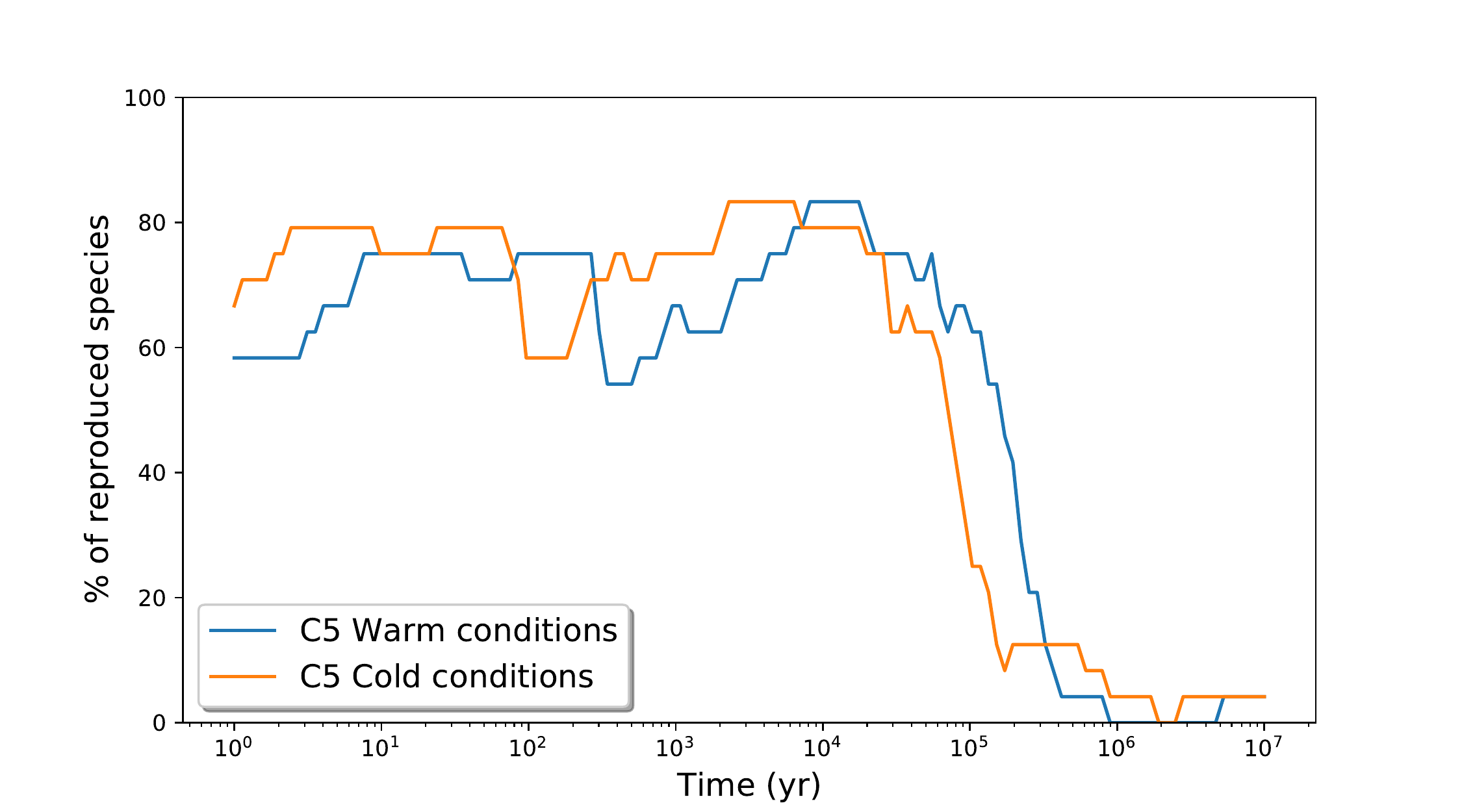}
\includegraphics[width=0.48\linewidth]{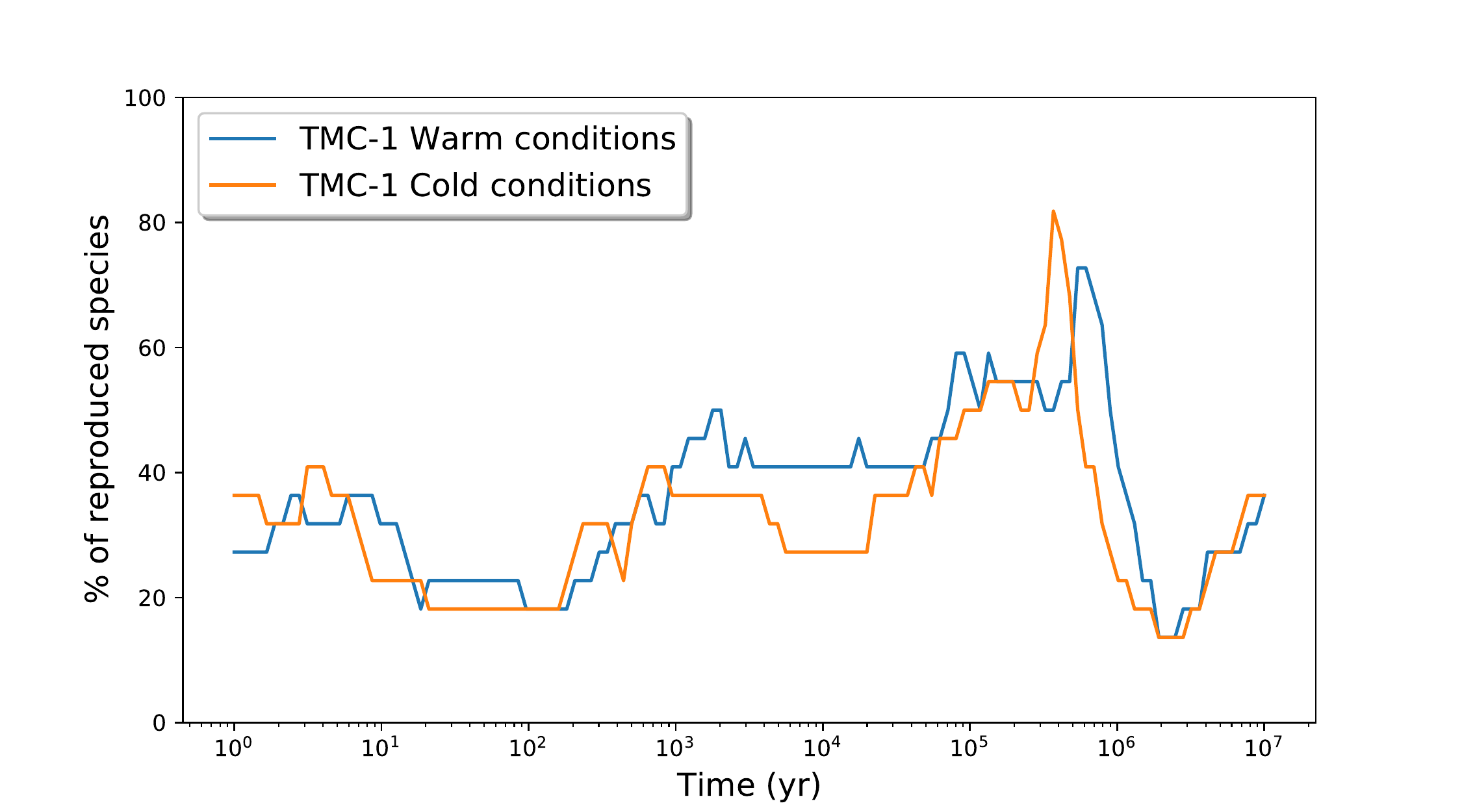}
\caption{Percentage of species reproduced by the different models for each source as a function of time for a cosmic-ray ionization rate of $6\times 10^{-17}$~s$^{-1}$. The labels 'warm conditions' and 'cold conditions' refer to the set of physical conditions as listed in Table~\ref{models}. \label{agreement_highZ}}
\end{figure*}

\begin{figure*}[htbp]
\centering
\includegraphics[width=0.48\linewidth]{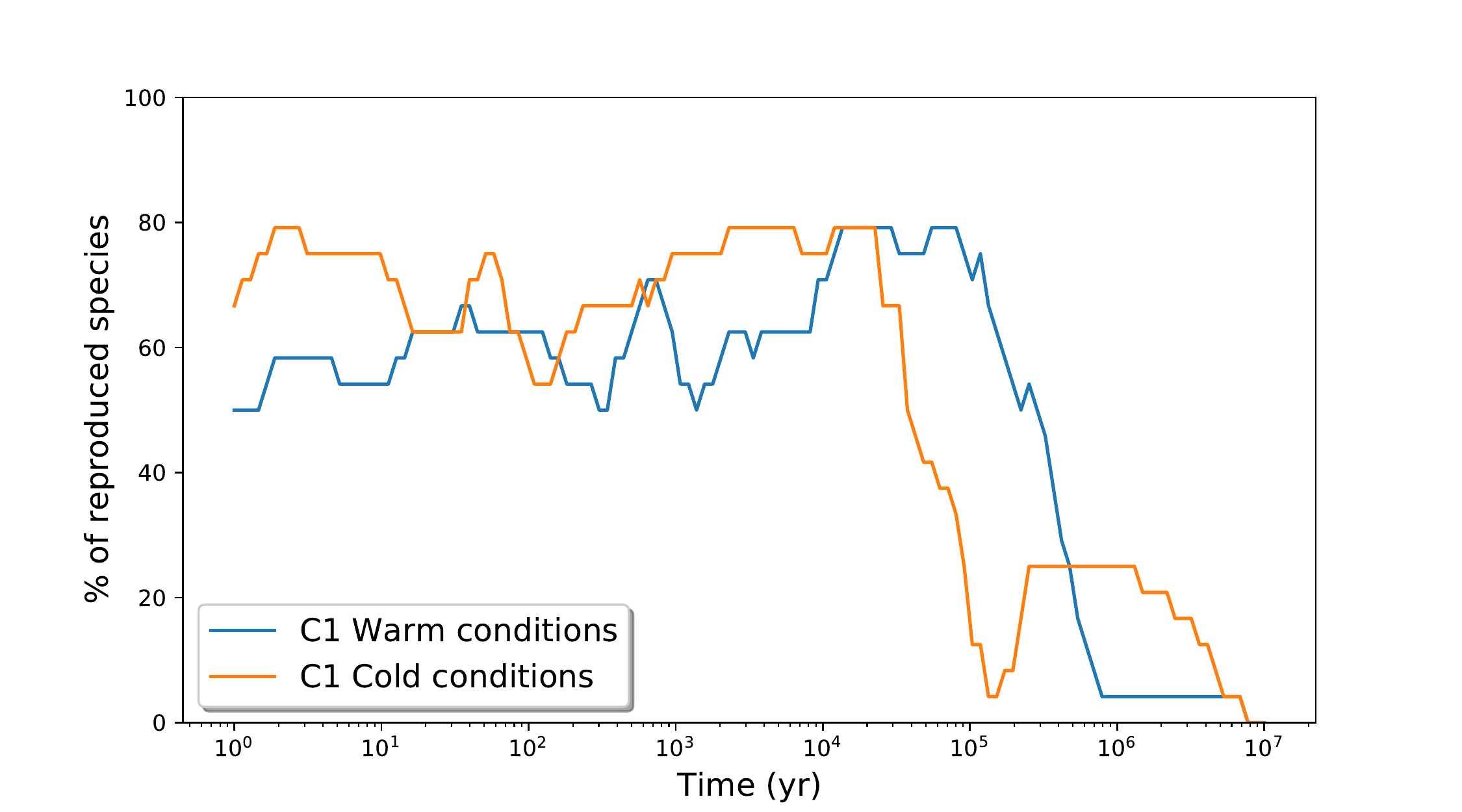}
\includegraphics[width=0.48\linewidth]{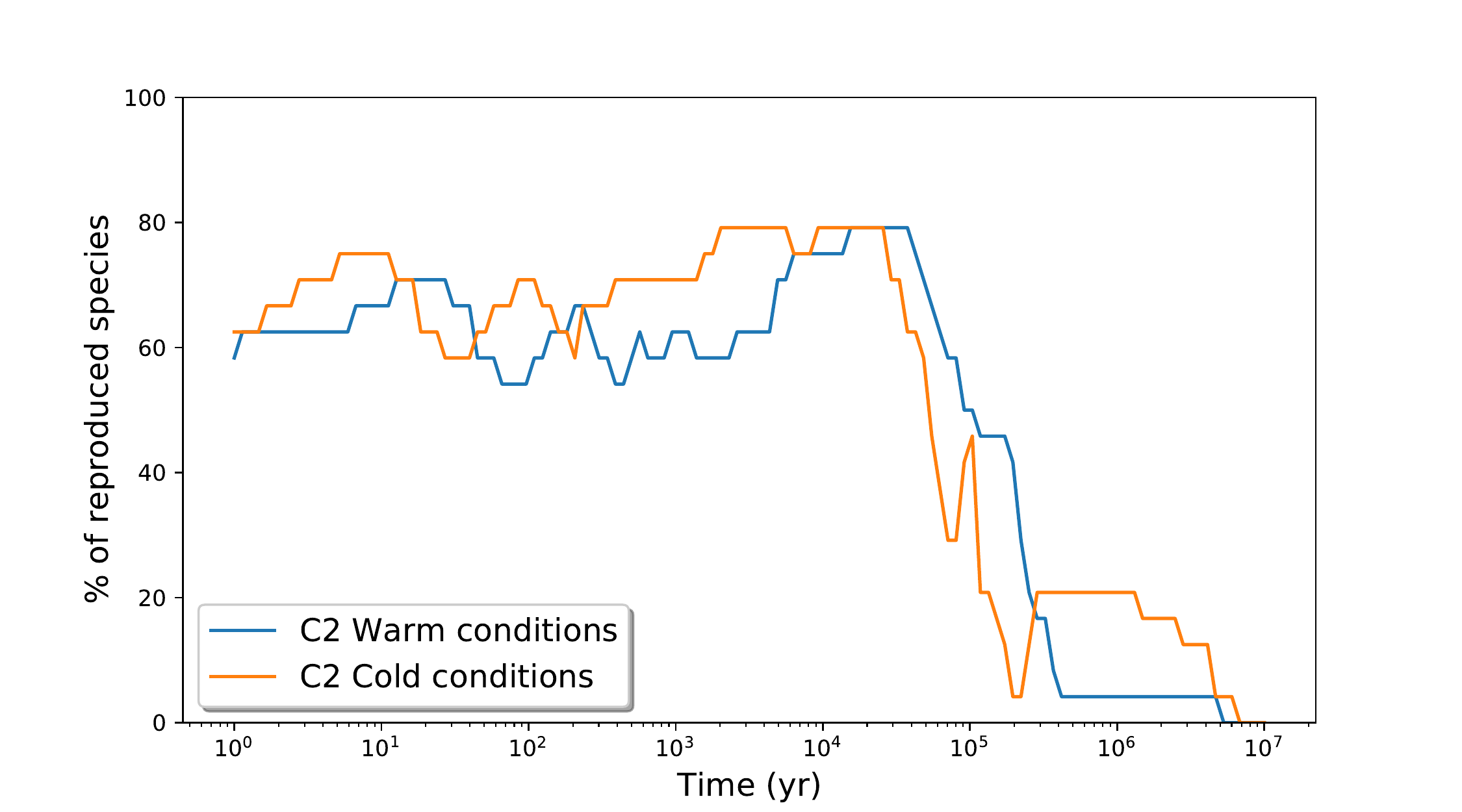}
\includegraphics[width=0.48\linewidth]{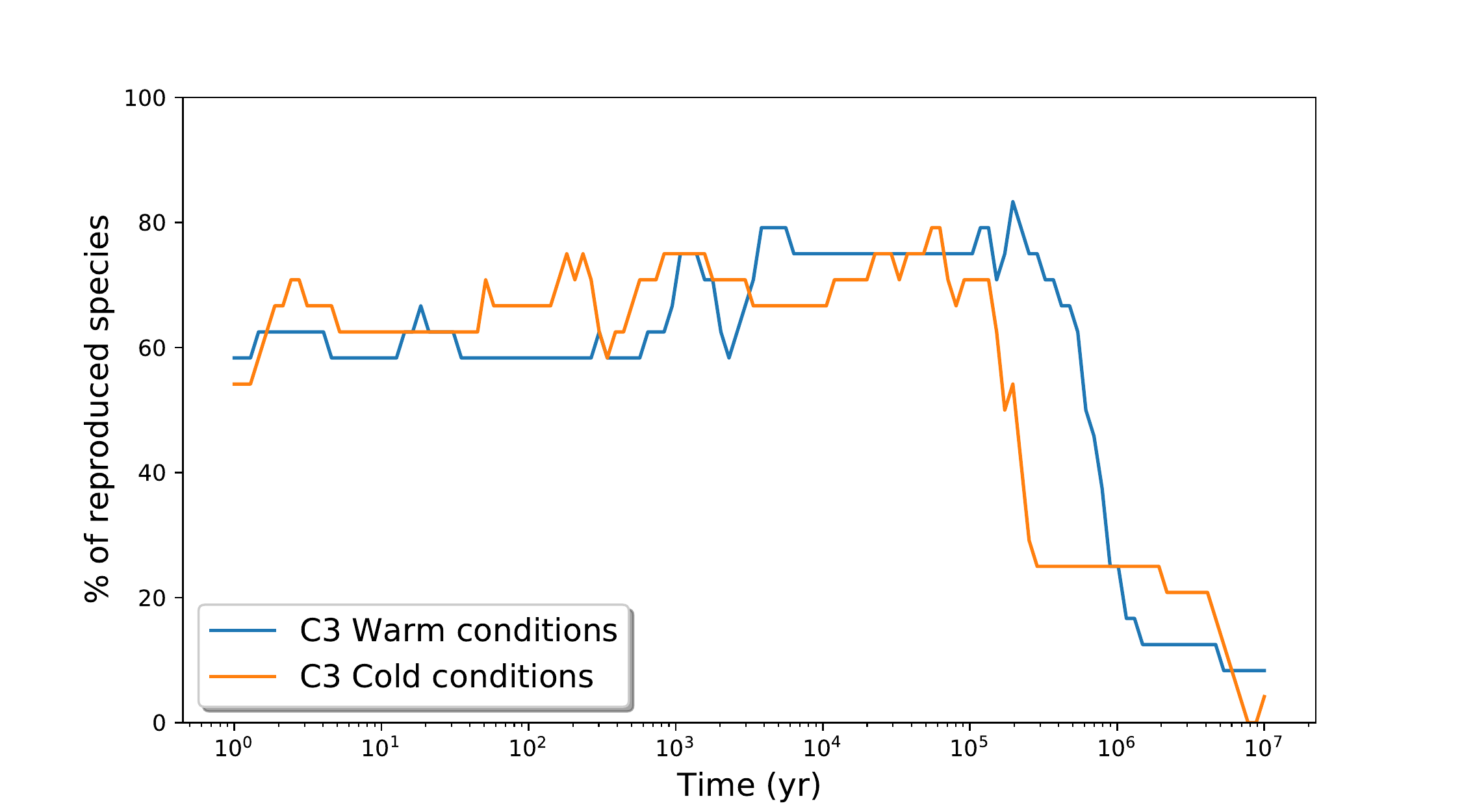}
\includegraphics[width=0.48\linewidth]{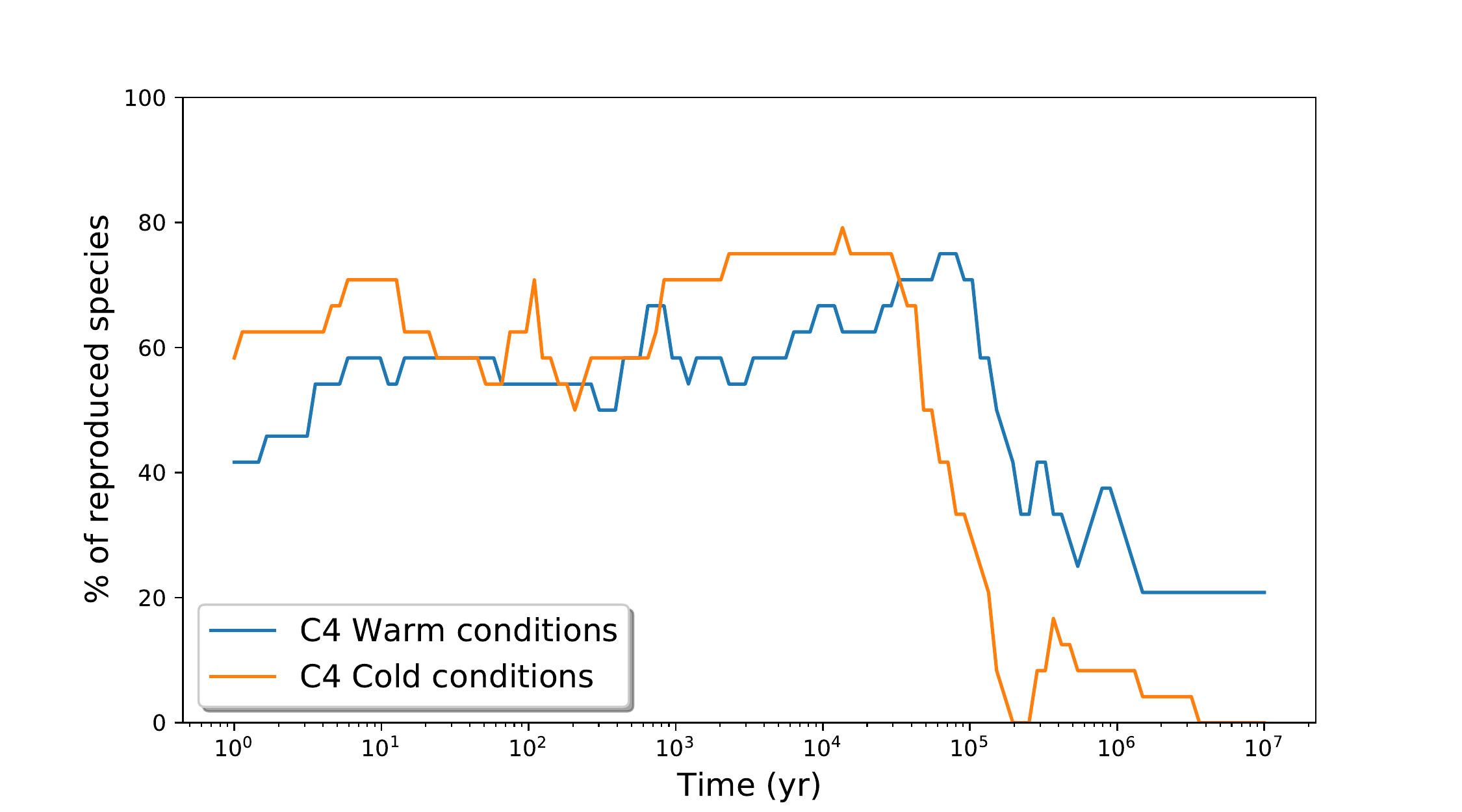}
\includegraphics[width=0.48\linewidth]{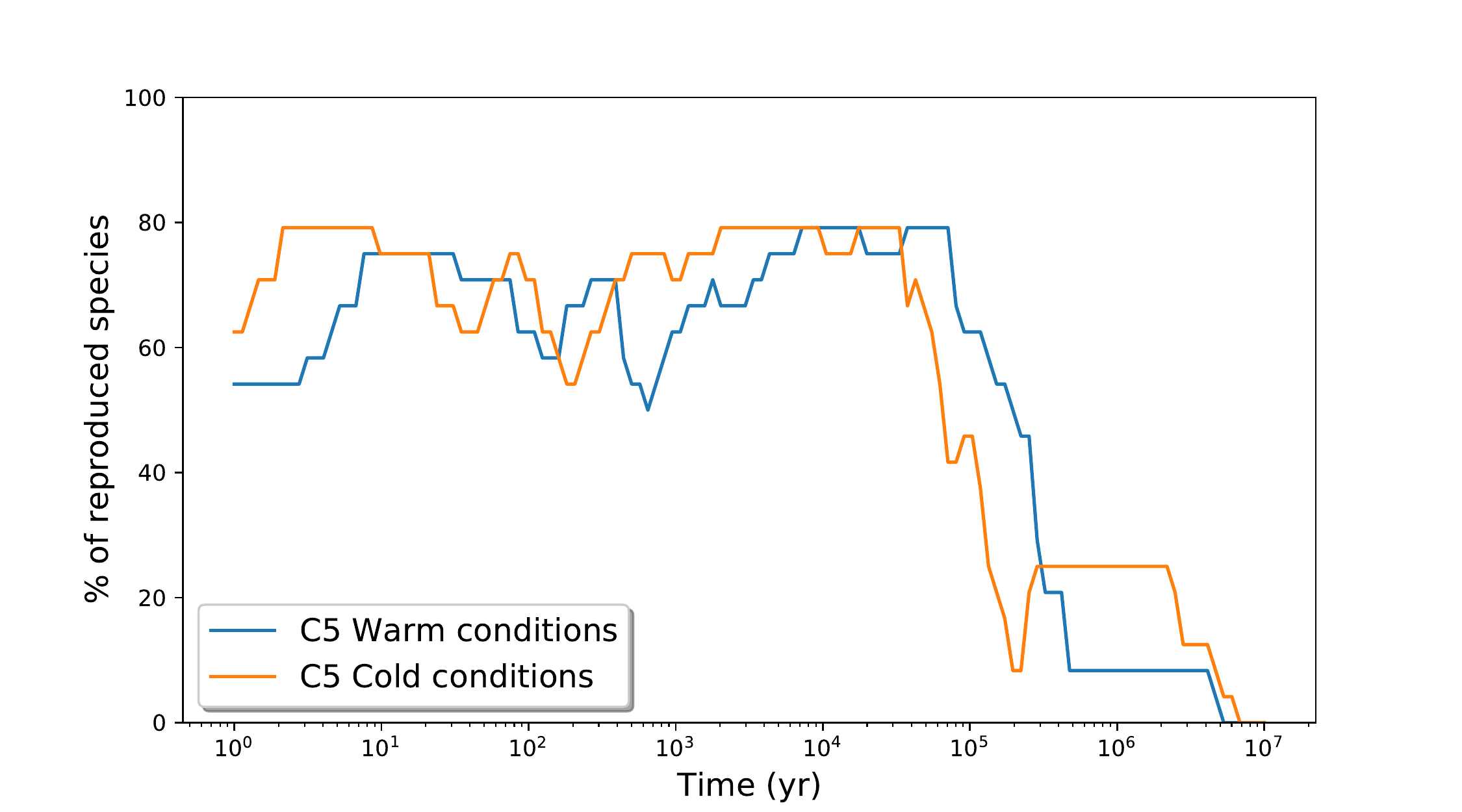}
\includegraphics[width=0.48\linewidth]{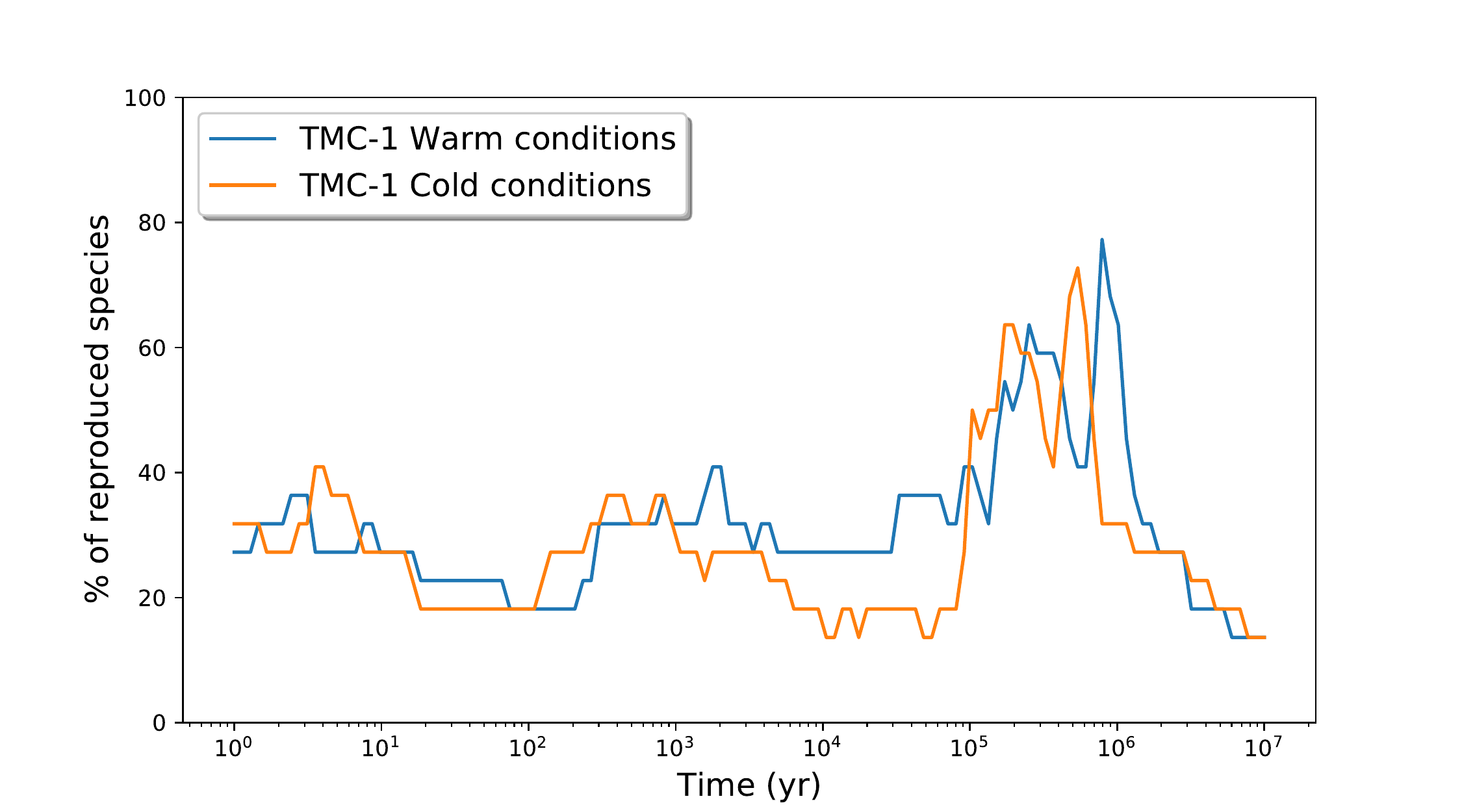}
\caption{Percentage of species reproduced by the different models for each source as a function of time with a 10 times depleted value of the elemental sulphur abundance. The labels 'warm conditions' and 'cold conditions' refer to the set of physical conditions as listed in Table~\ref{models}. \label{agreement_depS}}
\end{figure*}
\begin{figure*}[htbp]
\centering
\includegraphics[width=0.48\linewidth]{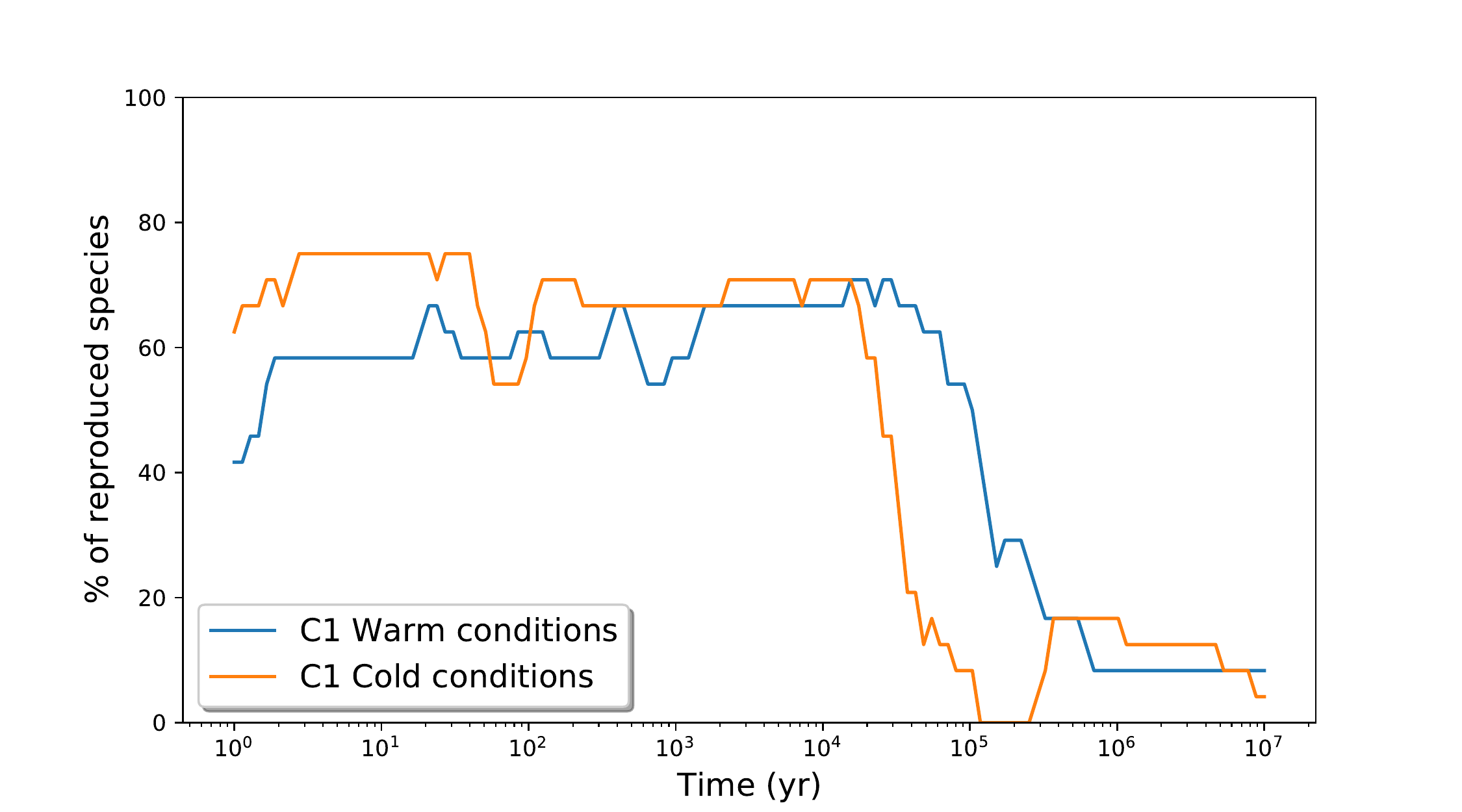}
\includegraphics[width=0.48\linewidth]{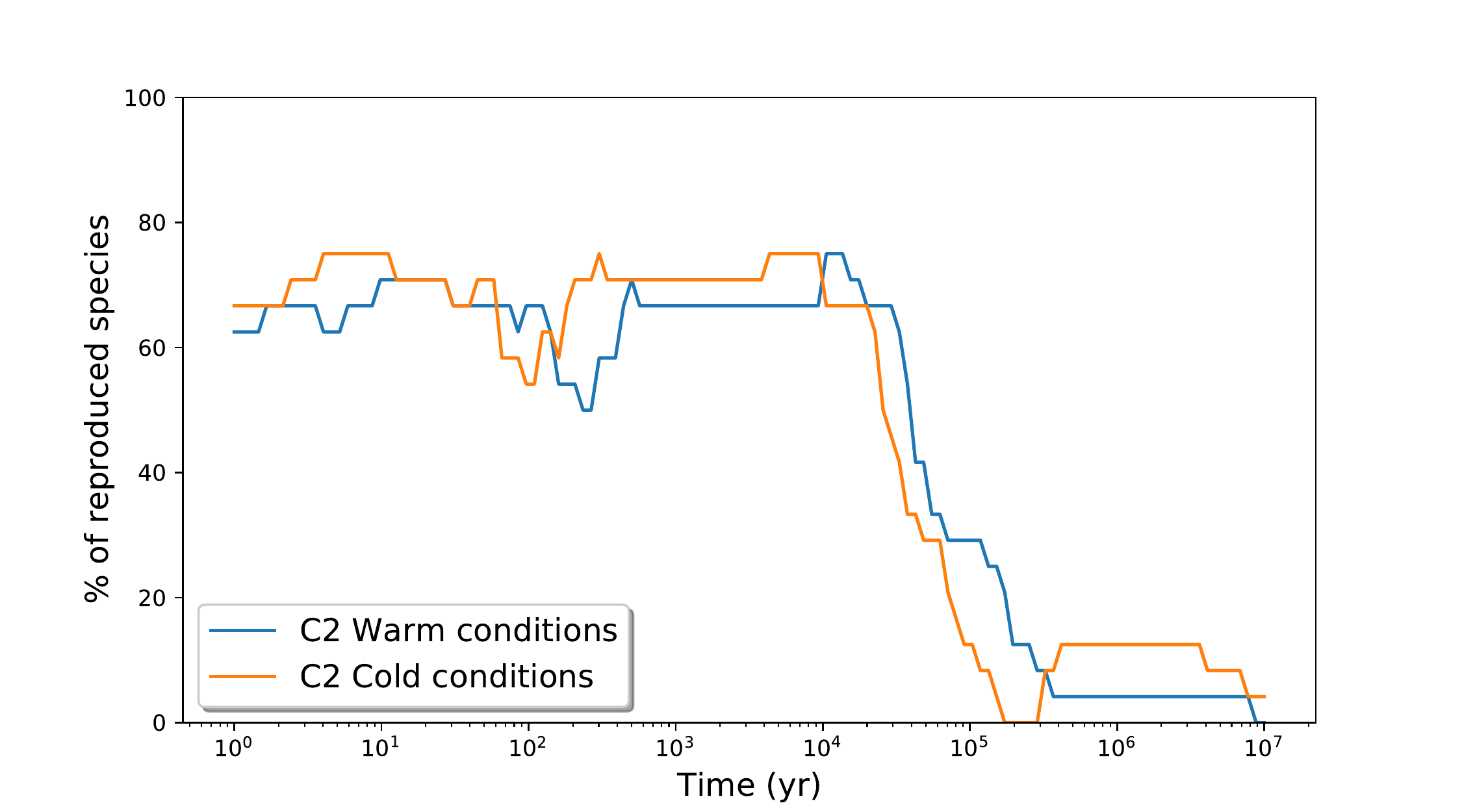}
\includegraphics[width=0.48\linewidth]{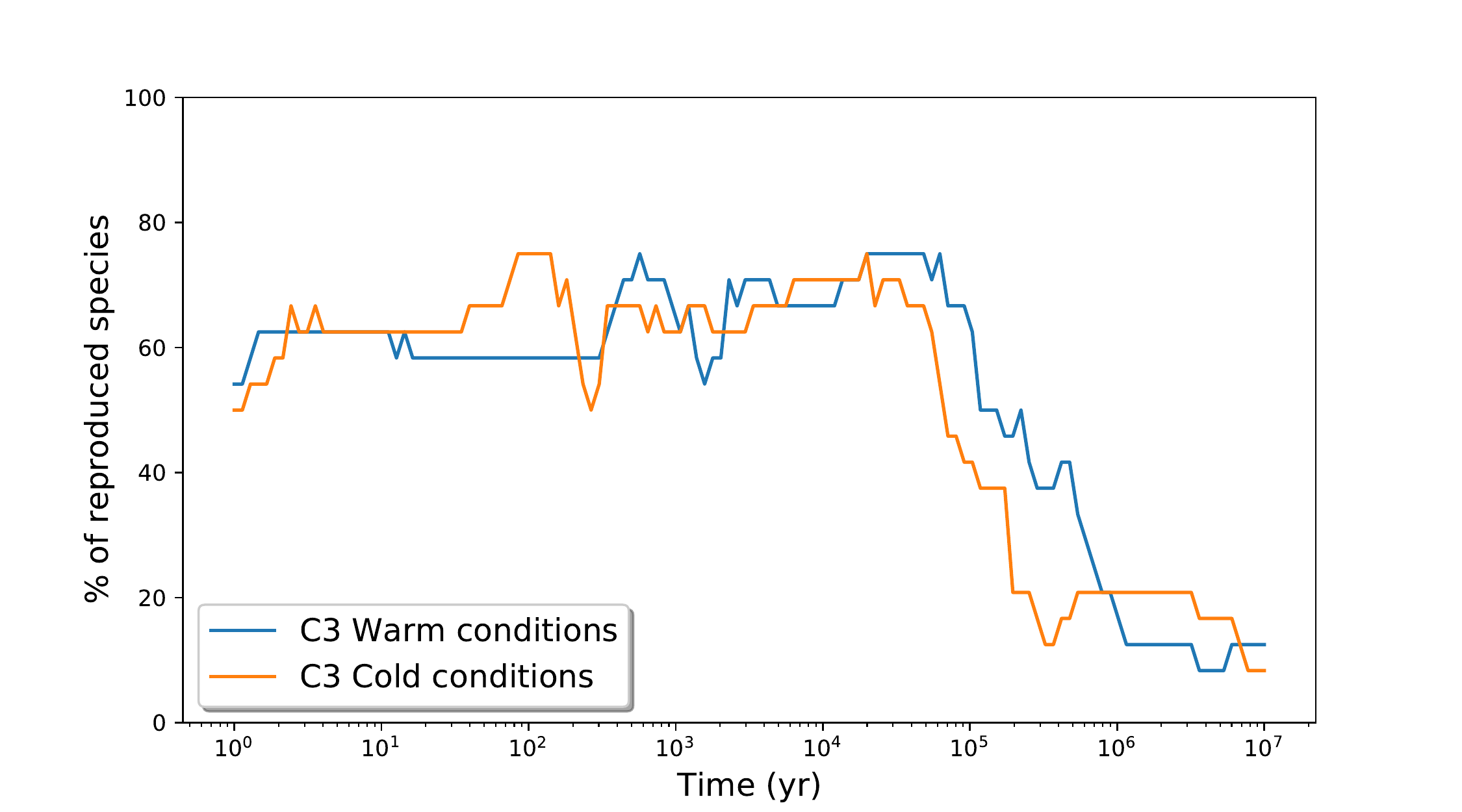}
\includegraphics[width=0.48\linewidth]{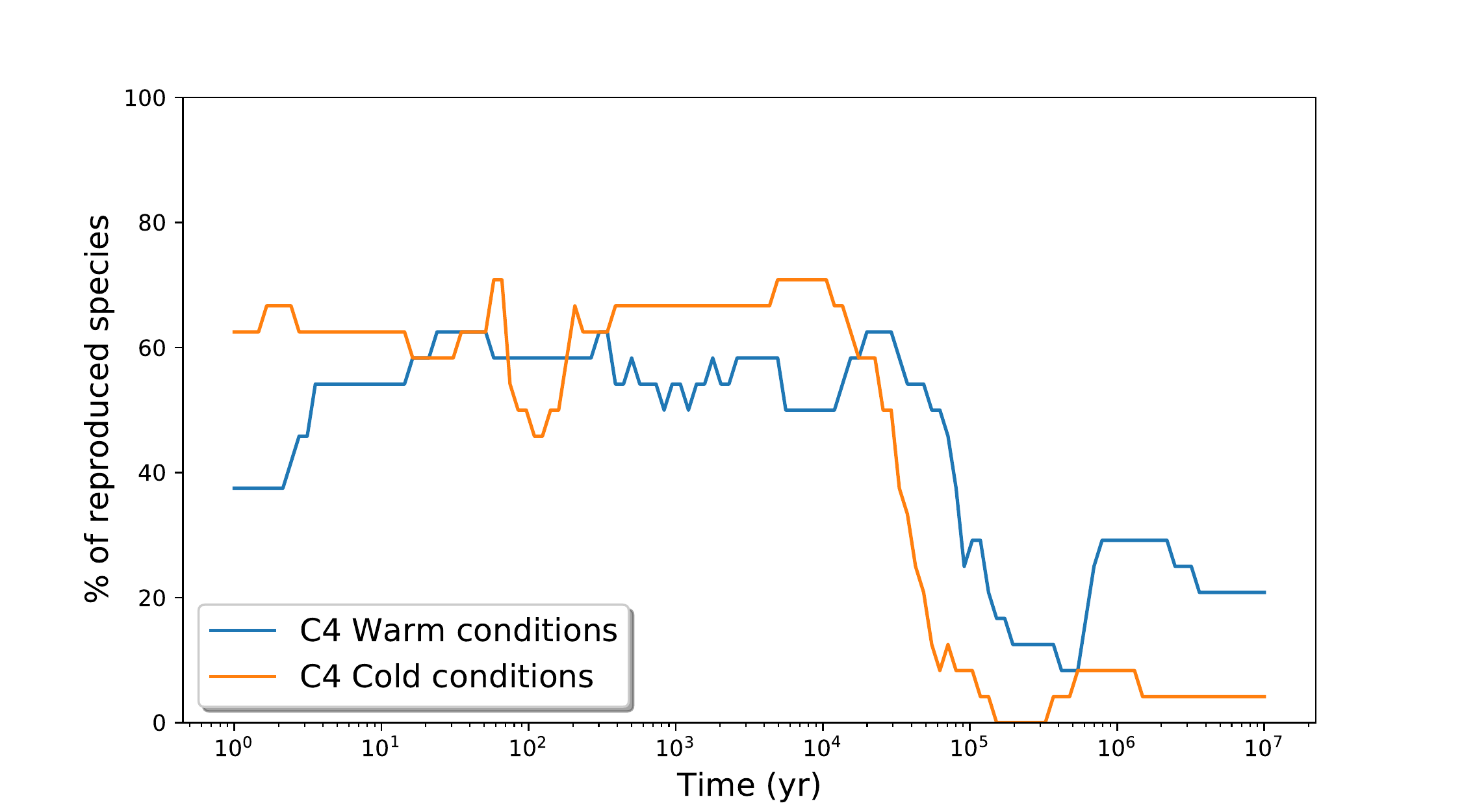}
\includegraphics[width=0.48\linewidth]{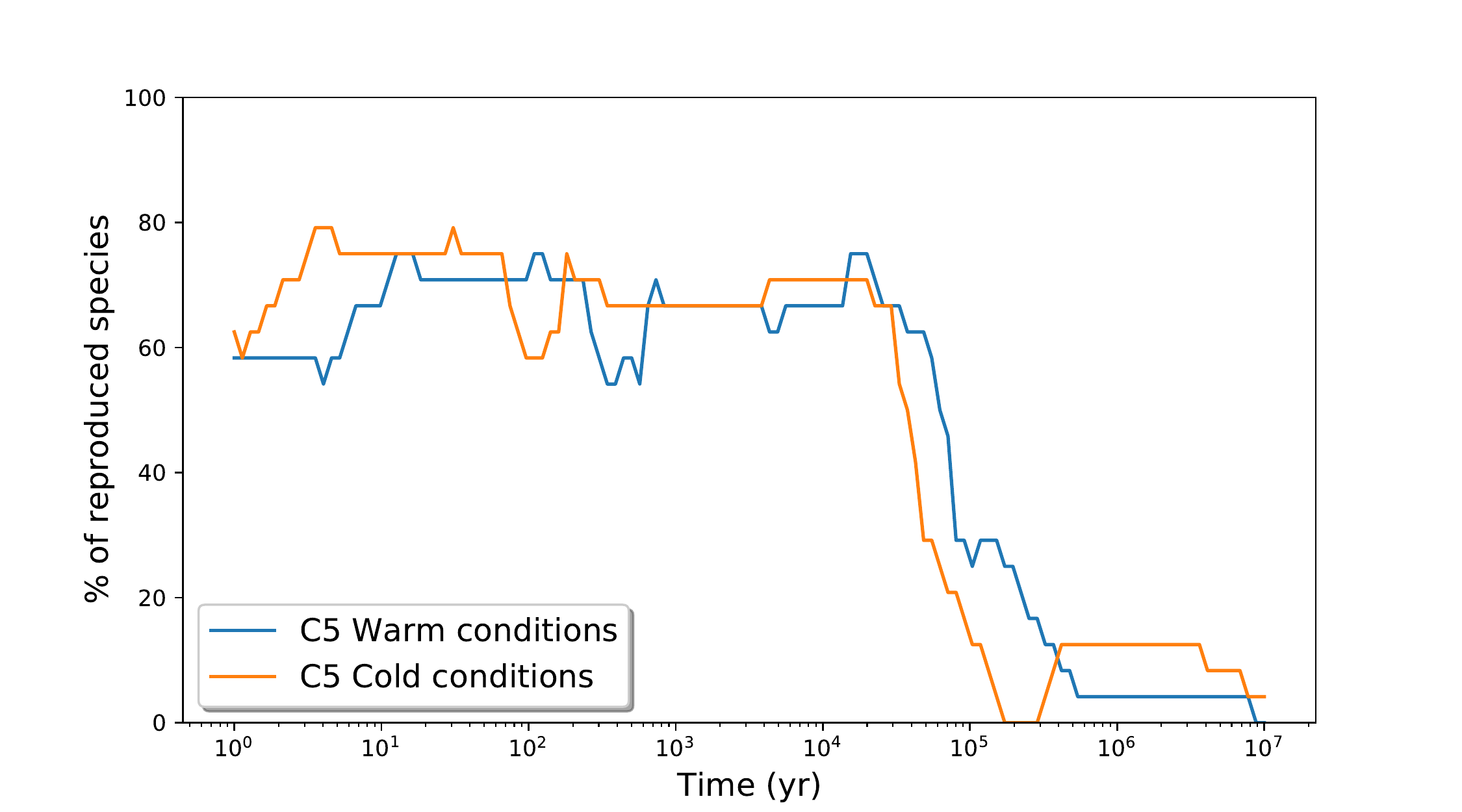}
\includegraphics[width=0.48\linewidth]{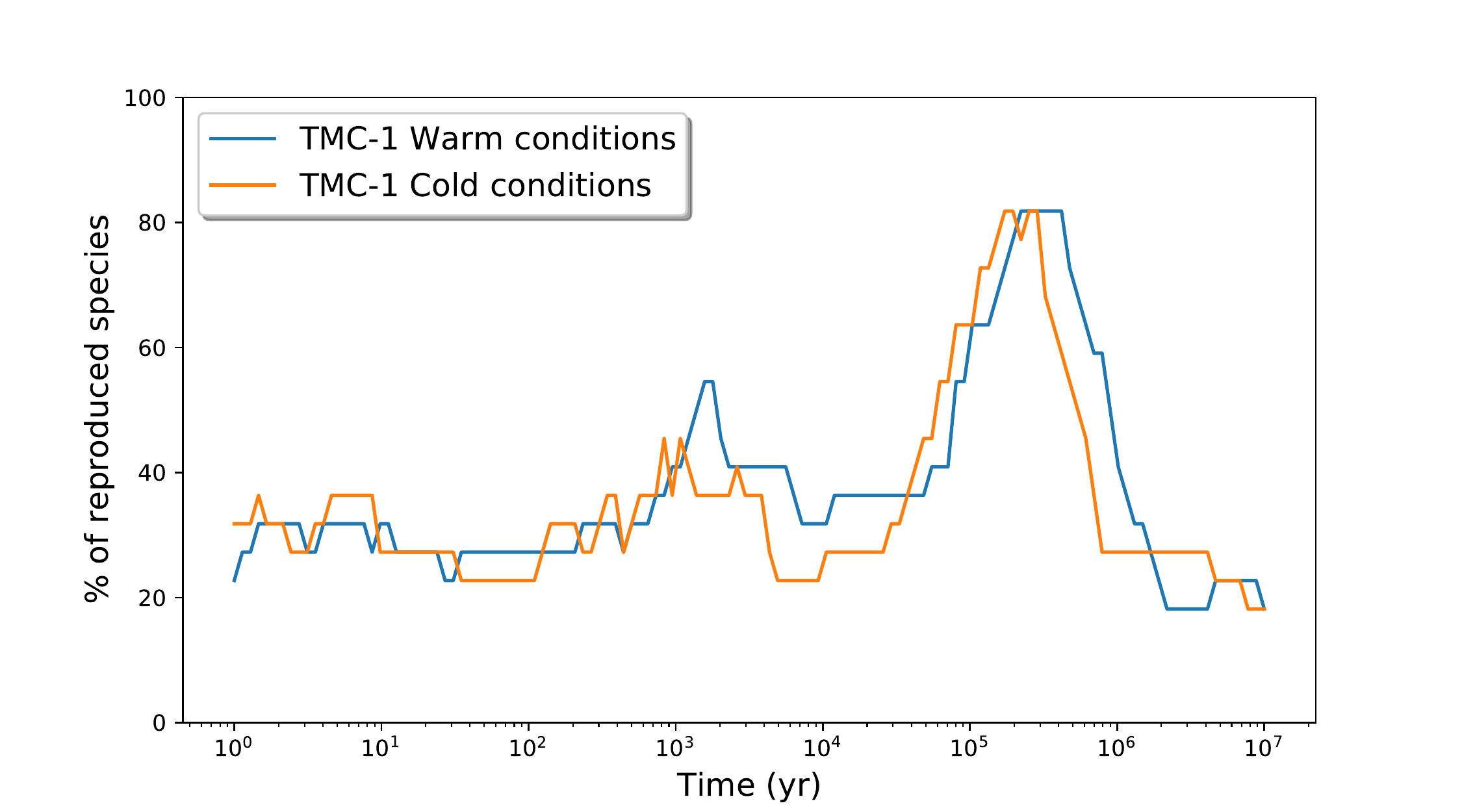}
\caption{Percentage of species reproduced by the different models for each source as a function of time for a C/O elemental ratio of 1.2. The labels 'warm conditions' and 'old conditions' refer to the set of physical conditions as listed in Table~\ref{models}. \label{agreement_C_O}}
\end{figure*}
\clearpage
\newpage
 
\section{Observed versus modelled abundances for the "best times"}\label{appendixD}

\begin{figure*}
\centering
\includegraphics[width=0.48\linewidth]{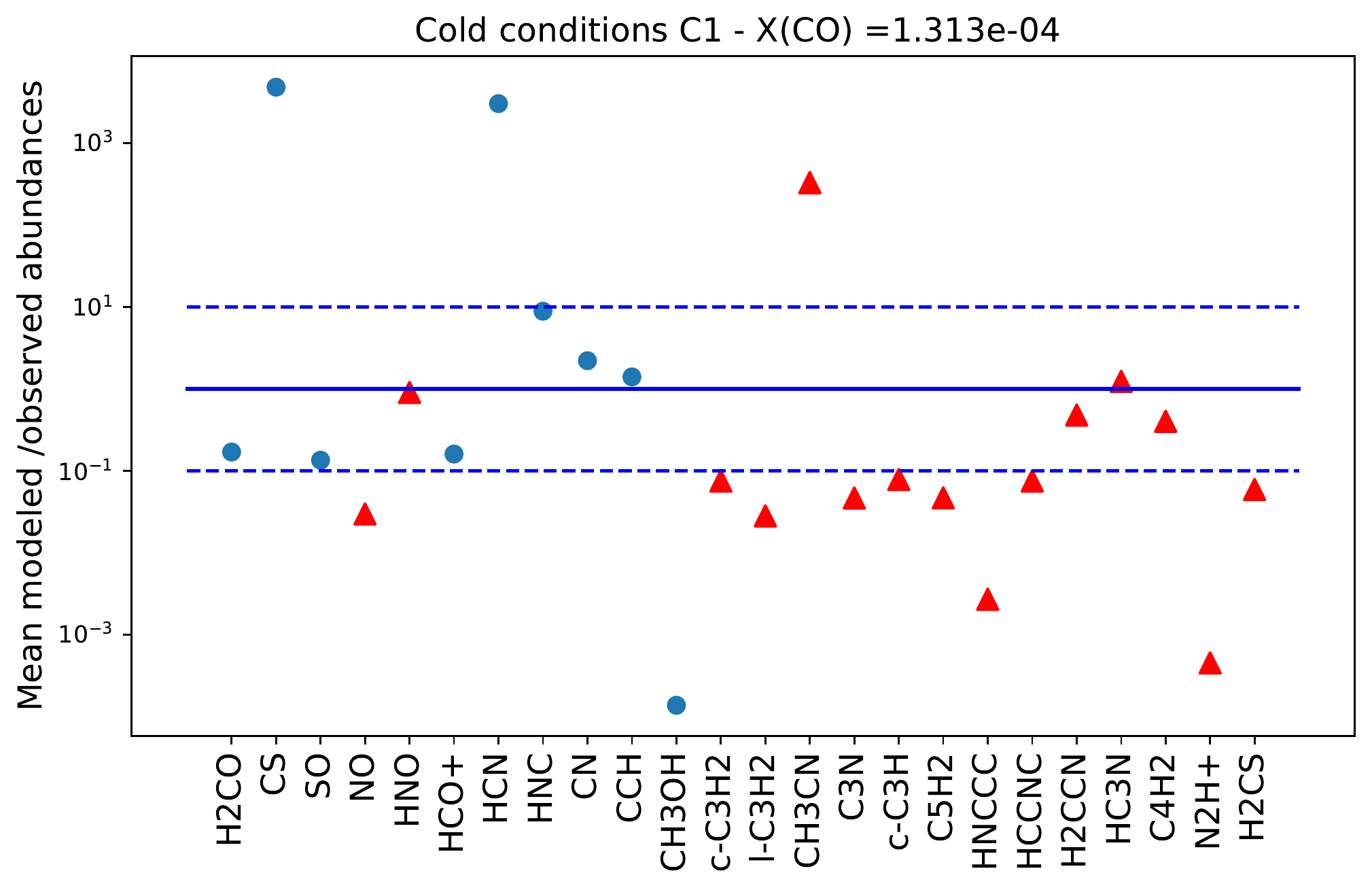}
\includegraphics[width=0.48\linewidth]{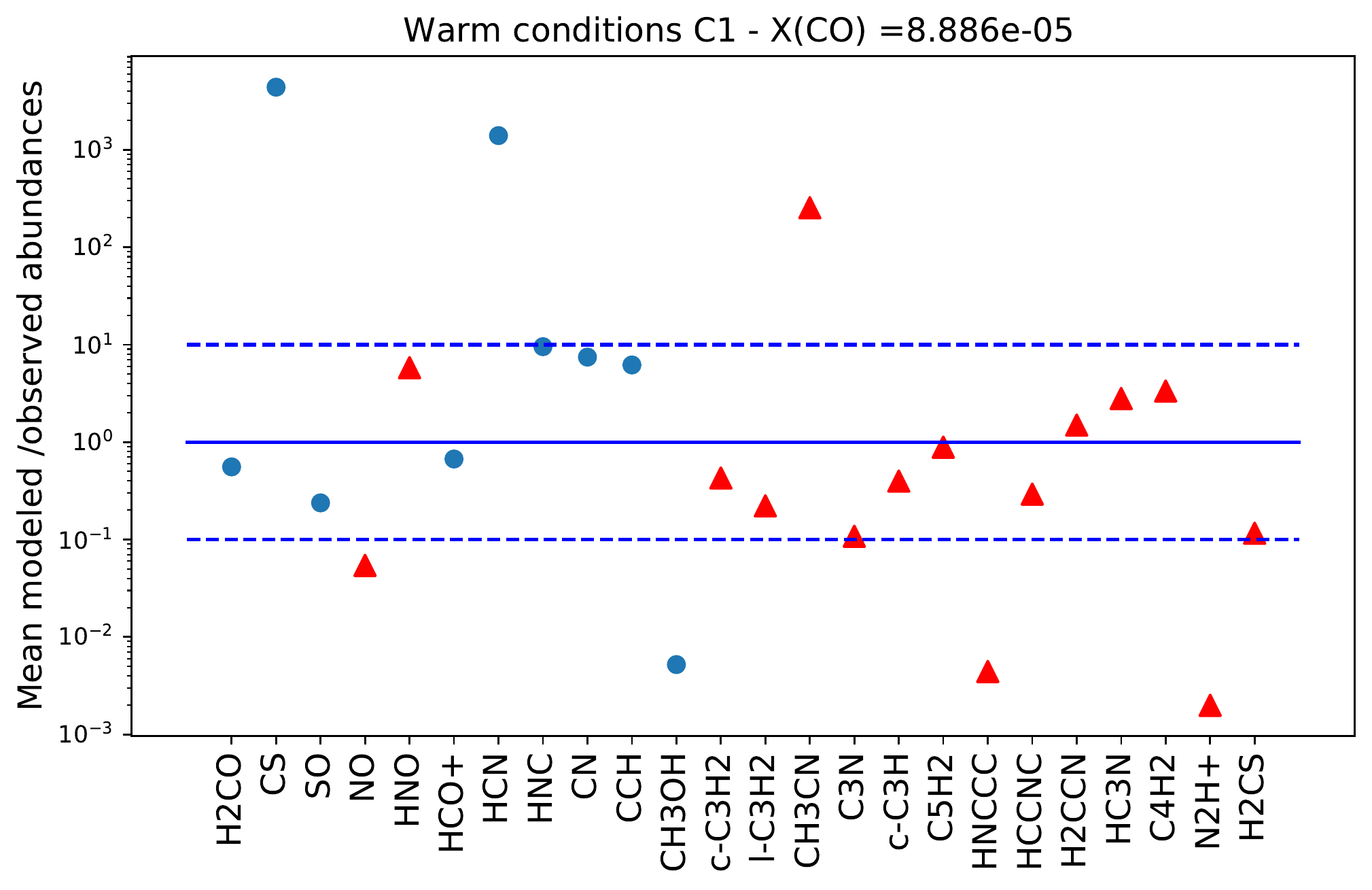}
\includegraphics[width=0.48\linewidth]{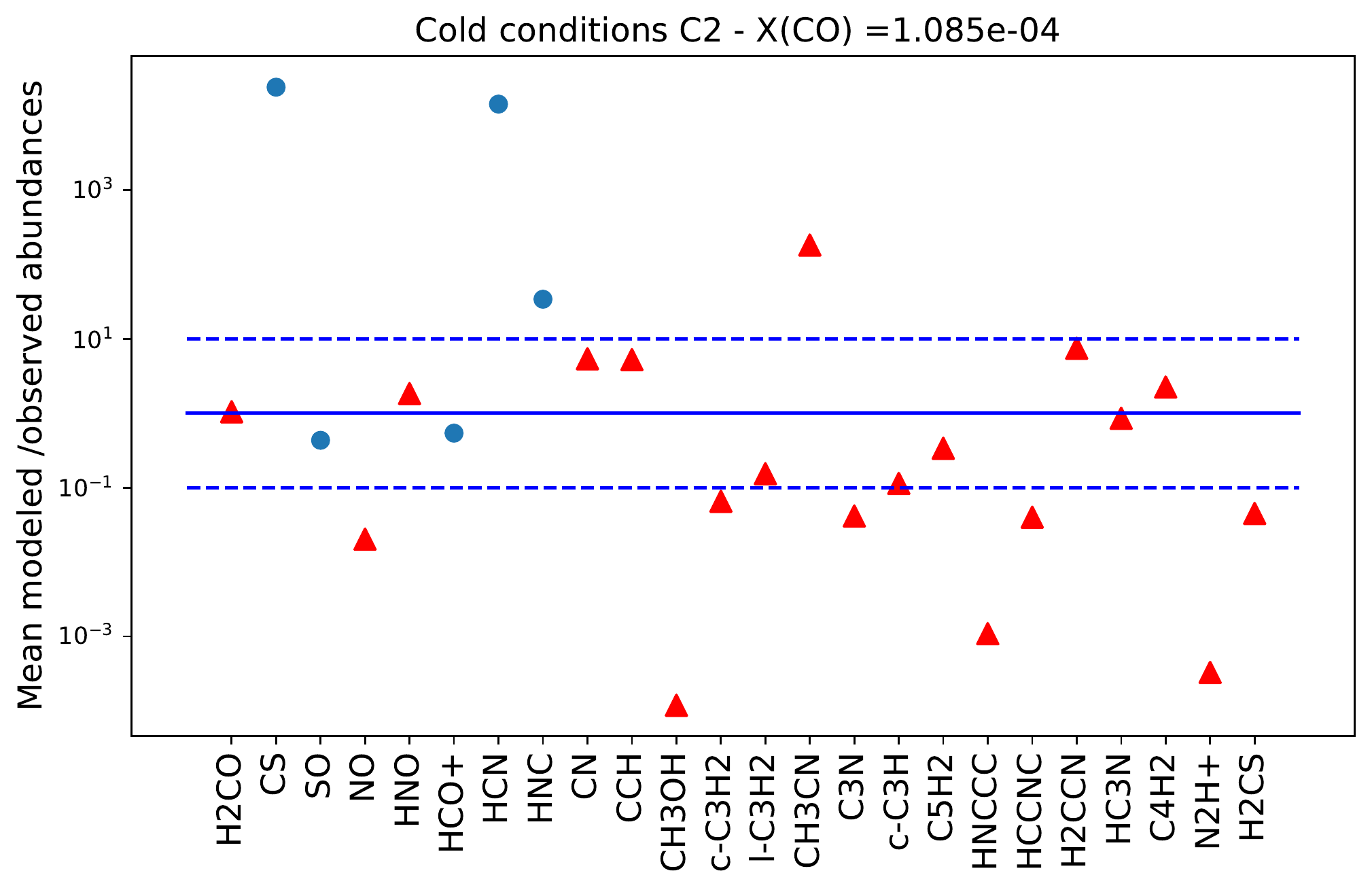}
\includegraphics[width=0.48\linewidth]{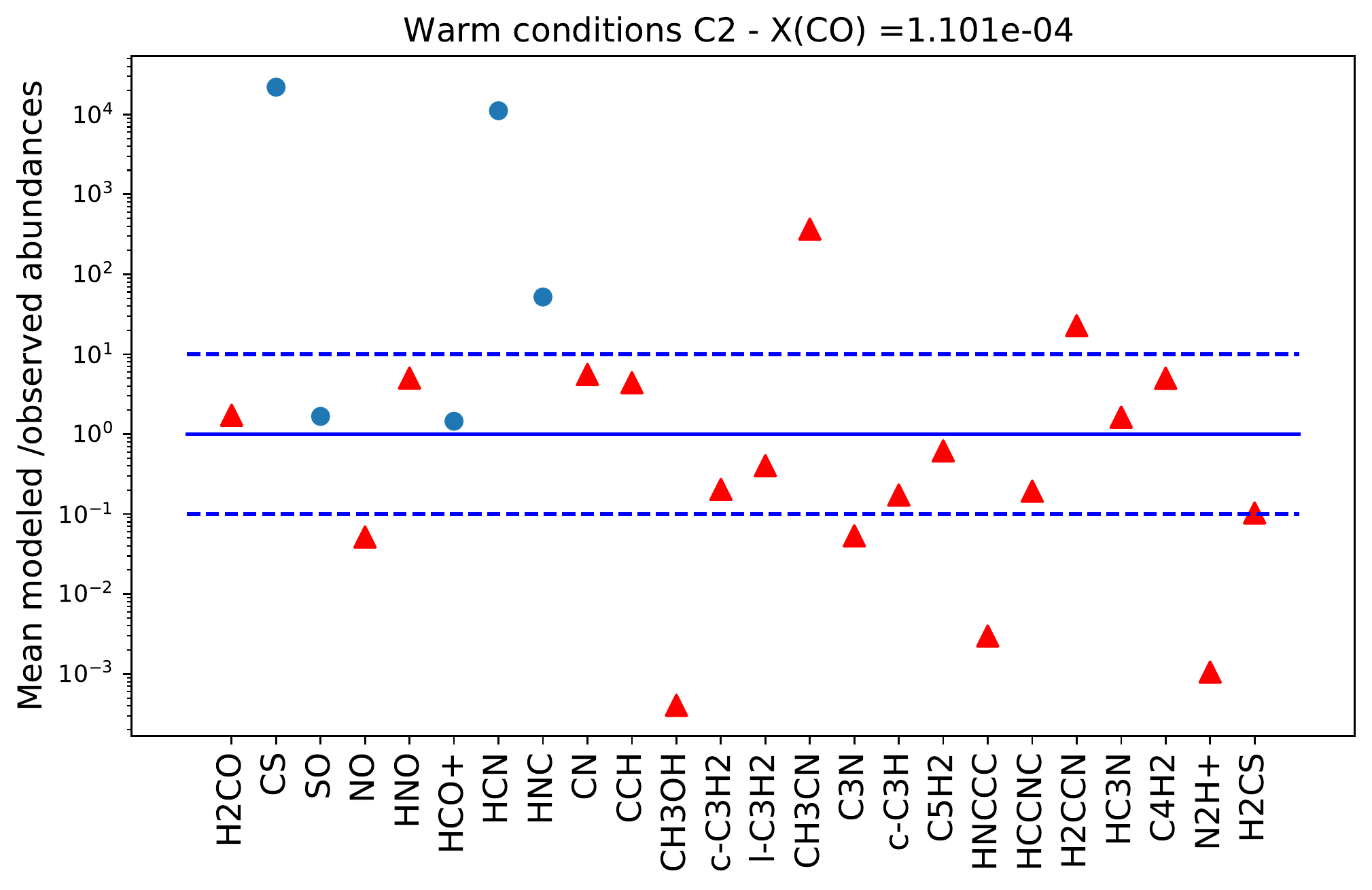}
\includegraphics[width=0.48\linewidth]{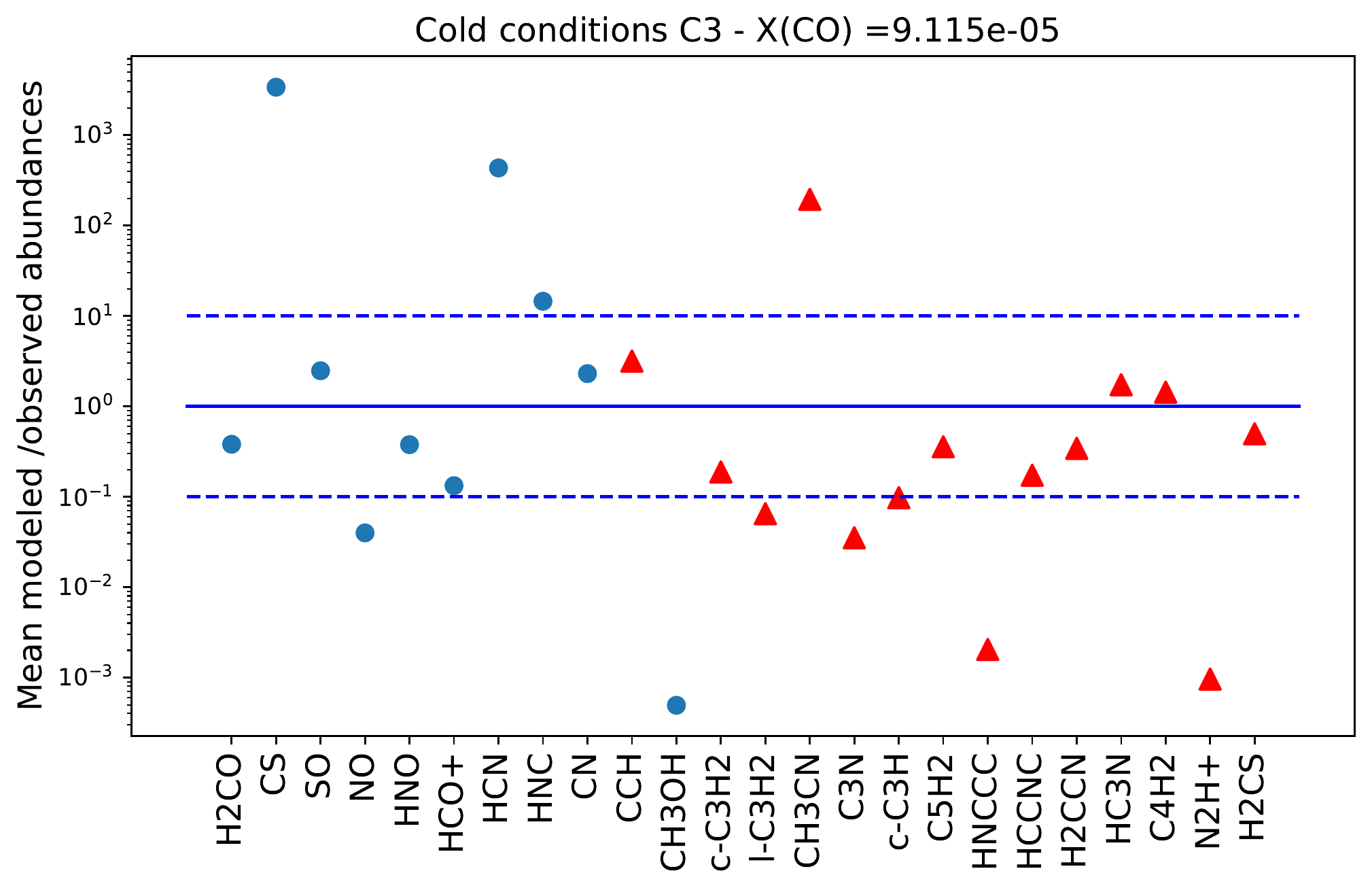}
\includegraphics[width=0.48\linewidth]{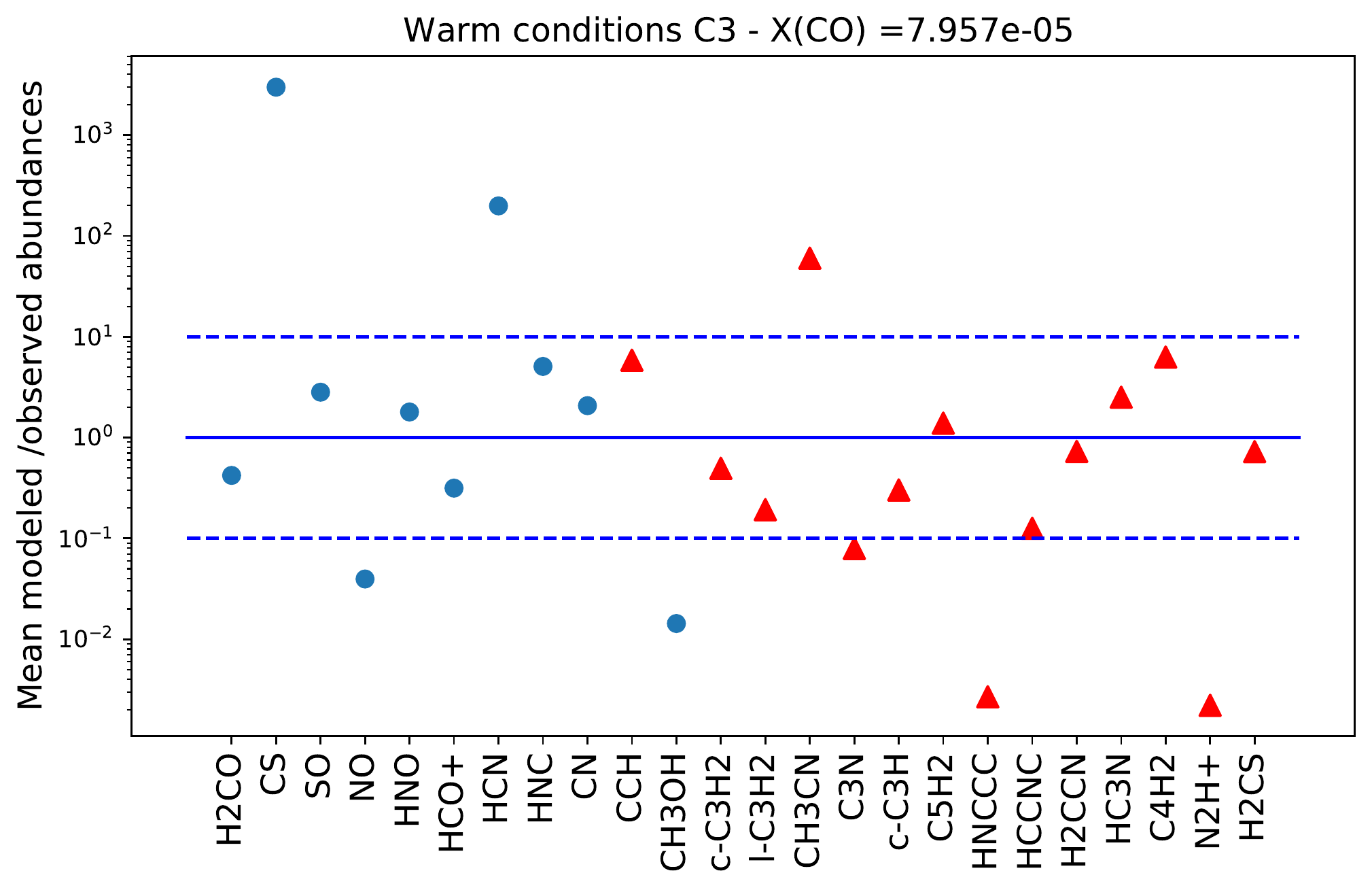}
\caption{Ratios between the modelled and observed abundances for each source and the two physical conditions. Horizontal lines are a ratio of 1 (solid), 10 (dashed), and 0.1 (dashed). Red dots are species for which we have only upper limits on the observed abundance. \label{modelled_observed_1}}
\end{figure*}

\begin{figure*}
\centering
\includegraphics[width=0.48\linewidth]{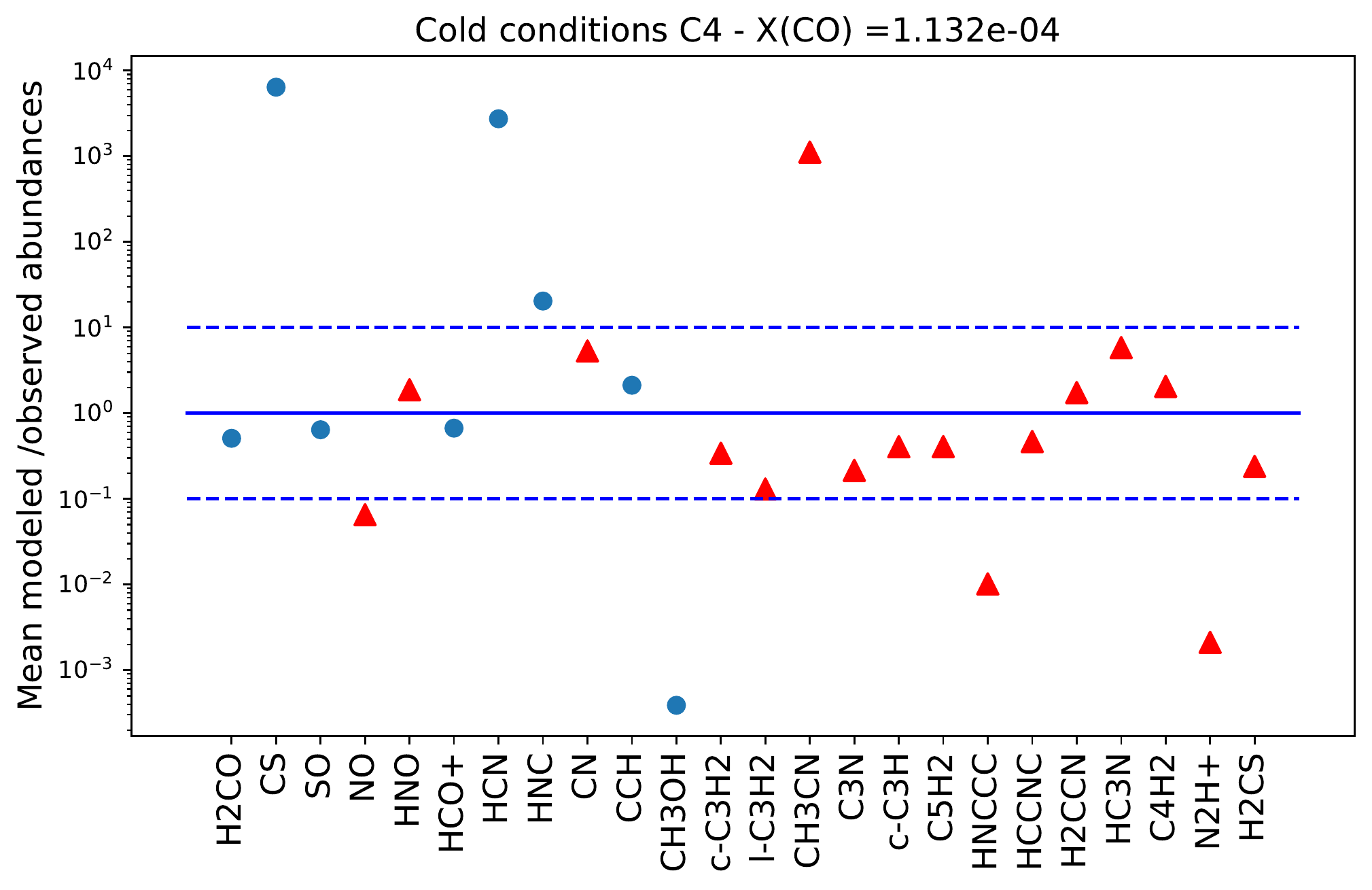}
\includegraphics[width=0.48\linewidth]{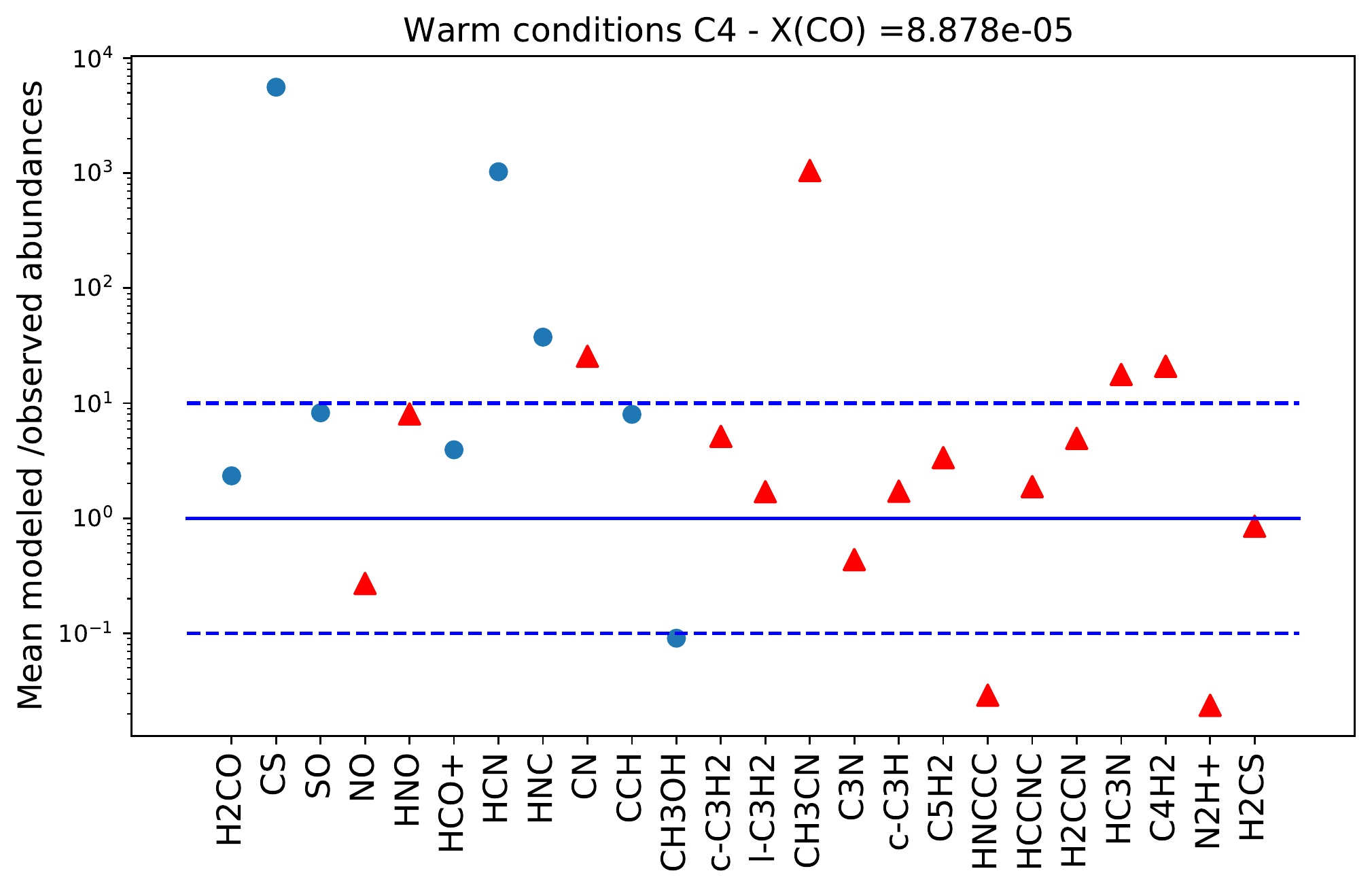}
\includegraphics[width=0.48\linewidth]{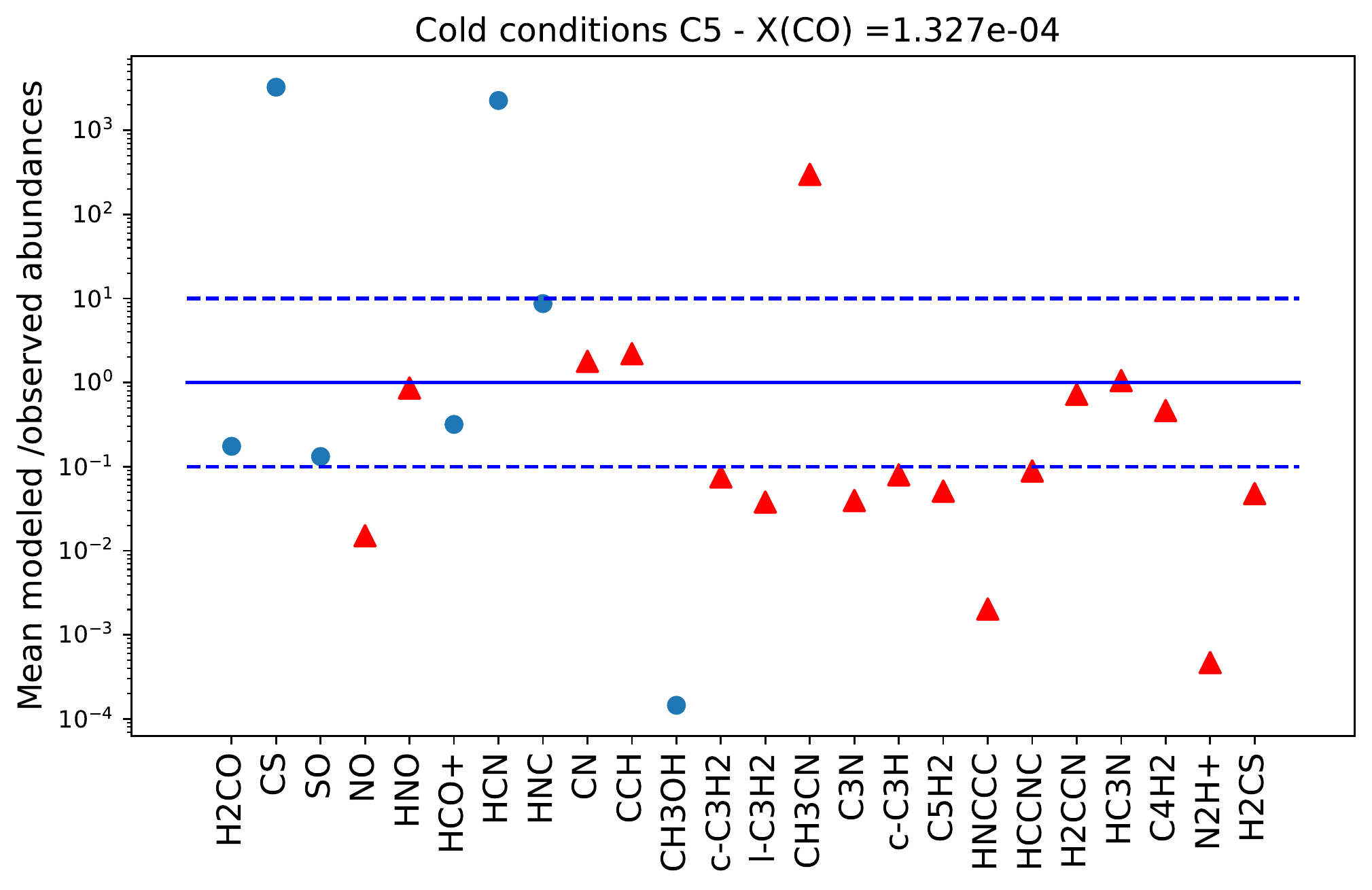}
\includegraphics[width=0.48\linewidth]{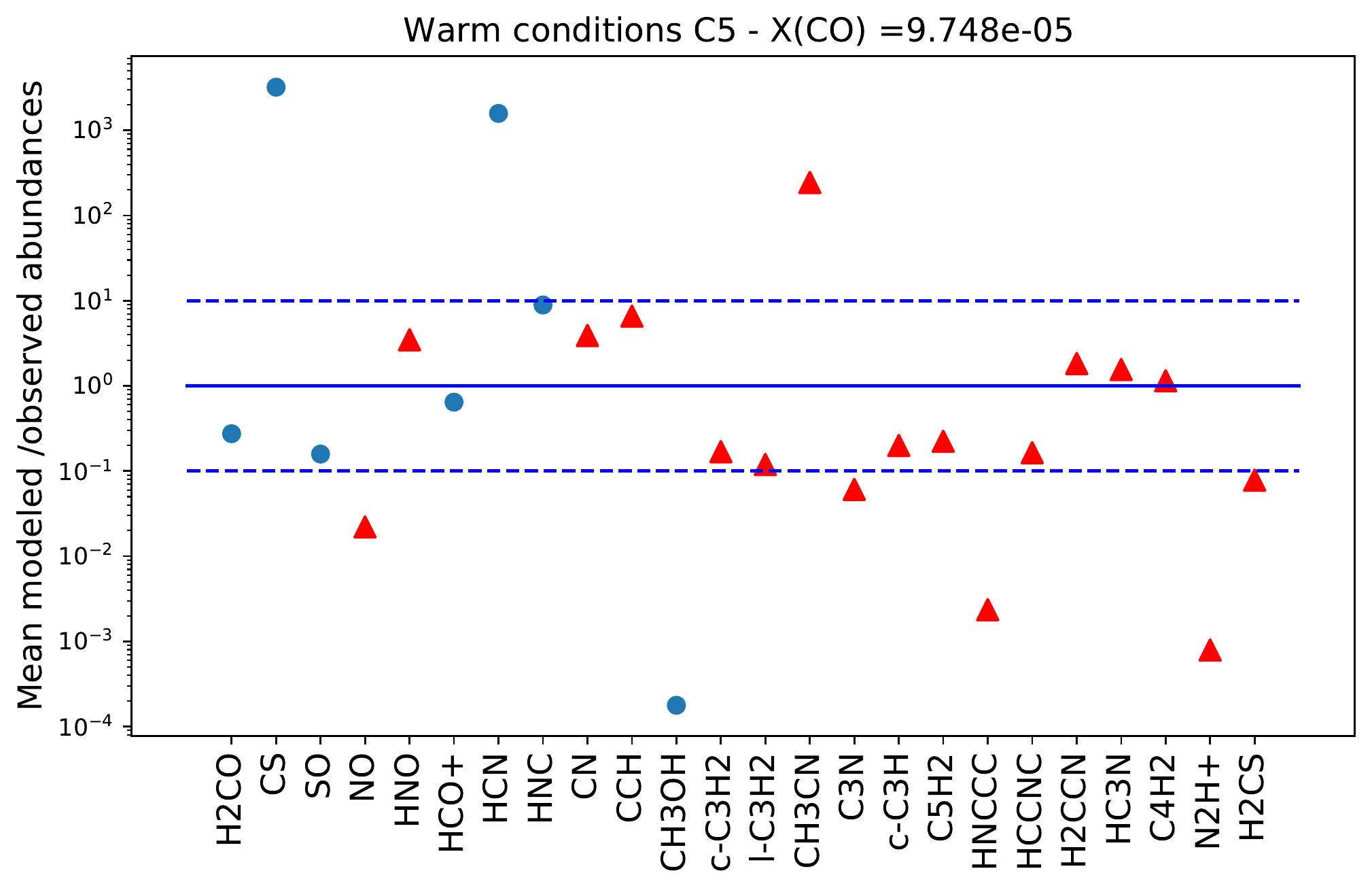}
\includegraphics[width=0.48\linewidth]{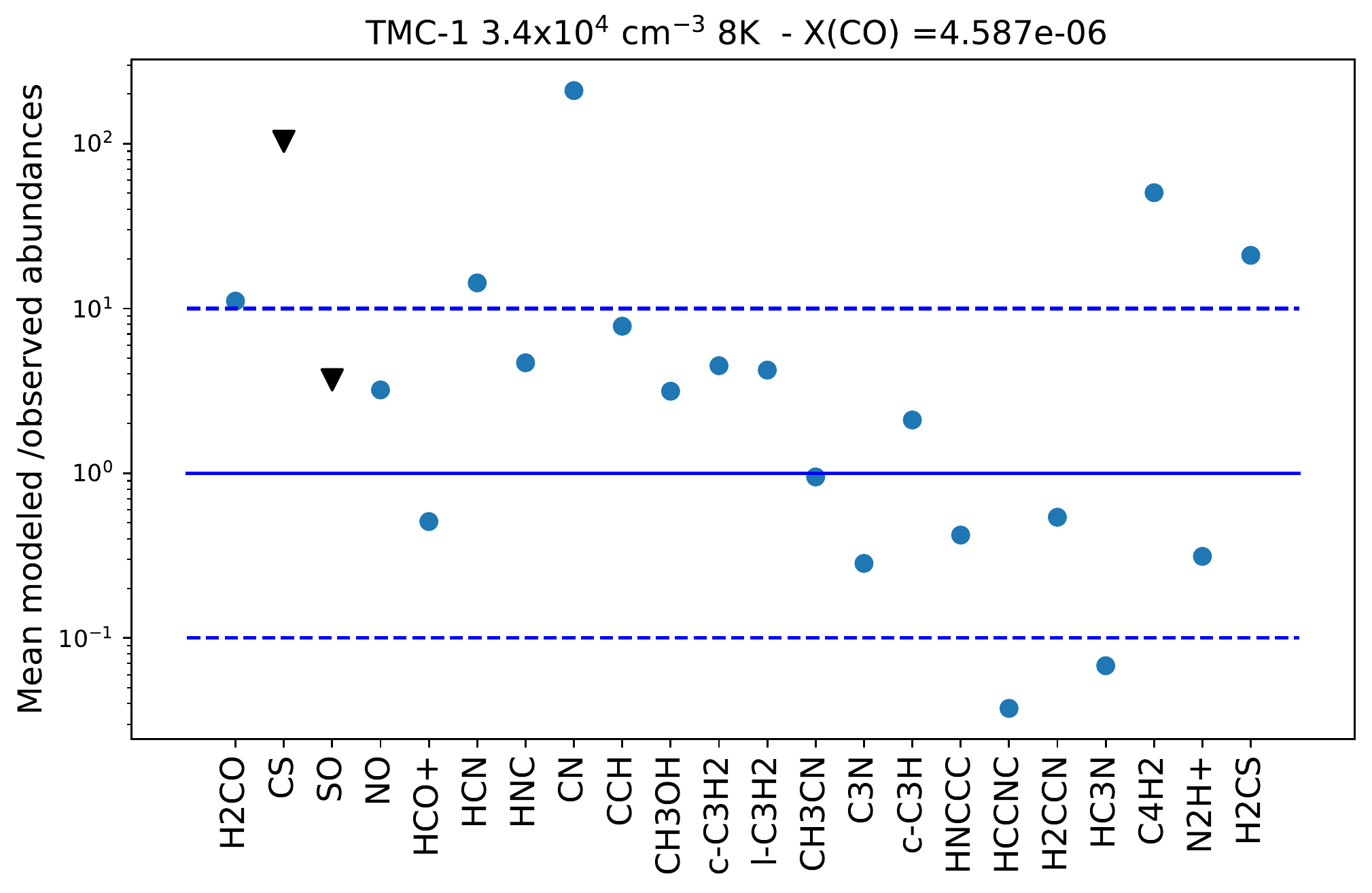}
\includegraphics[width=0.48\linewidth]{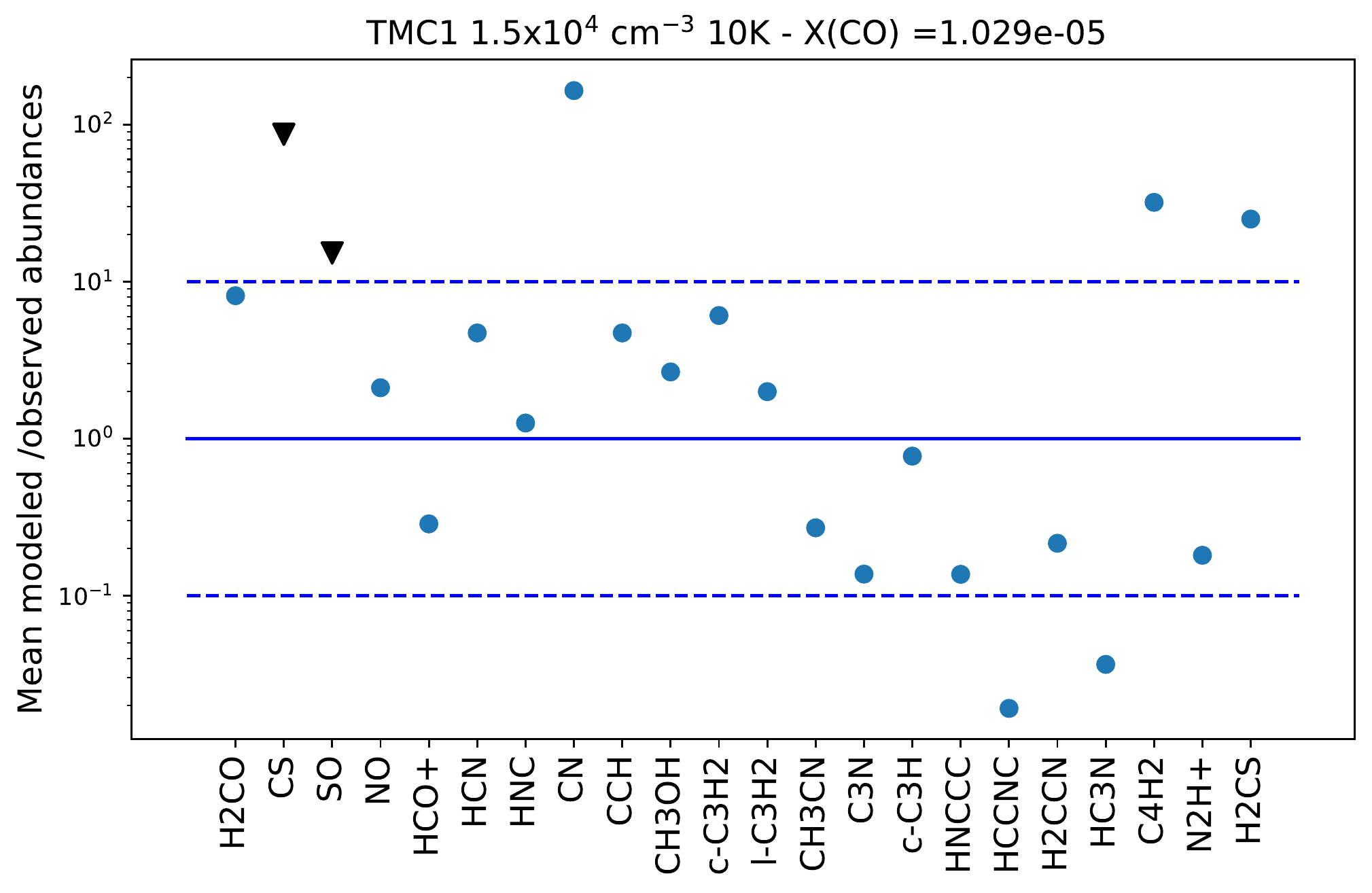}
\caption{Ratios between the modelled and observed abundances for each molecule, each source and the two physical conditions. Horizontal lines are a ratio of 1 (solid), 10 (dashed), and 0.1 (dashed). Red (black) dots are species for which we have only upper (lower) limits on the observed abundance. \label{modelled_observed_2}}
\end{figure*}

Figures~\ref{modelled_observed_1} and \ref{modelled_observed_2} show the ratio between the modelled and observed abundances (with respect to CO) for each source and models shown in section~\ref{chem_mod} at the best time of agreement. Since there may be several 'best' times, we used a mean value of abundance for all the 'best' times.  The mean CO abundance (with respect to H$_2$) for each plot is given in the plot title. The blue points within the dashed horizontal lines (a factor of 10) represent the species for which we have an agreement. To get an agreement, the red points (upper limits) have to be above the lower dashed line (divided by ten), while the black points (lower limits) have to below the upper dashed line (times 10).

\end{document}